\documentclass[11pt,graphicx,amsmath]{article}
\usepackage{amsmath}
\usepackage{graphicx}
\usepackage{CJK}
\usepackage{cancel}
\usepackage{bm}
\usepackage{xcolor}
\usepackage[numbers,sort&compress]{natbib}
\usepackage{caption}
\usepackage{subcaption}
\numberwithin{equation}{section}

\def\gsim{\;\rlap{\lower 2.5pt  \hbox{$\sim$}}\raise 1.5pt\hbox{$>$}\;}
\def\lsim{\;\rlap{\lower 2.5pt  \hbox{$\sim$}}\raise 1.5pt\hbox{$<$}\;}
\def\edth{\;\raise1.0pt\hbox{$'$}\hskip-6pt\partial\;}
\def\baredth{\;\overline{\raise1.0pt\hbox{$'$}\hskip-6pt \partial}\;}
\def\be{\begin{equation}}
\def\ee{\end{equation}}
\def\ba{\begin{eqnarray}}
\def\ea{\end{eqnarray}}
\def\nn{\nonumber}
\def\bt{\bm{\theta}}
\def\cf{\mathcal{F}}
\def\ch{\mathcal{H}}
\def\bl#1\el{\begin{align}#1\end{align}}
\def\l{\left}
\def\r{\right}

\title{ Second-order cosmological perturbations. IV. Produced by scalar-tensor and tensor-tensor couplings during the radiation dominated stage}

\author{\small   Bo Wang\thanks{ymwangbo@mail.ustc.edu.cn}\ \ ,
          Yang  Zhang\thanks{yzh@ustc.edu.cn}
           \\
 \small   Department of  Astronomy, Key Laboratory for Researches in Galaxies and Cosmology, \\
 \small    University of Science and Technology of China,   Hefei, Anhui, 230026,  China   }

 \date{}

\begin{document}
\maketitle

\def\bl#1\el{\begin{align}#1\end{align}}
\def\gsim{\;\rlap{\lower 2.5pt  \hbox{$\sim$}}\raise 1.5pt\hbox{$>$}\;}
\def\lsim{\;\rlap{\lower 2.5pt  \hbox{$\sim$}}\raise 1.5pt\hbox{$<$}\;}
\def\edth{\;\raise1.0pt\hbox{$'$}\hskip-6pt\partial\;}
\def\baredth{\;\overline{\raise1.0pt\hbox{$'$}\hskip-6pt \partial}\;}
\def\be{\begin{equation}}
\def\ee{\end{equation}}
\def\ba{\begin{eqnarray}}
\def\ea{\end{eqnarray}}
\def\nn{\nonumber}
\def\bt{\bm{\theta}}
\def\cf{\mathcal{F}}
\def\ch{\mathcal{H}}
\def\l{\left}
\def\r{\right}

\baselineskip=19truept
\Large


\begin{center}
\text{\large\bf Abstract}
\end{center}

We continue to study  the 2nd-order cosmological perturbations
 in synchronous coordinates
in the framework of the general relativity (GR)
during the radiation dominated (RD) stage,
and to focus on the scalar-tensor and tensor-tensor couplings.
The 1st-order curl velocity
and the associated 1st-order vector metric perturbations
are assumed to be vanishing.
By  analytically solving the 2nd-order Einstein equation and
the energy-momentum conservation equations,
we obtain the 2nd-order formal solutions (in the integral form)
of all the metric perturbations,
density contrast and  velocity;
perform the  transformation between the synchronous coordinates;
and identify the residual gauge modes in the 2nd-order solutions.
In addition, we present the 2nd-order gauge transformations
of the solutions from   synchronous to  Poisson coordinates.
To apply these formal solutions to concrete cosmological study,
one needs to choose proper  initial conditions
and do several numerical integrals.

\section{Introduction}

The cosmological perturbations in
a Friedmann-Robertson-Walker Universe
are of great importance for the study of cosmology.
In the past the  linear cosmological perturbations
have been extensively explored  \cite{Lifshitz1946,LifshitzKhalatnikov1963,
PressVishniac1980,Bardeen1980,BrandenbergerKahnPress1983,
KodamSasaki1984,Grishchuk1994}
and widely applied to the large scale structure \cite{ Peebles1980},
the cosmic microwave background (CMB)
    \cite{BaskoPolnarev1984,Polnarev1985,
MaBertschinger1995,Bertschinger,ZaldarriagaHarari1995,
Kosowsky1996,ZaldarriagaSeljak1997,
Kamionkowski1997,KeatingTimbie1998,
ZhaoZhang2006,Baskaran,Polnarevmiller2008},
and relic gravitational waves (RGW)
\cite{Grishchuk,FordParker1977GW,Starobinsky,
Rubakov,Fabbri,AbbottWise1984,
    Allen1988,Giovannini,Tashiro,zhangyang05,Morais2014,
    Weinberg2004,MiaoZhang2007,WangZhang2019RAA}, etc.
Nowadays, observations with increasing accuracy
\cite{PyneCarroll1996,MollerachHarariMatarrese2004,
JeongKomatsu2006,YangZhang2007,AnandaClarksonWands2007,Baumann2007,
Bartolo2010,Matarrese2007,Pietroni2008,Matsubara2008}
are capable of probing the nonlinear cosmological perturbations.
Throughout all cosmic expansion stages,
the nonlinear perturbations exist at small scales;
may accumulate substantially in the course of expansion;
and can leave possibly observable effects in the large scale structure,
the CMB anisotropies and polarization \cite{PyneCarroll1996, MollerachHarariMatarrese2004},
the non-Gaussianity of primordial perturbation  \cite{Bartolo2010},
primordial black holes \cite{primordialBH}, etc.

The metric perturbations in GR are usually  decomposed into scalars, vectors, and tensors.
In regard to cosmology,
the linear vector modes decrease with cosmic expansion and can be neglected
in most applications,
while the scalar and tensor modes are generated with comparable magnitudes
during inflation.
So three types of nonlinear couplings are more interesting in
cosmological studies:  scalar-scalar, scalar-tensor, and  tensor-tensor.
In the literature \cite{Tomita1967,MatarresePantanoSa'ez1994,
Russ1996,Salopek,MalikWands2004,Nakamura2003,Lu2008,Lu2009},
the studies of the 2nd-order  perturbations
mainly consider the scalar-scalar coupling,
whereas the scalar-tensor and the tensor-tensor couplings
have not been sufficiently investigated.
The authors of Ref.\cite{NohHwang2004} used the Arnowitt-Deser-Misner (ADM) method on
the  2nd-order perturbation with the tensor-tensor coupling
for a zero pressure dust \cite{HwangNoh2006}
and made comparisons between the comoving gauge
and the synchronous gauge \cite{HwangNoh2007}.
The authors of Ref.\cite{GongHwangNoh2017} presented a complete set of 2nd-order equations for scalar, vector and tensor perturbations
in the ADM formulation for a general gauge.
The authors of Ref.\cite{HwangJeongNoh2017} studied the gauge dependence of
the spectrum of 2nd-order tensor mode with scalar-scalar couplings
during the matter dominated (MD) stage
and showed that the 2nd-order tensor mode could dominate over the 1st-order tensor mode.
The authors of Refs.\cite{Bruni97,Matarrese98} studied the 2nd-order perturbed Einstein equation
during  the MD stage driven by the zero pressure dust
in synchronous coordinates,
including only the scalar-scalar coupling,
and derived the solutions of 2nd-order scalar and tensor perturbations.
In our previous works,
we extended the study of Refs.\cite{Bruni97,Matarrese98},
including the scalar-scalar \cite{WangZhang2017},
scalar-tensor, and tensor-tensor couplings \cite{ZhangQinWang2017};
obtained all the 2nd-order solutions of the scalar, vector, tensor,
and density perturbations during the MD stage;
identified the residual gauge modes under
synchronous-to-synchronous transformations,
and also presented the 2nd-order transformation
from synchronous to Poisson coordinates.
In Ref.\cite{WangZhang2ndRD2018},
we studied the 2nd-order perturbations in the RD stage driven
by a relativistic perfect fluid,
including only the scalar-scalar coupling;
derived the 2nd-order formal solutions of
the scalar, vector, tensor, density, and velocity perturbations
in the integral form;
and also identified the residual gauge modes within the synchronous coordinates.
In this paper as a continuation of Ref.\cite{WangZhang2ndRD2018},
we shall include the scalar-tensor and tensor-tensor couplings.
We are motivated not only
by nonlinearity within one type of metric perturbation,
such as the transfer of perturbation power between  $k$-modes, etc.,
but also by the transfer of perturbation power between different types of
perturbations, such as those between scalar and tensor modes, etc.
We shall derive the 2nd-order  solutions of scalar, vector, tensor,
 density, and velocity perturbations in the integral form,
perform the 2nd-order synchronous-to-synchronous transformation,
and identify the residual gauge modes.
In addition, we shall present the 2nd-order gauge transformations
of the solutions from the synchronous to the Poisson coordinates;
the latter is also commonly used in cosmological studies.
Thus, together with the results in Ref.\cite{WangZhang2ndRD2018},
a complete set of solutions of 2nd-order cosmological perturbations
of the RD stage are available.

In Sec. 2,
we give some basic setups of the metric perturbations
and the relativistic fluid model.
We also list the solutions of 1st-order perturbations of the RD stage
for use in this paper.

In Sec. 3, we decompose the 2nd-order perturbed Einstein equation
into the equations of 2nd-order scalar, vector, tensor metric perturbations
with scalar-tensor and tensor-tensor couplings as part of the effective sources,
and also derive the equations of the 2nd-order density contrast and velocity
from the covariant conservation of the stress tensor.

In Sec. 4,
we derive all the 2nd-order  solutions in the integral form.
We also explain how to do the time and momentum integrals
that occur in the solutions by two examples.

In Sec. 5,
we perform the synchronous-to-synchronous transformation,
and identify the residual gauge modes in the 2nd-order solutions.

In Section 6,
we perform transformations from synchronous coordinates to Poisson coordinates
of the 2nd-order solutions in Sec. 4,
and also of those with the scalar-scalar couplings
that have been derived in Ref.\cite{WangZhang2ndRD2018}.

 Section 7  gives  the  conclusions and discussions.

Appendix \ref{Ap:PertEinEq} gives a list of
the 2nd-order perturbed Einstein equations
and covariant conservation equations of the stress tensor
with the scalar-tensor and tensor-tensor couplings
in a general RW spacetime.
We use units with the speed of light $c=1$.

\section{Basic setups and 1st-order solutions}

Here we  give the basic setups  of this  paper,
using the same notations as in Refs. \cite {Matarrese98,ZhangQinWang2017,WangZhang2017,WangZhang2ndRD2018}.
A   flat Robertson-Walker (RW) metric in synchronous coordinates
is given by
\be \label{18q1}
ds^2=g_{\mu\nu}   dx^{\mu}dx^{\nu}=
a^2(\tau)[-d\tau^2
+\gamma_{ij}   dx^idx^j] ,
\ee
\be\label{eq1}
\gamma_{ij} =\delta_{ij} + \gamma_{ij}^{(1)} + \frac{1}{2} \gamma_{ij}^{(2)}
\ee
with $\gamma_{ij}^{(1)}$ and  $\gamma_{ij}^{(2)}$
being  the 1st- and 2nd-order metric perturbation, respectively.
Writing  $g^{ij}=a^{-2}\gamma^{ij}$,
one has
\be\label{metricUp2}
\gamma^{ij}=\delta^{ij} -\gamma^{(1)ij}
-\frac{1}{2}\gamma^{(2)ij}+\gamma^{(1)ik}\gamma^{(1)j}_{k} .
\ee
Raising and lowering the three-dimentional spatial indices
will be done by $\delta^{ij}$.
The metric perturbations can be further written as
\be  \label{gqamma1}
\gamma^{(A)}_{ij}=-2\phi^{(A)}\delta_{ij}  +\chi_{ij}^{(A)},
~~\text{with}~~ A=1,2,
\ee
where  $\phi^{(A)}$ is the trace part of  scalar perturbation,
and $\chi_{ij}^{(A)}$ is  traceless and can be further decomposed into
the following
\be \label{xqiij1}
\chi_{ij}^{(A)} =D_{ij}\chi^{||(A)}
               +\chi^{\perp(A)}_{ij}
               +\chi^{\top(A)}_{ij},
~~\text{with}~~ A=1,2,
\ee
where $D_{ij} \equiv  \partial_i\partial_j-\frac{1}{3}\delta_{ij}\nabla^2 $,
$\chi^{||(A)}$ is a scalar function,
and  $D_{ij}\chi^{||(A)}$ is the traceless part of the scalar perturbation.
The vector metric perturbation
satisfies a condition $\partial^i\partial^j  \chi^{\perp(A) }_{ij}=0$
and can be written as
\be\label{chiVec0}
\chi^{\perp(A) }_{ij}= \partial_i B^{(A)}_j+\partial_j B^{(A)}_i,
\,\,\,\,\,
\text{with}~~  \partial^i B^{(A)}_i =0,
~~ A=1,2 ,
\ee
where $B^{(A)}_i$ is a curl  vector and has two independent modes.
The tensor metric perturbation
satisfies the traceless and transverse  condition:
$\chi^{\top(A)i}\, _i=0$, $\partial^i\chi^{\top(A)}_{ij}=0$,
having two independent modes.

The RD stage of expansion is driven by a relativistic fluid  (without a shear stress),
whose   energy-momentum tensor is
$T_{\mu\nu}=(\rho+p)U_\mu U_\nu+g_{\mu\nu}p$,
where $\rho$ and $p$ are, respectively,
the energy density and pressure measured by a comoving observer
in the locally inertial frame,
and $U^\mu=\frac{d x^\mu}{d\lambda}$ with $d\lambda^2 =-ds^2$
is the fluid  4-velocity
with a normalization condition $g_{\mu\nu}U^\mu U^\nu=-1$.
We write
\be\label{rho}
\rho=\rho^{(0)} + \rho^{(1)}+\frac{1}{2}\rho^{(2)} ,
\ee
\be\label{p1}
p=p^{(0)}+p^{(1)}+\frac{1}{2}p^{(2)}.
\ee
where $ \rho^{(0)}$ is the background density
and $ \rho^{(1)}$, $ \rho^{(2)}$ are  the respective 1st-, 2nd-order density perturbations.
We introduce the  density contrast
\be\label{deltaA}
\delta^{(A)}\equiv\frac{\rho^{(A)}}{\rho^{(0)}}, ~~ A=1,2.
\ee{One can define the following parameters  of a fluid
\be \label{cn2}
c_s^2\equiv \frac {p^{(0)'}}{\rho^{(0)'}},
~~~~~~
c_L^2\equiv \frac {p^{(1)}}{\rho^{(1)}}
,
~~~~~~
c_N^2\equiv \frac {p^{(2)}}{\rho^{(2)}},
\ee
where $c_s$ is the sound speed.
For the relativistic  fluid,
$c_s^2=\frac13 $  which   also equals  the state of matter
$\omega= \frac {p^{(0)}}{\rho^{(0)}}=\frac13 $,
and $c_L^2=\frac13$ is taken \cite{Weinberg1972}.
In  this paper we assume $c_N^2=\frac13$ for computation convenience,
its actual value should be determined by future experiments.
The expansion of $U^\mu$ is also expanded up to 2nd order
\be
U^\mu\equiv U^{(0)\mu}+U^{(1)\mu}+\frac{1}{2}U^{(2)\mu} ,
\ee
where
\be\label{U0element}
U^{(0)0}=a^{-1},
~~~~~~
U^{(1)0}=0,
~~~~~~
U^{(2)0}=a^{-1} v^{(1)k}v^{(1)}_{k} ,
\ee
\be\label{Uielement}
U^{(0)i}=0,
~~~~~~
U^{(1)i}=a^{-1} v^{(1)i},
~~~~~~
U^{(2)i}= a^{-1}v^{(2)i} \, ,
\ee
with  the 3-velocity
 $v^i \equiv \frac{d x^i}{d\tau} =\frac{U^i}{U^0} =v^{(1)i}+\frac{1}{2}v^{(2)i}$
  \cite{Bertschinger}.
 $v^{(A)i}$ can be decomposed into noncurl and curl parts as
$v^{(A)i}=v^{||(A),\,i}+v^{\perp(A)i}$,
with $\partial_i v^{\perp(A)i}=0$ for $A=1,2$.
We also have
\be\label{Ulow0}
U_0=
-a \Big(1+\frac{1}{2}v^{(1)m}v^{(1)}_{m} \Big),
~~~
U_i=
a\Big(v^{(1)}_{i}  + \gamma_{ij}^{(1)}v^{(1)j} +\frac{1}{2}v^{(2)}_{i} \Big).
\ee

The  Einstein equation is expanded up to 2nd order of perturbations
\be\label{pertEinstein}
G^{(A)}_{\mu\nu}
\equiv R^{(A)}_{\mu\nu}-\frac{1}{2} \big[g_{\mu\nu}R\big]^{(A)}
=8\pi GT^{(A)}_{\mu\nu},
~~\text{with}~~ A= 0,1,2.
\ee
For each  order of  (\ref{pertEinstein}),
the (00) component is the energy constraint,
 $(0i)$ components are the momentum constraints,
and  $(ij)$ components are the evolution equations.
The 0th-order Einstein equation gives the Friedman equations
$\l(\frac{a'}{a}\r)^2
=\frac{8\pi G}{3} a^2\rho^{(0)}~$ and
$~-2\frac{a''}{a}
 + \l(\frac{a'}{a} \r)^2
 = 8\pi G   a^2  p^{(0)}$,
which have  a solution for  the   RD stage
$a(\tau)\propto\tau$ and
$\rho^{(0)}(\tau)=\frac{3}{8\pi G}\frac{a'^{\,2}(\tau)}{a^4(\tau)}
\propto\tau^{-4}$.
The covariant conservation of the stress tensor is
\be\label{covcons}
T^{ \mu\nu}\,_{; \, \nu}=0 .
\ee
The dynamics of gravitational systems is
determined by (\ref{pertEinstein}) and (\ref{covcons}).
The component $\mu=0$ of (\ref{covcons}) gives the energy conservation
\bl\label{EnConsv2}
&g^{00}p_{,\,0}
+\partial_0\big[(\rho+p)U^0 U^0\big]
+\partial_i\big[(\rho+p)U^0 U^i\big]
+\Gamma^{0}_{00}(\rho+p)U^0 U^0
+\Gamma^{0}_{ij}(\rho+p)U^i U^j
\nn\\
&
+\Gamma^{0}_{00}(\rho+p)U^0 U^0
+\Gamma^{k}_{k0}(\rho+p)U^0 U^0
+\Gamma^{k}_{km}(\rho+p)U^0 U^m =0 ,
\el
and the component $\mu=i$ gives  the momentum conservation,
\bl\label{MoConsv2}
&g^{ik}p_{,\,k}
+\partial_0\big[(\rho+p)U^i U^0\big]
+\partial_m\big[(\rho+p)U^i U^m\big]
+2\Gamma^{i}_{m0}(\rho+p)U^m U^0
+\Gamma^{i}_{ml}(\rho+p)U^m U^l
\nn\\
&
+\Gamma^{0}_{00}(\rho+p)U^i U^0
+\Gamma^{k}_{k0}(\rho+p)U^i U^0
+\Gamma^{k}_{kl}(\rho+p)U^i U^l =0 ,
\el
where the nonvanishing Christopher symbols are  listed in
(A1)--(A4) of Ref.\cite{WangZhang2ndRD2018}.
To each order of perturbation,
(\ref{EnConsv2}) and (\ref{MoConsv2}) will  determine $\rho$ and $U^\mu$.

The 1st-order perturbation in the RD stage is known in the literature,
and a complete list is given in Ref. \cite{WangZhang2ndRD2018}.
Here we quote  the 1st-order gauge-invariant modes for use in this paper.
The transverse part of 1st-order velocity $U^{\perp(1)i}$
and the vector mode $\chi^{\perp(1)}_{ij}$ of the metric
are decaying during the RD stage, so we take
\be \label{notrsv}
U^{\perp(1)i}= \chi^{\perp(1)}_{ij}= C_{1\,ij}= 0 \, ,
\ee
which amounts to assuming that
the relativistic fluid is irrotational  during the RD stage.
This will simplify the 2nd-order calculation
as the  coupling terms involving 1st-order vectors vanish.
The $k$-mode of the 1st-order longitudinal velocity
with a   constant $c_L^2$ is
\bl\label{v1||solgen}
v^{||(1)}_{\mathbf k}(\tau)=&
d_1  (\frac{c_L^{} k \tau}{2} )^{-\frac{3 c_L^2}{2}}
  \Gamma \left(\frac{3 c_L^2}{2}+1\right)
\left(J_{\frac{3 c_L^2}{2}}(c_L^{} k \tau)
+ (c_L^{} k \tau) J_{\frac{3 c_L^2}{2}+1}(c_L^{} k \tau)\right)
\nn\\
&
+  d_2   (\frac{c_L^{} k \tau}{2} )^3 \, _1F_2\left(2;\frac{5}{2},
\frac{3 c_L^2}{2}+\frac{5}{2}; -(\frac{c_L^{} k \tau}{2})^2\right)
\nn\\
&
 +d_3   (\frac{c_L^{} k \tau}{2})^{-3 c_L^2}  \,
   _1F_2\left(\frac{1}{2}-\frac{3 c_L^2}{2};
   -\frac{3 c_L^2}{2}-\frac{1}{2},1-\frac{3 c_L^2}{2};
   -(\frac{c_L^{} k \tau}{2})^2\right),
\el
where $d_1 ,d_2 ,d_3 $ are   $\bf k$-dependent integration constants;
$J_q(x)$ is the Bessel function;
$\Gamma(x)$ is the Gamma function;
and
$_pF_q$
is the generalized hypergeometric function.
From  the solution of  $v^{||(1)}_{\mathbf k}$,
other scalar perturbations can also be given  \cite{WangZhang2ndRD2018}.
For simplicity,
we take   $c_L^2 =\frac{1}{3}$  \cite{Grishchuk1994};
then   (\ref{v1||solgen}) reduces to
\be\label{vsol3}
 v^{||(1)}_{\mathbf k}
=
D_2(\mathbf k) \l(\frac{2}{k\tau}
    +\frac{i}{\sqrt3}\r)e^{-ik\tau/\sqrt3}
+D_3(\mathbf k) \l(\frac{2}{k\tau}
    -\frac{i}{\sqrt3}\r)e^{ik\tau/\sqrt3} \, ,
\ee
where $D_2=-\frac{3 \sqrt{3}}{4}d_2+\frac{i }{2} \sqrt{3} d_1$ and
$D_3 =-\frac{3 \sqrt{3}}{4}d_2+\frac{i }{2} \sqrt{3} d_1$
are   $\bf k$-dependent   constants.
The 1st-order density contrast is
\be\label{delta1sol3}
\delta^{(1)}_{\mathbf k}
=
D_2\l(\frac{8}{k\tau^2}
+ \frac{8 \,i }{\sqrt3 \,\tau}
-\frac{4 k}{3}\r)e^{-ik\tau/\sqrt3}
+D_3\l(\frac{8}{k\tau^2}
-\frac{8\,i}{\sqrt3\,\tau}
-\frac{4 k}{3}\r)e^{ik\tau/\sqrt3}
.
\ee
The two scalar modes  of the metric perturbation are
\bl
\phi^{(1)}_{\mathbf k}
=&
D_2\l(\frac{2}{k\tau^2}
+\frac{2 \,i }{\sqrt3 \,\tau}
    \r)e^{-ik\tau/\sqrt3}
+D_3\l(\frac{2}{k\tau^2}
-\frac{2\,i}{\sqrt3\,\tau}
    \r)e^{ik\tau/\sqrt3} \nn\\
& -\frac{2k}{3}\int^\tau\l[ D_2 e^{-ik\tau'/\sqrt3}
    +D_3e^{ik\tau'/\sqrt3}\r]\frac{d\tau'}{\tau'} ,   \label{phi1sol2}   \\
\chi^{||(1)}_{\mathbf k}
 =&
D_2\frac{4\sqrt3\,i}{k^2\tau}e^{-ik\tau/\sqrt3}
-D_3\frac{4\sqrt3\,i}{k^2\tau}e^{ik\tau/\sqrt3}
\nn\\
&
-\frac{4}{k}{}\int^\tau\l[ D_2 e^{-ik\tau'/\sqrt3}
    +D_3e^{ik\tau'/\sqrt3}\r]\frac{d\tau'}{\tau'}.
    \label{chi1sol3}
\el
And  the  equation for the  tensor mode is
\be  \label{evoEq1stRDtensor}
 \chi^{\top(1)''}_{ij}
+\frac{2}{\tau} \chi^{\top(1)'}_{ij}
- \nabla^2\chi^{\top(1) }_{ij}
=0 ,
\ee
which has the solution  written in terms of Fourier modes
\be  \label{Fourier}
\chi^{\top(1)}_{ij}  ( {\bf x},\tau)= \frac{1}{(2\pi)^{3/2}}
\int d^3k   e^{i \,\bf{k}\cdot\bf{x}}
\sum_{s={+,\times}} {\mathop \epsilon \limits^s}_{ij}({\bf k})  ~
{\mathop h\limits^s}_{ \bf k}  (\tau)
, \,\,\,\, {\bf k}=k\hat{k},
\ee
with two polarization tensors satisfying
\be\label{polarTensorSum}
{\mathop \epsilon  \limits^s}_{ij}({\bf k})  \delta^{ij}=0,\,\,\,\,
{\mathop \epsilon  \limits^s}_{ij}({\bf k})   k^i=0,\,\,\,
{\mathop \epsilon  \limits^s}_{ij}({\bf k})  {\mathop \epsilon  \limits^{s'}}{}^{ij}({\bf k})
=2\delta_{ss'}.
\ee
For RGW  generated during inflation
\cite{Grishchuk,Allen1988,zhangyang05},
the two polarization modes ${\mathop h\limits^s}_k $ with  $s= {+,\times}$
are usually assumed to be statistically equivalent,
the superscript $s$ can be dropped.
During the RD stage the mode is given by
\bl \label{GWmode}
h_{\bf k}  (\tau ) = &\frac{1}{a(\tau)}\sqrt{\frac{\pi}{2}}
\sqrt{\frac{\tau}{2}}
\big[b_1 (\mathbf k) H^{(1)}_{\frac{1}{2} } (k\tau )
+b_2 (\mathbf k) H^{(2)}_{\frac{1}{2} } (k\tau ) \big] \nn \\
= & \frac{1}{\tau}
\frac{i}{ \sqrt{2 k}}
\big[- b_1  e^{i k\tau}
+ b_2  e^{-i k\tau} \big],
\el
where  $b_1$, $b_2$
are $\bf k$-dependent coefficients,
to be determined by the initial condition during inflation
or a possible subsequent reheating  stage \cite{Grishchuk,zhangyang05}.
There are cosmic  processes occurring during the RD stage,
such as the QCD  transition and $e^+e^-$ annihilation \cite{WangZhang2008},
which  modify  only slightly the amplitude of RGW
and will be  neglected  in  this study.

The scalar modes   (\ref{phi1sol2}) and (\ref{chi1sol3}),
the density contrast  (\ref{delta1sol3}),
and the longitudinal velocity (\ref{vsol3})
all contain a factor $e^{\pm ik\tau/\sqrt3}$,
so that they are waves propagating at the sound speed  $\frac{1}{\sqrt 3}$
of  the relativistic fluid.
On the other hand,
the tensor modes (\ref{GWmode})  are waves
propagating at the speed of light.
In the  MD stage \cite{Matarrese98,WangZhang2017,ZhangQinWang2017},
  the scalar, and density contrast are not waves and do not propagate;
only the tensor modes still propagate at the speed of light.
Therefore,whether or not the scalar modes propagate
actually depends on the background  matter;
nevertheless,
the tensor modes always propagate at speed of light,
regardless of the background matter.
Thus,
the tensor modes are radiative as dynamic degrees of freedom,
differing from the scalar and vector modes.

\section{The 2nd-order perturbed equations }

\subsection{Equations with scalar-tensor couplings}

The 2nd-order perturbed Einstein equations
are  listed  in Appendix \ref{Ap:PertEinEq} for a general RW spacetime.
As said earlier in  (\ref{notrsv}),
the transverse vector mode and the curl velocity   are dropped,
there remain  only three types of couplings:
scalar-scalar, scalar-tensor, and tensor-tensor.
The scalar-scalar coupling has been studied in Ref.\cite{WangZhang2ndRD2018}.
Now we consider the scalar-tensor coupling.

The $(00)$ component of 2nd-order perturbed Einstein equation is
\be  \label{2nd00Einstein}
 G^{(2)}_{00}
 = 8\pi G T_{00}^{(2)} ,
\ee
where $G^{(2)}_{00}$ and $T_{00}^{(2)}$
are given by (A16) and (A22)
in Ref.\cite{WangZhang2ndRD2018}.
For the scalar-tensor coupling,
by using the Friedmann equation
and moving the  coupling terms to the rhs,
Eq.(\ref{2nd00Einstein}) gives the  2nd-order energy constraint
\be  \label{Ein2th003RD}
-\frac{6}{\tau} \phi^{(2)'}_{s(t)}
+2\nabla^2\phi^{(2)}_{s(t)}
+\frac{1}{3}\nabla^2\nabla^2\chi^{||(2)}_{s(t)}
=
\frac{3}{\tau^2}\delta^{(2)}_{s(t)}
+E_{s(t)}
,
\ee
where  a subscript ``s(t)" in $\phi^{(2)}_{s(t)}$, etc.,
indicates the case of scalar-tensor coupling,
and $E_{s(t)}$ is  the scalar-tensor coupling terms  as the following
\bl\label{ES1RD}
E_{s(t)}
\equiv  &
2\phi^{(1),\,lm}\chi^{\top(1)}_{lm}
+\frac{1}{2}\chi^{\top(1)'}_{lm}\chi^{||(1)',\,lm}
+\frac{2}{\tau} \chi^{\top(1)}_{lm}\chi^{||(1)',\,lm}
+\frac{2}{\tau} \chi^{\top(1)'}_{lm}\chi^{||(1),\,lm}
    \nn\\
    &
+\frac{1}{3}\chi^{\top(1)}_{lm}\nabla^2\chi^{||(1),\,lm}
-\chi^{||(1),\,lm}\nabla^2\chi^{\top(1)}_{lm}
-\frac{1}{2}\chi^{\top(1),\,l}_{mn}\chi^{||(1),\,mn}_{,\,l} ,
\el
which is part of the effective source
of the 2nd-order energy constraint.

The  $(0i)$ component of the 2nd-order perturbed Einstein equation is
\be \label{Ein2nd0i1}
 G^{(2)}_{0i}=  R^{(2)}_{0i}=8\pi G T_{0i}^{(2)},
\ee
where  $R^{(2) }_{0i}$
and $T^{(2)}_{0i}$ are given by (A13) and (A23)
in Ref.\cite{WangZhang2ndRD2018}.
This gives  the 2nd-order momentum constraint equation
for the scalar-tensor case
as the following
\be\label{MoConstr2ndv3RD}
2 \phi^{(2)'} _{s(t),\,i}
+\frac{1}{2}D_{ij}\chi^{||(2)',\,j}_{s(t)}
+\frac{1}{2}\chi^{\perp(2)',\,j}_{s(t)ij}
=
-\frac{4}{\tau^2}v^{(2)}_{s(t)i}
+M_{s(t)i},
\ee
where
\bl\label{MSi1RD}
M_{s(t)i}
\equiv&
-\frac{8}{\tau^2}\chi^{\top(1)}_{il}v^{||(1),\,l}
+\phi^{(1) ,\,l }\chi^{\top(1)' }_{il}
-2\phi^{(1)' ,\,l} \chi^{\top(1) }_{il}
\nn\\
&
-\chi^{\top(1)' }_{lm,  \, i}\chi^{||(1),\,lm}
- \frac{1}{2}\chi^{\top(1)'}_{lm }\chi^{||(1),\, lm}_{,i}
-\frac{1}{2} \chi^{\top(1)}_{lm,\,i}\chi^{||(1)',\,lm}
    \nn\\
&
+ \chi^{\top(1)'}_{il,\,m}\chi^{||(1),\,lm}
-\frac{1}{3}\chi^{\top(1)}_{li}\nabla^2\chi^{||(1)',\,l}
+\frac{2}{3}\chi^{\top (1)' }_{il}\nabla^2\chi^{||(1),\,l}
\el
is part of the effective  source.
Equation (\ref{MoConstr2ndv3RD}) can be further decomposed into two equations.
Applying $\nabla^{-2}\partial^i$ on (\ref{MoConstr2ndv3RD}) gives
the longitudinal  momentum constraint
\be \label{MoConstr2ndv3RD2}
2\phi^{(2)'}_{s(t)}
+\frac{1}{3}\nabla^2\chi^{||(2)'}_{s(t)}
=
-\frac{4}{\tau^2}v^{||(2)}_{s(t)}
+\nabla^{-2}M_{s(t)l}^{,\,l},
\ee
where
\bl\label{MSkkScalarRD}
M_{s(t)l}^{\,,\,l}
=&
-\frac{8}{\tau^2}\chi^{\top(1)}_{lm}v^{||(1),\,lm}
+\phi^{(1) ,\,lm}\chi^{\top(1)' }_{lm}
-2\phi^{(1)' ,\,lm} \chi^{\top(1) }_{lm}
\nn\\
&
- \frac{1}{2}\chi^{\top(1)'}_{lm ,n}\chi^{||(1),\, lmn}
+\frac{1}{6}\chi^{\top(1)'}_{lm}\nabla^2\chi^{||(1),\,lm}
-\chi^{||(1),\,lm}\nabla^2\chi^{\top(1)' }_{lm}
    \nn\\
&
-\frac{1}{2} \chi^{\top(1)}_{lm,\,n}\chi^{||(1)',\,lmn}
-\frac{1}{3}\chi^{\top(1)}_{lm}\nabla^2\chi^{||(1)',\,lm}
-\frac{1}{2} \chi^{||(1)',\,lm}\nabla^2\chi^{\top(1)}_{lm}
.
\el
A combination
[(\ref{MoConstr2ndv3RD})$-\partial_i$(\ref{MoConstr2ndv3RD2})]
gives the transverse  momentum constraint
\be \label{MoCons2ndCurlRD1}
\frac{1}{2}\chi^{\perp(2)',\,j}_{s(t)ij}
=
-\frac{4}{\tau^2}v^{\perp(2)}_{s(t)i}
+\l(
M_{s(t)i}
-\partial_i\nabla^{-2}M_{s(t)l}^{,\,l}
\r)
\ ,
\ee
where
\bl\label{MSiCurlRD}
&
\l( M_{s(t)i}-\partial_i\nabla^{-2}M_{s(t)l}^{\,,\,l} \r)
\nn\\
=&
-\frac{8}{\tau^2}\chi^{\top(1)}_{il}v^{||(1),\,l}
+\phi^{(1) ,\,l }\chi^{\top(1)' }_{il}
-2\phi^{(1)' ,\,l} \chi^{\top(1) }_{il}
-\frac{1}{2}\chi^{\top(1)' }_{lm,  \, i}\chi^{||(1),\,lm}
    \nn\\
&
-\frac{1}{2} \chi^{\top(1)}_{lm,\,i}\chi^{||(1)',\,lm}
+ \chi^{\top(1)'}_{il,\,m}\chi^{||(1),\,lm}
-\frac{1}{3}\chi^{\top(1)}_{li}\nabla^2\chi^{||(1)',\,l}
+\frac{2}{3}\chi^{\top (1)' }_{il}\nabla^2\chi^{||(1),\,l}
\nn\\
&
+\partial_i\nabla^{-2}\Big[
\frac{8}{\tau^2}\chi^{\top(1)}_{lm}v^{||(1),\,lm}
-\phi^{(1) ,\,lm}\chi^{\top(1)' }_{lm}
+2\phi^{(1)' ,\,lm} \chi^{\top(1) }_{lm}
\nn\\
&
- \frac{1}{2}\chi^{\top(1)'}_{lm ,n}\chi^{||(1),\, lmn}
-\frac{2}{3}\chi^{\top(1)'}_{lm}\nabla^2\chi^{||(1),\,lm}
+\frac{1}{2}\chi^{||(1),\,lm}\nabla^2\chi^{\top(1)' }_{lm}
    \nn\\
&
+\frac{1}{2} \chi^{\top(1)}_{lm,\,n}\chi^{||(1)',\,lmn}
+\frac{1}{3}\chi^{\top(1)}_{lm}\nabla^2\chi^{||(1)',\,lm}
+\frac{1}{2} \chi^{||(1)',\,lm}\nabla^2\chi^{\top(1)}_{lm}
\Big] .
\el

The $(ij)$ component of 2nd-order perturbed Einstein equation is
\be\label{Ein2ndijss1}
G^{(2)}_{ij}
=8\pi G T^{(2)}_{ij} ,
\ee
where  $G^{(2)}_{ij}$ and $T^{(2)}_{ij}$
are given in (A18) and(A24) of Ref.\cite{WangZhang2ndRD2018}.
For the scalar-tensor case,
(\ref{Ein2ndijss1}) gives the 2nd-order evolution equation
\bl\label{Evo2ndSs1RD}
&
2\phi^{(2)''}_{s(t)} \delta_{ij}
+\frac{4}{\tau}\phi^{(2)'}_{s(t)} \delta_{ij}
+\phi^{(2)}_{s(t),\,ij}
-\nabla^2\phi^{(2)}_{s(t)} \delta_{ij}
\nn\\
&
+\frac{1}{2}D_{ij}\chi^{||(2)''}_{s(t)}
+\frac{1}{\tau}D_{ij}\chi^{||(2)'}_{s(t)}
+\frac{1}{6}\nabla^2D_{ij}\chi^{||(2)}_{s(t)}
-\frac{1}{9}\delta_{ij}\nabla^2\nabla^2\chi^{||(2) }_{s(t)}
\nn\\
&
+\frac{1}{2}\chi^{\perp(2)''}_{s(t)ij}
+\frac{1}{\tau}\chi^{\perp(2)'}_{s(t)ij}
\nn\\&
+\frac{1}{2}\chi^{\top(2)''}_{s(t)ij}
+\frac{1}{\tau}\chi^{\top(2)'}_{s(t)ij}
-\frac{1}{2}\nabla^2\chi^{\top(2)}_{s(t)ij}
=
\frac{3 c_N^2}{\tau^2}\delta^{(2)}_{s(t)}\delta_{ij}
+S_{s(t)ij} ,
\el
where
\bl\label{Ss2ndij2RD}
S_{s(t)ij}
\equiv&
\frac{6}{\tau^2}c_L^2\chi^{\top(1)}_{ij}\delta^{(1)}
-6\phi^{(1)''}\chi^{\top(1)}_{ij}
-\frac{12}{\tau}\phi^{(1)'}\chi^{\top(1)}_{ij}
+4\chi^{\top(1)}_{ij}\nabla^2\phi^{(1)}
-\phi^{(1)'}\chi^{\top(1)'}_{ij}
\nn\\
&
-\phi^{(1),\,l}\chi^{\top(1)}_{l j,\,i}
-\phi^{(1),\,l}\chi^{\top(1)}_{l i,\,j}
+3\phi^{(1),\,l}\chi^{\top(1)}_{ij,\,l}
+2\phi^{(1)}\nabla^2\chi^{\top(1)}_{ij}
-2\phi^{(1),\,l}_{,\,i}\chi^{\top(1)}_{l j}
\nn\\
&
-2\phi^{(1),\,l}_{,\,j}\chi^{\top(1)}_{l i}
-\chi^{\top(1)''}_{lm}\chi^{||(1),\,lm}\delta_{ij}
-\chi^{\top(1)}_{lm}\chi^{||(1)'',\,lm}\delta_{ij}
-\frac{2}{\tau}\chi^{\top(1)'}_{lm}\chi^{||(1),\,lm}\delta_{ij}
\nn\\
&
-\frac{2}{\tau}\chi^{\top(1)}_{lm}\chi^{||(1)',\,lm}\delta_{ij}
+\chi^{\top(1)'}_{l i}\chi^{||(1)',\,l}_{,\,j}
+\chi^{\top(1)'}_{lj}\chi^{||(1)',\,l}_{,\,i}
-\frac{3}{2}\chi^{\top(1)'}_{lm}\chi^{||(1)',\,lm}\delta_{ij}
\nn\\
&
-\frac{1}{3}\chi^{\top(1)}_{li}\nabla^2\chi^{||(1),\,l}_{,\,j}
-\frac{1}{3}\chi^{\top(1)}_{lj}\nabla^2\chi^{||(1),\,l}_{,\,i}
-\chi^{\top(1)}_{lm,\,ij}\chi^{||(1),\,lm}
+\chi^{||(1),\,lm}\nabla^2\chi^{\top(1)}_{lm}\delta_{ij}
\nn\\
&
-\frac{1}{2}\chi^{\top(1)}_{lm,\,j}\chi^{||(1),\,lm}_{,\,i}
-\frac{1}{2}\chi^{\top(1)}_{lm,\,i}\chi^{||(1),\,lm}_{,\,j}
+\frac{1}{2}\chi^{||(1),\,nm l}\chi^{\top(1)}_{nm,\,l}\delta_{ij}
-\frac{2}{3}\chi^{\top(1)'}_{ij}\nabla^2\chi^{||(1)'}
\nn\\
&
+\frac{2}{3}\chi^{\top(1)}_{ij}\nabla^2\nabla^2\chi^{||(1)}
+\frac{1}{3}\nabla^2\chi^{\top(1)}_{ij}\nabla^2\chi^{||(1)}
+\frac{1}{3}\chi^{\top(1)}_{lj,\,i}\nabla^2\chi^{||(1),\,l}
+\frac{1}{3}\chi^{\top(1)}_{li,\,j}\nabla^2\chi^{||(1),\,l}
\nn\\
&
+\chi^{\top(1)}_{lj,\,im}\chi^{||(1),\,lm}
+\chi^{\top(1)}_{li,\,jm}\chi^{||(1),\,lm}
-\chi^{\top(1)}_{ij,\,lm}\chi^{||(1),\,lm}
\el
is part of the effective source.

We also need to decompose the evolution equation (\ref{Evo2ndSs1RD})
into the trace equation and the traceless equation.
First,
the trace equation from  (\ref{Evo2ndSs1RD}) is
\be\label{Evo2ndSsTr2RD}
2\phi^{(2)''}_{s(t)}
+\frac{4}{\tau}\phi^{(2)'}_{s(t)}
-\frac{2}{3}\nabla^2\phi^{(2)}_{s(t)}
-\frac{1}{9}\nabla^2\nabla^2\chi^{||(2) }_{s(t)}
=
\frac{3}{\tau^2}c_N^2\delta^{(2)}_{s(t)}
+\frac{1}{3}S_{s(t)l}^{\,l}
\,,
\ee
where
\bl\label{SijTraceRD}
S_{s(t)l}^{\,l}
=&
-4\phi^{(1),\,lm}\chi^{\top(1)}_{lm}
-3\chi^{\top(1)''}_{lm}\chi^{||(1),\,lm}
-3\chi^{\top(1)}_{lm}\chi^{||(1)'',\,lm}
-\frac{6}{\tau}\chi^{\top(1)'}_{lm}\chi^{||(1),\,lm}
\nn\\
&
-\frac{6}{\tau}\chi^{\top(1)}_{lm}\chi^{||(1)',\,lm}
-\frac{5}{2}\chi^{\top(1)'}_{lm}\chi^{||(1)',\,lm}
-\frac{2}{3}\chi^{\top(1)}_{lm}\nabla^2\chi^{||(1),\,lm}
+2\chi^{||(1),\,lm}\nabla^2\chi^{\top(1)}_{lm}
\nn\\
&
+\frac{1}{2}\chi^{\top(1)}_{lm,\,n}\chi^{||(1),\,lmn}
\ .
\el
The traceless equation from  (\ref{Evo2ndSs1RD})   is
\bl\label{Evo2ndSsNoTr1RD}
&
D_{ij}\phi^{(2)}_{s(t)}
+\frac{1}{2}D_{ij}\chi^{||(2)''}_{s(t)}
+\frac{1}{\tau} D_{ij}\chi^{||(2)'}_{s(t)}
+\frac{1}{6}\nabla^2D_{ij}\chi^{||(2)}_{s(t)}
\nn\\
&
+\frac{1}{2}\chi^{\perp(2)''}_{s(t)ij}
+\frac{1}{\tau} \chi^{\perp(2)'}_{s(t)ij}
\nn\\
&
+\frac{1}{2}\chi^{\top(2)''}_{s(t)ij}
+\frac{1}{\tau} \chi^{\top(2)'}_{s(t)ij}
-\frac{1}{2}\nabla^2\chi^{\top(2)}_{s(t)ij}
=
\bar S_{s(t)ij}
,
\el
where
\bl\label{SijTracelessRD}
\bar S_{s(t)ij}
\equiv&
S_{s(t)ij}-\frac{1}{3}S_{s(t)l}^l\delta_{ij}
\nn\\
=&
\frac{6}{\tau^2}c_L^2\chi^{\top(1)}_{ij}\delta^{(1)}
-6\phi^{(1)''}\chi^{\top(1)}_{ij}
-\frac{12}{\tau} \phi^{(1)'}\chi^{\top(1)}_{ij}
+4\chi^{\top(1)}_{ij}\nabla^2\phi^{(1)}
-\phi^{(1)'}\chi^{\top(1)'}_{ij}
\nn\\
&
-\phi^{(1),\,l}\chi^{\top(1)}_{l j,\,i}
-\phi^{(1),\,l}\chi^{\top(1)}_{l i,\,j}
+3\phi^{(1),\,l}\chi^{\top(1)}_{ij,\,l}
+2\phi^{(1)}\nabla^2\chi^{\top(1)}_{ij}
-2\phi^{(1),\,l}_{,\,i}\chi^{\top(1)}_{l j}
\nn\\
&
-2\phi^{(1),\,l}_{,\,j}\chi^{\top(1)}_{l i}
+\frac{4}{3}\phi^{(1),\,lm}\chi^{\top(1)}_{lm}\delta_{ij}
+\chi^{\top(1)'}_{l i}\chi^{||(1)',\,l}_{,\,j}
+\chi^{\top(1)'}_{lj}\chi^{||(1)',\,l}_{,\,i}
\nn\\
&
-\frac{2}{3}\chi^{\top(1)'}_{lm}\chi^{||(1)',\,lm}\delta_{ij}
-\frac{1}{3}\chi^{\top(1)}_{li}\nabla^2\chi^{||(1),\,l}_{,\,j}
-\frac{1}{3}\chi^{\top(1)}_{lj}\nabla^2\chi^{||(1),\,l}_{,\,i}
+\frac{2}{9}\chi^{\top(1)}_{lm}\nabla^2\chi^{||(1),\,lm}\delta_{ij}
\nn\\
&
-\chi^{\top(1)}_{lm,\,ij}\chi^{||(1),\,lm}
+\frac{1}{3}\chi^{||(1),\,lm}\nabla^2\chi^{\top(1)}_{lm}\delta_{ij}
-\frac{1}{2}\chi^{\top(1)}_{lm,\,j}\chi^{||(1),\,lm}_{,\,i}
-\frac{1}{2}\chi^{\top(1)}_{lm,\,i}\chi^{||(1),\,lm}_{,\,j}
\nn\\
&
+\frac{1}{3}\chi^{\top(1)}_{lm,\,n}\chi^{||(1),\,lmn}\delta_{ij}
-\frac{2}{3}\chi^{\top(1)'}_{ij}\nabla^2\chi^{||(1)'}
+\frac{2}{3}\chi^{\top(1)}_{ij}\nabla^2\nabla^2\chi^{||(1)}
+\frac{1}{3}\nabla^2\chi^{\top(1)}_{ij}\nabla^2\chi^{||(1)}
\nn\\
&
+\frac{1}{3}\chi^{\top(1)}_{lj,\,i}\nabla^2\chi^{||(1),\,l}
+\frac{1}{3}\chi^{\top(1)}_{li,\,j}\nabla^2\chi^{||(1),\,l}
+\chi^{\top(1)}_{lj,\,im}\chi^{||(1),\,lm}
+\chi^{\top(1)}_{li,\,jm}\chi^{||(1),\,lm}
\nn\\
&
-\chi^{\top(1)}_{ij,\,lm}\chi^{||(1),\,lm}
\ ,
\el
which still contains the scalar,  vector, and tensor components.
Applying
$3\nabla^{-2}\nabla^{-2}\partial_i\partial_j$ on (\ref{Evo2ndSsNoTr1RD})
gives the evolution equation for the scalar $\chi^{||(2)}_{s(t)}$
as follows:
\bl\label{Evo2ndSsChi1RD}
&
\chi^{||(2)''}_{s(t)}
+\frac{2}{\tau}\chi^{||(2)'}_{s(t)}
+\frac{1}{3}\nabla^2\chi^{||(2)}_{s(t)}
+2\phi^{(2)}_{s(t)}
=
3\nabla^{-2}\nabla^{-2}\bar S_{s(t)lm}^{\, ,\,lm}
,
\el
where
\bl\label{SijPijbarRD}
\bar S_{s(t)lm}^{\, ,\,lm}
=&
-\frac{2}{3}\nabla^2\nabla^2\Big[\chi^{||(1),\,lm}\chi^{\top(1)}_{lm}\Big]
+\nabla^2\Big[
\frac{4}{3}\phi^{(1),\,lm}\chi^{\top(1)}_{lm}
+\frac{1}{3}\chi^{\top(1)'}_{lm}\chi^{||(1)',\,lm}
\nn\\
&
+\frac{8}{9} \chi^{\top(1)}_{lm}\nabla^2\chi^{||(1),\,lm}
+\frac{7}{6}\chi^{\top(1)}_{lm,\,n}\chi^{||(1),\,lmn}
\Big]
\nn\\
&
+6c_L^2\l(\frac{a'}{a}\r)^2\chi^{\top(1)}_{lm}\delta^{(1),\,lm}
-6\phi^{(1)'',\,lm}\chi^{\top(1)}_{lm}
-\frac{12}{\tau}\phi^{(1)',\,lm}\chi^{\top(1)}_{lm}
\nn\\
&
-\phi^{(1)',\,lm}\chi^{\top(1)'}_{lm}
-3\phi^{(1),\,lmn}\chi^{\top(1)}_{lm,\,n}
-\chi^{||(1)',\,lm}\nabla^2\chi^{\top(1)'}_{lm}
+\frac{1}{3}\chi^{\top(1)'}_{lm}\nabla^2\chi^{||(1)',\,lm}
\nn\\
&
-\frac{1}{2}\chi^{\top(1)}_{lm,\,n}\nabla^2\chi^{||(1),\,lmn}
+\nabla^2\chi^{\top(1)}_{lm}\nabla^2\chi^{||(1),\,lm}
+\frac{3}{2}\chi^{||(1),\,lmn}\nabla^2\chi^{\top(1)}_{lm,n}
\ .
\el
By a combination
$\partial^i$\big[(\ref{Evo2ndSsNoTr1RD})$ -\frac{1}{2}D_{ij}$(\ref{Evo2ndSsChi1RD})\big],
one has
\be\label{Evo2ndSsVec1}
\frac{1}{2}\chi^{\perp(2)'',\,l}_{s(t)lj}
+\frac{1}{\tau} \chi^{\perp(2)',\,l}_{s(t)lj}
=
\bar S_{s(t)lj}^{\,,\,l}
-\partial_j\nabla^{-2}\bar S_{s(t)lm}^{\, ,\,lm}
.
\ee
By  $\nabla^{-2}\big[\partial_i(\ref{Evo2ndSsVec1})
+(i\leftrightarrow j)\big]$,
using (\ref{chiVec0}),
one gets the evolution equation for the vector mode
\be\label{Evo2ndSsVec2RD}
\frac{1}{2}\chi^{\perp(2)''}_{s(t)ij}
+\frac{1}{\tau} \chi^{\perp(2)'}_{s(t)ij}
=
V_{s(t)ij}
~,
\ee
where
{
\allowdisplaybreaks
\bl\label{SourceCurl1RD}
V_{s(t)ij}
\equiv
&
\nabla^{-2}\bar S_{s(t)lj,\,i}^{,\,l}
+\nabla^{-2}\bar S_{s(t)li,j}^{,\,l}
-2\nabla^{-2}\nabla^{-2}\bar S_{s(t)lm,\,ij}^{\, ,\,lm}
\nn\\
=&
\frac{2}{3}\partial_i\partial_j\Big[\chi^{||(1),\,lm}\chi^{\top(1)}_{lm}\Big]
+\partial_i\partial_j\nabla^{-2}\Big[
-\phi^{(1),\,lm}\chi^{\top(1)}_{lm}
-\frac{5}{6} \chi^{\top(1)}_{lm}\nabla^2\chi^{||(1),\,lm}
\nn\\
&
-\frac{1}{6}\chi^{||(1),\,lm}\nabla^2\chi^{\top(1)}_{lm}
-\frac{4}{3}\chi^{\top(1)}_{lm,\,n}\chi^{||(1),\,lmn}
\Big]
+\partial_i\nabla^{-2}\Big[
\frac{6}{\tau^2}c_L^2\,\chi^{\top(1)}_{lj}\delta^{(1),\,l}
-6\phi^{(1)'',\,l}\chi^{\top(1)}_{lj}
\nn\\
&
-\frac{12}{\tau} \phi^{(1)',\,l}\chi^{\top(1)}_{lj}
-\phi^{(1)',\,l}\chi^{\top(1)'}_{lj}
+2\chi^{\top(1)}_{lj}\nabla^2\phi^{(1),\,l}
+\phi^{(1),\,l}\nabla^2\chi^{\top(1)}_{lj}
-\phi^{(1),\,lm}_{,j}\chi^{\top(1)}_{lm}
\nn\\
&
+\chi^{\top(1)'}_{lj,\,m}\chi^{||(1)',\,lm}
-\chi^{\top(1)'}_{lm,j}\chi^{||(1)',\,lm}
+\frac{1}{3}\chi^{\top(1)'}_{lj}\nabla^2\chi^{||(1)',\,l}
-\frac{1}{6}\chi^{\top(1)}_{lm}\nabla^2\chi^{||(1),\,lm}_{,j}
\nn\\
&
+\frac{1}{3}\chi^{\top(1)}_{lj}\nabla^2\nabla^2\chi^{||(1),\,l}
+\frac{2}{3}\nabla^2\chi^{\top(1)}_{lj}\nabla^2\chi^{||(1),\,l}
+\chi^{||(1),\,lm}\nabla^2\chi^{\top(1)}_{lj,\,m}
-\frac{1}{2}\chi^{||(1),\,lm}\nabla^2\chi^{\top(1)}_{lm,j}
\Big]
\nn\\
&
+\partial_i\partial_j\nabla^{-2}\nabla^{-2}\Big[
-\frac{6}{\tau^2}c_L^2\,\chi^{\top(1)}_{lm}\delta^{(1),\,lm}
+6\phi^{(1)'',\,lm}\chi^{\top(1)}_{lm}
+\frac{12}{\tau} \phi^{(1)',\,lm}\chi^{\top(1)}_{lm}
\nn\\
&
+\phi^{(1)',\,lm}\chi^{\top(1)'}_{lm}
+3\phi^{(1),\,lmn}\chi^{\top(1)}_{lm,\,n}
+\chi^{||(1)',\,lm}\nabla^2\chi^{\top(1)'}_{lm}
-\frac{1}{3}\chi^{\top(1)'}_{lm}\nabla^2\chi^{||(1)',\,lm}
\nn\\
&
+\frac{1}{2}\chi^{\top(1)}_{lm,\,n}\nabla^2\chi^{||(1),\,lmn}
-\nabla^2\chi^{\top(1)}_{lm}\nabla^2\chi^{||(1),\,lm}
-\frac{3}{2}\chi^{||(1),\,lmn}\nabla^2\chi^{\top(1)}_{lm,n}
\Big]
\nn\\
&
+(i\leftrightarrow j)
\
\el
}
is the effective source of 2nd-order vector.
Finally,
[(\ref{Evo2ndSsNoTr1RD})$-\frac{1}{2}D_{ij}$(\ref{Evo2ndSsChi1RD})
$-$(\ref{Evo2ndSsVec2RD})] gives the
evolution equation for the 2nd-order tensor mode
\be \label{Evo2ndSsTen1RD}
\frac{1}{2}\chi^{\top(2)''}_{s(t)ij}
+\frac{1}{\tau} \chi^{\top(2)'}_{s(t)ij}
-\frac{1}{2}\nabla^2\chi^{\top(2)}_{s(t)ij}
=
J_{s(t)ij}
,
\ee
where
{
\allowdisplaybreaks
\bl\label{2ndTensorSourceRD}
J_{s(t)ij}
&
\equiv
\bar S_{s(t)ij}
-\frac{3}{2}D_{ij}\nabla^{-2}\nabla^{-2}\bar S_{s(t)lm}^{\, ,\,lm}
-\nabla^{-2}\bar S_{s(t)lj,\,i}^{,\,l}
-\nabla^{-2}\bar S_{s(t)li,j}^{,\,l}
+2\nabla^{-2}\nabla^{-2}\bar S_{s(t)lm,\,ij}^{\, ,\,lm}
\nn\\
=&
\Big[
\frac{6}{\tau^2}c_L^2\,\chi^{\top(1)}_{ij}\delta^{(1)}
-6\phi^{(1)''}\chi^{\top(1)}_{ij}
-\frac{12}{\tau} \phi^{(1)'}\chi^{\top(1)}_{ij}
+4\chi^{\top(1)}_{ij}\nabla^2\phi^{(1)}
-\phi^{(1)'}\chi^{\top(1)'}_{ij}
\nn\\
&
-\phi^{(1),\,l}\chi^{\top(1)}_{l j,\,i}
-\phi^{(1),\,l}\chi^{\top(1)}_{l i,\,j}
+3\phi^{(1),\,l}\chi^{\top(1)}_{ij,\,l}
+2\phi^{(1)}\nabla^2\chi^{\top(1)}_{ij}
-2\phi^{(1),\,l}_{,\,i}\chi^{\top(1)}_{l j}
\nn\\
&
-2\phi^{(1),\,l}_{,\,j}\chi^{\top(1)}_{l i}
+2\phi^{(1),\,lm}\chi^{\top(1)}_{lm}\delta_{ij}
+\chi^{\top(1)'}_{l i}\chi^{||(1)',\,l}_{,\,j}
+\chi^{\top(1)'}_{lj}\chi^{||(1)',\,l}_{,\,i}
-\frac{1}{2}\chi^{\top(1)'}_{lm}\chi^{||(1)',\,lm}\delta_{ij}
\nn\\
&
-\frac{1}{3}\chi^{\top(1)}_{li}\nabla^2\chi^{||(1),\,l}_{,\,j}
-\frac{1}{3}\chi^{\top(1)}_{lj}\nabla^2\chi^{||(1),\,l}_{,\,i}
+\frac{1}{3}\chi^{\top(1)}_{lm}\nabla^2\chi^{||(1),\,lm}\delta_{ij}
-\chi^{\top(1)}_{lm,\,ij}\chi^{||(1),\,lm}
\nn\\
&
-\frac{1}{2}\chi^{\top(1)}_{lm,\,j}\chi^{||(1),\,lm}_{,\,i}
-\frac{1}{2}\chi^{\top(1)}_{lm,\,i}\chi^{||(1),\,lm}_{,\,j}
+\frac{1}{4}\chi^{\top(1)}_{lm,\,n}\chi^{||(1),\,lmn}\delta_{ij}
-\frac{2}{3}\chi^{\top(1)'}_{ij}\nabla^2\chi^{||(1)'}
\nn\\
&
+\frac{2}{3}\chi^{\top(1)}_{ij}\nabla^2\nabla^2\chi^{||(1)}
+\frac{1}{3}\nabla^2\chi^{\top(1)}_{ij}\nabla^2\chi^{||(1)}
+\frac{1}{3}\chi^{\top(1)}_{lj,\,i}\nabla^2\chi^{||(1),\,l}
+\frac{1}{3}\chi^{\top(1)}_{li,\,j}\nabla^2\chi^{||(1),\,l}
\nn\\
&
+\chi^{\top(1)}_{lj,\,im}\chi^{||(1),\,lm}
+\chi^{\top(1)}_{li,\,jm}\chi^{||(1),\,lm}
-\chi^{\top(1)}_{ij,\,lm}\chi^{||(1),\,lm}
\Big]
+\delta_{ij}\nabla^{-2}\Big[
\frac{3}{\tau^2}c_L^2\chi^{\top(1)}_{lm}\delta^{(1),\,lm}
\nn\\
&
-3\phi^{(1)'',\,lm}\chi^{\top(1)}_{lm}
-\frac{6}{\tau} \phi^{(1)',\,lm}\chi^{\top(1)}_{lm}
-\frac{1}{2}\phi^{(1)',\,lm}\chi^{\top(1)'}_{lm}
-\frac{3}{2}\phi^{(1),\,lmn}\chi^{\top(1)}_{lm,\,n}
\nn\\
&
-\frac{1}{2}\chi^{||(1)',\,lm}\nabla^2\chi^{\top(1)'}_{lm}
+\frac{1}{6}\chi^{\top(1)'}_{lm}\nabla^2\chi^{||(1)',\,lm}
-\frac{1}{4}\chi^{\top(1)}_{lm,\,n}\nabla^2\chi^{||(1),\,lmn}
\nn\\
&
+\frac{1}{2}\nabla^2\chi^{\top(1)}_{lm}\nabla^2\chi^{||(1),\,lm}
+\frac{3}{4}\chi^{||(1),\,lmn}\nabla^2\chi^{\top(1)}_{lm,n}
\Big]
+\partial_i\partial_j\nabla^{-2}\Big[
-\frac{1}{2}\chi^{\top(1)'}_{lm}\chi^{||(1)',\,lm}
\nn\\
&
+\frac{1}{4}\chi^{\top(1)}_{lm,n}\chi^{||(1),\,lmn}
\Big]
+\partial_i\nabla^{-2}\Big[
-\frac{6}{\tau^2}c_L^2\,\chi^{\top(1)}_{lj}\delta^{(1),\,l}
+6\phi^{(1)'',\,l}\chi^{\top(1)}_{lj}
+\frac{12}{\tau} \phi^{(1)',\,l}\chi^{\top(1)}_{lj}
\nn\\
&
+\phi^{(1)',\,l}\chi^{\top(1)'}_{lj}
-2\chi^{\top(1)}_{lj}\nabla^2\phi^{(1),\,l}
-\phi^{(1),\,l}\nabla^2\chi^{\top(1)}_{lj}
+\phi^{(1),\,lm}_{,j}\chi^{\top(1)}_{lm}
-\chi^{\top(1)'}_{lj,\,m}\chi^{||(1)',\,lm}
\nn\\
&
+\chi^{\top(1)'}_{lm,j}\chi^{||(1)',\,lm}
-\frac{1}{3}\chi^{\top(1)'}_{lj}\nabla^2\chi^{||(1)',\,l}
+\frac{1}{6}\chi^{\top(1)}_{lm}\nabla^2\chi^{||(1),\,lm}_{,j}
-\frac{1}{3}\chi^{\top(1)}_{lj}\nabla^2\nabla^2\chi^{||(1),\,l}
\nn\\
&
-\frac{2}{3}\nabla^2\chi^{\top(1)}_{lj}\nabla^2\chi^{||(1),\,l}
-\chi^{||(1),\,lm}\nabla^2\chi^{\top(1)}_{lj,\,m}
+\frac{1}{2}\chi^{||(1),\,lm}\nabla^2\chi^{\top(1)}_{lm,j}
\Big]
\nn\\
&
+\partial_j\nabla^{-2}\Big[
-\frac{6}{\tau^2}c_L^2\chi^{\top(1)}_{li}\delta^{(1),\,l}
+6\phi^{(1)'',\,l}\chi^{\top(1)}_{li}
+\frac{12}{\tau}\phi^{(1)',\,l}\chi^{\top(1)}_{li}
+\phi^{(1)',\,l}\chi^{\top(1)'}_{li}
\nn\\
&
-2\chi^{\top(1)}_{li}\nabla^2\phi^{(1),\,l}
-\phi^{(1),\,l}\nabla^2\chi^{\top(1)}_{li}
+\phi^{(1),\,lm}_{,i}\chi^{\top(1)}_{lm}
-\chi^{\top(1)'}_{li,\,m}\chi^{||(1)',\,lm}
+\chi^{\top(1)'}_{lm,i}\chi^{||(1)',\,lm}
\nn\\
&
-\frac{1}{3}\chi^{\top(1)'}_{li}\nabla^2\chi^{||(1)',\,l}
+\frac{1}{6}\chi^{\top(1)}_{lm}\nabla^2\chi^{||(1),\,lm}_{,i}
-\frac{1}{3}\chi^{\top(1)}_{li}\nabla^2\nabla^2\chi^{||(1),\,l}
-\frac{2}{3}\nabla^2\chi^{\top(1)}_{li}\nabla^2\chi^{||(1),\,l}
\nn\\
&
-\chi^{||(1),\,lm}\nabla^2\chi^{\top(1)}_{li,\,m}
+\frac{1}{2}\chi^{||(1),\,lm}\nabla^2\chi^{\top(1)}_{lm,i}
\Big]
+\partial_i\partial_j\nabla^{-2}\nabla^{-2}\Big[
\frac{3}{\tau^2}c_L^2\chi^{\top(1)}_{lm}\delta^{(1),\,lm}
\nn\\
&
-3\phi^{(1)'',\,lm}\chi^{\top(1)}_{lm}
-\frac{6}{\tau} \phi^{(1)',\,lm}\chi^{\top(1)}_{lm}
-\frac{1}{2}\phi^{(1)',\,lm}\chi^{\top(1)'}_{lm}
-\frac{3}{2}\phi^{(1),\,lmn}\chi^{\top(1)}_{lm,\,n}
\nn\\
&
-\frac{1}{2}\chi^{||(1)',\,lm}\nabla^2\chi^{\top(1)'}_{lm}
+\frac{1}{6}\chi^{\top(1)'}_{lm}\nabla^2\chi^{||(1)',\,lm}
-\frac{1}{4}\chi^{\top(1)}_{lm,\,n}\nabla^2\chi^{||(1),\,lmn}
\nn\\
&
+\frac{1}{2}\nabla^2\chi^{\top(1)}_{lm}\nabla^2\chi^{||(1),\,lm}
+\frac{3}{4}\chi^{||(1),\,lmn}\nabla^2\chi^{\top(1)}_{lm,n}
\Big] ,
\el
}
which is the effective source of 2nd-order tensor.
(See also  (\ref{2ndTensorSource}) for a general RW spacetime.)
So far the 2nd-order perturbed Einstein equation has been decomposed
into separate equations for
the 2nd-order metric perturbations.

To solve these equations of  2nd-order metric perturbations,
we need also the 2nd-order energy-momentum conservation.
The general energy conservation
is given by (\ref{EnConsv2}).
Substituting $\Gamma^{\alpha}_{\mu\nu}$
in (A1)--(A4) of Ref.\cite{WangZhang2ndRD2018}
and $U^{\mu}$ in (\ref{U0element}) and (\ref{Uielement})
into (\ref{EnConsv2}) and keeping 2nd order,
one has the 2nd-order energy conservation for the scalar-tensor case
\be\label{enCons2ndRD2}
\delta^{(2)'}_{s(t)}
+\frac{3}{\tau}\l(c_N^2-\frac{1}{3}\r)\delta^{(2)}_{s(t)}
=
-\frac{4}{3}\nabla^2v^{||(2)}_{s(t)}
+4\phi^{(2)'}_{s(t)}
+A_{s(t)} ,
\ee
where
\be\label{AS}
A_{s(t)}\equiv
\frac{4}{3}\chi^{\top(1)'}_{lm}\chi^{||(1),\,lm}
+\frac{4}{3}\chi^{\top(1)}_{lm}\chi^{||(1)',\,lm}
.
\ee
Equation (\ref{enCons2ndRD2})   contains the scalar mode
$\phi^{(2)}_S$ and the velocity  $ v^{||(2)}_S$.
Note that $c_N$ defined by (\ref{cn2})
appears
in the 2nd-order energy conservation equation.

The general expression of momentum conservation
has been given in (\ref{MoConsv2}).
Substituting $\Gamma^{\alpha}_{\mu\nu}$
in (A1)--(A4)  in Ref.\cite{WangZhang2ndRD2018}
and $U^{\mu}$ in (\ref{U0element}) and (\ref{Uielement})
into (\ref{MoConsv2}) to 2nd order
gives the 2nd-order momentum conservation for the scalar-tensor case
\be\label{MoCons2ndRD2}
c_N^2 \delta^{(2)}_{s(t),\,i}
+\frac{4}{3} v^{||(2)'}_{s(t),i}
+\frac{4}{3} v^{\perp(2)'}_{s(t)i}
=
F_{s(t)i}
,
\ee
where
\be\label{FS}
F_{s(t)i}\equiv
2c_L^2\delta^{(1),\,l}\chi^{\top(1)}_{li}
-\frac{8}{3}v^{||(1),\,l}\chi^{\top(1)'}_{li}
.
\ee
Observe that  (\ref{MoCons2ndRD2}) involves the 2nd-order velocity
$v^{\perp(2)i}$,
yet does not involve the  2nd-order metric perturbations,
in contrast to the energy conservation (\ref{enCons2ndRD2}).
To proceed further,
(\ref{MoCons2ndRD2}) is decomposed into a longitudinal part
by [$\nabla^{-2}\partial^i$(\ref{MoCons2ndRD2})]   as
\be\label{MoC2ndNonCurl}
c_N^2 \delta_{s(t)}^{(2)}
+\frac{4}{3} v_{s(t)}^{||(2)'}
=
F^{||}_{s(t)}
\ee
with
\bl\label{FSnoncu}
F^{||}_{s(t)}
\equiv
\nabla^{-2}\partial^iF_{s(t)i}
=
\nabla^{-2}\Big[
2c_L^2\delta^{(1),\,lm}\chi^{\top(1)}_{lm}
-\frac{8}{3}v^{||(1),\,lm}\chi^{\top(1)'}_{lm}
\Big],
\el
and a transverse part
by [(\ref{MoCons2ndRD2})-$\partial_i$(\ref{MoC2ndNonCurl})]    as
\be\label{MoC2ndCurl}
\frac{4}{3} v^{\perp(2)'}_{s(t)i}
=
F^\perp_{s(t)i}
\ee
with
\bl\label{FScurl}
F^\perp_{s(t)i}
\equiv
&
F_{s(t)i}-\partial_i F^{||}_{s(t)}
\nn\\
=&
2c_L^2\delta^{(1),\,l}\chi^{\top(1)}_{li}
-\frac{8}{3}v^{||(1),\,l}\chi^{\top(1)'}_{li}
+\partial_i\nabla^{-2}\Big[
-2c_L^2\delta^{(1),\,lm}\chi^{\top(1)}_{lm}
+\frac{8}{3}v^{||(1),\,lm}\chi^{\top(1)'}_{lm}
\Big]
\el
being a transverse vector function.

We find that the 2nd-order trace evolution equation (\ref{Evo2ndSsTr2RD})
can be given as the combination
\be\label{relation1}
 (\ref{Evo2ndSsTr2RD})
 =
 -\frac{1}{3}(\ref{Ein2th003RD})
-\frac{\tau}{3}\frac{d}{d\tau}(\ref{Ein2th003RD})
+\frac{\tau}{3}\nabla^2(\ref{MoConstr2ndv3RD2})
-\frac{1}{\tau}(\ref{enCons2ndRD2})
,
\ee
and the 2nd-order  traceless  scalar evolution equation (\ref{Evo2ndSsChi1RD})
as  follows:
\be\label{relation3}
(\ref{Evo2ndSsChi1RD})
=
\nabla^{-2}\Big[
(\ref{Ein2th003RD})
+\tau\frac{d}{d\tau}(\ref{Ein2th003RD})
-\tau\nabla^2(\ref{MoConstr2ndv3RD2})
+3\frac{d}{d\tau}(\ref{MoConstr2ndv3RD2})
+\frac{6}{\tau}(\ref{MoConstr2ndv3RD2})
+\frac{3}{\tau}(\ref{enCons2ndRD2})
-\frac{9}{\tau^2}(\ref{MoC2ndNonCurl})
\Big]
 .
\ee
These mean that, for the scalars,
we can solve the equations of constraints and conservations,
and the solutions satisfy the evolution equations automatically.
(see in Sec.\ref{Sec:Solve2ndEq}.)

\subsection{Equations with tensor-tensor couplings}

Now we consider the  tensor-tensor couplings.
Moving the tensor-tensor coupling terms to the rhs
of the  $(00)$ component of the Einstein equation (\ref{2nd00Einstein}),
and using  the Friedmann equation,
one has the  2nd-order energy constraint equation as the following
\be  \label{TEin2th003RD}
-\frac{6}{\tau} \phi^{(2)'}_{T}
+2\nabla^2\phi^{(2)}_{T}
+\frac{1}{3}\nabla^2\nabla^2\chi^{||(2)}_{T}
=
\frac{3}{\tau^2}\delta^{(2)}_{T}
+E_{T}
,
\ee
where  a subscript $T$ in $\phi^{(2)}_{T}$ etc.
indicates the case of tensor-tensor coupling,
and $E_{T}$ in the above is
\bl\label{TES1RD}
E_{T}
\equiv
&
\frac{1}{4}\chi^{\top(1)'lm}\chi^{\top(1)'}_{lm}
+\frac{2}{\tau} \chi^{\top(1)lm}\chi^{\top(1)'}_{lm}
-\chi^{\top(1)lm}\nabla^2\chi^{\top(1)}_{lm}
    \nn\\
    &
-\frac{3}{4}\chi^{\top(1)lm,\,n}\chi^{\top(1)}_{lm,\,n}
+\frac{1}{2}\chi^{\top(1)lm,\,n}\chi^{\top(1)}_{ln,\,m}
.
\el

The  $(0i)$ component of the 2nd-order perturbed Einstein equation
(\ref{Ein2nd0i1}) gives the 2nd-order momentum constraint equation
as the following
\be\label{TMoConstr2ndv3RD}
2 \phi^{(2)'} _{T,\,i}
+\frac{1}{2}D_{ij}\chi^{||(2)',\,j}_{T}
+\frac{1}{2}\chi^{\perp(2)',\,j}_{T\,ij}
=
-\frac{4}{\tau^2}v^{(2)}_{T\,i}
+M_{T\,i},
\ee
with
\be\label{TMSi1RD}
M_{T\,i}
\equiv
\chi^{\top(1) lm} \chi^{\top(1)'}_{il,\,m}
-\chi^{\top(1)lm} \chi^{\top(1)' }_{lm,i}
-\frac{1}{2}  \chi^{\top(1)  lm}_{ ,\,i} \chi^{\top(1)'}_{lm}
\ .
\ee
As before,
[$\nabla^{-2}\partial^i$(\ref{TMoConstr2ndv3RD})] gives
the longitudinal  momentum constraint
\be \label{TMoConstr2ndv3RD2}
2\phi^{(2)'}_{T}
+\frac{1}{3}\nabla^2\chi^{||(2)'}_{T}
=
-\frac{4}{\tau^2}v^{||(2)}_{T}
+\nabla^{-2}M_{T\,l}^{,\,l},
\ee
where
\bl\label{TMSkkScalarRD}
\nabla^{-2}M_{T\,l}^{\,,\,l}
=&
\nabla^{-2}\Big[
\chi^{\top(1) lm,n} \chi^{\top(1)'}_{ln,\,m}
-\frac{1}{2} \chi^{\top(1)lm,n} \chi^{\top(1)' }_{lm,n}
-\frac{1}{2} \chi^{\top(1)lm} \nabla^2\chi^{\top(1)' }_{lm}
\Big]
-\frac{1}{2} \chi^{\top(1)lm} \chi^{\top(1)' }_{lm}
\el
is the effective source term.
A combination
[(\ref{TMoConstr2ndv3RD})-$\partial_i$(\ref{TMoConstr2ndv3RD2})]
gives the transverse   momentum constraint
\be \label{TMoCons2ndCurlRD1}
\frac{1}{2}\chi^{\perp(2)',\,j}_{T\,ij}
=
-\frac{4}{\tau^2}v^{\perp(2)}_{T\,i}
+\l(
M_{T\,i}
-\partial_i\nabla^{-2}M_{T\,l}^{,\,l}
\r)
\ ,
\ee
with
\bl\label{TMSiCurlRD}
&
\l( M_{T\,i}-\partial_i\nabla^{-2}M_{T\,l}^{,\,l} \r)
\nn\\
=&
\chi^{\top(1) lm} \chi^{\top(1)'}_{il,\,m}
-\frac{1}{2} \chi^{\top(1)lm} \chi^{\top(1)' }_{lm,i}
\nn\\
&
+\partial_i\nabla^{-2}\Big[
-\chi^{\top(1) lm,n} \chi^{\top(1)'}_{ln,\,m}
+\frac{1}{2} \chi^{\top(1)lm,n} \chi^{\top(1)' }_{lm,n}
+\frac{1}{2} \chi^{\top(1)lm} \nabla^2\chi^{\top(1)' }_{lm}
\Big] .
\el

The $(ij)$ component of 2nd-order perturbed Einstein equation
(\ref{Ein2ndijss1}) gives  the 2nd-order evolution equation
\bl\label{TEvo2ndSs1RD}
&
2\phi^{(2)''}_{T} \delta_{ij}
+\frac{4}{\tau}\phi^{(2)'}_{T} \delta_{ij}
+\phi^{(2)}_{T,\,ij}
-\nabla^2\phi^{(2)}_{T} \delta_{ij}
\nn\\
&
+\frac{1}{2}D_{ij}\chi^{||(2)''}_{T}
+\frac{1}{\tau}D_{ij}\chi^{||(2)'}_{T}
+\frac{1}{6}\nabla^2D_{ij}\chi^{||(2)}_{T}
-\frac{1}{9}\delta_{ij}\nabla^2\nabla^2\chi^{||(2) }_{T}
\nn\\
&
+\frac{1}{2}\chi^{\perp(2)''}_{T\,ij}
+\frac{1}{\tau}\chi^{\perp(2)'}_{T\,ij}
\nn\\&
+\frac{1}{2}\chi^{\top(2)''}_{T\,ij}
+\frac{1}{\tau}\chi^{\top(2)'}_{T\,ij}
-\frac{1}{2}\nabla^2\chi^{\top(2)}_{T\,ij}
=
\frac{3 c_N^2}{\tau^2}\delta^{(2)}_{T}\delta_{ij}
+S_{T\,ij} ,
\el
with
\bl\label{TSs2ndij2RD}
S_{Tij}\equiv
&
-\frac{2}{\tau}\chi^{\top(1)lm}\chi^{\top(1)'}_{lm}\delta_{ij}
-\chi^{\top(1)lm}\chi^{\top(1)}_{lm,\,ij}
+\chi^{\top(1)lm}\nabla^2\chi^{\top(1)}_{lm}\delta_{ij}
\nn\\
&
-\frac{1}{2}\chi^{\top(1)lm}_{,\,i}\chi^{\top(1)}_{lm,\,j}
-\chi^{\top(1)}_{li,\,m}\chi^{\top(1)l,\,m}_{j}
+\frac34\chi^{\top(1)lm,\,n}\chi^{\top(1)}_{lm,\,n}\delta_{ij}
\nn\\
&
+\chi^{\top(1)}_{l i,\,m}\chi^{\top(1)m,\,l}_{j}
-\frac{1}{2}\chi^{\top(1)lm,\,n}\chi^{\top(1)}_{ln,\,m}\delta_{ij}
+\chi^{\top(1)'l}_{i}\chi^{\top(1)'}_{lj}
-\frac34\chi^{\top(1)'\,lm}\chi^{\top(1)'}_{lm}\delta_{ij}
\nn\\
&
-\chi^{\top(1)lm}\chi^{\top(1)''}_{lm}\delta_{ij}
+\chi^{\top(1)lm}\chi^{\top(1)}_{lj,\,im}
+\chi^{\top(1)lm}\chi^{\top(1)}_{li,\,jm}
-\chi^{\top(1)lm}\chi^{\top(1)}_{ij,\,lm}
\
\el
as part of the effective source term.

We also need to decompose the evolution equation (\ref{TEvo2ndSs1RD})
into the trace  part and the  traceless part.
The trace part of 2nd-order evolution equation (\ref{TEvo2ndSs1RD}) is
\be\label{TEvo2ndSsTr2RD}
2\phi^{(2)''}_{T}
+\frac{4}{\tau}\phi^{(2)'}_{T}
-\frac{2}{3}\nabla^2\phi^{(2)}_{T}
-\frac{1}{9}\nabla^2\nabla^2\chi^{||(2) }_{T}
=
\frac{3}{\tau^2}c_N^2\delta^{(2)}_{T}
+\frac{1}{3}S_{Tl}^{\,l}
\,,
\ee
with
\bl\label{TSijTraceRD}
S_{Tl}^{\,l}
=&
-\frac{6}{\tau}\chi^{\top(1)lm}\chi^{\top(1)'}_{lm}
+2\chi^{\top(1)lm}\nabla^2\chi^{\top(1)}_{lm}
+\frac34\chi^{\top(1)lm,\,n}\chi^{\top(1)}_{lm,\,n}
\nn\\
&
-\frac{1}{2}\chi^{\top(1)lm,n}\chi^{\top(1)}_{ln,m}
-\frac54\chi^{\top(1)'\,lm}\chi^{\top(1)'}_{lm}
-3\chi^{\top(1)lm}\chi^{\top(1)''}_{lm}
\ .
\el
The traceless part of (\ref{TEvo2ndSs1RD}) is
\bl\label{TEvo2ndSsNoTr1RD}
&
D_{ij}\phi^{(2)}_{T}
+\frac{1}{2}D_{ij}\chi^{||(2)''}_{T}
+\frac{1}{\tau} D_{ij}\chi^{||(2)'}_{T}
+\frac{1}{6}\nabla^2D_{ij}\chi^{||(2)}_{T}
\nn\\
&
+\frac{1}{2}\chi^{\perp(2)''}_{T\,ij}
+\frac{1}{\tau} \chi^{\perp(2)'}_{T\,ij}
\nn\\
&
+\frac{1}{2}\chi^{\top(2)''}_{T\,ij}
+\frac{1}{\tau} \chi^{\top(2)'}_{T\,ij}
-\frac{1}{2}\nabla^2\chi^{\top(2)}_{T\,ij}
=
\bar S_{T\,ij}
,
\el
with
\bl\label{TSijTracelessRD}
\bar S_{T\,ij}
\equiv&
S_{T\,ij}-\frac{1}{3}S_{T\,l}^l\delta_{ij}
\nn\\
=&
-\chi^{\top(1)lm}\chi^{\top(1)}_{lm,\,ij}
+\frac{1}{3}\chi^{\top(1)lm}\nabla^2\chi^{\top(1)}_{lm}\delta_{ij}
-\frac{1}{2}\chi^{\top(1)lm}_{,\,i}\chi^{\top(1)}_{lm,\,j}
-\chi^{\top(1)}_{li,\,m}\chi^{\top(1)l,\,m}_{j}
\nn\\
&
+\frac{1}{2}\chi^{\top(1)lm,n}\chi^{\top(1)}_{lm,n}\delta_{ij}
+\chi^{\top(1)}_{l i,\,m}\chi^{\top(1)m,\,l}_{j}
-\frac{1}{3}\chi^{\top(1)lm,n}\chi^{\top(1)}_{ln,m}\delta_{ij}
+\chi^{\top(1)'l}_{i}\chi^{\top(1)'}_{lj}
\nn\\
&
-\frac{1}{3}\chi^{\top(1)'\,lm}\chi^{\top(1)'}_{lm}\delta_{ij}
+\chi^{\top(1)lm}\chi^{\top(1)}_{lj,\,im}
+\chi^{\top(1)lm}\chi^{\top(1)}_{li,\,jm}
-\chi^{\top(1)lm}\chi^{\top(1)}_{ij,\,lm}
\ .
\el
Equation (\ref{TEvo2ndSsNoTr1RD}) contains the scalar, the vector, and the tensor.
Applying
$3\nabla^{-2}\nabla^{-2}\partial_i\partial_j$ on (\ref{TEvo2ndSsNoTr1RD})
gives the evolution equation for the scalar $\chi^{||(2)}_{T}$
as follows:
\bl\label{TEvo2ndSsChi1RD}
&
\chi^{||(2)''}_{T}
+\frac{2}{\tau}\chi^{||(2)'}_{T}
+\frac{1}{3}\nabla^2\chi^{||(2)}_{T}
+2\phi^{(2)}_{T}
=
3\nabla^{-2}\nabla^{-2}\bar S_{T\,lm}^{\, ,\,lm}
,
\el
where
\bl\label{TSijPijbarRD}
\bar S_{T\,lm}^{\, ,\,lm}
=&
-\nabla^2\nabla^2\Big[
\frac{1}{8}\chi^{\top(1)lm}\chi^{\top(1)}_{lm}
\Big]
+\nabla^2\Big[
-\frac{1}{3}\chi^{\top(1)'\,lm}\chi^{\top(1)'}_{lm}
+\frac{1}{12}\chi^{\top(1)lm}\nabla^2\chi^{\top(1)}_{lm}
\nn\\
&
+\frac{1}{6}\chi^{\top(1)lm,n}\chi^{\top(1)}_{ln,m}
\Big]
+\chi^{\top(1)'lm,n}\chi^{\top(1)'}_{ln,m}
-\frac{1}{2}\chi^{\top(1)lm}\nabla^2\nabla^2\chi^{\top(1)}_{lm}
\nn\\
&
-\frac{1}{2}\chi^{\top(1)lm,n}\nabla^2\chi^{\top(1)}_{lm,n}
+\chi^{\top(1)lm,n}\nabla^2\chi^{\top(1)}_{ln,m}
\ .
\el
A combination
$\partial^i$\big[(\ref{TEvo2ndSsNoTr1RD})$ -\frac{1}{2}D_{ij}$(\ref{TEvo2ndSsChi1RD})\big]
gives
\be\label{TEvo2ndSsVec1}
\frac{1}{2}\chi^{\perp(2)'',\,l}_{T\,lj}
+\frac{1}{\tau} \chi^{\perp(2)',\,l}_{T\,lj}
=
\bar S_{T\,lj}^{\,,\,l}
-\partial_j\nabla^{-2}\bar S_{T\,lm}^{\, ,\,lm}
.
\ee
By  $\nabla^{-2}\big[\partial_i(\ref{TEvo2ndSsVec1})
+(i\leftrightarrow j)\big]$,
using (\ref{chiVec0}),
one gets the evolution equation for the vector mode
\be\label{TEvo2ndSsVec2RD}
\frac{1}{2}\chi^{\perp(2)''}_{T\,ij}
+\frac{1}{\tau} \chi^{\perp(2)'}_{T\,ij}
=
V_{T\,ij}
~,
\ee
where
{
\bl\label{TSourceCurl1RD}
V_{T\,ij}
\equiv
&
\nabla^{-2}\bar S_{T\,lj,\,i}^{,\,l}
+\nabla^{-2}\bar S_{T\,li,j}^{,\,l}
-2\nabla^{-2}\nabla^{-2}\bar S_{T\,lm,\,ij}^{\, ,\,lm}
\nn\\
=&
\partial_i\nabla^{-2}\Big[
\chi^{\top(1)'lm}\chi^{\top(1)'}_{jl,m}
-\frac{1}{2}\chi^{\top(1)lm}\nabla^2\chi^{\top(1)}_{lm,j}
+\chi^{\top(1)lm}\nabla^2\chi^{\top(1)}_{jl,\,m}
\Big]
\nn\\
&
+\partial_i\partial_j\nabla^{-2}\nabla^{-2}\Big[
-\chi^{\top(1)'lm,n}\chi^{\top(1)'}_{ln,m}
+\frac{1}{2}\chi^{\top(1)lm}\nabla^2\nabla^2\chi^{\top(1)}_{lm}
\nn\\
&
+\frac{1}{2}\chi^{\top(1)lm,n}\nabla^2\chi^{\top(1)}_{lm,n}
-\chi^{\top(1)lm,n}\nabla^2\chi^{\top(1)}_{ln,m}
\Big]
+(i\leftrightarrow j)
\el
}
is  the effective source.
Finally,
[(\ref{TEvo2ndSsNoTr1RD})$-\frac{1}{2}D_{ij}$(\ref{TEvo2ndSsChi1RD})
$-$(\ref{TEvo2ndSsVec2RD})] gives the
evolution equation for the 2nd-order tensor mode
\be \label{TEvo2ndSsTen1RD}
\frac{1}{2}\chi^{\top(2)''}_{T\,ij}
+\frac{1}{\tau} \chi^{\top(2)'}_{T\,ij}
-\frac{1}{2}\nabla^2\chi^{\top(2)}_{T\,ij}
=
J_{T\,ij}
,
\ee
where
{
\allowdisplaybreaks
\bl\label{T2ndTensorSourceRD}
J_{T\,ij}
\equiv
&
\bar S_{T\,ij}
-\frac{3}{2}D_{ij}\nabla^{-2}\nabla^{-2}\bar S_{T\,lm}^{\, ,\,lm}
-\nabla^{-2}\bar S_{T\,lj,\,i}^{,\,l}
-\nabla^{-2}\bar S_{T\,li,j}^{,\,l}
+2\nabla^{-2}\nabla^{-2}\bar S_{T\,lm,\,ij}^{\, ,\,lm}
\nn\\
=&
\Big[-\frac{5}{8}\chi^{\top(1)lm}\chi^{\top(1)}_{lm,ij}
+\frac{1}{4}\chi^{\top(1)lm}\nabla^2\chi^{\top(1)}_{lm}\delta_{ij}
-\frac{1}{8}\chi^{\top(1)lm}_{,i}\chi^{\top(1)}_{lm,j}
-\chi^{\top(1)}_{li,\,m}\chi^{\top(1)l,\,m}_{j}
\nn\\
&
+\frac{3}{8}\chi^{\top(1)lm,n}\chi^{\top(1)}_{lm,n}\delta_{ij}
+\chi^{\top(1)}_{l i,\,m}\chi^{\top(1)m,\,l}_{j}
-\frac{1}{4} \chi^{\top(1)lm,n}\chi^{\top(1)}_{ln,m}\delta_{ij}
+\chi^{\top(1)'l}_{i}\chi^{\top(1)'}_{lj}
\nn\\
&
-\frac{1}{2}\chi^{\top(1)'\,lm}\chi^{\top(1)'}_{lm}\delta_{ij}
+\chi^{\top(1)lm}\chi^{\top(1)}_{lj,\,im}
+\chi^{\top(1)lm}\chi^{\top(1)}_{li,\,jm}
-\chi^{\top(1)lm}\chi^{\top(1)}_{ij,\,lm}
\Big]
\nn\\
&
+\delta_{ij}\nabla^{-2}\Big[
\frac{1}{2}\chi^{\top(1)'lm,n}\chi^{\top(1)'}_{ln,m}
-\frac{1}{4}\chi^{\top(1)lm}\nabla^2\nabla^2\chi^{\top(1)}_{lm}
\nn\\
&
-\frac{1}{4}\chi^{\top(1)lm,n}\nabla^2\chi^{\top(1)}_{lm,n}
+\frac{1}{2}\chi^{\top(1)lm,n}\nabla^2\chi^{\top(1)}_{ln,m}
\Big]
\nn\\
&
+\partial_i\nabla^{-2}\Big[
-\chi^{\top(1)'lm}\chi^{\top(1)'}_{jl,m}
+\frac{1}{2}\chi^{\top(1)lm}\nabla^2\chi^{\top(1)}_{lm,j}
-\chi^{\top(1)lm}\nabla^2\chi^{\top(1)}_{jl,\,m}
\Big]
\nn\\
&
+\partial_j\nabla^{-2}\Big[
-\chi^{\top(1)'lm}\chi^{\top(1)'}_{il,m}
+\frac{1}{2}\chi^{\top(1)lm}\nabla^2\chi^{\top(1)}_{lm,i}
-\chi^{\top(1)lm}\nabla^2\chi^{\top(1)}_{il,\,m}
\Big]
\nn\\
&
+\partial_i\partial_j\nabla^{-2}\Big[
\frac{1}{2}\chi^{\top(1)'\,lm}\chi^{\top(1)'}_{lm}
-\frac{1}{8}\chi^{\top(1)lm}\nabla^2\chi^{\top(1)}_{lm}
-\frac{1}{4}\chi^{\top(1)lm,n}\chi^{\top(1)}_{ln,m}
\Big]
\nn\\
&
+\partial_i\partial_j\nabla^{-2}\nabla^{-2}\Big[
\frac{1}{2}\chi^{\top(1)'lm,n}\chi^{\top(1)'}_{ln,m}
-\frac{1}{4}\chi^{\top(1)lm}\nabla^2\nabla^2\chi^{\top(1)}_{lm}
\nn\\
&
-\frac{1}{4}\chi^{\top(1)lm,n}\nabla^2\chi^{\top(1)}_{lm,n}
+\frac{1}{2}\chi^{\top(1)lm,n}\nabla^2\chi^{\top(1)}_{ln,m}
\Big]
\el
is  the effective source.
}
So far the 2nd-order perturbed Einstein equation has been decomposed
into the separate equations for
the 2nd-order metric perturbations.

Next,
we  derive the 2nd-order energy-momentum conservation
with tensor-tensor couplings.
Expanding (\ref{EnConsv2}) to 2nd order,
one has the 2nd-order energy conservation for the tensor-tensor case as
\be\label{TenCons2ndRD2}
\delta^{(2)'}_{T}
+\frac{3}{\tau}\l(c_N^2-\frac{1}{3}\r)\delta^{(2)}_{T}
=
-\frac{4}{3}\nabla^2v^{||(2)}_{T}
+4\phi^{(2)'}_{T}
+A_{T} ,
\ee
where
\be\label{TAS}
A_{T}\equiv
\frac{4}{3}\chi^{\top(1)'}_{lm}\chi^{\top(1)lm}
.
\ee
Expanding  (\ref{MoConsv2}) to 2nd order,
one has the 2nd-order momentum conservation for the tensor-tensor case as
\be\label{TMoCons2ndRD2}
c_N^2 \delta^{(2)}_{T,\,i}
+\frac{4}{3} v^{||(2)'}_{T,i}
+\frac{4}{3} v^{\perp(2)'}_{T\,i}
=
0
,
\ee
which is homogeneous,  involving no tensor-tensor coupling terms.
Observe that  (\ref{TMoCons2ndRD2}) involves the 2nd-order velocity
$v^{\perp(2)}_{T\,i}$,
yet does not involve the  2nd-order metric perturbations,
in contrast to the energy conservation (\ref{enCons2ndRD2}).
To proceed further,
(\ref{TMoCons2ndRD2}) can be decomposed into a longitudinal part
by [$\nabla^{-2}\partial^i$(\ref{TMoCons2ndRD2})]   as
\be\label{TMoC2ndNonCurl}
c_N^2 \delta_{T}^{(2)}
+\frac{4}{3} v_{T}^{||(2)'}
=
0
\ee
and a transverse part
by [(\ref{TMoCons2ndRD2})-$\partial_i$(\ref{TMoC2ndNonCurl})]   as
\be\label{TMoC2ndCurl}
\frac{4}{3} v^{\perp(2)'}_{T\,i}
=
0.
\ee

Similar to the relations
(\ref{relation1}) and (\ref{relation3}),
the trace of the 2nd-order evolution equation (\ref{TEvo2ndSsTr2RD})
can be given by a combination as
\be\label{Trelation1}
 (\ref{TEvo2ndSsTr2RD})
 =
 -\frac{1}{3}(\ref{TEin2th003RD})
-\frac{\tau}{3}\frac{d}{d\tau}(\ref{TEin2th003RD})
+\frac{\tau}{3}\nabla^2(\ref{TMoConstr2ndv3RD2})
-\frac{1}{\tau}(\ref{TenCons2ndRD2})
,
\ee
and the scalar part of
the 2nd-order traceless evolution equation (\ref{TEvo2ndSsChi1RD})
is given by
\bl\label{Trelation3}
(\ref{TEvo2ndSsChi1RD})
=&
\nabla^{-2}\Big[
(\ref{TEin2th003RD})
+\tau\frac{d}{d\tau}(\ref{TEin2th003RD})
-\tau\nabla^2(\ref{TMoConstr2ndv3RD2})
+3\frac{d}{d\tau}(\ref{TMoConstr2ndv3RD2})
+\frac{6}{\tau}(\ref{TMoConstr2ndv3RD2})
\nn\\
&
+\frac{3}{\tau}(\ref{TenCons2ndRD2})
-\frac{9}{\tau^2}(\ref{TMoC2ndNonCurl})
\Big]
 .
\el
So we can use the equations of constraints
and conservations
to solve the scalars,
and the solutions satisfy the evolution equations automatically.

\section{Solution of the 2nd-order perturbations}
\label{Sec:Solve2ndEq}

\subsection{Solution for scalar-tensor couplings}

Given the 2nd-order perturbed equations,
we shall solve for the 2nd-order perturbations.
This subsection is for  the scalar-tensor case.
First,
the solution of  tensor  equation (\ref{Evo2ndSsTen1RD})
is
\be\label{solgw}
\chi^{\top(2)}_{s(t)ij}({\bf x},\tau)
=\int  \frac{d^3k}{(2\pi)^{\frac{3}{2}}}  e^{i{\bf k\cdot x}}
\Big(
\bar I_{s(t) ij}({\bf k}, \tau) +
\sum_{s={+,\times}} {\mathop \epsilon
\limits^s}_{ij}({\bf k})
\Big[-a_{1}^s \sqrt{\frac{2}{\pi}}\,\frac{i\, e^{i  k\tau}}{ k\tau}
+a_{2}^s \sqrt{\frac{2}{\pi}}\,\frac{i\, e^{-i  k\tau}}{ k\tau} \Big]
\Big),
\ee
where $a_{1}^s({\bf k})$ and  $a_{2}^s({\bf k})$
are  polarization-dependent and $\bf k$-dependent coefficients,
to be determined by    initial conditions;
their  associated term is   the  homogeneous solution of  (\ref{Evo2ndSsTen1RD})
and has   the same form as   (\ref{GWmode}).
In certain applications, they can be absorbed into   (\ref{GWmode}).
The integrand of the inhomogeneous solution in (\ref{solgw}) is given by
\be\label{GW2ndStepI}
\bar I_{s(t)ij}({\bf k},\tau)\equiv
\frac{ie^{-ik\tau}}{k\tau}\int^{\tau}\tau'e^{ik\tau'}\bar J_{s(t) ij}({\bf k},\tau')d\tau'
-\frac{ie^{ik\tau}}{k\tau}\int^{\tau}\tau'e^{-ik\tau'}\bar J_{s(t) ij}({\bf k},\tau')d\tau'
,
\ee
with $\bar J_{s(t) ij}$ being the Fourier transform of the source
$J_{s(t)ij}$ in (\ref{2ndTensorSourceRD})
that contains many terms of products of 1st-order solutions.

Next, the  vector solution  of evolution equation (\ref{Evo2ndSsVec2RD}) is
\be\label{chiVecSol}
\chi^{\perp(2)}_{s(t)ij}({\bf x},\tau)
=q_{1ij}({\bf x})
+\frac{q_{2ij}({\bf x})}{\tau}
+\int^\tau \frac{d\tau'}{\tau^{'2}}
    \int^{\tau'}2\tau^{''2}\,V_{s(t)ij}({\bf x},\tau'')d\tau''
\,,
\ee
where $q_{1ij}({\bf x})$ and $q_{2ij}({\bf x})$ are
two time-independent functions
determined by initial values and
 $V_{s(t)ij}$ is given by (\ref{SourceCurl1RD}).
Note that $q_{1ij}({\bf x})$ is a gauge mode
as shall be seen in Sec. 5.
Plugging the solution (\ref{chiVecSol}) into (\ref{MoCons2ndCurlRD1})
yields  the transverse part of the 2nd-order velocity
\bl\label{Vperp2ndSol}
v^{\perp(2)}_{s(t)i}
=&
\frac{q_{2ij}^{\,,\,j}({\bf x})}{8}
+\frac{\tau^2}{4}\l(
M_{s(t)i}
-\partial_i\nabla^{-2}M_{s(t)k}^{,\,k}
\r)
- \frac{1}{4}
    \int^{\tau}\tau^{'2}\,V_{s(t)ij}^{,j}({\bf x},\tau')d\tau'
\ ,
\el
where $(M_{s(t)i}-\partial_i\nabla^{-2}M_{s(t)k}^{,k})$
is in (\ref{MSiCurlRD}).
This solution can also be derived from integration of
the transverse  momentum conservation (\ref{MoC2ndCurl}),
as we have checked.
Thus, although  the 1st-order curl vector $v^{\perp(1)}_{i}$ is vanishing
by assumption,
nevertheless,
the 2nd-order curl vector $v^{\perp(2)}_{s(t)i}$ is generated by
the coupling according to (\ref{Vperp2ndSol}).

Next,
we solve the scalars.
From the longitudinal momentum conservation
(\ref{MoC2ndNonCurl}), one has
\be \label{deltaV2}
 \delta_{s(t)}^{(2)}
=
-\frac{4}{3c_N^2} v_{s(t)}^{||(2)'}
+\frac{1}{c_N^2}F^{||}_{s(t)}
\ .
\ee
Plugging the above $\delta^{(2)}_{s(t)}$ into
the energy conservation (\ref{enCons2ndRD2})
gives $\phi^{(2)'}_{s(t)}$ in terms of $v_{s(t)}^{||(2)}$,
$v_{s(t)}^{||(2)'}$, $v_{s(t)}^{||(2)''}$ as
\be\label{Vphi2}
\phi^{(2)'}_{s(t)}
=
-\frac{1}{3 c_N^2} v_{s(t)}^{||(2)''}
-\frac{1}{\tau}\frac{c_N^2-\frac{1}{3}}{c_N^2} v_{s(t)}^{||(2)'}
+\frac{1}{3}\nabla^2v^{||(2)}_{s(t)}
+\frac{1}{4 c_N^2}F^{||'}_{s(t)}
+\frac{3}{4\tau}\frac{c_N^2-\frac{1}{3}}{c_N^2}F^{||}_{s(t)}
-\frac{1}{4}A_{s(t)}
.
\ee

To use the energy constraint (\ref{Ein2th003RD}),
taking [$\frac{d}{d\tau}$(\ref{Ein2th003RD})] gives
\be
-\frac{6}{\tau} \phi^{(2)''}_{s(t)}
+\frac{6}{\tau^2} \phi^{(2)'}_{s(t)}
+\nabla^2 \Big[
2\phi^{(2)'}_{s(t)}
+\frac{1}{3}\nabla^2\chi^{||(2)'}_{s(t)}
\Big]
=
\frac{3}{\tau^2}\delta^{(2)'}_{s(t)}
-\frac{6}{\tau^3}\delta^{(2)}_{s(t)}
+E^{\,'}_{s(t)}
.
\ee
Plugging the momentum constraint (\ref{MoConstr2ndv3RD2}) into the above
to eliminate the $\nabla^2\nabla^2\chi^{||(2)'}_{s(t)}$ term leads to
\be
-\frac{6}{\tau} \phi^{(2)''}_{s(t)}
+\frac{6}{\tau^2} \phi^{(2)'}_{s(t)}
-\frac{4}{\tau^2}\nabla^2v^{||(2)}_{s(t)}
+M_{s(t)k}^{,\,k}
=
\frac{3}{\tau^2}\delta^{(2)'}_{s(t)}
-\frac{6}{\tau^3}\delta^{(2)}_{s(t)}
+E^{\,'}_{s(t)}
.
\ee
Then,
plugging $\delta^{(2)}_{s(t)}$  of  (\ref{deltaV2}) and
$\phi^{(2)'}_{s(t)} $ of  (\ref{Vphi2}) into the above
yields a 3rd-order differential equation of $v^{||(2)}_{s(t)}$
as follows:
\be\label{V2ndeq1}
v_{s(t)}^{||(2)'''}
+\frac{3c_N^2}{\tau} v_{s(t)}^{||(2)''}
-\frac{6c_N^2+2}{\tau^2}v_{s(t)}^{||(2)'}
-\frac{c_N^2}{\tau}\nabla^2v^{||(2)}_{s(t)}
-c_N^2\nabla^2v^{||(2)'}_{s(t)}
=Z_{s(t)}
,
\ee
with
{\allowdisplaybreaks
\bl\label{ZSall}
Z_{s(t)}
\equiv &
\frac{3}{4}F^{||''}_{s(t)}
+\frac{9c_N^2}{4\tau}F^{||'}_{s(t)}
-\frac{9c_N^2+3}{2\tau^2}F^{||}_{s(t)}
-\frac{3c_N^2}{4}A_{s(t)}^{\,'}
+\frac{3c_N^2}{4\tau}A_{s(t)} \nn \\
& -\frac{\tau}{2}c_N^2M_{s(t)l}^{,\,l}
+\frac{\tau}{2}c_N^2E^{\,'}_{s(t)} \nn\\
 =&
\bigg[
\frac{4}{3\tau}\chi^{\top(1)}_{lm}v^{||(1),\,lm}
+\frac{2\tau}{3}\phi^{(1)',\,lm}\chi^{\top(1)}_{lm}
+\frac{\tau}{6}\phi^{(1),\,lm}\chi^{\top(1)'}_{lm}
+\frac{\tau}{12}\chi^{\top(1)''}_{lm}\chi^{||(1)',\,lm}
\nn\\
&
+\frac{\tau}{12}\chi^{\top(1)'}_{lm}\chi^{||(1)'',\,lm}
+\frac{\tau}{36}\chi^{\top(1)'}_{lm}\nabla^2\chi^{||(1),\,lm}
+\frac{\tau}{9}\chi^{\top(1)}_{lm}\nabla^2\chi^{||(1)',\,lm}
\nn\\
&
-\frac{\tau}{12}\chi^{||(1)',\,lm}\nabla^2\chi^{\top(1)}_{lm}
\bigg]
+\nabla^{-2}\bigg[
\frac{3}{2}c_L^2\delta^{(1)'',\,lm}\chi^{\top(1)}_{lm}
+\frac{3}{2}c_L^2\delta^{(1),\,lm}\chi^{\top(1)''}_{lm}
\nn\\
&
+\frac{3}{2\tau}c_L^2\delta^{(1)',\,lm}\chi^{\top(1)}_{lm}
+\frac{3}{2\tau}c_L^2\delta^{(1),\,lm}\chi^{\top(1)'}_{lm}
-\frac{6}{\tau^2}c_L^2\delta^{(1),\,lm}\chi^{\top(1)}_{lm}
+3c_L^2\delta^{(1)',\,lm}\chi^{\top(1)'}_{lm}
\nn\\
&
-2v^{||(1)'',\,lm}\chi^{\top(1)'}_{lm}
-2v^{||(1),\,lm}\chi^{\top(1)'''}_{lm}
-4v^{||(1)',\,lm}\chi^{\top(1)''}_{lm}
-\frac{2}{\tau}v^{||(1),\,lm}\chi^{\top(1)''}_{lm}
\nn\\
&
-\frac{2}{\tau}v^{||(1)',lm}\chi^{\top(1)'}_{lm}
+\frac{8}{\tau^2}v^{||(1),\,lm}\chi^{\top(1)'}_{lm}
\bigg]
\, ,
\el
}
being the effective source,
formed from the products of 1st-order scalar and tensor modes.
Written in  the $\bf k$-space,
(\ref{V2ndeq1})  is
\be\label{V2eqFourier1}
v_{s(t)\mathbf k}^{||(2)'''}
+\frac{3c_N^2}{\tau} v_{s(t)\mathbf k}^{||(2)''}
+\l(
c_N^2k^2
-\frac{6c_N^2+2}{\tau^2}
\r)v_{s(t)\mathbf k}^{||(2)'}
+\frac{c_N^2}{\tau}k^2v_{s(t)\mathbf k}^{||(2)}
=Z_{s(t)\mathbf k}(\tau)
,
\ee
where
{\allowdisplaybreaks
\be\label{ZSallFourier}
Z_{s(t)\mathbf k}(\tau)
 =
\frac{1}{(2\pi)^{3}}
    \int d^3k_2
    \bigg\{
\bigg[
 k^l k^m\sum_{s={+,\times}}
{\mathop \epsilon \limits^s}_{lm}({{\bf k}_2})
\bigg]
\widetilde{Z}(\tau; {\mathbf k}, {\mathbf k_2})
\bigg\}
\,,
\ee
is the  Fourier transform  of (\ref{ZSall}) with
\bl\label{ZTildeDecom2}
\widetilde{Z}(\tau; {\mathbf k}, {\mathbf k_2})
=&
-\frac{4}{3\tau}h_{ {\bf k}_2}v^{||(1)}_{({\bf k} -{\bf k}_2)}
-\frac{2\tau}{3}h_{ {\bf k}_2}\phi^{(1)'}_{({\bf k} -{\bf k}_2)}
-\frac{\tau}{6}h'_{ {\bf k}_2}\phi^{(1)}_{({\bf k} -{\bf k}_2)}
-\frac{\tau}{12}h''_{ {\bf k}_2}\chi^{||(1)'}_{({\bf k} -{\bf k}_2)}
\nn\\
&
-\frac{\tau}{12}h'_{ {\bf k}_2}\chi^{||(1)''}_{({\bf k} -{\bf k}_2)}
+\frac{\tau}{36}|{\bf k} -{\bf k}_2|^2h'_{ {\bf k}_2}\chi^{||(1)}_{({\bf k} -{\bf k}_2)}
+\frac{\tau}{9}|{\bf k} -{\bf k}_2|^2h_{ {\bf k}_2}\chi^{||(1)'}_{({\bf k} -{\bf k}_2)}
\nn\\
&
-\frac{\tau}{12}( k_2)^2 h_{ {\bf k}_2}\chi^{||(1)'}_{({\bf k} -{\bf k}_2)}
+\frac{3}{2}(k)^{-2}c_L^2h_{ {\bf k}_2}\delta^{(1)''}_{({\bf k} -{\bf k}_2)}
+\frac{3}{2}(k)^{-2}c_L^2h''_{ {\bf k}_2}\delta^{(1)}_{({\bf k} -{\bf k}_2)}
\nn\\
&
+\frac{3}{2\tau}(k)^{-2}c_L^2h_{ {\bf k}_2}\delta^{(1)'}_{({\bf k} -{\bf k}_2)}
+\frac{3}{2\tau}(k)^{-2}c_L^2h'_{ {\bf k}_2}\delta^{(1)}_{({\bf k} -{\bf k}_2)}
-\frac{6}{\tau^2}(k)^{-2}c_L^2h_{ {\bf k}_2}\delta^{(1)}_{({\bf k} -{\bf k}_2)}
\nn\\
&
+3(k)^{-2}c_L^2 h'_{ {\bf k}_2}\delta^{(1)'}_{({\bf k} -{\bf k}_2)}
-2 (k)^{-2}h'_{ {\bf k}_2}v^{||(1)''}_{({\bf k} -{\bf k}_2)}
-2 (k)^{-2}h'''_{ {\bf k}_2}v^{||(1)}_{({\bf k} -{\bf k}_2)}
\nn\\
&
-4 (k)^{-2}h''_{ {\bf k}_2}v^{||(1)'}_{({\bf k} -{\bf k}_2)}
-\frac{2}{\tau}(k)^{-2}h''_{ {\bf k}_2}v^{||(1)}_{({\bf k} -{\bf k}_2)}
-\frac{2}{\tau}(k)^{-2}h'_{ {\bf k}_2}v^{||(1)'}_{({\bf k} -{\bf k}_2)}
\nn\\
&
+\frac{8}{\tau^2}(k)^{-2}h'_{ {\bf k}_2}v^{||(1)}_{({\bf k} -{\bf k}_2)}
\ ,
\el
}
where   ${\mathop \epsilon  \limits^s}_{ij}({\bf k})k^i=0$ has been  used.
Plugging the 1st-order solutions (\ref{vsol3})--(\ref{GWmode})
into the above yields
{\allowdisplaybreaks
\bl\label{ZTildeDecom3}
\widetilde{Z}(\tau; {\mathbf k}, {\mathbf k_2})
=&
 b_1({\bf k}_2)D_2({\bf k}-{\bf k}_2)\frac{i}{ \sqrt{2 k_2}}
   e^{i( k_2-|\mathbf k-\mathbf k_2|/\sqrt3\,\,)\tau}
\bigg[
-\frac{40}{|\mathbf k-\mathbf k_2|(k)^2\tau^5}
\nn\\
&
-\frac{40 i}{\sqrt3 (k)^2\tau^4}
+\frac{40 i k_2}{|\mathbf k-\mathbf k_2|(k)^2\tau^4}
-\frac{4i}{\sqrt3|\mathbf k-\mathbf k_2|^2\tau^4}
+\frac{20 |\mathbf k-\mathbf k_2|}{3 (k)^2\tau^3}
\nn\\
&
+\frac{20(k_2)^2}{|\mathbf k-\mathbf k_2|(k)^2\tau^3}
-\frac{4 k_2}{\sqrt3|\mathbf k-\mathbf k_2|^2\tau^3}
-\frac{40 k_2}{\sqrt3(k)^2\tau^3}
-\frac{4i(k_2)^3}{|\mathbf k-\mathbf k_2|(k)^2\tau^2}
\nn\\
&
+\frac{16 i(k_2)^2}{\sqrt3(k)^2\tau^2}
+\frac{8i |\mathbf k-\mathbf k_2|^2}{3\sqrt3(k)^2\tau^2}
-\frac{20i |\mathbf k-\mathbf k_2|k_2}{3(k)^2\tau^2}
-\frac{2|\mathbf k-\mathbf k_2|^3}{9(k)^2\tau}
\nn\\
&
+\frac{2|\mathbf k-\mathbf k_2|^2 k_2}{\sqrt3(k)^2\tau}
-\frac{2|\mathbf k-\mathbf k_2|(k_2)^2}{(k)^2\tau}
+\frac{2(k_2)^3}{\sqrt3(k)^2\tau}
\bigg]
\nn\\
&
+ b_1({\bf k}_2)D_3({\bf k}-{\bf k}_2)\frac{i}{ \sqrt{2 k_2}}
   e^{i( k_2+|\mathbf k-\mathbf k_2|/\sqrt3\,\,)\tau}
\bigg[
-\frac{40}{|\mathbf k-\mathbf k_2|(k)^2\tau^5}
\nn\\
&
+\frac{40i}{\sqrt3(k)^2\tau^4}
+\frac{40ik_2}{|\mathbf k-\mathbf k_2|(k)^2\tau^4}
+\frac{4i}{\sqrt3|\mathbf k-\mathbf k_2|^2\tau^4}
+\frac{20|\mathbf k-\mathbf k_2|}{3(k)^2\tau^3}
\nn\\
&
+\frac{20(k_2)^2}{|\mathbf k-\mathbf k_2|(k)^2\tau^3}
+\frac{4 k_2}{\sqrt3|\mathbf k-\mathbf k_2|^2\tau^3}
+\frac{40k_2}{\sqrt3(k)^2\tau^3}
-\frac{4i(k_2)^3}{|\mathbf k-\mathbf k_2|(k)^2\tau^2}
\nn\\
&
-\frac{16i(k_2)^2}{\sqrt3(k)^2\tau^2}
-\frac{8i|\mathbf k-\mathbf k_2|^2}{3\sqrt3(k)^2\tau^2}
-\frac{20i|\mathbf k-\mathbf k_2|k_2}{3(k)^2\tau^2}
-\frac{2|\mathbf k-\mathbf k_2|^3}{9(k)^2\tau}
\nn\\
&
-\frac{2|\mathbf k-\mathbf k_2|^2k_2}{\sqrt3(k)^2\tau}
-\frac{2|\mathbf k-\mathbf k_2|(k_2)^2}{(k)^2\tau}
-\frac{2(k_2)^3}{\sqrt3(k)^2\tau}
\bigg]
\nn\\
&
+ b_2({\bf k}_2)D_2({\bf k}-{\bf k}_2)\frac{i}{ \sqrt{2 k_2}}
    e^{-i (k_2+|\mathbf k-\mathbf k_2|/\sqrt3\,\,)\tau}
\bigg[
\frac{40}{|\mathbf k-\mathbf k_2|(k)^2\tau^5}
\nn\\
&
+\frac{40 i}{\sqrt3 (k)^2\tau^4}
+\frac{40i k_2}{|\mathbf k-\mathbf k_2|(k)^2\tau^4}
+\frac{4i}{\sqrt3|\mathbf k-\mathbf k_2|^2\tau^4}
-\frac{20 |\mathbf k-\mathbf k_2|}{3(k)^2 \tau^3}
\nn\\
&
- \frac{20(k_2)^2}{|\mathbf k-\mathbf k_2|(k)^2\tau^3}
-\frac{4k_2}{\sqrt3|\mathbf k-\mathbf k_2|^2\tau^3}
-\frac{40k_2}{\sqrt3 (k)^2\tau^3}
-\frac{4i(k_2)^3}{|\mathbf k-\mathbf k_2|(k)^2\tau^2}
\nn\\
&
-\frac{16i(k_2)^2}{\sqrt3(k)^2\tau^2}
-\frac{8i|\mathbf k-\mathbf k_2|^2}{3\sqrt3(k)^2\tau^2}
-\frac{20i|\mathbf k-\mathbf k_2|k_2}{3(k)^2\tau^2}
+\frac{2|\mathbf k-\mathbf k_2|^3}{9(k)^2\tau}
\nn\\
&
+\frac{2|\mathbf k-\mathbf k_2|^2 k_2}{\sqrt3(k)^2\tau}
    +\frac{2|\mathbf k-\mathbf k_2|(k_2)^2}{(k)^2\tau}
+\frac{2(k_2)^3}{\sqrt3(k)^2\tau}
\bigg]
\nn\\
&
+ b_2({\bf k}_2)D_3({\bf k}-{\bf k}_2)\frac{i}{ \sqrt{2 k_2}}
    e^{-i (k_2-|\mathbf k-\mathbf k_2|/\sqrt3\,)\tau}
\bigg[
\frac{40}{|\mathbf k-\mathbf k_2|(k)^2\tau^5}
\nn\\
&
-\frac{40i}{\sqrt3(k)^2\tau^4}
+\frac{40i k_2}{|\mathbf k-\mathbf k_2|(k)^2\tau^4}
-\frac{4i}{\sqrt3|\mathbf k-\mathbf k_2|^2\tau^4}
-\frac{20|\mathbf k-\mathbf k_2|}{3(k)^2\tau^3}
\nn\\
&
-\frac{20(k_2)^2}{|\mathbf k-\mathbf k_2|(k)^2\tau^3}
+\frac{4 k_2}{\sqrt3|\mathbf k-\mathbf k_2|^2\tau^3}
+\frac{40k_2}{\sqrt3(k)^2\tau^3}
-\frac{4 i(k_2)^3}{|\mathbf k-\mathbf k_2|(k)^2 \tau^2}
\nn\\
&
+\frac{16 i(k_2)^2}{\sqrt3(k)^2\tau^2}
+\frac{8i|\mathbf k-\mathbf k_2|^2}{3\sqrt3(k)^2\tau^2}
-\frac{20i|\mathbf k-\mathbf k_2|k_2}{3(k)^2\tau^2}
+\frac{2|\mathbf k-\mathbf k_2|^3}{9(k)^2\tau}
\nn\\
&
-\frac{2|\mathbf k-\mathbf k_2|^2 k_2}{\sqrt3(k)^2\tau}
+\frac{2|\mathbf k-\mathbf k_2|(k_2)^2}{(k)^2\tau}
-\frac{2(k_2)^3}{\sqrt3(k)^2 \tau}
\bigg]
.
\el
}
Notice that (\ref{ZTildeDecom3}) contains no time-integration terms.

The homogeneous solution of (\ref{V2eqFourier1}) for
 a general value of $c_N$
is similar to  Eq.(\ref{v1||solgen}) just by a replacement of
$c_L\rightarrow c_N$,
and the inhomogeneous solution of (\ref{V2eqFourier1}) is complicated.
For the case  $c_N^2 =\frac{1}{3}$ and a general $c_L$,
(\ref{V2eqFourier1}) becomes
\be\label{V2eqFourier2}
v_{s(t)\mathbf k}^{||(2)'''}
+\frac{1}{\tau} v_{s(t)\mathbf k}^{||(2)''}
+\l(\frac{k^2}{3}
-\frac{4}{\tau^2}\r)v_{s(t)\mathbf k}^{||(2)'}
+\frac{k^2}{3\tau}v_{s(t)\mathbf k}^{||(2)}
=Z_{s(t)\mathbf k}(\tau)
.
\ee
The solution of (\ref{V2eqFourier2}) is
{\allowdisplaybreaks
\bl\label{v2ndsol}
v^{||(2)}_{s(t)\mathbf k}
&=
\frac{P_1(\mathbf k)}{k\tau}
+P_2(\mathbf k)\l(\frac{2}{k\tau}
    +\frac{i}{\sqrt3}\r)e^{-ik\tau/\sqrt3}
+P_3(\mathbf k)\l(\frac{2}{k\tau}
    -\frac{i}{\sqrt3}\r)e^{ik\tau/\sqrt3}
\nn\\
&
-\left(
\frac{2}{k\tau} \cos (\frac{k \tau}{\sqrt{3}})
+\frac{1}{\sqrt{3}}\sin (\frac{k \tau}{\sqrt{3}})
\right)
\int^\tau
\bigg(
9\cos (\frac{k \tau'}{\sqrt{3}})
+3\sqrt{3}k \tau' \sin (\frac{k \tau'}{\sqrt{3}})
\bigg)
\frac{  Z_{s(t)\mathbf k}(\tau') }{ k^3 \tau'} \, d\tau'
\nn\\
&
-\left(
\frac{ 2}{k\tau} \sin (\frac{k \tau}{\sqrt{3}})
-\frac{1}{\sqrt{3}} \cos (\frac{k\tau}{\sqrt{3}})
\right)
    \int^\tau
    \bigg( 9 \sin (\frac{k \tau'}{\sqrt{3}})
- 3\sqrt{3} k \tau' \cos (\frac{k \tau'}{\sqrt{3}})\bigg)
    \frac{Z_{s(t)\mathbf k}(\tau')}{k^3 \tau'}d\tau'
    \nn\\
&
+\frac{1}{ k\tau}
\int^\tau \frac{
    3\left(k^2 \tau'^2+6\right)}{k^3 \tau'} Z_{s(t)\mathbf k}(\tau')d\tau'
,
\el
}
where $P_1(\mathbf k)$, $P_2(\mathbf k)$, $P_3(\mathbf k)$
are arbitrary time-independent functions,
and determined by initial conditions,
and their associated  terms
are the homogeneous solution.
The $P_1(\mathbf k)$ terms are gauge modes as shall be seen in Sec. 5.

The solution of $\delta_{s(t){\bf k}}^{(2)}$ is directly
given by (\ref{deltaV2}) in
 $\bf k$-space  as
{\allowdisplaybreaks
\bl\label{deltaV2solu}
 \delta_{s(t){\bf k}}^{(2)}=&
-4 v_{s(t){\bf k}}^{||(2)'}
+3 F^{||}_{s(t){\bf k}}
\nn\\
=&
\frac{4 P_1(\mathbf k)}{k\tau^2}
+P_2(\mathbf k)
\l(
\frac{8}{k\tau^2}
+\frac{8 i}{\sqrt3\,\tau}
-\frac{4 k}{3}
\r)e^{-ik\tau/\sqrt3}
\nn\\
    &
+P_3(\mathbf k)\l(
\frac{8}{k\tau^2}
-\frac{8 i}{\sqrt3\,\tau}
-\frac{4 k}{3}
\r)e^{ik\tau/\sqrt3}
+3 F^{||}_{s(t){\bf k}}
        \nn\\
        &
+\bigg(
-\frac{8}{k\tau^2} \cos (\frac{k \tau}{\sqrt{3}})
-\frac{8}{\sqrt{3}\tau} \sin (\frac{k \tau}{\sqrt{3}})
+\frac{4k}{3}\cos (\frac{k \tau}{\sqrt{3}})
\bigg)
\int^\tau
\bigg(
9\cos (\frac{k \tau'}{\sqrt{3}})
\nn\\
&
+3\sqrt{3}k \tau' \sin (\frac{k \tau'}{\sqrt{3}})
\bigg)
\frac{  Z_{s(t)\mathbf k}(\tau') }{ k^3 \tau'} \, d\tau'
\nn\\
&
+\bigg(
-\frac{ 8}{k\tau^2} \sin (\frac{k \tau}{\sqrt{3}})
+\frac{ 8}{\sqrt{3}\,\tau} \cos (\frac{k \tau}{\sqrt{3}})
+\frac{4k}{3} \sin (\frac{k\tau}{\sqrt{3}})
\bigg)
    \int^\tau
    \bigg( 9 \sin (\frac{k \tau'}{\sqrt{3}})
\nn\\
&
- 3\sqrt{3} k \tau' \cos (\frac{k \tau'}{\sqrt{3}})\bigg)
    \frac{Z_{s(t)\mathbf k}(\tau')}{k^3 \tau'}d\tau'
    \nn\\
&
+\frac{1}{ k\tau^2}
\int^\tau \frac{
    12(k^2 \tau'^2+6)}{k^3 \tau'} Z_{s(t)\mathbf k}(\tau')d\tau'
.
\el
}
Integrating the $k$-mode equation of (\ref{Vphi2}) yields
the scalar solution
{\allowdisplaybreaks
\bl\label{phi2SkSol}
\phi^{(2)}_{s(t){\bf k}}
=&
-v_{s(t){\bf k}}^{||(2)'}
-\int^\tau \frac{k^2}{3}v^{||(2)}_{s(t){\bf k}} d\tau'
+\frac{3}{4}F^{||}_{s(t){\bf k}}
- \frac{1}{4}\int^\tau A_{s(t){\bf k}} d\tau'
+P_4(\mathbf k)(\tau)
\nn\\
=&
P_1(\mathbf k)\l(
\frac{1}{k\tau^2}
-\frac{k\ln\tau}{3}
\r)
+P_2(\mathbf k)
\l(
\frac{2}{k\tau^2}
+\frac{2 i}{\sqrt3\,\tau}
\r)e^{-ik\tau/\sqrt3}
\nn\\
&
+P_3(\mathbf k)\l(
\frac{2}{k\tau^2}
-\frac{2 i}{\sqrt3\,\tau}
\r)e^{ik\tau/\sqrt3}
+P_4(\mathbf k)
        \nn\\
        &
-\frac{2 k}{3}\int^\tau
\l[
P_2(\mathbf k)e^{-ik\tau'/\sqrt3}
+P_3(\mathbf k)e^{ik\tau'/\sqrt3}
\r]
\frac{d\tau'}{\tau'}
+\frac{3}{4}F^{||}_{s(t){\bf k}}(\tau)
\nn\\
&
- \frac{1}{4}\int^\tau A_{s(t){\bf k}}(\tau') d\tau'
+\int^\tau\frac{(k^2 \tau'\,^2+6)\ln\tau'+3}{k^2 \tau'} Z_{s(t)\mathbf k}(\tau')d\tau'
\nn\\
&
+\l(
\frac{1}{ k\tau^2}
-\frac{ k\ln\tau}{3}
\r)\int^\tau \frac{
    3(k^2 \tau'^2+6)}{k^3 \tau'} Z_{s(t)\mathbf k}(\tau')d\tau'
\nn\\
&
-\bigg(
\frac{2}{k\tau^2} \cos (\frac{k \tau}{\sqrt{3}})
+\frac{2}{\sqrt{3}\tau} \sin (\frac{k \tau}{\sqrt{3}})
\bigg)
\int^\tau
\bigg(
9\cos (\frac{k \tau'}{\sqrt{3}})
\nn\\
&
+3\sqrt{3}k \tau' \sin (\frac{k \tau'}{\sqrt{3}})
\bigg)
\frac{  Z_{s(t)\mathbf k}(\tau') }{ k^3 \tau'} \, d\tau'
\nn\\
&
-\bigg(
\frac{ 2}{k\tau^2} \sin (\frac{k \tau}{\sqrt{3}})
-\frac{ 2}{\sqrt{3}\,\tau} \cos (\frac{k \tau}{\sqrt{3}})
\bigg)
    \int^\tau
    \bigg( 9 \sin (\frac{k \tau'}{\sqrt{3}})
\nn\\
&
- 3\sqrt{3} k \tau' \cos (\frac{k \tau'}{\sqrt{3}})\bigg)
    \frac{Z_{s(t)\mathbf k}(\tau')}{k^3 \tau'}d\tau'
\nn\\
&
+\int^\tau
\bigg[
\frac{2}{k\tau''} \cos (\frac{k \tau''}{\sqrt{3}})
\int^{\tau''}
\bigg(
3\cos (\frac{k \tau'}{\sqrt{3}})
+\sqrt{3}k \tau' \sin (\frac{k \tau'}{\sqrt{3}})
\bigg)
\frac{  Z_{s(t)\mathbf k}(\tau') }{ k \tau'} \, d\tau'
\bigg]d\tau''
\nn\\
&
+\int^\tau
\bigg[
\frac{ 2}{k\tau''} \sin (\frac{k \tau''}{\sqrt{3}})
    \int^{\tau''}
    \bigg(
    3 \sin (\frac{k \tau'}{\sqrt{3}})
- \sqrt{3} k \tau' \cos (\frac{k \tau'}{\sqrt{3}})\bigg)
    \frac{Z_{s(t)\mathbf k}(\tau')}{k \tau'}d\tau'
\bigg]d\tau''
,
\el
}
where  $A_{s(t)}\equiv
\frac{d}{d\tau}\l[
\frac{4}{3}\chi^{\top(1)}_{lm}\chi^{||(1),\,lm}\r]$   by (\ref{AS}),
and
{\allowdisplaybreaks
\bl
\int^\tau A_{s(t){\bf k}}& d\tau'
=
\frac{1}{(2\pi)^{3}}\int d^3k_2\bigg\{
\Big[
 k^l k^m\sum_{s={+,\times}}
{\mathop \epsilon \limits^s}_{lm}({{\bf k}_2})
\Big]
\,
\Big[
-\frac{4}{3}h_{ {\bf k}_2}\chi^{||(1)}_{({\bf k} -{\bf k}_2)}
\Big]
\bigg\}
\nn\\
=&
\frac{1}{(2\pi)^{3}}\int d^3k_2\bigg\{
\Big[
 k^l k^m\sum_{s={+,\times}}
{\mathop \epsilon \limits^s}_{lm}({{\bf k}_2})
\Big]
\frac{16 i}{3\sqrt{2 k_2}|{\bf k}-{\bf k}_2|\tau}
\nn\\
&
\times
\Big[
b_1({\bf k}_2)  D_2({\bf k}-{\bf k}_2)
\Big(
\frac{\sqrt3\,ie^{i(k_2-|{\bf k}-{\bf k}_2|/\sqrt3\,\,)\tau}}{|{\bf k}-{\bf k}_2|\tau}
- e^{i k_2\tau}\int^\tau \frac{e^{-i|{\bf k}-{\bf k}_2|\tau'/\sqrt3}}{\tau'}d\tau'
\Big)
\nn\\
&
+b_1({\bf k}_2)  D_3({\bf k}-{\bf k}_2)
\Big(
-\frac{\sqrt3\,i e^{i( k_2+|{\bf k}-{\bf k}_2|/\sqrt3\,\,)\tau}}{|{\bf k}-{\bf k}_2|\tau}
-e^{i k_2\tau}\int^\tau \frac{e^{i|{\bf k}-{\bf k}_2|\tau'/\sqrt3}}{\tau'}d\tau'
\Big)
\nn\\
&
+ b_2({\bf k}_2)D_2({\bf k}-{\bf k}_2)
\Big(
-\frac{\sqrt3\,ie^{-i( k_2+|{\bf k}-{\bf k}_2|/\sqrt3\,\,)\tau}}{|{\bf k}-{\bf k}_2|\tau}
+ e^{-i k_2\tau}\int^\tau \frac{e^{-i|{\bf k}-{\bf k}_2|\tau'/\sqrt3}}{\tau'}d\tau'
\Big)
\nn\\
&
+ b_2({\bf k}_2)  D_3({\bf k}-{\bf k}_2)
\Big(
\frac{\sqrt3\,i e^{-i( k_2-|{\bf k}-{\bf k}_2|/\sqrt3\,\,)\tau}}{|{\bf k}-{\bf k}_2|\tau}
+e^{-i k_2\tau}\int^\tau \frac{e^{i|{\bf k}-{\bf k}_2|\tau'/\sqrt3}}{\tau'}d\tau'
\Big)
\Big]
\bigg\}.
\el
}
Notice that in the above,
$P_4(\mathbf k)$ terms in (\ref{phi2SkSol})
are gauge modes,
as shall be seen in Sec. 5.

Finally,
plugging (\ref{deltaV2solu}) and (\ref{phi2SkSol}) into
(\ref{Ein2th003RD}) in $\bf k$-space  gives the scalar solution
{\allowdisplaybreaks
\bl\label{chi2Ssol}
\chi^{||(2)}_{s(t){\bf k}}
=&
\frac{18}{k^4\tau} \phi^{(2)'}_{s(t){\bf k}}
+\frac{6}{k^2}\phi^{(2)}_{s(t){\bf k}}
+\frac{9}{k^4\tau^2}\delta^{(2)}_{s(t){\bf k}}
+\frac{3}{k^4} E_{s(t){\bf k}}
\nn\\
=&
-P_1(\mathbf k)\frac{2\ln\tau}{k}
+P_2(\mathbf k)\frac{4\sqrt3 \,i}{k^2\tau}e^{-ik\tau/\sqrt3}
-P_3(\mathbf k)\frac{4\sqrt3 \,i}{k^2\tau}e^{ik\tau/\sqrt3}
+\frac{6 P_4(\mathbf k)}{k^2}
        \nn\\
        &
-\frac{4}{k}\int^\tau
\l[
P_2(\mathbf k)e^{-ik\tau'/\sqrt3}
+P_3(\mathbf k)e^{ik\tau'/\sqrt3}
\r]
\frac{d\tau'}{\tau'}
\nn\\
&
-\frac{2 \ln\tau}{k }\int^\tau \frac{
    3(k^2 \tau'^2+6)}{k^3 \tau'} Z_{s(t)\mathbf k}(\tau')d\tau'
+\int^\tau\frac{6(k^2 \tau'\,^2+6)\ln\tau'+18}{k^4 \tau'} Z_{s(t)\mathbf k}(\tau')d\tau'
    \nn\\
&
-\frac{4\sqrt3}{k^2\tau} \sin (\frac{k \tau}{\sqrt{3}})
\int^\tau
\bigg(
9\cos (\frac{k \tau'}{\sqrt{3}})
+3\sqrt{3}k \tau' \sin (\frac{k \tau'}{\sqrt{3}})
\bigg)
\frac{  Z_{s(t)\mathbf k}(\tau') }{ k^3 \tau'} \, d\tau'
\nn\\
&
+\frac{4\sqrt3}{k^2\tau} \cos (\frac{k \tau}{\sqrt{3}})
    \int^\tau
    \bigg( 9 \sin (\frac{k \tau'}{\sqrt{3}})
- 3\sqrt{3} k \tau' \cos (\frac{k \tau'}{\sqrt{3}})\bigg)
    \frac{Z_{s(t)\mathbf k}(\tau')}{k^3 \tau'}d\tau'
\nn\\
&
+\int^\tau
\bigg[
\frac{12}{k^3\tau''} \cos (\frac{k \tau''}{\sqrt{3}})
\int^{\tau''}
\bigg(
3\cos (\frac{k \tau'}{\sqrt{3}})
+\sqrt{3}k \tau' \sin (\frac{k \tau'}{\sqrt{3}})
\bigg)
\frac{  Z_{s(t)\mathbf k}(\tau') }{ k \tau'} \, d\tau'
\bigg]d\tau''
\nn\\
&
+\int^\tau
\bigg[
\frac{12}{k^3\tau''} \sin (\frac{k \tau''}{\sqrt{3}})
    \int^{\tau''}
    \bigg(
    3 \sin (\frac{k \tau'}{\sqrt{3}})
- \sqrt{3} k \tau' \cos (\frac{k \tau'}{\sqrt{3}})\bigg)
    \frac{Z_{s(t)\mathbf k}(\tau')}{k \tau'}d\tau'
\bigg]d\tau''
        \nn\\
        &
+\frac{3}{k^4} E_{s(t){\bf k}}
+\frac{27}{2k^4\tau}F^{||'}_{s(t){\bf k}}(\tau)
+\frac{27}{k^4\tau^2} F^{||}_{s(t){\bf k}}
+\frac{9}{2k^2}F^{||}_{s(t){\bf k}}(\tau)
        \nn\\
        &
- \frac{9}{2k^4\tau} A_{s(t){\bf k}}(\tau)
-\frac{3}{2k^2}\int^\tau A_{s(t){\bf k}}(\tau') d\tau'
.
\el
}
We have checked that the scalar solutions
(\ref{v2ndsol})--(\ref{phi2SkSol}) and (\ref{chi2Ssol})
satisfy the scalar parts of the evolution equation,
(\ref{Evo2ndSsTr2RD}) and (\ref{Evo2ndSsChi1RD}).

The above 2nd-order solutions  involve many   integrals
 $\int d^3 k$  or  $\int d\tau$ of
the scalar-tensor coupling terms.
In the $\bf k$-integrations,
the four functions $b_1({\bf k})$, $b_2({\bf k})$,
$D_2({\bf k})$, and  $D_3({\bf k})$
depend upon the concrete initial conditions
and  inflation models.
Moreover, in actually doing integration,
one should avoid     IR and UV divergences
which may arise from the lower and upper limits of  $\int d^3 k$
  \cite{WangZhangChen2016,ZhangWangJCAP2018}.
As an illustration,
suppose $b_1({\bf k}) \propto k^{N_1}$, $b_2({\bf k}) \propto k^{N_2}$,
$D_2({\bf k}) \propto k^{N_3}$, and $D_3({\bf k}) \propto k^{N_4}$.
Then we shall have the following typical integration terms
\be\label{integrl}
\int  d^3k_2\,
\Big[
e^{-i (k_2+|\mathbf k-\mathbf k_2|/\sqrt3\,\,)\tau}
(k_2)^{n_1} \big|{\bf k -k_2}\big|^{n_2}
k^l k^m\sum_{s={+,\times}}
{\mathop \epsilon \limits^{s}}_{lm}({{\bf k}_2})
\Big],
\ee
where $n_1$ and $n_2$ are linearly related to $N_1$, $N_2$, $N_3$ and $N_4$.
Let  $\bf k$ be along the $z$-axis
and $\theta$ be  the angle between $\mathbf k_2$ and $\bf k$.
The unit vector along $\mathbf k_2$ is
\be \label{direction1}
\hat{k}_2=\cos\phi\sin\theta\ \hat{x}
+\sin\phi\sin\theta\ \hat{y}+\cos\theta\ \hat{z},
\ee
the orthogonal unit vectors normal to  $\mathbf k_2$ are
\be \label{direction2}
\hat{u}=\sin\phi\ \hat{x}-\cos\phi\ \hat{y},
\ee
\be\label{direction3}
\hat{v}=\cos\phi\cos\theta\ \hat{x}
+\sin\phi\cos\theta\ \hat{y}-\sin\theta\ \hat{z} ,
\ee
and the corresponding polarization tensors of the 1st-order tensor mode
are as follows:
\ba\label{basis tensor 1}
{\mathop \epsilon \limits^{\times}}_{ij}({{\bf k}_2})
&=& \hat{u}_i\hat{v}_j+\hat{v}_i\hat{u}_j \nn \\
&=& \left(       \begin{array}{ccc}
           2\sin\phi\cos\phi\cos\theta & \sin^2\phi\cos\theta-\cos^2\phi\cos\theta & -\sin\phi\sin\theta \\
            \sin^2\phi\cos\theta-\cos^2\phi\cos\theta & -2\sin\phi\cos\phi\cos\theta & \cos\phi\sin\phi \\
           -\sin\phi\sin\theta & \cos\phi\sin\theta & 0 \\
                 \end{array}
            \right), \nn \\
\ea
\ba\label{basis tensor 2}
{\mathop \epsilon \limits^{+}}_{ij}({{\bf k}_2})
& =& \hat{u}_{i}\hat{u}_j -\hat{v}_i\hat{v}_j \nn \\
&=& \left(     \begin{array}{ccc}
     \sin^2\phi-\cos^2\phi\cos^2\theta & - \sin\phi\cos\phi (1+\cos^2\theta) & \cos\phi\cos\theta\sin\theta \\
       -\sin\phi\cos\phi (1+\cos^2\theta) & \cos^2\phi-\sin^2\phi\cos^2\theta & \sin\phi\cos\theta\sin\theta \\
     \cos\phi\cos\theta\sin\theta & \sin\phi\cos\theta\sin\theta & -\sin^2\theta \\
              \end{array}
            \right). \nn \\
\ea
So  one has
$k^l k^m\sum_{s={+,\times}}{\mathop \epsilon \limits^{s}}_{lm}({{\bf k}_2})
=-(k_2)^2\sin^2\theta$.
The angular integration in (\ref{integrl}) can be carried out,
yielding
\bl
&
- 2\pi\int_{K_1}^{K_2} dk_2\int_0^\pi d\theta
\Big[
e^{-i \left(k_2+\sqrt{(k)^2+(k_2)^2-2k_2k\cos\theta}\left/\sqrt3\right.\right)\tau}
(k_2)^{n_1+2}
\nn\\
&
\hspace{4.8cm}
\times\left((k)^2+(k_2)^2-2k_2k\cos\theta\right)^{\frac{n_2}{2}}
(k)^2\sin^3\theta
\Big]
\nn\\
=&
-\pi\int_{K_1}^{K_2} dk_2
\Bigg\{
\frac{3^{\frac{n_2}{2}+1}}{2k \tau ^6}
(k_2)^{n_1-1} e^{-i k_2 \tau }
\tau^{-n_2}\left(-i\right)^{n_2}
 \bigg[
\tau ^4 \left((k)^2-(k_2)^2\right)^2
    \Gamma \big(n_2+2,\frac{i |k-k_2| \tau }{\sqrt{3}}\big)
\nn\\
&
+6 \tau ^2 \left((k)^2+(k_2)^2\right)
    \Gamma \big(n_2+4,\frac{i |k-k_2| \tau }{\sqrt{3}}\big)
+9 \Gamma \big(n_2+6,\frac{i |k-k_2| \tau }{\sqrt{3}}\big)
\nn\\
&
-\tau ^4 \left((k)^2-(k_2)^2\right)^2
    \Gamma \big(n_2+2,\frac{i (k+k_2) \tau }{\sqrt{3}}\big)
-6 \tau ^2 \left((k)^2+(k_2)^2\right)
    \Gamma \big(n_2+4,\frac{i (k+k_2) \tau }{\sqrt{3}}\big)
\nn\\
&
-9 \Gamma \big(n_2+6,\frac{i (k+k_2) \tau }{\sqrt{3}}\big)
\bigg]
\Bigg\}
,
\nn
\el
where the integration limits $K_1$ and $K_2$
are   cutoffs to ensure IR and UV convergence,
and the remaining integration over $k_2$ can be done numerically.
All other  $\int d^3 k$ terms can be treated similarly.

The time integrations can also be carried out.
$Z_{s(t)\bf k}$
in (\ref{ZSallFourier}) and (\ref{ZTildeDecom3})
only has one type of  term:
$ \frac{1}{\tau^{N}} e^{ -i \frac{ k\tau}{\sqrt{3}}}$,
so that the single time integrations of $Z_{s(t)\bf k}$
have the nontrivial terms,
\be\label{nonTriv1}
 \int^\tau\frac{d\tau'}{\tau'^n} \exp \Big[- i \frac{ k_1 \tau'}{\sqrt{3}}\Big]
 \propto  k ^{n-1} \Gamma (1-n,\frac{i k  \tau }{\sqrt{3}} ),
\ee
\bl\label{nonTriv2}
 \int^\tau \frac{d\tau'}{\tau'^n}\ln\tau'
 \exp \Big[- i \frac{ k_1 \tau'}{\sqrt{3}}\Big]
 \propto \
&
    3^{\frac{1-n}{2}}  (i k   )^{n-1}  \ln\tau
     \Gamma  (1-n,0,\frac{i k \tau }{\sqrt{3}} ) \nn \\
&   -\frac{\tau ^{1-n}}{(1-n)^{2} }
       \, _2F_2  (1-n,1-n;2-n,2-n;-\frac{i k \tau }{\sqrt{3}} )    ,
\el
and the double time integrations of
$Z_{s(t)\bf k}$ have the following nontrivial term,
\be\label{nonTriv3}
 \int^{\tau}  \frac{ d\tau'}{\tau'} \exp \Big[ i \frac{ k_1 \tau'}{\sqrt{3}}\Big]
 \int^{\tau'}  \frac{ d\tau''}{\tau''\, ^n} \exp \Big[ i \frac{ k_2\tau''}{\sqrt{3}}\Big]
 \equiv  z_5 (\tau;n; k_1,k_2 ) ,
\ee
which, in actual computing,
 can be defined as a function
and recalled \cite{WangZhang2ndRD2018}.
As an illustration, we plot the real part of  $z_5$  in Fig.\ref{z5}.
\begin{figure}[htb]
\centering
 \includegraphics[width=0.5\linewidth]{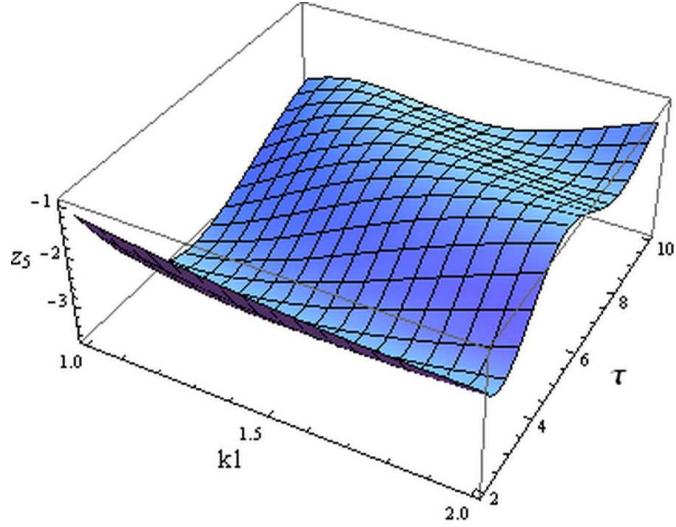}
\caption{Real part of $z_5(\tau; 1; k_1,k_2)$ at fixed $k_2$.
   }\label{z5}
\end{figure}

\subsection{Solution for tensor-tensor couplings}

Now
we solve the 2nd-order perturbations with tensor-tensor couplings.
The tensor part of
the traceless part of evolution equation (\ref{TEvo2ndSsTen1RD})
has the general solution
\be\label{Tsolgw}
\chi^{\top(2)}_{T\,ij}({\bf x},\tau)
=\int  \frac{d^3k}{(2\pi)^{\frac{3}{2}}}  e^{i{\bf k\cdot x}}
\Big(
\bar I_{T\, ij}({\bf k}, \tau) +
\sum_{s={+,\times}} {\mathop \epsilon
\limits^s}_{ij}({\bf k})
\Big[-f_{1}^s \sqrt{\frac{2}{\pi}}\,\frac{i\, e^{i  k\tau}}{ k\tau}
+f_{2}^s \sqrt{\frac{2}{\pi}}\,\frac{i\, e^{-i  k\tau}}{ k\tau} \Big]
\Big),
\ee
where $f_{1}^s({\bf k})$ and  $f_{2}^s({\bf k})$
terms represent  a homogeneous solution,
similar to (\ref{solgw}).
The integrand of the inhomogeneous solution in (\ref{Tsolgw}) is given by
\be\label{TGW2ndStepI}
\bar I_{T\,ij}({\bf k},\tau)\equiv
\frac{ie^{-ik\tau}}{k\tau}\int^{\tau}\tau'e^{ik\tau'}\bar J_{T\, ij}({\bf k},\tau')d\tau'
-\frac{ie^{ik\tau}}{k\tau}\int^{\tau}\tau'e^{-ik\tau'}\bar J_{T\, ij}({\bf k},\tau')d\tau'
,
\ee
with $\bar J_{T\, ij}$ being the Fourier transform of the source
$J_{T\,ij}$ in (\ref{T2ndTensorSourceRD})
that contains many terms of products of 1st-order solutions.
The solution (\ref{Tsolgw}) has a similar structure to the solution (\ref{solgw})
except for the inhomogeneous part with a different source term.

The   vector  solution of Eq.(\ref{TEvo2ndSsVec2RD}) is
\be\label{TchiVecSol}
\chi^{\perp(2)}_{T\,ij}({\bf x},\tau)
=q_{3ij}({\bf x})
+\frac{q_{4ij}({\bf x})}{\tau}
+\int^\tau \frac{d\tau'}{\tau^{'2}}
    \int^{\tau'}2\tau^{''2}\,V_{T\,ij}({\bf x},\tau'')d\tau''
\,,
\ee
with $V_{T\,ij}$ given by (\ref{TSourceCurl1RD})
and $q_{3ij}({\bf x})$ and $q_{4ij}({\bf x})$ are
two coefficients to be  determined by initial conditions.
Note that $q_{3ij}({\bf x})$ is a gauge mode
as shall be seen in the next section.
Also the solution (\ref{TchiVecSol}) has
a similar structure to the solution (\ref{chiVecSol}).
Plugging the solution (\ref{TchiVecSol}) into (\ref{TMoCons2ndCurlRD1}),
one has the transverse part of the 2nd-order velocity as
\bl\label{TVperp2ndSol}
v^{\perp(2)}_{T\,i}
=&
\frac{q_{4ij}^{\,,\,j}({\bf x})}{8}
+\frac{\tau^2}{4}\l(
M_{T\,i}
-\partial_i\nabla^{-2}M_{T\,k}^{,\,k}
\r)
- \frac{1}{4}
    \int^{\tau}\tau^{'2}\,V_{T\,ij}^{,j}({\bf x},\tau')d\tau'
\ ,
\el
where $(M_{T\,i}-\partial_i\nabla^{-2}M_{T\,k}^{,k})$
is in (\ref{TMSiCurlRD}).
This solution satisfies
the transverse  momentum conservation (\ref{TMoC2ndCurl}),
i.e.,
$v^{\perp(2)'}_{Ti} =0$,
as we have checked.
 According to (\ref{TVperp2ndSol}),
the 2nd-order curl vector $v^{\perp(2)}_{i}$ is generated by the coupling,
although  the 1st-order curl vector $v^{\perp(1)}_{i}$ is zero
by assumption.

Next,
we solve the scalars.
From the longitudinal  momentum conservation
(\ref{TMoC2ndNonCurl}),
\be \label{TdeltaV2}
 \delta_{T}^{(2)}
=
-\frac{4}{3c_N^2} v_{T}^{||(2)'}
\ .
\ee
Plugging the above $\delta^{(2)}_{T}$ into
the energy conservation (\ref{TenCons2ndRD2}),
gives $\phi^{(2)'}_{T}$ in terms of $v_{T}^{||(2)}$,
$v_{T}^{||(2)'}$, $v_{T}^{||(2)''}$ as
\be\label{TVphi2}
\phi^{(2)'}_{T}
=
-\frac{1}{3 c_N^2} v_{T}^{||(2)''}
-\frac{1}{\tau}\frac{c_N^2-\frac{1}{3}}{c_N^2} v_{T}^{||(2)'}
+\frac{1}{3}\nabla^2v^{||(2)}_{T}
-\frac{1}{4}A_{T}
.
\ee
To use the energy constraint (\ref{TEin2th003RD}),
taking [$\frac{d}{d\tau}$(\ref{TEin2th003RD})] gives
\be
-\frac{6}{\tau} \phi^{(2)''}_{T}
+\frac{6}{\tau^2} \phi^{(2)'}_{T}
+\nabla^2 \Big[
2\phi^{(2)'}_{T}
+\frac{1}{3}\nabla^2\chi^{||(2)'}_{T}
\Big]
=
\frac{3}{\tau^2}\delta^{(2)'}_{T}
-\frac{6}{\tau^3}\delta^{(2)}_{T}
+E^{\,'}_{T}
.
\ee
Plugging the momentum constraint (\ref{TMoConstr2ndv3RD2}) into the above to
 eliminate  $\nabla^2\nabla^2\chi^{||(2)'}_{T}$,
 one has
\be
-\frac{6}{\tau} \phi^{(2)''}_{T}
+\frac{6}{\tau^2} \phi^{(2)'}_{T}
-\frac{4}{\tau^2}\nabla^2v^{||(2)}_{T}
+M_{T\,k}^{,\,k}
=
\frac{3}{\tau^2}\delta^{(2)'}_{T}
-\frac{6}{\tau^3}\delta^{(2)}_{T}
+E^{\,'}_{T}
.
\ee
Then,
plugging $\delta^{(2)}_{T}$  of  (\ref{TdeltaV2}) and
$\phi^{(2)'}_{T} $ of  (\ref{TVphi2}) into the above
yields a 3rd-order differential equation of $v^{||(2)}_{T}$
as
\be\label{TV2ndeq1}
v_{T}^{||(2)'''}
+\frac{3c_N^2}{\tau} v_{T}^{||(2)''}
-\frac{6c_N^2+2}{\tau^2}v_{T}^{||(2)'}
-\frac{c_N^2}{\tau}\nabla^2v^{||(2)}_{T}
-c_N^2\nabla^2v^{||(2)'}_{T}
=Z_{T}
,
\ee
with
\be\label{TZScn1}
Z_{T}
\equiv
-\frac{3c_N^2}{4}A_{T}^{\,'}
+\frac{3c_N^2}{4\tau}A_{T}
-\frac{\tau}{2}c_N^2M_{T\,l}^{,\,l}
+\frac{\tau}{2}c_N^2E^{\,'}_{T}
\, ,
\ee
which can also be written as
\be\label{TZSall}
Z_{T}
=
-\frac{1}{6}\chi^{\top(1)'\,lm}\chi^{\top(1)'}_{lm}
\,,
\ee
where  the 1st-order GW equation (\ref{evoEq1stRDtensor}) is used.
Equation (\ref{TV2ndeq1}) in the $\bf k$-space is
\be\label{TV2eqFourier1}
v_{T\mathbf k}^{||(2)'''}
+\frac{3c_N^2}{\tau} v_{T\mathbf k}^{||(2)''}
+\l(
c_N^2k^2
-\frac{6c_N^2+2}{\tau^2}
\r)v_{T\mathbf k}^{||(2)'}
+\frac{c_N^2}{\tau}k^2v_{T\mathbf k}^{||(2)}
=Z_{T\mathbf k}(\tau)
,
\ee
where the source
\bl
Z_{T{\bf k}}
=&
\frac1{(2\pi)^{3/2}}\int d^3x
\l[
-\frac{1}{6}\chi^{\top(1)'\,lm}\chi^{\top(1)'}_{lm}
\r]
e^{-i{\bf k\cdot x}}
\nn\\
=&
-\frac{1}{6(2\pi)^{3}}\int d^3k_2\,
\Big[
\sum_{s_1={+,\times}}  \sum_{s_2={+,\times}}
{\mathop \epsilon \limits^{s_1}}_{lm}({\bf k}-{{\bf k}_2})
{\mathop \epsilon \limits^{s_2}}{}^{lm}({{\bf k}_2})
\Big]
{h'}_{ {\bf k}-{{\bf k}_2}} {h'}_{ {\bf k}_2}
\el
is the Fourier transform of
$Z_{T}({\bf x},\tau)$ in (\ref{TZScn1}).
Plugging (\ref{GWmode}) into the above,
one has
{\allowdisplaybreaks
\bl\label{ZTk}
Z_{T{\bf k}}
=&
\frac{1}{12(2\pi)^{3}}\int d^3k_2\,
\bigg[
\sum_{s_1={+,\times}}  \sum_{s_2={+,\times}}
{\mathop \epsilon \limits^{s_1}}_{lm}({\bf k}-{{\bf k}_2})
{\mathop \epsilon \limits^{s_2}}{}^{lm}({{\bf k}_2})
\frac{1}{\sqrt{k_2|{\bf k}-{{\bf k}_2}|}}
\nn\\
&
\times
\Big\{
- b_1({{\bf k}_2})\ b_2({\bf k}-{{\bf k}_2})\
e^{i (k_2-|{\bf k}-{{\bf k}_2}|)\tau}
\Big(\frac{1}{\tau^4}
    +\frac{i (|{\bf k}-{{\bf k}_2}|-k_2)}{\tau^3}
    +\frac{k_2 |{\bf k}-{{\bf k}_2}|}{\tau^2}
    \Big)
\nn\\
&
+b_1({{\bf k}_2})\ b_1({\bf k}-{{\bf k}_2})\
e^{i (k_2+|{\bf k}-{{\bf k}_2}|)\tau}
\Big(\frac{1}{\tau^4}
    -\frac{i (|{\bf k}-{{\bf k}_2}|+k_2)}{\tau^3}
    -\frac{ k_2|{\bf k}-{{\bf k}_2}|}{\tau^2}
    \Big)
\nn\\
&
+b_2({{\bf k}_2})\ b_2({\bf k}-{{\bf k}_2})\
 e^{-i (k_2+|{\bf k}-{{\bf k}_2}|)\tau}\Big(
    \frac{1}{\tau^4}
    +\frac{i (|{\bf k}-{{\bf k}_2}|+k_2)}{\tau^3}
    -\frac{ k_2|{\bf k}-{{\bf k}_2}|}{\tau^2}
    \Big)
\nn\\
&
-b_2({{\bf k}_2})\ b_1({\bf k}-{{\bf k}_2})\
e^{-i (k_2-|{\bf k}-{{\bf k}_2}|)\tau}
\Big(\frac{1}{\tau^4}
    -\frac{i(|{\bf k}-{{\bf k}_2}|-k_2)}{\tau^3}
    +\frac{k_2|{\bf k}-{{\bf k}_2}|}{\tau^2}
    \Big)
 \Big\}
\bigg]
.
\el
}
The homogeneous solution of (\ref{TV2eqFourier1}) for
 a general value of $c_N$
is similar to the 1st order solution  (\ref{v1||solgen}) just by a replacement of
$c_L\rightarrow c_N$
and a new set of integration constants  $d_1,d_2,d_3$.
For simplicity,
we take $c_N^2 =\frac{1}{3}$, and
$c_L$ can be a general value
in the following calculations.
Then, (\ref{TV2eqFourier1}) becomes
\be\label{TV2eqFourier2}
v_{T\mathbf k}^{||(2)'''}
+\frac{1}{\tau} v_{T\mathbf k}^{||(2)''}
+\l(\frac{k^2}{3}
-\frac{4}{\tau^2}\r)v_{T\mathbf k}^{||(2)'}
+\frac{k^2}{3\tau}v_{T\mathbf k}^{||(2)}
=Z_{T\mathbf k}(\tau)
.
\ee
The solution of (\ref{TV2eqFourier2}) is
\bl\label{Tv2ndsol}
v^{||(2)}_{T\mathbf k}
=&
\frac{Q_1(\mathbf k)}{k\tau}
+Q_2(\mathbf k)\l(\frac{2}{k\tau}
    +\frac{i}{\sqrt3}\r)e^{-ik\tau/\sqrt3}
+Q_3(\mathbf k)\l(\frac{2}{k\tau}
    -\frac{i}{\sqrt3}\r)e^{ik\tau/\sqrt3}
\nn\\
&
-\left(
\frac{2}{k\tau} \cos (\frac{k \tau}{\sqrt{3}})
+\frac{1}{\sqrt{3}}\sin (\frac{k \tau}{\sqrt{3}})
\right)
\int^\tau
\Big(
9\cos (\frac{k \tau'}{\sqrt{3}})
+3\sqrt{3}k \tau' \sin (\frac{k \tau'}{\sqrt{3}})
\Big)
\frac{  Z_{T\mathbf k}(\tau') }{ k^3 \tau'} \, d\tau'
\nn\\
&
-\left(
\frac{ 2}{k\tau} \sin (\frac{k \tau}{\sqrt{3}})
-\frac{1}{\sqrt{3}} \cos (\frac{k\tau}{\sqrt{3}})
\right)
    \int^\tau
    \Big( 9 \sin (\frac{k \tau'}{\sqrt{3}})
- 3\sqrt{3} k \tau' \cos (\frac{k \tau'}{\sqrt{3}})\Big)
    \frac{Z_{T\mathbf k}(\tau')}{k^3 \tau'}d\tau'
    \nn\\
&
+\frac{1}{ k\tau}
\int^\tau \frac{
    3\left(k^2 \tau'^2+6\right)}{k^3 \tau'} Z_{T\mathbf k}(\tau')d\tau'
,
\el
where $Q_1(\mathbf k)$, $Q_2(\mathbf k)$, $Q_3(\mathbf k)$
are time-independent coefficients,
and determined by initial conditions,
corresponding to the homogeneous terms of the solution.
Note that the $Q_1(\mathbf k)$ term is a gauge mode
as shall be seen in Sec. 5.
The solution of $\delta_{T{\bf k}}^{(2)}$ is directly
given by (\ref{TdeltaV2}) in
 $\bf k$-space  as
{\allowdisplaybreaks
\bl\label{TdeltaV2solu}
 \delta_{T{\bf k}}^{(2)}=&
-4 v_{T{\bf k}}^{||(2)'}
\nn\\
=&
\frac{4 Q_1(\mathbf k)}{k\tau^2}
+Q_2(\mathbf k)
\l(
\frac{8}{k\tau^2}
+\frac{8 i}{\sqrt3\,\tau}
-\frac{4 k}{3}
\r)e^{-ik\tau/\sqrt3}
\nn\\
    &
+Q_3(\mathbf k)\l(
\frac{8}{k\tau^2}
-\frac{8 i}{\sqrt3\,\tau}
-\frac{4 k}{3}
\r)e^{ik\tau/\sqrt3}
        \nn\\
        &
+\Big(
-\frac{8}{k\tau^2} \cos (\frac{k \tau}{\sqrt{3}})
-\frac{8}{\sqrt{3}\tau} \sin (\frac{k \tau}{\sqrt{3}})
+\frac{4k}{3}\cos (\frac{k \tau}{\sqrt{3}})
\Big)
\int^\tau
\Big(
9\cos (\frac{k \tau'}{\sqrt{3}})
\nn\\
&
+3\sqrt{3}k \tau' \sin (\frac{k \tau'}{\sqrt{3}})
\Big)
\frac{  Z_{T\mathbf k}(\tau') }{ k^3 \tau'} \, d\tau'
\nn\\
&
+\Big(
-\frac{ 8}{k\tau^2} \sin (\frac{k \tau}{\sqrt{3}})
+\frac{ 8}{\sqrt{3}\,\tau} \cos (\frac{k \tau}{\sqrt{3}})
+\frac{4k}{3} \sin (\frac{k\tau}{\sqrt{3}})
\Big)
    \int^\tau
    \Big( 9 \sin (\frac{k \tau'}{\sqrt{3}})
\nn\\
&
- 3\sqrt{3} k \tau' \cos (\frac{k \tau'}{\sqrt{3}})\Big)
    \frac{Z_{T\mathbf k}(\tau')}{k^3 \tau'}d\tau'
  +\frac{1}{ k\tau^2}
\int^\tau \frac{
    12(k^2 \tau'^2+6)}{k^3 \tau'} Z_{T\mathbf k}(\tau')d\tau'
.
\el
}
Integrating (\ref{TVphi2}) yields
the solution of $\phi^{(2)}_{T{\bf k}}$  as
{\allowdisplaybreaks
\bl\label{Tphi2SkSol}
\phi^{(2)}_{T{\bf k}}
&=
-v_{T{\bf k}}^{||(2)'}
-\int^\tau \frac{k^2}{3}v^{||(2)}_{T{\bf k}} d\tau'
- \frac{1}{4}\int^\tau A_{T{\bf k}} d\tau'
+Q_4(\mathbf k)(\tau)
\nn\\
&=
Q_1(\mathbf k)\l(
\frac{1}{k\tau^2}
-\frac{k\ln\tau}{3}
\r)
+Q_2(\mathbf k)
\l(
\frac{2}{k\tau^2}
+\frac{2 i}{\sqrt3\,\tau}
\r)e^{-ik\tau/\sqrt3}
\nn\\
&
+Q_3(\mathbf k)\l(
\frac{2}{k\tau^2}
-\frac{2 i}{\sqrt3\,\tau}
\r)e^{ik\tau/\sqrt3}
+Q_4(\mathbf k)
        \nn\\
        &
-\frac{2 k}{3}\int^\tau
\l[
Q_2(\mathbf k)e^{-ik\tau'/\sqrt3}
+Q_3(\mathbf k)e^{ik\tau'/\sqrt3}
\r]
\frac{d\tau'}{\tau'}
- \frac{1}{4}\int^\tau A_{T{\bf k}}(\tau') d\tau'
\nn\\
&
+\int^\tau\frac{(k^2 \tau'\,^2+6)\ln\tau'+3}{k^2 \tau'} Z_{T\mathbf k}(\tau')d\tau'
+\l(
\frac{1}{ k\tau^2}
-\frac{ k\ln\tau}{3}
\r)\int^\tau \frac{
    3(k^2 \tau'^2+6)}{k^3 \tau'} Z_{T\mathbf k}(\tau')d\tau'
\nn\\
&
-\Big(
\frac{2}{k\tau^2} \cos (\frac{k \tau}{\sqrt{3}})
+\frac{2}{\sqrt{3}\tau} \sin (\frac{k \tau}{\sqrt{3}})
\Big)
\int^\tau
\Big(
9\cos (\frac{k \tau'}{\sqrt{3}})
+3\sqrt{3}k \tau' \sin (\frac{k \tau'}{\sqrt{3}})
\Big)
\frac{  Z_{T\mathbf k}(\tau') }{ k^3 \tau'} \, d\tau'
\nn\\
&
-\Big(
\frac{ 2}{k\tau^2} \sin (\frac{k \tau}{\sqrt{3}})
-\frac{ 2}{\sqrt{3}\,\tau} \cos (\frac{k \tau}{\sqrt{3}})
\Big)
    \int^\tau
    \Big( 9 \sin (\frac{k \tau'}{\sqrt{3}})
- 3\sqrt{3} k \tau' \cos (\frac{k \tau'}{\sqrt{3}})\Big)
    \frac{Z_{T\mathbf k}(\tau')}{k^3 \tau'}d\tau'
\nn\\
&
+\int^\tau
\Big[
\frac{2}{k\tau''} \cos (\frac{k \tau''}{\sqrt{3}})
\int^{\tau''}
\Big(
3\cos (\frac{k \tau'}{\sqrt{3}})
+\sqrt{3}k \tau' \sin (\frac{k \tau'}{\sqrt{3}})
\Big)
\frac{  Z_{T\mathbf k}(\tau') }{ k \tau'} \, d\tau'
\Big]d\tau''
\nn\\
&
+\int^\tau
\Big[
\frac{ 2}{k\tau''} \sin (\frac{k \tau''}{\sqrt{3}})
    \int^{\tau''}
    \Big(
    3 \sin (\frac{k \tau'}{\sqrt{3}})
- \sqrt{3} k \tau' \cos (\frac{k \tau'}{\sqrt{3}})\Big)
    \frac{Z_{T\mathbf k}(\tau')}{k \tau'}d\tau'
\Big]d\tau''
,
\el
}
where (by  $A_{T}=\frac{d}{d\tau}\l[
\frac{2}{3}\chi^{\top(1)}_{lm}\chi^{\top(1)lm}\r]$ in (\ref{TAS}))
\bl\label{TintAS}
\int^\tau A_{T{\bf k}} d\tau'
=&
\frac{2}{3(2\pi)^{3}}\int d^3k_2\,
\Big[
\sum_{s_1={+,\times}}  \sum_{s_2={+,\times}}
{\mathop \epsilon \limits^{s_1}}_{lm}({\bf k}-{{\bf k}_2})
{\mathop \epsilon \limits^{s_2}}{}^{lm}({{\bf k}_2})
{h}_{ {\bf k}-{{\bf k}_2}} {h}_{ {\bf k}_2}
\Big]
\nn\\
=&
\frac{1}{3(2\pi)^{3}}\int d^3k_2\,
\bigg[
\frac{1}{\tau^2 \sqrt{k_2 |{\bf k}-{{\bf k}_2}|}}
\sum_{s_1={+,\times}}  \sum_{s_2={+,\times}}
{\mathop \epsilon \limits^{s_1}}_{lm}({\bf k}-{{\bf k}_2})
{\mathop \epsilon \limits^{s_2}}{}^{lm}({{\bf k}_2})
\nn\\
&
\times
\Big(
b_1({{\bf k}_2})b_2({\bf k}-{{\bf k}_2})e^{i(k_2-|{\bf k}-{{\bf k}_2}|)\tau}
-b_1({{\bf k}_2})b_1({\bf k}-{{\bf k}_2})e^{i(k_2+|{\bf k}-{{\bf k}_2}|)\tau}
\nn\\
&
-b_2({{\bf k}_2}) b_2({\bf k}-{{\bf k}_2})e^{-i(k_2+|{\bf k}-{{\bf k}_2}|)\tau}
+b_2({{\bf k}_2}) b_1({\bf k}-{{\bf k}_2})e^{-i(k_2-|{\bf k}-{{\bf k}_2}|)\tau}
\Big)
\bigg]
.
\el
In the above and the following solutions,
$Q_4(\mathbf k)$ terms are gauge modes,
as shall be seen in Sec. 5.

Finally,
plugging (\ref{TdeltaV2solu}) and (\ref{Tphi2SkSol}) into
(\ref{TEin2th003RD}) in $\bf k$-space,
one has the scalar solution
{\allowdisplaybreaks
\bl\label{Tchi2Ssol}
\chi^{||(2)}_{T{\bf k}}
=&
\frac{18}{k^4\tau} \phi^{(2)'}_{T{\bf k}}
+\frac{6}{k^2}\phi^{(2)}_{T{\bf k}}
+\frac{9}{k^4\tau^2}\delta^{(2)}_{T{\bf k}}
+\frac{3}{k^4} E_{T{\bf k}}
\nn\\
=&
-Q_1(\mathbf k)\frac{2\ln\tau}{k}
+Q_2(\mathbf k)\frac{4\sqrt3 \,i}{k^2\tau}e^{-ik\tau/\sqrt3}
-Q_3(\mathbf k)\frac{4\sqrt3 \,i}{k^2\tau}e^{ik\tau/\sqrt3}
+\frac{6 Q_4(\mathbf k)}{k^2}
        \nn\\
        &
-\frac{4}{k}\int^\tau
\l[
Q_2(\mathbf k)e^{-ik\tau'/\sqrt3}
+Q_3(\mathbf k)e^{ik\tau'/\sqrt3}
\r]
\frac{d\tau'}{\tau'}
\nn\\
&
-\frac{2 \ln\tau}{k }\int^\tau \frac{
    3(k^2 \tau'^2+6)}{k^3 \tau'} Z_{T\mathbf k}(\tau')d\tau'
+\int^\tau\frac{6(k^2 \tau'\,^2+6)\ln\tau'+18}{k^4 \tau'} Z_{T\mathbf k}(\tau')d\tau'
    \nn\\
&
-\frac{4\sqrt3}{k^2\tau} \sin (\frac{k \tau}{\sqrt{3}})
\int^\tau
\Big(
9\cos (\frac{k \tau'}{\sqrt{3}})
+3\sqrt{3}k \tau' \sin (\frac{k \tau'}{\sqrt{3}})
\Big)
\frac{  Z_{T\mathbf k}(\tau') }{ k^3 \tau'} \, d\tau'
\nn\\
&
+\frac{4\sqrt3}{k^2\tau} \cos (\frac{k \tau}{\sqrt{3}})
    \int^\tau
    \Big( 9 \sin (\frac{k \tau'}{\sqrt{3}})
- 3\sqrt{3} k \tau' \cos (\frac{k \tau'}{\sqrt{3}})\Big)
    \frac{Z_{T\mathbf k}(\tau')}{k^3 \tau'}d\tau'
\nn\\
&
+\int^\tau
\Big[
\frac{12}{k^3\tau''} \cos (\frac{k \tau''}{\sqrt{3}})
\int^{\tau''}
\Big(
3\cos (\frac{k \tau'}{\sqrt{3}})
+\sqrt{3}k \tau' \sin (\frac{k \tau'}{\sqrt{3}})
\Big)
\frac{  Z_{T\mathbf k}(\tau') }{ k \tau'} \, d\tau'
\Big]d\tau''
\nn\\
&
+\int^\tau
\Big[
\frac{12}{k^3\tau''} \sin (\frac{k \tau''}{\sqrt{3}})
    \int^{\tau''}
    \Big(
    3 \sin (\frac{k \tau'}{\sqrt{3}})
- \sqrt{3} k \tau' \cos (\frac{k \tau'}{\sqrt{3}})\Big)
    \frac{Z_{T\mathbf k}(\tau')}{k \tau'}d\tau'
\Big]d\tau''
        \nn\\
        &
+\frac{3}{k^4} E_{T{\bf k}}
- \frac{9}{2k^4\tau} A_{T{\bf k}}(\tau)
-\frac{3}{2k^2}\int^\tau A_{T{\bf k}}(\tau') d\tau'
.
\el
}
We have checked that the scalar solutions
(\ref{Tv2ndsol})--(\ref{Tphi2SkSol}) and (\ref{Tchi2Ssol})
satisfy the scalar parts of the evolution equation,
(\ref{TEvo2ndSsTr2RD}) and (\ref{TEvo2ndSsChi1RD}).
So far,
the solutions for the 2nd-order perturbations have all been given.
The expressions of the 2nd-order solutions for the RD stage
generally  contain many  more terms
than those for the MD stage  \cite{WangZhang2017,ZhangQinWang2017}.

The 2nd-order solutions contain several terms of integrals of
the tensor-tensor coupling.
The   $\int d^3 k$  integral
can be treated in a similar way to the paragraph  below (\ref{chi2Ssol}).
As an illustration,
suppose $b_1({\bf k}) \propto k^{N_1}$ and $b_2({\bf k}) \propto k^{N_2}$.
Then we shall have the following typical integration
\bl\label{2ndIntK2}
&
\int  d^3k_2\,
\Big[
e^{-i (k_2+|\mathbf k-\mathbf k_2|)\tau}
(k_2)^{n_3} \big|{\bf k -k_2}\big|^{n_4}
\sum_{s_1={+,\times}}  \sum_{s_2={+,\times}}
{\mathop \epsilon \limits^{s_1}}_{lm}({\bf k}-{{\bf k}_2})
{\mathop \epsilon \limits^{s_2}}{}^{lm}({{\bf k}_2})
\Big]
\nn\\
=&
\int_{K_3}^{K_4}  dk_2\int_0^\pi d\phi \int_0^\pi\sin\theta d\theta\,
\Big[
e^{-i (k_2+|\mathbf k-\mathbf k_2|)\tau}
(k_2)^{n_3+2} \big|{\bf k -k_2}\big|^{n_4}
\nn\\
&
\hspace{6.3cm}
\times\sum_{s_1={+,\times}}  \sum_{s_2={+,\times}}
{\mathop \epsilon \limits^{s_1}}_{lm}({\bf k}-{{\bf k}_2})
{\mathop \epsilon \limits^{s_2}}{}^{lm}({{\bf k}_2})
\Big]
,
\el
where $n_3$ and $n_4$ are linearly related to $N_1$ and $N_2$, and
$K_3$ and $K_4$ are   cutoffs to ensure IR and UV convergence.
Take $\bf k$ as the $z$-axis
and $\theta$ as the angle between $\mathbf k_2$ and $\bf k$.
The  polarization tensors
${\mathop \epsilon \limits^{s}}_{ij}({{\bf k}_2})$
are the same as (\ref{basis tensor 1}) and (\ref{basis tensor 2}),
and ${\mathop \epsilon \limits^{s}}_{ij}(\mathbf k-\mathbf k_2)$
are given by a replacement of
$\theta\rightarrow \arctan\l(\frac{k_2\sin\theta}{k-k_2\cos\theta}\r)$
and $\phi\rightarrow\phi+\pi$
in (\ref{basis tensor 1}) and (\ref{basis tensor 2}).
With these relations,
the integration over $\mathbf k_2$ can be treated
similarly to the paragraph below (\ref{chi2Ssol}).
As for the time integrations  $\int d\tau$,
since $Z_{T {\bf k}}(\tau)$ in (\ref{ZTk}) only has one type of term,
$ \frac{1}{\tau^{N'}} e^{ -i  k\tau}$,
the time integrations in the tensor-tensor coupling case
have similar forms as in (\ref{nonTriv1})--(\ref{nonTriv3})
and can be done numerically.

\section{2nd-order residual gauge modes in synchronous coordinates}

The 2nd-order perturbation solutions in the last section
contain the gauge modes.
In this section,
we shall give the 2nd-order residual gauge transformations
for scalar-tensor and tensor-tensor coupling cases,
and eliminate the 2nd-order residual gauge modes in the solutions.

Consider  the coordinate transformation up to 2nd order
 \cite{Matarrese98,WangZhang2017,WangZhang2ndRD2018}:
\be\label{xmutransf}
x^\mu \rightarrow  \bar x^\mu = x^\mu +\xi^{(1)\mu}
+ \frac{1}{2}\xi^{(1)\mu}_{,\alpha}\xi^{(1)\alpha}
+ \frac{1}{2}\xi ^{(2)\mu},
\ee
where $\xi^{(1)\mu}$ is a  1st-order vector field
and  $\xi^{(2)\mu}$ is a 2nd-order vector field,
  and they  can be written in terms of
their respective parameters,
\bl \label{alpha_r}
&\xi^{(A)0}=\alpha^{(A)},
~~~~\text{with}~~ A=1,2,
\\
&\xi^{(A)i}=\partial^i\beta^{(A)}+d^{ (A)i}
~~~~\text{with}~~ \partial_i d^{(A)i}=0
.    \label{xi_r}
\el
For the RD stage with  $a(\tau)\propto \tau$,
\be \label{xi0trans}
\xi^{(1)0}(\tau, {\bf x}) = \frac{A^{(1)}(\mathbf x)}{\tau},
\ee
\be  \label{gi0}
 \xi^{(1)i} (\tau, {\bf x})= A^{(1),i} \ln\tau
    +C^{||(1),i}(\mathbf x)  +C^{\perp(1) i}(\mathbf x)  ,
\ee
where   $A^{(1)}$ and $C^{||(1)}$
are   $\tau$-independent scalar functions,
$C^{\perp(1)}_{\,i}$ is a $\tau$-independent curl vector,
and all of them are of 1st-order.
[See   (C12) and (C13) of Ref.\cite{WangZhang2ndRD2018}.]
The 1st-order gauge transformations of metric perturbations
between two synchronous coordinate systems
are given in (3.37)--(3.50) of Ref.\cite{WangZhang2ndRD2018}.

The   general 2nd-order synchronous-to-synchronous gauge transformations
in a general RW spacetime
are given in  Appendix C in  Ref.\cite{WangZhang2ndRD2018}.
Here  we apply them to the case of  $a(\tau)\propto \tau$.
First we   shall give the 2nd-order gauge transformations
for the scalar-tensor coupling.
From (C27), (C29),  and (C30) of Ref.\cite{WangZhang2ndRD2018},
keeping only the   $\chi^{\top(1)}_{ij} $-linear-dependent terms,
the 2nd-order vector field  $\xi^{(2)\mu } $ is given
as follows:
\be  \label{STalpha2_3}
\alpha^{(2)}
=\frac{ A^{(2)}(\mathbf x)}{\tau} \, ,
\ee
\be\label{STbeta2_1}
\beta^{(2)}=
\nabla^{-2}\Big[
-A^{(1) ,\, lm}\int^\tau\frac{2\chi^{\top(1)}_{lm}(\tau',{\bf x})}{\tau'}d\tau'
\Big]
+A^{(2)}(\mathbf x) \ln\tau
+C^{||(2)}(\mathbf x)
\,,
\ee
\be \label{STd2_2}
d^{(2)}_i
=
\partial_i\nabla^{-2}\Big[
2A^{(1) ,\,lm}\int^\tau \frac{ \chi^{\top(1)}_{lm}(\tau',{\bf x})}{\tau'}d\tau'
\Big]
-2A^{(1) ,\,l}\int^\tau  \frac{\chi^{\top(1)}_{li}(\tau',{\bf x})}{\tau'}d\tau'
+ C^{\perp(2)}_i (\mathbf x)
,
\ee
where   $A^{(2)}$ and $C^{||(2)}$ are   $\tau$-independent scalar functions,
$C^{\perp(2)}_{\,i}$ is  a $\tau$-independent  curl vector,
all of them are of 2nd-order.
From (C31) (C32) (C33) (C34) of Ref.\cite{WangZhang2ndRD2018}
applied to the RD stage and for the scalar-tensor coupling,
we have the 2nd-order transformation of metric perturbations
\bl\label{phi2TransRD}
\bar \phi^{(2)}_{s(t)} = &
\phi^{(2)}_{s(t)}
+\frac{2\ln\tau}{3}\chi^{\top(1)}_{\,lm}A^{(1),\,lm}
-A^{(1),\,lm}\int^\tau\frac{2\chi^{\top(1)}_{\,lm}(\tau',{\bf x})}{3\tau'}d\tau'
\nn\\
&
+\frac{1}{\tau^2}A^{(2)}
+\frac{\ln\tau}{3}\nabla^2A^{(2)}
+\frac{1}{3}\nabla^2C^{||(2)}
,
\el
\bl\label{chi||2transRD}
 \bar\chi^{||(2)}
=&
 \chi^{||(2)}
- \frac{6}{\tau^2} \nabla^{-2}\nabla^{-2}\big(
\chi^{\top(1)}_{lm}A^{(1),\,lm}
\big)
-  \frac{3}{\tau}   \nabla^{-2} \nabla^{-2}\big(
\chi^{\top(1)'}_{lm}A^{(1),\,lm}
\big)
\nn\\
&
+ \big[\ln\tau\big] \bigg\{
2\nabla^{-2}\big(
\chi^{\top(1)}_{lm}A^{(1),\,lm}
\big)
-3\nabla^{-2}\nabla^{-2}\big(
3\chi^{\top(1)}_{lm,n}A^{(1),\,lmn}
+2\chi^{\top(1)}_{lm}\nabla^2A^{(1),\,lm}
\big)
\bigg\}
\nn\\
&
+   \bigg\{
2\nabla^{-2}\big(
\chi^{\top(1)}_{lm}C^{||(1),\,lm}
\big)
-3\nabla^{-2}\nabla^{-2}\big(
3\chi^{\top(1)}_{lm,n}C^{||(1),\,lmn}
+2\chi^{\top(1)}_{lm}\nabla^2C^{||(1),\,lm}
\big)
\bigg\}
\nn\\
&
-4 \nabla^{-2}\bigg\{
- A^{(1),\,lm} \int^\tau \frac{\chi^{\top(1)}_{lm}(\tau',{\bf x})}{\tau'}d\tau'
\bigg\}
-2\big[\ln\tau\big]A^{(2)}
-2C^{||(2)}
    ,
\el
{
\allowdisplaybreaks
\bl\label{chiPerp2TransRD}
\bar\chi^{\perp(2)}_{ij}
=&
\chi^{\perp(2)}_{ij}
+\frac{2}{\tau^2} \bigg\{
-2\partial_i\nabla^{-2}\big(
\chi^{\top(1)}_{lj}A^{(1),\,l}
\big)
+2\partial_i\partial_j\nabla^{-2}\nabla^{-2}\big(
\chi^{\top(1)}_{lm}A^{(1),\,lm}
\big)
\bigg\}
\nn \\
&
- \frac{1}{\tau} \bigg\{
2\partial_i\nabla^{-2}\big(
\chi^{\top(1)'}_{lj}A^{(1),\,l} \big)
-2\partial_i\partial_j\nabla^{-2}\nabla^{-2}\big(
\chi^{\top(1)'}_{lm}A^{(1),\,lm}
\big)
\bigg\}
\nn \\
&
- \big[\ln\tau \big]\bigg\{
2\partial_i\nabla^{-2}\big(
2\chi^{\top(1)}_{lj,m}A^{(1),\,lm}
+\chi^{\top(1)}_{lm}A^{(1),\,lm}_{,j}
+\chi^{\top(1)}_{lj}\nabla^2A^{(1),\,l}
\big)
\nn\\
&
-2\partial_i\partial_j\nabla^{-2}\nabla^{-2}\big(
3\chi^{\top(1)}_{lm,n}A^{(1),\,lmn}
+2\chi^{\top(1)}_{lm}\nabla^2A^{(1),\,lm}
\big)
\bigg\}
\nn\\
&
+\bigg\{
-2\partial_i\nabla^{-2}\big(
2\chi^{\top(1)}_{lj,m}C^{||(1),\,lm}
+\chi^{\top(1)}_{lm}C^{||(1),\,lm}_{,j}
+\chi^{\top(1)}_{lj}\nabla^2C^{||(1),\,l}
\big)
\nn\\
&
+2\partial_i\partial_j\nabla^{-2}\nabla^{-2}\big(
+3\chi^{\top(1)}_{lm,n}C^{||(1),\,lmn}
+2\chi^{\top(1)}_{lm}\nabla^2C^{||(1),\,lm}
\big)
\bigg\}
\nn\\
&
+\partial_i\bigg\{
A^{(1),\,l}\int^\tau \frac{2\chi^{\top(1)}_{lj}(\tau',{\bf x})}{\tau'}d\tau'
\bigg\}
+\partial_i\partial_j\nabla^{-2}\bigg\{
-A^{(1),\,lm}\int^\tau \frac{2\chi^{\top(1)}_{lm}(\tau' ,{\bf x})}{\tau'}d\tau
\bigg\}
\nn\\
&
  -C^{\perp(2)}_{i,j}
+(i \leftrightarrow j ) \ ,
\el
}
{\allowdisplaybreaks
\bl\label{chiT2transRD}
\bar\chi^{\top(2)}_{ij}
=&
\chi^{\top(2)}_{ij}
+\frac{2}{\tau^2}\bigg\{
-\delta_{ij}\nabla^{-2}\big(
\chi^{\top(1)}_{lm}A^{(1),\,lm}
\big)
-2\chi^{\top(1)}_{ij}A^{(1)}
+2\partial_i\nabla^{-2}\big(
\chi^{\top(1)}_{lj}A^{(1),\,l}
\big)
\nn\\
&
+2\partial_j\nabla^{-2}\big(
\chi^{\top(1)}_{li}A^{(1),\,l}
\big)
-\partial_i\partial_j\nabla^{-2}\nabla^{-2}\big(
\chi^{\top(1)}_{lm}A^{(1),\,lm}
\big)
\bigg\}
\nn\\
&
+\frac{1}{\tau}\bigg\{
-\delta_{ij}\nabla^{-2}\big(
\chi^{\top(1)'}_{lm}A^{(1),\,lm}
\big)
-2\chi^{\top(1)'}_{ij}A^{(1)}
+2\partial_i\nabla^{-2}\big(
\chi^{\top(1)'}_{lj}A^{(1),\,l}
\big)
\nn\\
&
+2\partial_j\nabla^{-2}\big(
\chi^{\top(1)'}_{li}A^{(1),\,l}
\big)
-\partial_i\partial_j\nabla^{-2}\nabla^{-2}\big(
\chi^{\top(1)'}_{lm}A^{(1),\,lm}
\big) \bigg\}
\nn\\
&
+\big[\ln\tau\big]\bigg\{
\delta_{ij}\nabla^{-2}\big(
\chi^{\top(1)}_{lm,n}A^{(1),\,lmn}
+2 A^{(1),\,lm}\nabla^2\chi^{\top(1)}_{lm}
\big)
-2\chi^{\top(1)}_{ij,\,l}A^{(1),\,l}
\nn\\
&
-2\chi^{\top(1)}_{li}A^{(1),\,l}_{,j}
-2\chi^{\top(1)}_{lj}A^{(1),\,l}_{,i}
    \nn\\
    &
+2\partial_i\nabla^{-2}\big(
2\chi^{\top(1)}_{lj,m}A^{(1),\,lm}
+\chi^{\top(1)}_{lm}A^{(1),\,lm}_{,j}
+\chi^{\top(1)}_{lj}\nabla^2A^{(1),\,l}
\big)
    \nn\\
    &
+2\partial_j\nabla^{-2}\big(
2\chi^{\top(1)}_{li,m}A^{(1),\,lm}
+\chi^{\top(1)}_{lm}A^{(1),\,lm}_{,i}
+\chi^{\top(1)}_{li}\nabla^2A^{(1),\,l}
\big)
\nn\\
&
-\partial_i\partial_j\nabla^{-2}\nabla^{-2}\big(
7\chi^{\top(1)}_{lm,n}A^{(1),\,lmn}
+4\chi^{\top(1)}_{lm}\nabla^2A^{(1),\,lm}
+2 A^{(1),\,lm}\nabla^2\chi^{\top(1)}_{lm}
\big)
\bigg\}
\nn\\
&
+\bigg\{
\delta_{ij}\nabla^{-2}\big(
\chi^{\top(1)}_{lm,n}C^{||(1),\,lmn}
+2C^{||(1),\,lm}\nabla^2\chi^{\top(1)}_{lm}
\big)
\nn\\
&
-2 \chi^{\top(1)}_{ij,\,l}C^{||(1),\,l}
-2\chi^{\top(1)}_{li}C^{||(1),\,l}_{,j}
-2\chi^{\top(1)}_{lj}C^{||(1),\,l}_{,i}
\nn\\
&
+2 \partial_i\nabla^{-2}\big(
2\chi^{\top(1)}_{lj,m}C^{||(1),\,lm}
+\chi^{\top(1)}_{lm}C^{||(1),\,lm}_{,j}
+\chi^{\top(1)}_{lj}\nabla^2C^{||(1),\,l}
\big)
\nn\\
&
+2 \partial_j\nabla^{-2}\big(
2\chi^{\top(1)}_{li,m}C^{||(1),\,lm}
+\chi^{\top(1)}_{lm}C^{||(1),\,lm}_{,i}
+\chi^{\top(1)}_{li}\nabla^2C^{||(1),\,l}
\big)
\nn\\
&
-\partial_i\partial_j\nabla^{-2}\nabla^{-2}\big(
7\chi^{\top(1)}_{lm,n}C^{||(1),\,lmn}
+4\chi^{\top(1)}_{lm}\nabla^2C^{||(1),\,lm}
+2 C^{||(1),\,lm}\nabla^2\chi^{\top(1)}_{lm}
\big)
\bigg\}
   .
\el
}
Equation (\ref {chiT2transRD}) tells that
the transformation of 2nd-order tensor
involves only  $\xi^{(1)\mu}$,
 independent of  the 2nd-order vector field $\xi^{(2)\mu}$.
So the 2nd-order tensor is effectively gauge invariant
under the 2nd-order gauge transformations.
This is similar to the case of the MD stage in Ref.\cite{WangZhang2017}.

From (C38), (C39), and (C42)--(C44) in  Ref.\cite{WangZhang2ndRD2018},
applied to the  RD stage and for the scalar-tensor coupling,
we have the  2nd-order transformation for the density contrast and the velocity
as follows:
\be\label{delta2TransRD}
\bar\delta^{(2)}_{s(t)}
=
\delta^{(2)}_{s(t)}
+\frac{4}{\tau^2}A^{(2)}(\mathbf x)
,
\ee
\be\label{v||2TransRD}
\bar v^{||(2)}_{s(t)}
=
v^{||(2)}_{s(t)}
+\frac{1}{\tau}
\nabla^{-2}
\big(
-2 A^{(1),\,lm}\chi^{\top(1)}_{lm}
\big)
+\frac{A^{(2)}(\mathbf x)}{\tau}
,
\ee
\be\label{vperp2TransRD}
\bar v^{\perp(2)}_{s(t)i}
=
v^{\perp(2)}_{s(t)i}
+\frac{1}{\tau}
\bigg[
-2 A^{(1),\,l}\chi^{\top(1)}_{li}
+\partial^i\nabla^{-2}
\big(
2 A^{(1),\,lm}\chi^{\top(1)}_{lm}
\big)
\bigg]
.
\ee

The above   general synchronous-to-synchronous transformations contain
two vector fields $\xi^{(1)\mu}$ and $\xi^{(2)\mu}$.
However, in applications,
 distinctions should be made between $\xi^{(1)\mu}$ and $\xi^{(2)\mu}$.
If one sets $\xi^{(2)\mu}=0$  \cite{Abramo1997,HwangNoh2012},
only  $\xi^{(1)\mu}$ remains,
which ensures $\bar g^{(1)}_{00}=0$, $\bar g^{(1)}_{0i}=0$
and keeps the obtained 1st-order solutions as gauge invariant;
one has no freedom
to make  $\bar g^{(2)}_{00}=0$ and  $\bar g^{(2)}_{0i}=0$ anymore,
because   $\xi^{(1)\mu}$ has been already fixed.
Thus,  this kind of 2nd-order transformations
is not   effective  when $\xi^{(2)\mu}=0$.
The more interesting case of 2nd-order gauge transformation
is   when the 1st-order solutions are held fixed
and only the 2nd-order $\xi^{(2)\mu}$  is allowed \cite{Gleiser1996}.
In this case
one simply sets $\xi^{(1)\mu}=0$ but $\xi^{(2)\mu} \ne 0$ in
 (\ref{STalpha2_3})--(\ref{STd2_2}),
which reduce to
\be\label{alpha2RD2}
\alpha^{(2)}(\tau,\mathbf x)
=\frac{ A^{(2)}(\mathbf x)}{\tau} \, ,
\ee
\be\label{beta2RD2}
\beta^{(2)}(\tau,\mathbf x)=
 A^{(2)}({\bf x})  \ln\tau
+C^{||(2)} ({\bf x})  \,,
\ee
\be
\label{d2RD2}
d^{(2)}_i(\mathbf x)
=C^{\perp(2)}_i ({\bf x}) \, ,
\ee
and (\ref{phi2TransRD})--(\ref{vperp2TransRD})  reduce to
\be\label{phi2TransRD2}
\bar \phi^{(2)}_{s(t)} (\tau,\mathbf x) =
\phi^{(2)}_{s(t)}(\tau,\mathbf x)
+\frac{A^{(2)}({\bf x})}{\tau^2}
+\frac{\ln\tau}{3}\nabla^2A^{(2)}({\bf x})
+\frac{1}{3}\nabla^2C^{||(2)}({\bf x})
,
\ee
\be\label{chi||2transRD2}
 \bar\chi^{||(2)}_{s(t)}(\tau,\mathbf x)
=
 \chi^{||(2)}_{s(t)}(\tau,\mathbf x)
 -2   A^{(2)}({\bf x}) \ln\tau
-2C^{||(2)}({\bf x})
    ,
\ee
\be\label{chiPerp2TransRD2}
\bar\chi^{\perp(2)}_{s(t)ij}(\tau,\mathbf x)
=
\chi^{\perp(2)}_{s(t)ij}(\tau,\mathbf x)
 -\partial_{j}C^{\perp(2)}_{i}({\bf x})
-\partial_{i}C^{\perp(2)}_{j}({\bf x})\ ,
\ee
\bl\label{chiT2transRD2}
\bar\chi^{\top(2)}_{s(t)ij}(\tau,\mathbf x)
=
\chi^{\top(2)}_{s(t)ij}(\tau,\mathbf x),
\el
\be\label{delta2TransRD2}
\bar\delta^{(2)}_{s(t)}(\tau,\mathbf x)
=
\delta^{(2)}_{s(t)}(\tau,\mathbf x)
+\frac{4}{\tau^2}A^{(2)}({\bf x})
,
\ee
\be\label{v||2TransRD2}
\bar v^{||(2)}_{s(t)}(\tau,\mathbf x)
=
v^{||(2)}_{s(t)}(\tau,\mathbf x)
+\frac{A^{(2)}({\bf x})}{\tau}
,
\ee
\be\label{vperp2TransRD2}
\bar v^{\perp(2)}_{s(t)i}(\tau,\mathbf x)
=
v^{\perp(2)}_{s(t)i}(\tau,\mathbf x)
,
\ee
which has the same structure as the 1st-order residual transforms
\cite{WangZhang2ndRD2018}.
It is seen that $\chi^{\top(2)}_{s(t)ij}$
and $v^{\perp(2)i}_{s(t)}$
are invariant within synchronous coordinates.
Equation (\ref{chiPerp2TransRD2}) tells
that the  $\tau$-independent term  $q_{1ij}({\bf x})$ in
the vector solution   (\ref{chiVecSol})
is a gauge term and can be eliminated by
a choice of $C^{\perp(2)}_{j}$,
so that the gauge-invariant vector solution  is
\be\label{chiVecSolphy}
\chi^{\perp(2)}_{s(t)ij}({\bf x},\tau)
=\frac{q_{2ij}({\bf x})}{\tau}
+\int^\tau \frac{d\tau'}{\tau^{'2}}
    \int^{\tau'}2\tau^{''2}\,V_{s(t)ij}({\bf x},\tau'')d\tau''
\,.
\ee

To identify the residual gauge modes in the 2nd-order scalar solutions,
we write (\ref{phi2TransRD2}), (\ref{chi||2transRD2}),
(\ref{delta2TransRD2}), and (\ref{v||2TransRD2})
 in $\bf k$-space:
\be\label{phi2TransRD2F}
\bar \phi^{(2)}_{s(t){\bf k}} (\tau) =
\phi^{(2)}_{s(t){\bf k}} (\tau)
+A^{(2)}_{\bf k}\l(
\frac{1}{\tau^2}
-\frac{k^2}{3}\ln\tau
\r)
-\frac{k^2}{3}C^{||(2)}_{\bf k}
,
\ee
\be\label{chi||2transRD2F}
 \bar\chi^{||(2)}_{s(t){\bf k}} (\tau)
=
 \chi^{||(2)}_{s(t){\bf k}} (\tau)
 -2A^{(2)}_{\bf k}\ln\tau
-2C^{||(2)}_{\bf k}
    ,
\ee
\be\label{delta2TransRD2F}
\bar\delta^{(2)}_{s(t){\bf k}} (\tau)
=
\delta^{(2)}_{s(t){\bf k}} (\tau)
+\frac{4}{\tau^2}A^{(2)}_{\bf k}
,
\ee
\be\label{v||2TransRD2F}
\bar v^{||(2)}_{s(t){\bf k}} (\tau)
=
v^{||(2)}_{s(t){\bf k}} (\tau)
+\frac{A^{(2)}_{\bf k}}{\tau}
.
\ee
Comparing (\ref{phi2TransRD2F})--(\ref{v||2TransRD2F})
with the solutions (\ref{phi2SkSol}),  (\ref{chi2Ssol}),
(\ref{deltaV2solu}), and (\ref{v2ndsol}), respectively,
tells us that $P_1(\mathbf k)$ and $P_4(\mathbf k)$ terms
in  (\ref{phi2SkSol}),   (\ref{chi2Ssol}), (\ref{deltaV2solu}), and (\ref{v2ndsol})
are gauge terms,
which can be eliminated  simultaneously by choosing
\be\label{A2Transk}
 A^{(2)}_{\mathbf k}
=-\frac{ P_1(\mathbf k)}{k},
\ee
\be\label{C2||Transk}
C^{||(2)}_{\mathbf k}=\frac{3 P_4(\mathbf k)}{k^2} .
\ee
Thus,  the gauge-invariant  modes of the 2nd-order scalar
perturbations are
{\allowdisplaybreaks
\bl\label{phi2SkSolphy}
\phi&^{(2)}_{s(t){\bf k}}
=
P_2(\mathbf k)
\l(
\frac{2}{k\tau^2}
+\frac{2 i}{\sqrt3\,\tau}
\r)e^{-ik\tau/\sqrt3}
+P_3(\mathbf k)\l(
\frac{2}{k\tau^2}
-\frac{2 i}{\sqrt3\,\tau}
\r)e^{ik\tau/\sqrt3}
        \nn\\
        &
-\frac{2 k}{3}\int^\tau
\l[
P_2(\mathbf k)e^{-ik\tau'/\sqrt3}
+P_3(\mathbf k)e^{ik\tau'/\sqrt3}
\r]
\frac{d\tau'}{\tau'}
+\frac{3}{4}F^{||}_{s(t){\bf k}}(\tau)
- \frac{1}{4}\int^\tau A_{s(t){\bf k}}(\tau') d\tau'
\nn\\
&
+\int^\tau\frac{(k^2 \tau'\,^2+6)\ln\tau'+3}{k^2 \tau'} Z_{s(t)\mathbf k}(\tau')d\tau'
+\l(
\frac{1}{ k\tau^2}
-\frac{ k\ln\tau}{3}
\r)\int^\tau \frac{
    3(k^2 \tau'^2+6)}{k^3 \tau'} Z_{s(t)\mathbf k}(\tau')d\tau'
\nn\\
&
-\bigg(
\frac{2}{k\tau^2} \cos (\frac{k \tau}{\sqrt{3}})
+\frac{2}{\sqrt{3}\tau} \sin (\frac{k \tau}{\sqrt{3}})
\bigg)
\int^\tau
\bigg(
9\cos (\frac{k \tau'}{\sqrt{3}})
+3\sqrt{3}k \tau' \sin (\frac{k \tau'}{\sqrt{3}})
\bigg)
\frac{  Z_{s(t)\mathbf k}(\tau') }{ k^3 \tau'} \, d\tau'
\nn\\
&
-\bigg(
\frac{ 2}{k\tau^2} \sin (\frac{k \tau}{\sqrt{3}})
-\frac{ 2}{\sqrt{3}\,\tau} \cos (\frac{k \tau}{\sqrt{3}})
\bigg)
    \int^\tau
    \bigg( 9 \sin (\frac{k \tau'}{\sqrt{3}})
- 3\sqrt{3} k \tau' \cos (\frac{k \tau'}{\sqrt{3}})\bigg)
    \frac{Z_{s(t)\mathbf k}(\tau')}{k^3 \tau'}d\tau'
\nn\\
&
+\int^\tau
\bigg[
\frac{2}{k\tau''} \cos (\frac{k \tau''}{\sqrt{3}})
\int^{\tau''}
\bigg(
3\cos (\frac{k \tau'}{\sqrt{3}})
+\sqrt{3}k \tau' \sin (\frac{k \tau'}{\sqrt{3}})
\bigg)
\frac{  Z_{s(t)\mathbf k}(\tau') }{ k \tau'} \, d\tau'
\bigg]d\tau''
\nn\\
&
+\int^\tau
\bigg[
\frac{ 2}{k\tau''} \sin (\frac{k \tau''}{\sqrt{3}})
    \int^{\tau''}
    \bigg(
    3 \sin (\frac{k \tau'}{\sqrt{3}})
- \sqrt{3} k \tau' \cos (\frac{k \tau'}{\sqrt{3}})\bigg)
    \frac{Z_{s(t)\mathbf k}(\tau')}{k \tau'}d\tau'
\bigg]d\tau''
,
\el
\bl\label{chi2Ssolphy}
\chi^{||(2)}_{s(t){\bf k}}
=&
P_2(\mathbf k)\frac{4\sqrt3 \,i}{k^2\tau}e^{-ik\tau/\sqrt3}
-P_3(\mathbf k)\frac{4\sqrt3 \,i}{k^2\tau}e^{ik\tau/\sqrt3}
        \nn\\
        &
-\frac{4}{k}\int^\tau
\l[
P_2(\mathbf k)e^{-ik\tau'/\sqrt3}
+P_3(\mathbf k)e^{ik\tau'/\sqrt3}
\r]
\frac{d\tau'}{\tau'}
\nn\\
&
-\frac{2 \ln\tau}{k }\int^\tau \frac{
    3(k^2 \tau'^2+6)}{k^3 \tau'} Z_{s(t)\mathbf k}(\tau')d\tau'
+\int^\tau\frac{6(k^2 \tau'\,^2+6)\ln\tau'+18}{k^4 \tau'} Z_{s(t)\mathbf k}(\tau')d\tau'
    \nn\\
&
-\frac{4\sqrt3}{k^2\tau} \sin (\frac{k \tau}{\sqrt{3}})
\int^\tau
\bigg(
9\cos (\frac{k \tau'}{\sqrt{3}})
+3\sqrt{3}k \tau' \sin (\frac{k \tau'}{\sqrt{3}})
\bigg)
\frac{  Z_{s(t)\mathbf k}(\tau') }{ k^3 \tau'} \, d\tau'
\nn\\
&
+\frac{4\sqrt3}{k^2\tau} \cos (\frac{k \tau}{\sqrt{3}})
    \int^\tau
    \bigg( 9 \sin (\frac{k \tau'}{\sqrt{3}})
- 3\sqrt{3} k \tau' \cos (\frac{k \tau'}{\sqrt{3}})\bigg)
    \frac{Z_{s(t)\mathbf k}(\tau')}{k^3 \tau'}d\tau'
\nn\\
&
+\int^\tau
\bigg[
\frac{12}{k^3\tau''} \cos (\frac{k \tau''}{\sqrt{3}})
\int^{\tau''}
\bigg(
3\cos (\frac{k \tau'}{\sqrt{3}})
+\sqrt{3}k \tau' \sin (\frac{k \tau'}{\sqrt{3}})
\bigg)
\frac{  Z_{s(t)\mathbf k}(\tau') }{ k \tau'} \, d\tau'
\bigg]d\tau''
\nn\\
&
+\int^\tau
\bigg[
\frac{12}{k^3\tau''} \sin (\frac{k \tau''}{\sqrt{3}})
    \int^{\tau''}
    \bigg(
    3 \sin (\frac{k \tau'}{\sqrt{3}})
- \sqrt{3} k \tau' \cos (\frac{k \tau'}{\sqrt{3}})\bigg)
    \frac{Z_{s(t)\mathbf k}(\tau')}{k \tau'}d\tau'
\bigg]d\tau''
        \nn\\
        &
+\frac{3}{k^4} E_{s(t){\bf k}}
+\frac{27}{2k^4\tau}F^{||'}_{s(t){\bf k}}(\tau)
+\frac{27}{k^4\tau^2} F^{||}_{s(t){\bf k}}
+\frac{9}{2k^2}F^{||}_{s(t){\bf k}}(\tau)
        \nn\\
        &
- \frac{9}{2k^4\tau} A_{s(t){\bf k}}(\tau)
-\frac{3}{2k^2}\int^\tau A_{s(t){\bf k}}(\tau') d\tau'
,
\el
\bl\label{deltaV2soluphy}
\delta_{s(t){\bf k}}^{(2)}
=&
P_2(\mathbf k)
\l(
\frac{8}{k\tau^2}
+\frac{8 i}{\sqrt3\,\tau}
-\frac{4 k}{3}
\r)e^{-ik\tau/\sqrt3}
+P_3(\mathbf k)\l(
\frac{8}{k\tau^2}
-\frac{8 i}{\sqrt3\,\tau}
-\frac{4 k}{3}
\r)e^{ik\tau/\sqrt3}
        \nn\\
        &
+3 F^{||}_{s(t){\bf k}}
+\bigg(
-\frac{8}{k\tau^2} \cos (\frac{k \tau}{\sqrt{3}})
-\frac{8}{\sqrt{3}\tau} \sin (\frac{k \tau}{\sqrt{3}})
+\frac{4k}{3}\cos (\frac{k \tau}{\sqrt{3}})
\bigg)
\int^\tau
\bigg(
9\cos (\frac{k \tau'}{\sqrt{3}})
\nn\\
&
+3\sqrt{3}k \tau' \sin (\frac{k \tau'}{\sqrt{3}})
\bigg)
\frac{  Z_{s(t)\mathbf k}(\tau') }{ k^3 \tau'} \, d\tau'
\nn\\
&
+\bigg(
-\frac{ 8}{k\tau^2} \sin (\frac{k \tau}{\sqrt{3}})
+\frac{ 8}{\sqrt{3}\,\tau} \cos (\frac{k \tau}{\sqrt{3}})
+\frac{4k}{3} \sin (\frac{k\tau}{\sqrt{3}})
\bigg)
    \int^\tau
    \bigg( 9 \sin (\frac{k \tau'}{\sqrt{3}})
\nn\\
&
- 3\sqrt{3} k \tau' \cos (\frac{k \tau'}{\sqrt{3}})\bigg)
    \frac{Z_{s(t)\mathbf k}(\tau')}{k^3 \tau'}d\tau'
    \nn\\
&
+\frac{1}{ k\tau^2}
\int^\tau \frac{
    12(k^2 \tau'^2+6)}{k^3 \tau'} Z_{s(t)\mathbf k}(\tau')d\tau'
,
\el
\bl\label{v2ndsolphy}
v&^{||(2)}_{s(t)\mathbf k}
=
P_2(\mathbf k)\l(\frac{2}{k\tau}
    +\frac{i}{\sqrt3}\r)e^{-ik\tau/\sqrt3}
+P_3(\mathbf k)\l(\frac{2}{k\tau}
    -\frac{i}{\sqrt3}\r)e^{ik\tau/\sqrt3}
\nn\\
&
-\left(
\frac{2}{k\tau} \cos (\frac{k \tau}{\sqrt{3}})
+\frac{1}{\sqrt{3}}\sin (\frac{k \tau}{\sqrt{3}})
\right)
\int^\tau
\bigg(
9\cos (\frac{k \tau'}{\sqrt{3}})
+3\sqrt{3}k \tau' \sin (\frac{k \tau'}{\sqrt{3}})
\bigg)
\frac{  Z_{s(t)\mathbf k}(\tau') }{ k^3 \tau'} \, d\tau'
\nn\\
&
-\left(
\frac{ 2}{k\tau} \sin (\frac{k \tau}{\sqrt{3}})
-\frac{1}{\sqrt{3}} \cos (\frac{k\tau}{\sqrt{3}})
\right)
    \int^\tau
    \bigg( 9 \sin (\frac{k \tau'}{\sqrt{3}})
- 3\sqrt{3} k \tau' \cos (\frac{k \tau'}{\sqrt{3}})\bigg)
    \frac{Z_{s(t)\mathbf k}(\tau')}{k^3 \tau'}d\tau'
    \nn\\
&
+\frac{1}{ k\tau}
\int^\tau \frac{
    3\left(k^2 \tau'^2+6\right)}{k^3 \tau'} Z_{s(t)\mathbf k}(\tau')d\tau'
,
\el
}
where $Z_{s(t)}$, $E_{s(t)}$, $A_{s(t)}$, $F^{||}_{s(t)}$
are in (\ref{ZSall}) (\ref{ES1RD}) (\ref{AS}) (\ref{FSnoncu}).

Next is the case of  tensor-tensor coupling.
The analysis is  similar to the above paragraphs.
In particular,
the 2nd-order residual gauge transformation
is effectively implemented only by the 2nd-order  vector field $\xi^{(2)\mu}$,
and the gauge transformations are
\be\label{Tphi2TransRD2}
\bar \phi^{(2)}_{T} (\tau,\mathbf x) =
\phi^{(2)}_{T}(\tau,\mathbf x)
+\frac{A^{(2)}({\bf x})}{\tau^2}
+\frac{\ln\tau}{3}\nabla^2A^{(2)}({\bf x})
+\frac{1}{3}\nabla^2C^{||(2)}({\bf x})
,
\ee
\be\label{Tchi||2transRD2}
 \bar\chi^{||(2)}_{T}(\tau,\mathbf x)
=
 \chi^{||(2)}_{T}(\tau,\mathbf x)
 -2   A^{(2)}({\bf x}) \ln\tau
-2C^{||(2)}({\bf x})
    ,
\ee
\be\label{TchiPerp2TransRD2}
\bar\chi^{\perp(2)}_{T\,ij}(\tau,\mathbf x)
=
\chi^{\perp(2)}_{T\,ij}(\tau,\mathbf x)
 -\partial_{j}C^{\perp(2)}_{i}({\bf x})
-\partial_{i}C^{\perp(2)}_{j}({\bf x})\ ,
\ee
\bl\label{TchiT2transRD2}
\bar\chi^{\top(2)}_{T\,ij}(\tau,\mathbf x)
=
\chi^{\top(2)}_{T\,ij}(\tau,\mathbf x),
\el
\be\label{Tdelta2TransRD2}
\bar\delta^{(2)}_{T}(\tau,\mathbf x)
=
\delta^{(2)}_{T}(\tau,\mathbf x)
+\frac{4}{\tau^2}A^{(2)}({\bf x})
,
\ee
\be\label{Tv||2TransRD2}
\bar v^{||(2)}_{T}(\tau,\mathbf x)
=
v^{||(2)}_{T}(\tau,\mathbf x)
+\frac{A^{(2)}({\bf x})}{\tau}
,
\ee
\be\label{Tvperp2TransRD2}
\bar v^{\perp(2)}_{T\,i}(\tau,\mathbf x)
=
v^{\perp(2)}_{T\,i}(\tau,\mathbf x)
.
\ee
Again,
$\chi^{\top(2)}_{T\,ij}$ in (\ref{Tsolgw})
and $v^{\perp(2)}_{T\,i}$  in (\ref{TVperp2ndSol})
are invariant within synchronous coordinates.
By  (\ref{TchiPerp2TransRD2}),
the  $q_{3ij}({\bf x})$ term in
the solution  (\ref{TchiVecSol})
is a gauge term and can be eliminated,
so that  the gauge-invariant vector solution   is
\be\label{TchiVecSolphy}
\chi^{\perp(2)}_{T\,ij}({\bf x},\tau)
=\frac{q_{4ij}({\bf x})}{\tau}
+\int^\tau \frac{d\tau'}{\tau^{'2}}
    \int^{\tau'}2\tau^{''2}\,V_{T\,ij}({\bf x},\tau'')d\tau''
\,.
\ee

To identify the residual gauge modes in the 2nd-order scalar solutions,
we write (\ref{Tphi2TransRD2}), (\ref{Tchi||2transRD2}),
(\ref{Tdelta2TransRD2}), and (\ref{Tv||2TransRD2})
 in $\bf k$-space:
\be\label{Tphi2TransRD2F}
\bar \phi^{(2)}_{T{\bf k}} (\tau) =
\phi^{(2)}_{T{\bf k}} (\tau)
+A^{(2)}_{\bf k}\l(
\frac{1}{\tau^2}
-\frac{k^2}{3}\ln\tau
\r)
-\frac{k^2}{3}C^{||(2)}_{\bf k}
,
\ee
\be\label{Tchi||2transRD2F}
 \bar\chi^{||(2)}_{T{\bf k}} (\tau)
=
 \chi^{||(2)}_{T{\bf k}} (\tau)
 -2A^{(2)}_{\bf k}\ln\tau
-2C^{||(2)}_{\bf k}
    ,
\ee
\be\label{Tdelta2TransRD2F}
\bar\delta^{(2)}_{T{\bf k}} (\tau)
=
\delta^{(2)}_{T{\bf k}} (\tau)
+\frac{4}{\tau^2}A^{(2)}_{\bf k}
,
\ee
\be\label{Tv||2TransRD2F}
\bar v^{||(2)}_{T{\bf k}} (\tau)
=
v^{||(2)}_{T{\bf k}} (\tau)
+\frac{A^{(2)}_{\bf k}}{\tau}
.
\ee
Comparing (\ref{Tphi2TransRD2F})--(\ref{Tv||2TransRD2F})
with the solutions (\ref{Tphi2SkSol}), (\ref{Tchi2Ssol}),
(\ref{TdeltaV2solu}), and (\ref{Tv2ndsol}), respectively,
tells us that the $Q_1(\mathbf k)$ and $Q_4(\mathbf k)$ terms
in  (\ref{Tphi2SkSol}),   (\ref{Tchi2Ssol}), (\ref{TdeltaV2solu}), and (\ref{Tv2ndsol})
are gauge terms,
which can be removed simultaneously by choosing
\be\label{TA2Transk}
 A^{(2)}_{\mathbf k}
=-\frac{ Q_1(\mathbf k)}{k},
\ee
\be\label{TC2||Transk}
C^{||(2)}_{\mathbf k}=\frac{3 Q_4(\mathbf k)}{k^2} .
\ee
Thus,  the gauge-invariant solutions of the 2nd-order scalar
perturbations are
{\allowdisplaybreaks
\bl\label{Tphi2SkSolphy}
\phi^{(2)}_{T{\bf k}}
&=
Q_2(\mathbf k)
\l(
\frac{2}{k\tau^2}
+\frac{2 i}{\sqrt3\,\tau}
\r)e^{-ik\tau/\sqrt3}
+Q_3(\mathbf k)\l(
\frac{2}{k\tau^2}
-\frac{2 i}{\sqrt3\,\tau}
\r)e^{ik\tau/\sqrt3}
        \nn\\
        &
-\frac{2 k}{3}\int^\tau
\l[
Q_2(\mathbf k)e^{-ik\tau'/\sqrt3}
+Q_3(\mathbf k)e^{ik\tau'/\sqrt3}
\r]
\frac{d\tau'}{\tau'}
- \frac{1}{4}\int^\tau A_{T{\bf k}}(\tau') d\tau'
\nn\\
&
+\int^\tau\frac{(k^2 \tau'\,^2+6)\ln\tau'+3}{k^2 \tau'} Z_{T\mathbf k}(\tau')d\tau'
+\l(
\frac{1}{ k\tau^2}
-\frac{ k\ln\tau}{3}
\r)\int^\tau \frac{
    3(k^2 \tau'^2+6)}{k^3 \tau'} Z_{T\mathbf k}(\tau')d\tau'
\nn\\
&
-\Big(
\frac{2}{k\tau^2} \cos (\frac{k \tau}{\sqrt{3}})
+\frac{2}{\sqrt{3}\tau} \sin (\frac{k \tau}{\sqrt{3}})
\Big)
\int^\tau
\Big(
9\cos (\frac{k \tau'}{\sqrt{3}})
+3\sqrt{3}k \tau' \sin (\frac{k \tau'}{\sqrt{3}})
\Big)
\frac{  Z_{T\mathbf k}(\tau') }{ k^3 \tau'} \, d\tau'
\nn\\
&
-\Big(
\frac{ 2}{k\tau^2} \sin (\frac{k \tau}{\sqrt{3}})
-\frac{ 2}{\sqrt{3}\,\tau} \cos (\frac{k \tau}{\sqrt{3}})
\Big)
    \int^\tau
    \Big( 9 \sin (\frac{k \tau'}{\sqrt{3}})
- 3\sqrt{3} k \tau' \cos (\frac{k \tau'}{\sqrt{3}})\Big)
    \frac{Z_{T\mathbf k}(\tau')}{k^3 \tau'}d\tau'
\nn\\
&
+\int^\tau
\Big[
\frac{2}{k\tau''} \cos (\frac{k \tau''}{\sqrt{3}})
\int^{\tau''}
\Big(
3\cos (\frac{k \tau'}{\sqrt{3}})
+\sqrt{3}k \tau' \sin (\frac{k \tau'}{\sqrt{3}})
\Big)
\frac{  Z_{T\mathbf k}(\tau') }{ k \tau'} \, d\tau'
\Big]d\tau''
\nn\\
&
+\int^\tau
\Big[
\frac{ 2}{k\tau''} \sin (\frac{k \tau''}{\sqrt{3}})
    \int^{\tau''}
    \Big(
    3 \sin (\frac{k \tau'}{\sqrt{3}})
- \sqrt{3} k \tau' \cos (\frac{k \tau'}{\sqrt{3}})\Big)
    \frac{Z_{T\mathbf k}(\tau')}{k \tau'}d\tau'
\Big]d\tau''
,
\el
\bl\label{Tchi2Ssolphy}
\chi^{||(2)}_{T{\bf k}}
=&
Q_2(\mathbf k)\frac{4\sqrt3 \,i}{k^2\tau}e^{-ik\tau/\sqrt3}
-Q_3(\mathbf k)\frac{4\sqrt3 \,i}{k^2\tau}e^{ik\tau/\sqrt3}
        \nn\\
        &
-\frac{4}{k}\int^\tau
\l[
Q_2(\mathbf k)e^{-ik\tau'/\sqrt3}
+Q_3(\mathbf k)e^{ik\tau'/\sqrt3}
\r]
\frac{d\tau'}{\tau'}
\nn\\
&
-\frac{2 \ln\tau}{k }\int^\tau \frac{
    3(k^2 \tau'^2+6)}{k^3 \tau'} Z_{T\mathbf k}(\tau')d\tau'
+\int^\tau\frac{6(k^2 \tau'\,^2+6)\ln\tau'+18}{k^4 \tau'} Z_{T\mathbf k}(\tau')d\tau'
    \nn\\
&
-\frac{4\sqrt3}{k^2\tau} \sin (\frac{k \tau}{\sqrt{3}})
\int^\tau
\Big(
9\cos (\frac{k \tau'}{\sqrt{3}})
+3\sqrt{3}k \tau' \sin (\frac{k \tau'}{\sqrt{3}})
\Big)
\frac{  Z_{T\mathbf k}(\tau') }{ k^3 \tau'} \, d\tau'
\nn\\
&
+\frac{4\sqrt3}{k^2\tau} \cos (\frac{k \tau}{\sqrt{3}})
    \int^\tau
    \Big( 9 \sin (\frac{k \tau'}{\sqrt{3}})
- 3\sqrt{3} k \tau' \cos (\frac{k \tau'}{\sqrt{3}})\Big)
    \frac{Z_{T\mathbf k}(\tau')}{k^3 \tau'}d\tau'
\nn\\
&
+\int^\tau
\Big[
\frac{12}{k^3\tau''} \cos (\frac{k \tau''}{\sqrt{3}})
\int^{\tau''}
\Big(
3\cos (\frac{k \tau'}{\sqrt{3}})
+\sqrt{3}k \tau' \sin (\frac{k \tau'}{\sqrt{3}})
\Big)
\frac{  Z_{T\mathbf k}(\tau') }{ k \tau'} \, d\tau'
\Big]d\tau''
\nn\\
&
+\int^\tau
\Big[
\frac{12}{k^3\tau''} \sin (\frac{k \tau''}{\sqrt{3}})
    \int^{\tau''}
    \Big(
    3 \sin (\frac{k \tau'}{\sqrt{3}})
- \sqrt{3} k \tau' \cos (\frac{k \tau'}{\sqrt{3}})\Big)
    \frac{Z_{T\mathbf k}(\tau')}{k \tau'}d\tau'
\Big]d\tau''
        \nn\\
        &
+\frac{3}{k^4} E_{T{\bf k}}
- \frac{9}{2k^4\tau} A_{T{\bf k}}(\tau)
-\frac{3}{2k^2}\int^\tau A_{T{\bf k}}(\tau') d\tau'
,
\el
\bl\label{TdeltaV2soluphy}
\delta_{T{\bf k}}^{(2)}
=&
Q_2(\mathbf k)
\l(
\frac{8}{k\tau^2}
+\frac{8 i}{\sqrt3\,\tau}
-\frac{4 k}{3}
\r)e^{-ik\tau/\sqrt3}
+Q_3(\mathbf k)\l(
\frac{8}{k\tau^2}
-\frac{8 i}{\sqrt3\,\tau}
-\frac{4 k}{3}
\r)e^{ik\tau/\sqrt3}
        \nn\\
        &
+\Big(
-\frac{8}{k\tau^2} \cos (\frac{k \tau}{\sqrt{3}})
-\frac{8}{\sqrt{3}\tau} \sin (\frac{k \tau}{\sqrt{3}})
+\frac{4k}{3}\cos (\frac{k \tau}{\sqrt{3}})
\Big)
\int^\tau
\Big(
9\cos (\frac{k \tau'}{\sqrt{3}})
\nn\\
&
+3\sqrt{3}k \tau' \sin (\frac{k \tau'}{\sqrt{3}})
\Big)
\frac{  Z_{T\mathbf k}(\tau') }{ k^3 \tau'} \, d\tau'
\nn\\
&
+\Big(
-\frac{ 8}{k\tau^2} \sin (\frac{k \tau}{\sqrt{3}})
+\frac{ 8}{\sqrt{3}\,\tau} \cos (\frac{k \tau}{\sqrt{3}})
+\frac{4k}{3} \sin (\frac{k\tau}{\sqrt{3}})
\Big)
    \int^\tau
    \Big( 9 \sin (\frac{k \tau'}{\sqrt{3}})
\nn\\
&
- 3\sqrt{3} k \tau' \cos (\frac{k \tau'}{\sqrt{3}})\Big)
    \frac{Z_{T\mathbf k}(\tau')}{k^3 \tau'}d\tau'
    \nn\\
&
+\frac{1}{ k\tau^2}
\int^\tau \frac{
    12(k^2 \tau'^2+6)}{k^3 \tau'} Z_{T\mathbf k}(\tau')d\tau'
,
\el
\bl\label{Tv2ndsolphy}
v^{||(2)}_{T\mathbf k}
&=
Q_2(\mathbf k)\l(\frac{2}{k\tau}
    +\frac{i}{\sqrt3}\r)e^{-ik\tau/\sqrt3}
+Q_3(\mathbf k)\l(\frac{2}{k\tau}
    -\frac{i}{\sqrt3}\r)e^{ik\tau/\sqrt3}
\nn\\
&
-\left(
\frac{2}{k\tau} \cos (\frac{k \tau}{\sqrt{3}})
+\frac{1}{\sqrt{3}}\sin (\frac{k \tau}{\sqrt{3}})
\right)
\int^\tau
\Big(
9\cos (\frac{k \tau'}{\sqrt{3}})
+3\sqrt{3}k \tau' \sin (\frac{k \tau'}{\sqrt{3}})
\Big)
\frac{  Z_{T\mathbf k}(\tau') }{ k^3 \tau'} \, d\tau'
\nn\\
&
-\left(
\frac{ 2}{k\tau} \sin (\frac{k \tau}{\sqrt{3}})
-\frac{1}{\sqrt{3}} \cos (\frac{k\tau}{\sqrt{3}})
\right)
    \int^\tau
    \Big( 9 \sin (\frac{k \tau'}{\sqrt{3}})
- 3\sqrt{3} k \tau' \cos (\frac{k \tau'}{\sqrt{3}})\Big)
    \frac{Z_{T\mathbf k}(\tau')}{k^3 \tau'}d\tau'
    \nn\\
&
+\frac{1}{ k\tau}
\int^\tau \frac{
    3\left(k^2 \tau'^2+6\right)}{k^3 \tau'} Z_{T\mathbf k}(\tau')d\tau'
,
\el
}
where $Z_{T}$ is in (\ref{TZSall}),
$E_{T}$ is in (\ref{TES1RD}),
and $A_{T}$ is in (\ref{TAS}).

Thus,  we have obtained all the gauge-invariant solutions of
the 2nd-order perturbations
with scalar-tensor couplings
in Eqs.(\ref{phi2SkSolphy})--(\ref{v2ndsolphy}), (\ref{chiVecSolphy}),
(\ref{Vperp2ndSol}), and (\ref{solgw})
and with tensor-tensor couplings
in Eqs.(\ref{Tphi2SkSolphy})--(\ref{Tv2ndsolphy}), (\ref{TchiVecSolphy}),
(\ref{TVperp2ndSol}), and (\ref{Tsolgw}).

To illustrate the physical picture of the solved 2nd-order perturbations,
we now calculate the power spectrum of
the 2nd-order scalar perturbation $\phi^{(2)}_{T{\bf k}}$ in (\ref{Tphi2SkSol})
generated by the tensor-tensor coupling.
Equation (\ref{Tphi2SkSol}) contains many terms.
For an illustration, here we consider only
 the term $- \frac{1}{4}\int^\tau A_{T{\bf k}}(\tau') d\tau'$
in the first line in (\ref{Tphi2SkSol}) as follows:
\bl\label{ph2kA}
\phi^{(2)}_{T{\bf k}}
&=
- \frac{1}{4}\int^\tau A_{T{\bf k}}(\tau') d\tau'
\nn\\
&=
-\frac{1}{6(2\pi)^{3}}\int d^3k_2\,
\Big[
\sum_{s_1, s_2={+,\times}}
{\mathop \epsilon \limits^{s_1}}_{lm}({\bf k}-{{\bf k}_2})
{\mathop \epsilon \limits^{s_2}}{}^{lm}({{\bf k}_2})
{h}_{ {\bf k}-{{\bf k}_2}} {h}_{ {\bf k}_2}
\Big] ,
\el
where  ${h}_{\bf k}$ is the 1st-order tensor mode given by (\ref{GWmode}).
As  a complex random variable,
${h}_{\bf k}$  has a zero mean
\be \label{0h}
\langle{h}_{\bf k}\rangle=0  ~~~ \text{ for each }  \bf k,
\ee
and its real  and imaginary parts have the same Gaussian probability distribution,
with no correlation between them,
i.e., the phase of ${h}_{\bf k}$ is random
with a uniform probability distribution \cite{LiddleLythbook}.
Furthermore, $h_{\bf k}$ are statistically independent for different $\bf k$,
and
the power spectrum of the 1st-order tensor is defined
in terms of the ensemble average as follows
\cite{Allen & Romano,zhangyang05}:
\be\label{spectrum1stGW}
\langle{h}_{{\bf k}}^*{h}_{{\bf k}'}\rangle
\equiv\frac{2\pi^2}{ k^{3}}\delta({\bf k}-{\bf k}')
\Delta^2_t(k,\tau) \, .
\ee
Also
\be\label{0hh}
\langle {h}_{{\bf k}} {h}_{{\bf k}}\rangle=0
\ee
holds since  the phase of ${h}_{{\bf k}}$
is random  \cite{LiddleLythbook, Cornish & Larson}.
By these, the statistical properties of $h_{\bf k}$ are completely determined.
[In parallel,  one can write down the properties similar to
(\ref{0h})--(\ref{0hh}),
for the 1st-order scalar and vector metric perturbations,
as well as for  the 1st-order density  and velocity perturbations.]
For inflation models described by a power law $a(\tau)\propto  |\tau|^{1+\beta}$,
the 1st-order power spectrum is known for the RD stage
resulting from the cosmic evolution from inflation \cite{zhangyang05}.
We define  the  2nd-order power spectrum in terms of the covariance
as follows:
\bl\label{spectrum2ndTTS1}
\Big\langle
\Big(\phi^{(2)*}_{T{\bf k}}-\langle\phi^{(2)*}_{T{\bf k}}\rangle\Big)
\Big(\phi^{(2)}_{T{\bf k}'}-\langle\phi^{(2)}_{T{\bf k}'}\rangle\Big)
\Big\rangle
=&
\Big\langle
\phi^{(2)*}_{T{\bf k}}
\phi^{(2)}_{T{\bf k}'}
\Big\rangle
-
\langle\phi^{(2)*}_{T{\bf k}}\rangle
\langle\phi^{(2)}_{T{\bf k}'}\rangle
\nn\\
\equiv&
\frac{2\pi^2}{ k^{3}}\delta(\mathbf k-\mathbf k')P_{\phi_T}(k,\tau)
,
\el
where the mean of the 2nd-order scalar perturbation
  is given by
\bl
\langle\phi^{(2)}_{T{\bf k}}\rangle
=&
-\frac{1}{6(2\pi)^{3}}\int d^3k_2\,
\Big[
\sum_{s_1,s_2={+,\times}}
{\mathop \epsilon \limits^{s_1}}_{lm}({\bf k}-{{\bf k}_2})
{\mathop \epsilon \limits^{s_2}}{}^{lm}({{\bf k}_2})
\langle{h}_{ {\bf k}-{{\bf k}_2}} {h}_{ {\bf k}_2}\rangle
\Big]
 =0,
\el
where the second equality follows from  (\ref{0hh}).
In fact,
a 2nd-order perturbation, either scalar, vector,  or tensor,
 always has a vanishing mean
since it is formed from the products of the 1st-order perturbations
with the  statistical properties (\ref{spectrum1stGW}) and (\ref{0hh}).
The covariance (\ref{spectrum2ndTTS1})
is always proportional to the Dirac delta function as we have checked.
This is consistent with
Refs.\cite{Lu2008} and \cite{HwangJeongNoh2017} in other coordinates.
(We remark that
the 2nd-order perturbations, such as $\phi^{(2)}_{T{\bf k}}$,
are statistically independent for different  $\bf k$
since their  covariances are  zero for ${\bf k} \ne {\bf k'}$.
However,  $\phi^{(2)}_{T{\bf k}}$ for different  $\bf k$
are   dynamically  related,
since  it contains products of the 1st-order perturbations
with different $\bf k$.)
By using the   ``factorization"  property \cite{Isserlis1918,Allen & Romano,Lu2008}
\be\label{factorization}
\langle x_1x_2x_3x_4\rangle
=\langle x_1x_2\rangle \langle x_3x_4\rangle
+\langle x_1 x_3\rangle \langle x_2x_4\rangle
+\langle x_1x_4\rangle \langle x_2 x_3\rangle \, ,
\ee
which is valid for generic Gaussian random variables $x_1,x_2,x_3,x_4$,
each having zero mean;
using the property  (\ref{polarTensorSum}) of the polarization tensors;
and using (\ref{ph2kA}), (\ref{spectrum1stGW}), and (\ref{0hh}),
the covariance is written as
\bl
\langle
\phi^{(2)*}_{T{\bf k}}\phi^{(2)}_{T{\bf k}'}
\rangle
=&
\delta({\bf k}-{\bf k}')
\frac{(2\pi^2)^2}{36(2\pi)^{6}}
\int d^3k_2\,
\frac{2}{ (k_2)^{3}|{\bf k}-{{\bf k}_2}|^{3}}
\Big(
\sum_{s_1, s_2={+,\times}}
{\mathop \epsilon \limits^{s_1}}_{lm}({\bf k}-{{\bf k}_2})
{\mathop \epsilon \limits^{s_2}}{}^{lm}({{\bf k}_2})
\Big)^2
\nn\\
&
\times
\Delta^2_t(|{\bf k}-{{\bf k}_2}|,\tau)
\Delta^2_t( k_2,\tau)
.
\el
So  by the definition (\ref{spectrum2ndTTS1}),
 we read off the 2nd-order power  spectrum
\bl\label{2powersp}
P_{\phi_T}(k,\tau)
=&
\frac{k^{3}}{36(2\pi)^{4}}
\int \frac{d^3k'}{k^{\prime\,3}}\,
\bigg[\frac{1}{ |{\bf k}-{{\bf k}'}|^{3}}
\Big(
\sum_{s_1, s_2 ={+,\times}}
{\mathop \epsilon \limits^{s_1}}_{lm}({\bf k}-{{\bf k}'})
{\mathop \epsilon \limits^{s_2}}{}^{lm}({{\bf k}'})
\Big)^2
\nn\\
&
\times
\Delta^2_t(|{\bf k}-{{\bf k}'}|,\tau)
\Delta^2_t( k',\tau)
\bigg]
.
\el
Given $\Delta^2_t(k,\tau)$ in a cosmological model \cite{zhangyang05},
the power spectrum $P_{\phi_T}(k,\tau)$ is  determined
by the integration (\ref{2powersp}).
The frequency  at a  conformal  time $\tau$ is related to $k$
via $f(\tau) =  k/ 2\pi a(\tau)$.
By the normalization of $a(\tau)$  in  Ref.\cite{ZhangWangJCAP2018},
the present frequency is related to $k$ as
$f  \simeq   1.7\times10^{-19}\,  k ~ {\rm Hz}$.
We plot $\Delta_t^2(f)$ and $P_{\phi_ T}(f)$
in Fig.\ref{PowerT1st} for  different values of cosmic redshift $z$
in the low frequency range,
which may affect the CMB anisotropies and polarization.
It is seen that, at low $f$,
$\Delta_t^2(f) \sim 10^{-10}$ in magnitude,
and $P_{\phi_ T}(f)\sim[\Delta_t^2(f)]^2  \sim 10^{-20}$,
so that the amplitude of 2nd-order spectrum is extremely small
in this case.
As expected,
the amplitudes of the power spectra are decreasing as  $z$
becomes smaller,
since the amplitude $h_{\bf k}$ of RGW
is decreasing inside the horizon  \cite{zhangyang05}.
So far in the above,
we present  only the contribution of
 one term $- \frac{1}{4}\int^\tau A_{T{\bf k}}  d\tau'$,
as an illustration.
The whole contribution to the 2nd-order spectrum of $\phi^{(2)}_{T{\bf k}}$
will involve  hundreds of terms like this.
Besides, there are
the 2nd-order scalars  $\phi^{(2)}_{s(t){\bf k}}$,
$\chi^{||(2)}_{s(t){\bf k}}$,
generated by the scalar-tensor coupling, etc.
All these  require  many more efforts of computation
and will be left for future study.

\begin{figure}[htbp]
\centering
\subcaptionbox{}
    {%
 \includegraphics[width=0.65\linewidth]{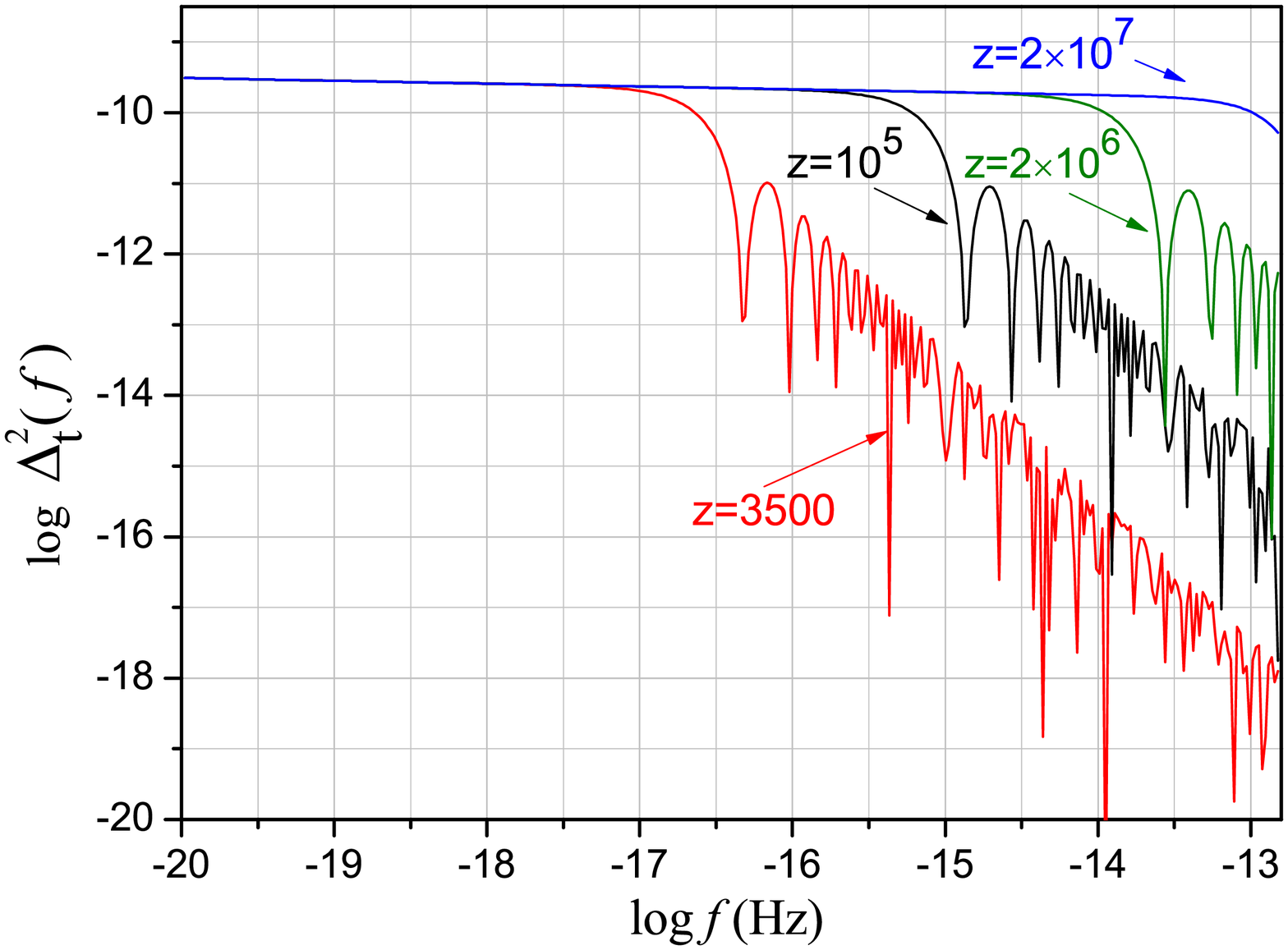}
}
\subcaptionbox{}
    {%
 \includegraphics[width=0.65\linewidth]{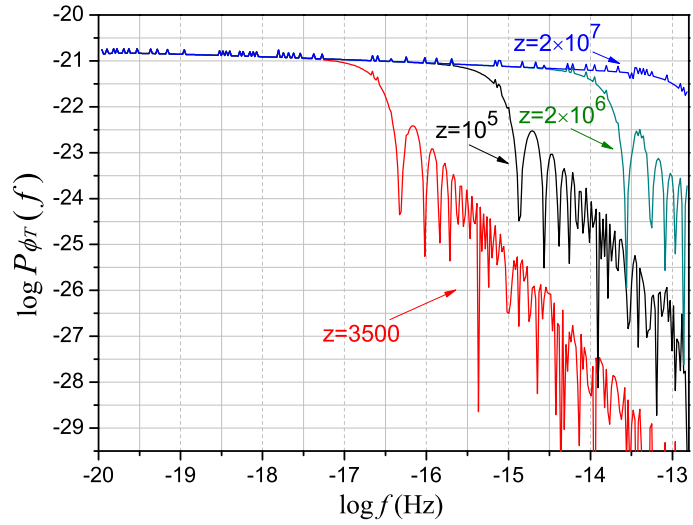}
}
\caption{(a) The power spectrum of 1st-order tensor.
(b) The power spectrum (\ref{2powersp})
   of 2nd-order scalar $\phi^{(2)}_{T{\bf k}}$
   generated by the 1st-order tensor-tensor coupling term (\ref{ph2kA}).
   }\label{PowerT1st}
\end{figure}

\section{Transformation of 2nd-order solutions
from synchronous to Poisson coordinates }

The Poisson coordinates are also commonly used in cosmological study,
so in this section we perform transformation of the 2nd-order solutions
from synchronous coordinates
into  Poisson coordinates.
A perturbed metric up to 2nd order in the Poisson coordinates  is
generally written as
\cite{Matarrese98,Bertschinger,MaBertschinger1995}
\be\label{g00poiss}
g_{00}=-a^2\bigg[
1+\psi^{(1)}_{P}
+\frac{1}{2}\psi^{(2)}_{P}\bigg],
\ee
\be
g_{0i}=a^2\bigg[
w^{(1)}_{P\,i}
+\frac{1}{2}w^{(2)}_{P\,i}\bigg],
\ee
\be \label{gijpoiss}
g_{ij}=a^2\bigg[
\delta_{ij}
-2\bigg(\phi^{(1)}_{P}+\frac{1}{2}\phi^{(2)}_{P}\bigg)\delta_{ij}
+\chi^{\top(1)}_{P\,ij}
+\frac{1}{2}\chi^{\top(2)}_{P\,ij}\bigg].
\ee
with  the vectors satisfying
\be \label{stpoissonCon}
\partial^i w^{(A)}_{P\,i}=0,  \,\,\, A=1,2
\ee
and  the tensor satisfying
\be  \label{stpoissonConTensor}
\chi^{\top(A)i}_{P\,i}=0,
~~~~~~
\partial^i\chi^{\top(A)}_{P\,ij}=0 .
\ee
[When the vector and tensor modes are zero,
the coordinate system  (\ref{g00poiss})--(\ref{gijpoiss})
is also called the longitudinal (conformal-Newtonian) coordinate.]
Consider a transformation of the solutions of metric perturbations
from a synchronous coordinate to a Poisson coordinate.
The 1st- and 2nd-order transformation vectors
$\xi^{(1)\mu}$ and $\xi^{(2)\mu}$  are introduced,
respectively,
as (\ref{alpha_r}) and (\ref{xi_r}).
Let the 1st-order metric perturbations,
$\phi^{(1)},D_{ij}\chi^{||(1)},\chi^{\top(1)}_{ij}$
 be given as (\ref{phi1sol2}), (\ref{chi1sol3}), and (\ref{Fourier})
 in a synchronous coordinate without vector mode.
By   the coordinate transformation  (C4) in Ref.\cite{WangZhang2ndRD2018},
one gets  the 1st-order   metric perturbation in Poisson coordinates
as follows \cite{Matarrese98}:
\be\label{stpsi1P}
\psi^{(1)}_P=-\alpha^{(1)'}-\frac{a'}{a}\alpha^{(1)},
\ee
\be\label{stwi1P}
w^{(1)}_{P\,i}=\alpha^{(1)}_{,\,i}
-\beta^{(1)'}_{,\,i}
-d^{(1)'}_i,
\ee
\be\label{stphi1P}
\phi^{(1)}_{P}=\phi^{(1)}
+\frac{1}{3}\nabla^2\beta^{(1)}
+\frac{a'}{a}\alpha^{(1)},
\ee
\be\label{stchi1P}
\chi^{\top(1)}_{P\,ij}=
D_{ij}\chi^{||(1)}
+\chi^{\top(1)}_{ij}
-2D_{ij}\beta^{(1)}
-d^{(1)}_{i,j}
-d^{(1)}_{j,\,i}.
\ee
By the constraints (\ref{stpoissonCon}) and (\ref{stpoissonConTensor})
in the Poisson coordinate,
the transformation parameters satisfy  the following constraints
\be
\alpha^{(1)}=\beta^{(1)'},
\ee
\be
D_{ij}(\chi^{||(1)}-2\beta^{(1)})=0,
\ee
\be
d^{(1)}_i = 0
\ee
in the absence of the 1st-order vector mode.
Also, their  solutions are given by  $\alpha^{(1)} =\frac{1}{2}\chi^{||(1)'} $,
$\beta^{(1)} =\frac{1}{2}\chi^{||(1)}$.
In $\bf k$-space,
the solution  $\chi^{||(1)}_{\bf k}$ is known in  (\ref{chi1sol3}),
by which one obtains  the 1st-order transformation parameters
\be\label{stalpha1Po}
\alpha^{(1)}_{\bf k}=\frac{1}{2}\chi^{||(1)'}_{\bf k}
=
-D_2\frac{2\sqrt3\,i}{k^2\tau^2}e^{-ik\tau/\sqrt3}
+D_3\frac{2\sqrt3\,i}{k^2\tau^2}e^{ik\tau/\sqrt3}
,
\ee
\bl\label{stbeta1Po}
\beta^{(1)}_{\bf k}=\frac{1}{2}\chi^{||(1)}_{\bf k}
=&
D_2\frac{2\sqrt3\,i}{k^2\tau}e^{-ik\tau/\sqrt3}
-D_3\frac{2\sqrt3\,i}{k^2\tau}e^{ik\tau/\sqrt3}
\nn\\
&
-\frac{2}{k}\int^\tau\l[ D_2 e^{-ik\tau'/\sqrt3}
    +D_3e^{ik\tau'/\sqrt3}\r]\frac{d\tau'}{\tau'}
.
\el
Plugging these and  (\ref{phi1sol2}), (\ref{chi1sol3})
into (\ref{stpsi1P})--(\ref{stchi1P})
yields the  metric perturbations in Poisson coordinates
\be\label{Poisson1stSol}
\psi^{(1)}_{P\,\bf k}=
\phi^{(1)}_{P\,\bf k}
=
D_2\Big(
\frac{2}{k\tau^2}
-\frac{2\sqrt3\,i}{k^2\tau^3}
\Big)
e^{-ik\tau/\sqrt3}
+D_3\Big(
\frac{2}{k\tau^2}
+\frac{2\sqrt3\,i}{k^2\tau^3}
\Big)
e^{ik\tau/\sqrt3},
\ee
\be
w^{(1)}_{P\,i}=0,
\ee
\be
\chi^{\top(1)}_{P\,ij}=  \chi^{\top(1)}_{\,ij}   .
\ee
Our result  (\ref{Poisson1stSol}) is the general 1st-order scalar solution.
By taking $D_2=(-A_r+i B_r)/(4 \sqrt{3} k)$ and
$D_3=(-A_r-i B_r)/(4 \sqrt{3} k)$,
Eq.(\ref{Poisson1stSol}) is the same as Eq.(43) of Ref.\cite{Lu2008}.
The 1st-order scalar mode $\Phi$ in Ref.\cite{Lu2008}
is related to our $\chi^{||(1)}$  as
$\Phi=-\frac{1}{2}\chi^{||(1)''}-\frac{1}{2\tau}\chi^{||(1)'}$.
Also note that
 Ref.\cite{Lu2008} ignored certain terms as decaying modes
in (\ref{Poisson1stSol}) in their actual subsequent computing.
We see  that in the Poisson coordinate
the 1st-order vector  mode is still zero and
the 1st-order tensor mode is the same as in synchronous coordinates,
and the two 1st-order scalar modes are equal,
a fact that can be checked
using (\ref{stwi1P}), (\ref{stphi1P}), (\ref{stalpha1Po}), and (\ref{stbeta1Po}),
\[
\phi^{(1)}_{P}-\psi^{(1)}_P=
 \frac{1}{2}\chi^{||(1)''}
+\frac{1}{\tau}\chi^{||(1)'}
+\frac{1}{6}\nabla^2\chi^{||(1)} + \phi^{(1) }
=0,
\]
where in the last step
the evolution equation (3.11) in Ref.\cite{WangZhang2ndRD2018} is used.
That is why the Poisson coordinate is often used
when only the scalar metric perturbation is present.

Next,
we turn to the 2nd-order metric perturbations
with scalar-tensor couplings in the Poisson  coordinates.
By (C5) in Ref.\cite{WangZhang2ndRD2018} with the scalar-tensor coupling,
one has
\be\label{stpsi2PChi2}
\psi^{(2)}_{P_{s(t)}}=
-\alpha^{(2)'}_{s(t)}
-\frac{a'}{a}\alpha^{(2)}_{s(t)}
,
\ee
\be\label{stwi2PChi2}
w^{(2)}_{P_{s(t)}i}
=
W_{s(t)i}
+\alpha^{(2)}_{s(t),\,i}
-\beta^{(2)'}_{s(t),\,i}
-d^{(2)'}_{s(t)i},
\ee
\be\label{stphi2PChi2}
\phi^{(2)}_{P_{s(t)}}=
\phi^{(2)}_{s(t)}
+\frac{1}{3}\chi^{\top(1)}_{lm}\chi^{||,\,lm}
+\frac{a'}{a}\alpha^{(2)}_{s(t)}
+\frac{1}{3}\nabla^2\beta^{(2)}_{s(t)},
\ee
\be\label{stchi2PChi2}
\chi^{\top(2)}_{P_{s(t)}ij}=
D_{ij}\chi^{||(2)}_{s(t)}
+\chi^{\perp(2)}_{s(t)ij}
+\chi^{\top(2)}_{s(t)ij}
+Y_{s(t)ij}
-\big(d^{(2)}_{s(t)i,j}
+d^{(2)}_{s(t)j,i}
+2 D_{ij}\beta^{(2)}_{s(t)}\big)
,
\ee
where
\be\label{stXiChi2}
W_{s(t)i}\equiv
-\chi^{\top(1)}_{il}\chi^{||(1)',\,l},
\ee
\bl\label{stYijChi2}
Y_{s(t)ij}\equiv&
-\frac{2a'}{a}\chi^{\top(1)}_{ij}\chi^{||(1)'}
-\chi^{\top(1)'}_{ij}\chi^{||(1)'}
-\chi^{\top(1)}_{ij,\,l}\chi^{||(1),\,l}
\nn\\
&
-\chi^{\top(1)}_{il}\chi^{||(1),\,l}_{,j}
-\chi^{\top(1)}_{jl}\chi^{||(1),\,l}_{,i}
+\frac{2}{3}\chi^{\top(1)}_{lm}\chi^{||(1),\,lm}\delta_{ij}
.
\el
In order to get perturbations in the Poisson  coordinates,
one needs to solve for the parameters
$\alpha^{(2)}_{s(t)}$,   $\beta^{(2)}_{s(t)}$ and $d^{(2)}_{s(t)i}$.
We  decompose $W_{s(t)i}$ in (\ref{stwi2PChi2})
into  two  parts,
$W_{s(t)i}=W^{||}_{s(t),\,i}+W^{\perp}_{s(t)i}$,
with
\be\label{stW||}
W^{||}_{s(t)}=
\nabla^{-2}\partial^iW_{s(t)i}
=
\nabla^{-2}\Big[
-\chi^{\top(1)}_{lm}\chi^{||(1)',\,lm}
\Big]
,
\ee
\bl\label{stWperp}
W^{\perp}_{s(t)i}=&
W_{s(t)i}-\partial_i W^{||}_{s(t)}
\nn\\
=&
-\chi^{\top(1)}_{il}\chi^{||(1)',\,l}
+\partial_i\nabla^{-2}\Big[
\chi^{\top(1)}_{lm}\chi^{||(1)',\,lm}
\Big]
.
\el
Since $w^{(2)}_{P_{s(t)}i}$  contains only the curl part,
the gradient part of Eq.(\ref{stwi2PChi2})  gives an equation
\be \label{stwi2Pncurl}
W^{||}_{s(t)} +\alpha^{(2)}_{s(t)} -\beta^{(2)'}_{s(t)}
=0.
\ee
Also $\chi^{\top(2)}_{P_{s(t)}ij}$ is traceless and transverse;
i.e., the rhs of (\ref{stchi2PChi2}) should contain no  scalar, vector modes.
We  decompose $Y_{s(t)ij}$
into scalar, vector, tensor modes as
\be
Y_{s(t)ij}=D_{ij}Y^{||}_{s(t)}+2Y^{\perp}_{s(t)(i,j)}+Y^{\top}_{s(t)ij}  ,
\ee
where $Y^{\perp}_{s(t)(i,j)}=\frac{1}{2}(Y^{\perp}_{s(t)i,j}+Y^{\perp}_{s(t)j,i})$,
$Y^{\perp,\,i}_{s(t)i}=0$, and $Y^{\top,\,i}_{s(t)ij}=0$.
We  get
\bl\label{stY||}
Y^{||}_{s(t)}=&
\frac{3}{2}\nabla^{-2}\nabla^{-2}Y_{s(t)lm}^{,\,lm}
\nn\\
=&
\nabla^{-2}\Big[
\chi^{\top(1)}_{lm}\chi^{||(1),\,lm}
\Big]
+\nabla^{-2}\nabla^{-2}\Big[
-\frac{3a'}{a}\chi^{\top(1)}_{lm}\chi^{||(1)',\,lm}
-\frac{3}{2}\chi^{\top(1)'}_{lm}\chi^{||(1)',\,lm}
\nn\\
&
-\frac{9}{2}\chi^{\top(1)}_{lm,n}\chi^{||(1),\,lmn}
-3\chi^{\top(1)}_{lm}\nabla^2\chi^{||(1),\,lm}
\Big]
,
\el
\bl\label{stYperp}
Y^{\perp}_{s(t)j}=&
\nabla^{-2}Y_{s(t)lj}^{,\,l}
-\frac{2}{3}Y^{||}_{s(t),j}
\nn\\
=&
\nabla^{-2}\Big[
-\frac{2a'}{a}\chi^{\top(1)}_{lj}\chi^{||(1)',\,l}
-\chi^{\top(1)'}_{lj}\chi^{||(1)',\,l}
-2\chi^{\top(1)}_{jl,m}\chi^{||(1),\,lm}
-\chi^{\top(1)}_{lm}\chi^{||(1),\,lm}_{,j}
\nn\\
&
-\chi^{\top(1)}_{lj}\nabla^2\chi^{||(1),\,l}
\Big]
+\partial_j\nabla^{-2}\nabla^{-2}\Big[
\frac{2a'}{a}\chi^{\top(1)}_{lm}\chi^{||(1)',\,lm}
+\chi^{\top(1)'}_{lm}\chi^{||(1)',\,lm}
\nn\\
&
+3\chi^{\top(1)}_{lm,n}\chi^{||(1),\,lmn}
+2\chi^{\top(1)}_{lm}\nabla^2\chi^{||(1),\,lm}
\Big]
,
\el
{\allowdisplaybreaks
\bl\label{stYtop1}
Y^{\top}_{s(t)ij}=&
Y_{s(t)ij}-D_{ij}Y^{||}_{s(t)}-2Y^{\perp}_{s(t)(i,j)}
\nn\\
=&
-\frac{2a'}{a}\chi^{\top(1)}_{ij}\chi^{||(1)'}
-\chi^{\top(1)'}_{ij}\chi^{||(1)'}
-\chi^{\top(1)}_{ij,\,l}\chi^{||(1),\,l}
-\chi^{\top(1)}_{il}\chi^{||(1),\,l}_{,j}
-\chi^{\top(1)}_{jl}\chi^{||(1),\,l}_{,i}
\nn\\
&
+\chi^{\top(1)}_{lm}\chi^{||(1),\,lm}\delta_{ij}
+\delta_{ij}\nabla^{-2}\Big[
-\frac{a'}{a}\chi^{\top(1)}_{lm}\chi^{||(1)',\,lm}
-\frac{1}{2}\chi^{\top(1)'}_{lm}\chi^{||(1)',\,lm}
\nn\\
&
-\frac{3}{2}\chi^{\top(1)}_{lm,n}\chi^{||(1),\,lmn}
-\chi^{\top(1)}_{lm}\nabla^2\chi^{||(1),\,lm}
\Big]
+\partial_i\nabla^{-2}\Big[
\frac{2a'}{a}\chi^{\top(1)}_{lj}\chi^{||(1)',\,l}
+\chi^{\top(1)'}_{lj}\chi^{||(1)',\,l}
\nn\\
&
+2\chi^{\top(1)}_{jl,m}\chi^{||(1),\,lm}
+\chi^{\top(1)}_{lm}\chi^{||(1),\,lm}_{,j}
+\chi^{\top(1)}_{lj}\nabla^2\chi^{||(1),\,l}
\Big]
+\partial_j\nabla^{-2}\Big[
\frac{2a'}{a}\chi^{\top(1)}_{li}\chi^{||(1)',\,l}
\nn\\
&
+\chi^{\top(1)'}_{li}\chi^{||(1)',\,l}
+2\chi^{\top(1)}_{il,m}\chi^{||(1),\,lm}
+\chi^{\top(1)}_{lm}\chi^{||(1),\,lm}_{,i}
+\chi^{\top(1)}_{li}\nabla^2\chi^{||(1),\,l}
\Big]
\nn\\
&
-\partial_i\partial_j\nabla^{-2}\Big[
\chi^{\top(1)}_{lm}\chi^{||(1),\,lm}
\Big]
+\partial_i\partial_j\nabla^{-2}\nabla^{-2}\Big[
-\frac{a'}{a}\chi^{\top(1)}_{lm}\chi^{||(1)',\,lm}
-\frac{1}{2}\chi^{\top(1)'}_{lm}\chi^{||(1)',\,lm}
\nn\\
&
-\frac{3}{2}\chi^{\top(1)}_{lm,n}\chi^{||(1),\,lmn}
-\chi^{\top(1)}_{lm}\nabla^2\chi^{||(1),\,lm}
\Big]
.
\el
}
Requiring the scalar and  vector modes in Eq.(\ref{stchi2PChi2})
to be  zero  leads to the two equations
\be\label{stchi2Pscalar}
\chi^{||(2)}_{s(t)}
+Y^{||}_{s(t)}
-2\beta^{(2)}_{s(t)}
=0
,
\ee
\be\label{stchi2Pvector}
\nabla^{-2}\partial^j\chi^{\perp(2)}_{s(t)ij}
+Y_{s(t)i}^{\perp}
-d^{(2)}_{s(t)i}
=0
.
\ee
The solutions of (\ref{stwi2Pncurl}), (\ref{stchi2Pscalar}), and (\ref{stchi2Pvector})
are follows:
\be\label{stwi2PncurlSol1}
\alpha^{(2)}_{s(t)}
=\frac{1}{2}\chi^{||(2)\,'}_{s(t)}
+\frac{1}{2}Y^{||\,'}_{s(t)}-W^{||}_{s(t)},
\ee
\be\label{stchi2PscalarSol1}
\beta^{(2)}_{s(t)}=\frac{1}{2}\chi^{||(2)}_{s(t)}
+\frac{1}{2}Y^{||}_{s(t)}
,
\ee
\be\label{stchi2PvectorSol1}
d^{(2)}_{s(t)i}=\nabla^{-2}\chi^{\perp(2),j}_{s(t)ij}
+Y_{s(t)i}^{\perp}
.
\ee
Plugging the solutions $\chi^{\perp(2)}_{s(t)ij}$  (\ref{chiVecSolphy}) and
$\chi^{||(2)}_{s(t)}$   (\ref{chi2Ssolphy})  into the above equations,
one obtains the parameters
{\allowdisplaybreaks
\bl\label{stalpha2bar}
\alpha^{(2)}_{s(t)\bf k}
=&
-P_2 \frac{2\sqrt3 \,i}{k^2\tau^2}e^{-ik\tau/\sqrt3}
+P_3 \frac{2\sqrt3 \,i}{k^2\tau^2}e^{ik\tau/\sqrt3}
        \nn\\
        &
-\frac{1}{k\tau}\int^\tau \frac{
    3(k^2 \tau'^2+6)}{k^3 \tau'} Z_{s(t)\mathbf k}(\tau')d\tau'
-\frac{9}{k^4\tau}  Z_{s(t)\mathbf k}
    \nn\\
&
+\frac{2\sqrt3}{k^2\tau^2} \sin (\frac{k \tau}{\sqrt{3}})
\int^\tau
\bigg(
9\cos (\frac{k \tau'}{\sqrt{3}})
+3\sqrt{3}k \tau' \sin (\frac{k \tau'}{\sqrt{3}})
\bigg)
\frac{  Z_{s(t)\mathbf k}(\tau') }{ k^3 \tau'} \, d\tau'
\nn\\
&
-\frac{2\sqrt3}{k^2\tau^2} \cos (\frac{k \tau}{\sqrt{3}})
    \int^\tau
    \bigg( 9 \sin (\frac{k \tau'}{\sqrt{3}})
- 3\sqrt{3} k \tau' \cos (\frac{k \tau'}{\sqrt{3}})\bigg)
    \frac{Z_{s(t)\mathbf k}(\tau')}{k^3 \tau'}d\tau'
        \nn\\
        &
+\frac{3}{2k^4} E^{\,'}_{s(t){\bf k}}
+\frac{27}{4k^4\tau}F^{||''}_{s(t){\bf k}}
+\frac{27}{4k^4\tau^2}F^{||'}_{s(t){\bf k}}
-\frac{27}{k^4\tau^3} F^{||}_{s(t){\bf k}}
+\frac{9}{4k^2}F^{||'}_{s(t){\bf k}}
        \nn\\
        &
- \frac{9}{4k^4\tau} A^{'}_{s(t){\bf k}}
+ \frac{9}{4k^4\tau^2} A_{s(t){\bf k}}
-\frac{3}{4k^2} A_{s(t){\bf k}}
+\frac{1}{2}Y^{||\,'}_{s(t)\bf k}
-W^{||}_{s(t)\bf k},
\el
\bl\label{stbeta2bar}
\beta^{(2)}_{s(t)\bf k}
=&
P_2 \frac{2\sqrt3 \,i}{k^2\tau}e^{-ik\tau/\sqrt3}
-P_3 \frac{2\sqrt3 \,i}{k^2\tau}e^{ik\tau/\sqrt3}
        \nn\\
        &
-\frac{2}{k}\int^\tau
\l[
P_2 e^{-ik\tau'/\sqrt3}
+P_3 e^{ik\tau'/\sqrt3}
\r]
\frac{d\tau'}{\tau'}
\nn\\
&
-\frac{ \ln\tau}{k }\int^\tau \frac{
    3(k^2 \tau'^2+6)}{k^3 \tau'} Z_{s(t)\mathbf k}(\tau')d\tau'
+\int^\tau\frac{3(k^2 \tau'^2+6)\ln\tau'+9}{k^4 \tau'} Z_{s(t)\mathbf k}(\tau')d\tau'
    \nn\\
&
-\frac{2\sqrt3}{k^2\tau} \sin (\frac{k \tau}{\sqrt{3}})
\int^\tau
\bigg(
9\cos (\frac{k \tau'}{\sqrt{3}})
+3\sqrt{3}k \tau' \sin (\frac{k \tau'}{\sqrt{3}})
\bigg)
\frac{  Z_{s(t)\mathbf k}(\tau') }{ k^3 \tau'} \, d\tau'
\nn\\
&
+\frac{2\sqrt3}{k^2\tau} \cos (\frac{k \tau}{\sqrt{3}})
    \int^\tau
    \bigg( 9 \sin (\frac{k \tau'}{\sqrt{3}})
- 3\sqrt{3} k \tau' \cos (\frac{k \tau'}{\sqrt{3}})\bigg)
    \frac{Z_{s(t)\mathbf k}(\tau')}{k^3 \tau'}d\tau'
\nn\\
&
+\int^\tau
\bigg[
\frac{6}{k^3\tau''} \cos (\frac{k \tau''}{\sqrt{3}})
\int^{\tau''}
\bigg(
3\cos (\frac{k \tau'}{\sqrt{3}})
+\sqrt{3}k \tau' \sin (\frac{k \tau'}{\sqrt{3}})
\bigg)
\frac{  Z_{s(t)\mathbf k}(\tau') }{ k \tau'} \, d\tau'
\bigg]d\tau''
\nn\\
&
+\int^\tau
\bigg[
\frac{6}{k^3\tau''} \sin (\frac{k \tau''}{\sqrt{3}})
    \int^{\tau''}
    \bigg(
    3 \sin (\frac{k \tau'}{\sqrt{3}})
- \sqrt{3} k \tau' \cos (\frac{k \tau'}{\sqrt{3}})\bigg)
    \frac{Z_{s(t)\mathbf k}(\tau')}{k \tau'}d\tau'
\bigg]d\tau''
        \nn\\
        &
+\frac{3}{2k^4} E_{s(t){\bf k}}
+\frac{27}{4k^4\tau}F^{||'}_{s(t){\bf k}}(\tau)
+\frac{27}{2k^4\tau^2} F^{||}_{s(t){\bf k}}
+\frac{9}{4k^2}F^{||}_{s(t){\bf k}}(\tau)
\nn\\
        &
- \frac{9}{4k^4\tau} A_{s(t){\bf k}}(\tau)
-\frac{3}{4k^2}\int^\tau A_{s(t){\bf k}}(\tau') d\tau'
+\frac{1}{2}Y^{||}_{s(t)\bf k}
,
\el
\be\label{std2bar2}
d^{(2)}_{s(t)i}
=
\frac{ \nabla^{-2}q_{2ij}^{,\,j}}{\tau}
 + 2\int^\tau \frac{d\tau'}{\tau^{'2}}
    \int^{\tau'}\tau^{''2}\,\nabla^{-2}V_{s(t)ij}^{,j}({\bf x},\tau'')d\tau''
+Y^{\perp}_{s(t)i},
\ee
}
where $E_{s(t)}$, $F_{s(t)}$, $A_{s(t)}$, $Z_{s(t)}$,
and $V_{s(t)ij}$ are given by
Eqs.(\ref{ES1RD}), (\ref{FSnoncu}), (\ref{AS}), (\ref{ZSall}), and
 (\ref{SourceCurl1RD}).

Plugging Eqs.(\ref{phi2SkSolphy}) and
 (\ref{stalpha2bar})--(\ref{std2bar2})
into (\ref{stpsi2PChi2})--(\ref{stphi2PChi2})
with $a(\tau)\propto\tau$,
one obtains the 2nd-order metric perturbations in the Poisson coordinate as
\be\label{stpsi2Psol}
\psi^{(2)}_{P_{s(t)}}=
\int  \frac{d^3k}{(2\pi)^{3/2}}\Big[
-\alpha^{(2)'}_{s(t)\bf k}
-\frac{1}{\tau}\alpha^{(2)}_{s(t)\bf k}
\Big]e^{i{\bf k\cdot x}}
,
\ee
where
{\allowdisplaybreaks
\bl\label{stalphaTaualpha}
&\ \ \
-\alpha^{(2)'}_{s(t)\bf k}
-\frac{1}{\tau}\alpha^{(2)}_{s(t)\bf k}
\nn\\
&=
P_2 (
    -\frac{2\sqrt3 \,i}{k^2\tau^3}
    +\frac{2}{k\tau^2})e^{-ik\tau/\sqrt3}
+P_3 (
   -\frac{2\sqrt3 \,i}{k^2\tau^3}
    +\frac{2 }{k\tau^2})e^{ik\tau/\sqrt3}
    \nn\\
    &
+\Big(\frac{2\sqrt3}{k^2\tau^3} \sin (\frac{k \tau}{\sqrt{3}})
-\frac{2}{k\tau^2} \cos (\frac{k \tau}{\sqrt{3}})\Big)
\int^\tau
\bigg(
9\cos (\frac{k \tau'}{\sqrt{3}})
+3\sqrt{3}k \tau' \sin (\frac{k \tau'}{\sqrt{3}})
\bigg)
\frac{  Z_{s(t)\mathbf k}(\tau') }{ k^3 \tau'} \, d\tau'
\nn\\
&
+\Big(
-\frac{2\sqrt3}{k^2\tau^3} \cos (\frac{k \tau}{\sqrt{3}})
-\frac{2}{k\tau^2} \sin (\frac{k \tau}{\sqrt{3}})
\Big)
    \int^\tau
    \bigg( 9 \sin (\frac{k \tau'}{\sqrt{3}})
- 3\sqrt{3} k \tau' \cos (\frac{k \tau'}{\sqrt{3}})\bigg)
    \frac{Z_{s(t)\mathbf k}(\tau')}{k^3 \tau'}d\tau'
        \nn\\
        &
+\frac{3}{k^2} Z_{s(t)\mathbf k}
+\frac{9}{k^4\tau}  Z^{\,'}_{s(t)\mathbf k}
-\frac{3}{2k^4\tau} E^{\,'}_{s(t){\bf k}}
-\frac{3}{2k^4} E^{\,''}_{s(t){\bf k}}
-\frac{27}{4k^4\tau}F^{||'''}_{s(t){\bf k}}
+\frac{135}{4k^4\tau^3}F^{||'}_{s(t){\bf k}}
        \nn\\
        &
-\frac{27}{4k^4\tau^2}F^{||''}_{s(t){\bf k}}
-\frac{54}{k^4\tau^4} F^{||}_{s(t){\bf k}}
-\frac{9}{4k^2\tau}F^{||'}_{s(t){\bf k}}
-\frac{9}{4k^2}F^{||''}_{s(t){\bf k}}
+ \frac{9}{4k^4\tau} A^{''}_{s(t){\bf k}}
+ \frac{9}{4k^4\tau^3} A_{s(t){\bf k}}
\nn\\
&
- \frac{9}{4k^4\tau^2} A^{'}_{s(t){\bf k}}
+\frac{3}{4k^2\tau} A_{s(t){\bf k}}
+\frac{3}{4k^2} A^{'}_{s(t){\bf k}}
-\frac{1}{2\tau}Y^{||\,'}_{s(t)\bf k}
-\frac{1}{2}Y^{||\,''}_{s(t)\bf k}
+\frac{1}{\tau}W^{||}_{s(t)\bf k}
+W^{||\,'}_{s(t)\bf k}
,
\el
}
and
\be\label{stphi2Psol}
\phi^{(2)}_{P_{s(t)}}=
\int  \frac{d^3k}{(2\pi)^{3/2}}\Big[
\phi^{(2)}_{s(t)\bf k}
+\frac{1}{\tau}\alpha^{(2)}_{s(t)\bf k}
-\frac{k^2}{3}\beta^{(2)}_{s(t)\bf k}
\Big]e^{i{\bf k\cdot x}}
+\frac{1}{3}\chi^{\top(1)}_{lm}\chi^{||,\,lm}
,
\ee
where
{\allowdisplaybreaks
\bl
&\ \ \ \
\phi^{(2)}_{s(t)\bf k}
+\frac{1}{\tau}\alpha^{(2)}_{s(t)\bf k}
-\frac{k^2}{3}\beta^{(2)}_{s(t)\bf k}
\nn\\
&=
P_2
\l(-\frac{2\sqrt3 \,i}{k^2\tau^3}
+\frac{2}{k\tau^2}
\r)e^{-ik\tau/\sqrt3}
+P_3\l(
\frac{2\sqrt3 \,i}{k^2\tau^3}
+\frac{2}{k\tau^2}
\r)e^{ik\tau/\sqrt3}
\nn\\
&
+\bigg(
\frac{2\sqrt3}{k^2\tau^3} \sin (\frac{k \tau}{\sqrt{3}})
-\frac{2}{k\tau^2} \cos (\frac{k \tau}{\sqrt{3}})
\bigg)
\int^\tau
\bigg(
9\cos (\frac{k \tau'}{\sqrt{3}})
+3\sqrt{3}k \tau' \sin (\frac{k \tau'}{\sqrt{3}})
\bigg)
\frac{  Z_{s(t)\mathbf k}(\tau') }{ k^3 \tau'} \, d\tau'
    \nn\\
&
+\bigg(
-\frac{2\sqrt3}{k^2\tau^3} \cos (\frac{k \tau}{\sqrt{3}})
-\frac{ 2}{k\tau^2} \sin (\frac{k \tau}{\sqrt{3}})
\bigg)
    \int^\tau
    \bigg( 9 \sin (\frac{k \tau'}{\sqrt{3}})
- 3\sqrt{3} k \tau' \cos (\frac{k \tau'}{\sqrt{3}})\bigg)
    \frac{Z_{s(t)\mathbf k}(\tau')}{k^3 \tau'}d\tau'
\nn\\
&
-\frac{9}{k^4\tau^2}  Z_{s(t)\mathbf k}
- \frac{9}{4k^4\tau^2} A^{'}_{s(t){\bf k}}
+ \frac{9}{4k^4\tau^3} A_{s(t){\bf k}}
+\frac{3}{2k^4\tau} E^{\,'}_{s(t){\bf k}}
+\frac{27}{4k^4\tau^2}F^{||''}_{s(t){\bf k}}
+\frac{27}{4k^4\tau^3}F^{||'}_{s(t){\bf k}}
        \nn\\
        &
-\frac{27}{k^4\tau^4} F^{||}_{s(t){\bf k}}
-\frac{1}{2k^2} E_{s(t){\bf k}}
-\frac{9}{2k^2\tau^2} F^{||}_{s(t){\bf k}}
-\frac{k^2}{6}Y^{||}_{s(t)\bf k}
+\frac{1}{2\tau}Y^{||\,'}_{s(t)\bf k}
-\frac{1}{\tau}W^{||}_{s(t)\bf k}
,
\el
}
and
\bl\label{stwi2Psol}
w^{(2)}_{P_{s(t)}i}
=&
\frac{ \nabla^{-2}q_{2ij}^{,\,j}}{\tau^2}
-\frac{2}{\tau^{2}}
    \int^{\tau}\tau^{'2}\,\nabla^{-2}V_{s(t)ij}^{,j}({\bf x},\tau')d\tau'
-Y^{\perp'}_{s(t)i}
+W^{\perp}_{s(t)i}
,
\el
with $W^{||}_{s(t)}$, $Y^{||}_{s(t)}$, $Y^{\perp}_{s(t)i}$, $W^{\perp}_{s(t)i}$ given in
(\ref{stW||})--(\ref{stYperp}).
Also, the tensor
$\chi^{\top(2)}_{P_{s(t)}ij}$ is given from (\ref{stchi2PChi2}) as
\be
\chi^{\top(2)}_{P_{s(t)}ij}=
\chi^{\top(2)}_{s(t)ij}
+Y^\top_{s(t)ij}
,
\ee
with $\chi^{\top(2)}_{s(t)ij}$ in (\ref{solgw})
and $Y^{\top}_{s(t)ij}$ in (\ref{stYtop1}).
From the above, it is seen that in the Poisson coordinate
the two 2nd-order scalar modes are no longer equal,
that  the 2nd-order vector mode is nonzero
and is contributed  by the scalar-tensor couplings,
and that the 2nd-order tensor mode is different
from the one in synchronous coordinates
by the coupling term $Y^\top_{s(t)ij}$.

Next,
we calculate the density in the Poisson coordinate.
Equation (C6) in Ref.\cite{WangZhang2ndRD2018}
gives that the 0th-order mass-energy density
 as
\be\label{stPoiRho0}
\rho^{(0)}_{P}=\rho^{(0)}=\frac{3}{8\pi G}\frac{a'^{\,2}(\tau)}{a^4(\tau)}
\propto\tau^{-4}.
\ee
By   Eq.(C7) in Ref.\cite{WangZhang2ndRD2018},
the  1st-order $\rho^{(1)}_{P}$ is given by
\be
  \rho ^{(1)}_{P}
=    \rho ^{(1)} - \rho ^{(0)}_{,\, \mu} \xi^{(1)\mu}
.
\ee
For the RD stage,
from $\delta^{(1)}_{\bf k}$ in (\ref{delta1sol3})
and $\alpha^{(1)}_{\bf k}$ in (\ref{stalpha1Po}),
the above gives
\bl
\delta^{(1)}_{P\bf k}
= &
D_2\l(
-\frac{8\sqrt3\,i}{k^2\tau^3}
+\frac{8}{k\tau^2}
+ \frac{8 \,i }{\sqrt3 \,\tau}
-\frac{4 k}{3}\r)e^{-ik\tau/\sqrt3}
\nn\\
&
+D_3\l(
\frac{8\sqrt3\,i}{k^2\tau^3}
+\frac{8}{k\tau^2}
-\frac{8\,i}{\sqrt3\,\tau}
-\frac{4 k}{3}\r)e^{ik\tau/\sqrt3}
\,  .
\el

To 2nd  order,
by using (C8) in Ref.\cite{WangZhang2ndRD2018},
one has
\bl\label{stf2Trans}
\rho^{(2)}_{P}
=&
\rho^{(2)}
-2\rho^{(1)}_{,\, 0}\xi^{(1)0}
-2\rho^{(1)}_{,\, k}\xi^{(1)k}
+\rho^{(0)}_{,\, 0}\xi^{(1)0}_{,\, 0}\xi^{(1)0}
+\rho^{(0)}_{,\, 0}\xi^{(1)0}_{,\, k}\xi^{(1)k}
\nn\\
&
+\rho^{(0)}_{,\, 00}\xi^{(1)0}\xi^{(1)0}
-\rho^{(0)}_{,\, 0}\xi^{(2)0}
\el
which can be written in terms of the 2nd-order density contrast
contributed by scalar-tensor couplings as
\be
\delta^{(2)}_{P_{s(t)}}
=
\delta^{(2)}_{s(t)}
+\l[
-2\frac{a''(\tau)}{a'(\tau)}
+4\frac{a'(\tau)}{a(\tau)}
\r]\alpha^{(2)}_{s(t)} .
\ee
For the RD stage,
using the solution
$\delta_{s(t){\bf k}}^{(2)}$ in (\ref{deltaV2soluphy})
and
$\alpha^{(2)}_{s(t)\bf k}$ in (\ref{stalpha2bar}),
the above becomes
\be
\delta^{(2)}_{P_{s(t)}}
=
\int  \frac{d^3k}{(2\pi)^{3/2}}
\Big[
\delta_{s(t){\bf k}}^{(2)}
+\frac{4}{\tau}\alpha^{(2)}_{s(t)\bf k}
\Big]
e^{i{\bf k\cdot x}} ,
\ee
where
{\allowdisplaybreaks
\bl
&
\delta_{s(t){\bf k}}^{(2)}
+\frac{4}{\tau}\alpha^{(2)}_{s(t)\bf k}
\nn\\
=&
P_2
\l(
-\frac{8\sqrt3 \,i}{k^2\tau^3}
+\frac{8}{k\tau^2}
+\frac{8 i}{\sqrt3\,\tau}
-\frac{4 k}{3}
\r)e^{-ik\tau/\sqrt3}
\nn\\
&
+P_3 \l(
\frac{8\sqrt3 \,i}{k^2\tau^3}
+\frac{8}{k\tau^2}
-\frac{8 i}{\sqrt3\,\tau}
-\frac{4 k}{3}
\r)e^{ik\tau/\sqrt3}
        \nn\\
        &
+\bigg(
\frac{8\sqrt3}{k^2\tau^3} \sin (\frac{k \tau}{\sqrt{3}})
-\frac{8}{k\tau^2} \cos (\frac{k \tau}{\sqrt{3}})
-\frac{8}{\sqrt3\,\tau} \sin (\frac{k \tau}{\sqrt{3}})
\nn\\
&
+\frac{4k}{3}\cos (\frac{k \tau}{\sqrt{3}})
\bigg)
\int^\tau
\bigg(
9\cos (\frac{k \tau'}{\sqrt{3}})
+3\sqrt{3}k \tau' \sin (\frac{k \tau'}{\sqrt{3}})
\bigg)
\frac{  Z_{s(t)\mathbf k}(\tau') }{ k^3 \tau'} \, d\tau'
\nn\\
&
+\bigg(
-\frac{8\sqrt3}{k^2\tau^3} \cos (\frac{k \tau}{\sqrt{3}})
-\frac{ 8}{k\tau^2} \sin (\frac{k \tau}{\sqrt{3}})
+\frac{ 8}{\sqrt{3}\,\tau} \cos (\frac{k \tau}{\sqrt{3}})
\nn\\
&
+\frac{4k}{3} \sin (\frac{k\tau}{\sqrt{3}})
\bigg)
    \int^\tau
    \bigg( 9 \sin (\frac{k \tau'}{\sqrt{3}})
- 3\sqrt{3} k \tau' \cos (\frac{k \tau'}{\sqrt{3}})\bigg)
    \frac{Z_{s(t)\mathbf k}(\tau')}{k^3 \tau'}d\tau'
\nn\\
&
-\frac{36}{k^4\tau^2}  Z_{s(t)\mathbf k}
+\frac{6}{k^4\tau} E^{\,'}_{s(t){\bf k}}
+\frac{27}{k^4\tau^2}F^{||''}_{s(t){\bf k}}
+\frac{27}{k^4\tau^3}F^{||'}_{s(t){\bf k}}
-\frac{108}{k^4\tau^4} F^{||}_{s(t){\bf k}}
+\frac{9}{k^2\tau}F^{||'}_{s(t){\bf k}}
        \nn\\
        &
+3 F^{||}_{s(t){\bf k}}
- \frac{9}{k^4\tau^2} A^{'}_{s(t){\bf k}}
+ \frac{9}{k^4\tau^3} A_{s(t){\bf k}}
-\frac{3}{k^2\tau} A_{s(t){\bf k}}
+\frac{2}{\tau}Y^{||\,'}_{s(t)\bf k}
-\frac{4}{\tau}W^{||}_{s(t)\bf k}
 .
\el
}
It is seen that only $\xi^{(2)\mu}$
contributes to $\delta^{(2)}_{P_{s(t)}}$,
yet $\xi^{(1)\mu}$ does not contribute.

Next,
we solve   for
the 4-velocity in the Poisson coordinate.
By (C9) in Ref.\cite{WangZhang2ndRD2018},
the 0th-order 4-velocity is
$U^{(0)0}_{P}=U^{(0)0}=a^{-1}$ and
$U^{(0)i}_{P}=U^{(0)i}=0$.
By (C10) in Ref.\cite{WangZhang2ndRD2018},
the 1st-order velocity transforms as
\be\label{stU10P}
U^{(1)0}_{P}
=
U^{(1)0}
+\frac{a'}{a^2}\alpha^{(1)}
+\frac{1}{a}\alpha^{(1)'} ,
\ee
\be\label{stU1iP}
v^{(1)i}_{P}
=
v^{(1)i}
+\beta^{(1)',i}
+d^{(1)i\,'} ,
\ee
where the definition
$U^{(1)i}_{P}=a^{-1} v^{(1)i}_{P}$ is used.
From $\alpha^{(1)}_{\bf k}$
in (\ref{stalpha1Po})
for the RD stage,
and using $U^{(1)0}=0$ in (\ref{U0element}),
Eq.(\ref{stU10P}) becomes
\be
U^{(1)0}_{P}
=
D_2(
\frac{2\sqrt3\,i}{k^2\tau^4}
-\frac{2}{k\tau^3}
)e^{-ik\tau/\sqrt3}
+D_3(
-\frac{2\sqrt3\,i}{k^2\tau^4}
-\frac{2}{k\tau^3}
)e^{ik\tau/\sqrt3}
 ,
\ee
which is nonzero,
even in the synchronous coordinate $U^{(1)0}=0$.
By using $v^{||(1)}$ in  (\ref{vsol3}) and
$\beta^{(1)}$ in (\ref{stbeta1Po}),
and taking $v^{\perp(1)i}=0$ and $d^{(1)i}=0$,
Eq.(\ref{stU1iP})  gives
\be
 v^{\perp(1)i}_{P}
=  0,
\ee
\bl\label{stv1||P}
v^{||(1)}_{P\bf k}
=&
D_2\l(
-\frac{2\sqrt3\,i}{k^2\tau^2}
+\frac{2}{k\tau}
    +\frac{i}{\sqrt3}\r)e^{-ik\tau/\sqrt3}
+D_3\l(
\frac{2\sqrt3\,i}{k^2\tau^2}
+\frac{2}{k\tau}
    -\frac{i}{\sqrt3}\r)e^{ik\tau/\sqrt3}
.
\el
Thus the 1st-order velocity remains noncurl in the Poisson coordinate.

To 2nd order,
from (C11) in Ref.\cite{WangZhang2ndRD2018}
and (\ref{U0element}), (\ref{Uielement}),
(\ref{stalpha1Po}), and (\ref{stbeta1Po}),
one has the 0-component of the 4-velocity as
\be
U^{(2)0}_{P_{s(t)}}
        =
-\frac{1}{\tau}\int  \frac{d^3k}{(2\pi)^{3/2}}\Big[
-\alpha^{(2)'}_{s(t)\bf k}
-\frac{1}{\tau}\alpha^{(2)}_{s(t)\bf k}
\Big]e^{i{\bf k\cdot x}}
\,,
\ee
where $-\alpha^{(2)'}_{s(t)\bf k}
-\frac{1}{\tau}\alpha^{(2)}_{s(t)\bf k}$
is given in (\ref{stalphaTaualpha}).
By a definition  $U^{(2)i}_{P_{s(t)}}=a^{-1}v^{(2)i}_{P_{s(t)}}$,
one has the $i$-component as
\be\label{stvP2i4}
v^{(2)}_{P_{s(t)}i}
    =
v^{(2)}_{s(t)i}
+\beta^{(2)'}_{s(t),i}
+d^{(2)'}_{s(t)i}
\, ,
\ee
where  only $\xi^{(2)\mu}$
contributes to $U^{(2)\mu}_{P_{s(t)}}$.
By writing $v^{(2)}_{s(t)i}=v^{||(2)}_{s(t),i}+v^{\perp(2)}_{s(t)i}$
with $v^{\perp(2),i}_{s(t)i}=0$,
 [$\nabla^{-2}\partial^i$(\ref{stvP2i4})]
gives the noncurl part of $v^{(2)}_{P_{s(t)}i}$ as
\bl\label{stvP||2i2}
v^{||(2)}_{P_{s(t)}}
    =&
\int  \frac{d^3k}{(2\pi)^{3/2}}\Big[
v^{||(2)}_{s(t){\bf k}}
+\beta^{(2)'}_{s(t){\bf k}}
\Big]e^{i{\bf k\cdot x}}
\,,
\el
where
{\allowdisplaybreaks
\bl
v^{||(2)}_{s(t){\bf k}}
+\beta^{(2)'}_{s(t){\bf k}}
=&
P_2 \l(
    -\frac{2\sqrt3 \,i}{k^2\tau^2}
    +\frac{2}{k\tau}
    +\frac{i}{\sqrt3}
\r)    e^{-ik\tau/\sqrt3}
+P_3 \l(
     \frac{2\sqrt3 \,i}{k^2\tau^2}
    +\frac{2}{k\tau}
    -\frac{i}{\sqrt3}
    \r)e^{ik\tau/\sqrt3}
\nn\\
&
+\left(
\frac{2\sqrt3}{k^2\tau^2} \sin (\frac{k \tau}{\sqrt{3}})
-\frac{2}{k\tau} \cos (\frac{k \tau}{\sqrt{3}})
-\frac{1}{\sqrt{3}}\sin (\frac{k \tau}{\sqrt{3}})
\right)
\int^\tau
\bigg(
9\cos (\frac{k \tau'}{\sqrt{3}})
\nn\\
&
+3\sqrt{3}k \tau' \sin (\frac{k \tau'}{\sqrt{3}})
\bigg)
\frac{  Z_{s(t)\mathbf k}(\tau') }{ k^3 \tau'} \, d\tau'
\nn\\
&
+\left(
-\frac{2\sqrt3}{k^2\tau^2} \cos (\frac{k \tau}{\sqrt{3}})
-\frac{2}{k\tau} \sin (\frac{k \tau}{\sqrt{3}})
+\frac{1}{\sqrt{3}} \cos (\frac{k\tau}{\sqrt{3}})
\right)
    \int^\tau
    \bigg( 9 \sin (\frac{k \tau'}{\sqrt{3}})
\nn\\
&
- 3\sqrt{3} k \tau' \cos (\frac{k \tau'}{\sqrt{3}})\bigg)
    \frac{Z_{s(t)\mathbf k}(\tau')}{k^3 \tau'}d\tau'
        \nn\\
        &
-\frac{9}{k^4\tau}  Z_{s(t)\mathbf k}
+\frac{3}{2k^4} E^{\,'}_{s(t){\bf k}}
+\frac{27}{4k^4\tau}F^{||''}_{s(t){\bf k}}
+\frac{27}{4k^4\tau^2}F^{||'}_{s(t){\bf k}}
-\frac{27}{k^4\tau^3} F^{||}_{s(t){\bf k}}
        \nn\\
        &
+\frac{9}{4k^2}F^{||'}_{s(t){\bf k}}
- \frac{9}{4k^4\tau} A^{'}_{s(t){\bf k}}
+ \frac{9}{4k^4\tau^2} A_{s(t){\bf k}}
-\frac{3}{4k^2} A_{s(t){\bf k}}
+\frac{1}{2}Y^{||'}_{s(t){\bf k}}
,
\el
}
with $E_{s(t)}$, $F_{s(t)}$, $A_{s(t)}$, $Z_{s(t)}$,
and $Y^{||}_{s(t)}$ given by
Eqs.(\ref{ES1RD}), (\ref{FSnoncu}), (\ref{AS}), (\ref{ZSall}), and
(\ref{stY||}).

The curl part of $v^{(2)}_{P_{s(t)}i}$ is given by
[(\ref{stvP2i4})$-\partial_i$(\ref{stvP||2i2})] as follows:
\bl
v^{\perp(2)}_{P_{s(t)}i}
    =&
\frac{q_{2ij}^{\,,\,j}({\bf x})}{8}
-\frac{ \nabla^{-2}q_{2ij}^{\,,\,j}}{\tau^2}
- \frac{1}{4}
    \int^{\tau}\tau^{'2}\,V_{s(t)ij}^{,j}({\bf x},\tau')d\tau'
\nn\\
&
+ \frac{2}{\tau^{2}}
    \int^{\tau}\tau^{'2}\,\nabla^{-2}V_{s(t)ij}^{,j}({\bf x},\tau')d\tau'
+\frac{\tau^2}{4}\l(
M_{s(t)i}
-\partial_i\nabla^{-2}M_{s(t)k}^{,\,k}
\r)
+Y_{s(t)i}^{\perp'}
\,.
\el
where  $(M_{s(t)i}-\partial_i\nabla^{-2}M_{s(t)k}^{,k})$,
$V_{s(t)ij}$,
and $Y^{\perp}_{s(t)i}$ are given in
(\ref{MSiCurlRD}), (\ref{SourceCurl1RD}), and (\ref{stYperp}),
respectively.
Thus the  2nd-order curl part of the 3-velocity is nonzero
even if $v^{\perp(1)}_{P\,i}=0$.

Next, we will calculate the 2nd-order perturbations with
tensor-tensor couplings in the Poisson coordinate,
by  similar procedures to the above.
By  (C5)  with the coupling of tensor-tensor  in Ref.\cite{WangZhang2ndRD2018},
one has
\be\label{Tpsi2PChi2}
\psi^{(2)}_{P_{T}}=
-\alpha^{(2)'}_{T}
-\frac{a'}{a}\alpha^{(2)}_{T}
,
\ee
\be\label{Twi2PChi2}
w^{(2)}_{P_{T}\,i}
=
\alpha^{(2)}_{T,\,i}
-\beta^{(2)'}_{T,\,i}
-d^{(2)'}_{T\,i},
\ee
\be\label{Tphi2PChi2}
\phi^{(2)}_{P_{T}}=
\phi^{(2)}_{T}
+\frac{a'}{a}\alpha^{(2)}_{T}
+\frac{1}{3}\nabla^2\beta^{(2)}_{T},
\ee
\be\label{Tchi2PChi2}
\chi^{\top(2)}_{P_{T}\,ij}=
D_{ij}\chi^{||(2)}_{T}
+\chi^{\perp(2)}_{Tij}
+\chi^{\top(2)}_{Tij}
-\big(d^{(2)}_{T\,i,j}
+d^{(2)}_{T\,j,i}
+2 D_{ij}\beta^{(2)}_{T}\big)
,
\ee
where only $\xi^{(2)\mu}$
contributes to the above transformations,
as $\xi^{(1)\mu}$ does not contain the tensor mode.

We  need to solve for
$\alpha^{(2)}_{T}$,  $\beta^{(2)}_{T}$ and $d^{(2)}_{T\,i}$.
To do this,
we use the transverse condition of
$w^{(2)}_{P_{T}\,i}$ and  the traceless and transverse condition of
$\chi^{\top(2)}_{P_{T}\,ij}$,  which lead to
\be\label{Twi2Pncurl}
\alpha^{(2)}_{T}
-\beta^{(2)'}_{T}
=0,
\ee
\be\label{Tchi2Pscalar}
\chi^{||(2)}_{T}
-2\beta^{(2)}_{T}
=0
,
\ee
\be\label{Tchi2Pvector}
\nabla^{-2}\partial^j\chi^{\perp(2)}_{Tij}
-d^{(2)}_{T\,i}
=0 \, ,
\ee
and the solutions
\be\label{Twi2PncurlSol1}
\alpha^{(2)}_{T}
=\frac{1}{2}\chi^{||(2)\,'}_{T},
\ee
\be\label{Tchi2PscalarSol1}
\beta^{(2)}_{T}=\frac{1}{2}\chi^{||(2)}_{T}
,
\ee
\be\label{Tchi2PvectorSol1}
d^{(2)}_{T\,i}=\nabla^{-2}\chi^{\perp(2),j}_{Tij}
.
\ee
Plugging the solutions $\chi^{\perp(2)}_{Tij}$  (\ref{TchiVecSolphy}) and
$\chi^{||(2)}_{T}$  (\ref{Tchi2Ssolphy}) into the above equations,
one has
{\allowdisplaybreaks
\bl\label{Talpha2bar}
\alpha^{(2)}_{T\bf k}
=&
-Q_2 \frac{2\sqrt3 \,i}{k^2\tau^2}e^{-ik\tau/\sqrt3}
+Q_3 \frac{2\sqrt3 \,i}{k^2\tau^2}e^{ik\tau/\sqrt3}
        \nn\\
        &
-\frac{1}{k\tau}\int^\tau \frac{
    3(k^2 \tau'^2+6)}{k^3 \tau'} Z_{T\mathbf k}(\tau')d\tau'
-\frac{9}{k^4\tau}  Z_{T\mathbf k}
    \nn\\
&
+\frac{2\sqrt3}{k^2\tau^2} \sin (\frac{k \tau}{\sqrt{3}})
\int^\tau
\bigg(
9\cos (\frac{k \tau'}{\sqrt{3}})
+3\sqrt{3}k \tau' \sin (\frac{k \tau'}{\sqrt{3}})
\bigg)
\frac{  Z_{T\mathbf k}(\tau') }{ k^3 \tau'} \, d\tau'
\nn\\
&
-\frac{2\sqrt3}{k^2\tau^2} \cos (\frac{k \tau}{\sqrt{3}})
    \int^\tau
    \bigg( 9 \sin (\frac{k \tau'}{\sqrt{3}})
- 3\sqrt{3} k \tau' \cos (\frac{k \tau'}{\sqrt{3}})\bigg)
    \frac{Z_{T\mathbf k}(\tau')}{k^3 \tau'}d\tau'
        \nn\\
        &
+\frac{3}{2k^4} E^{\,'}_{T{\bf k}}
- \frac{9}{4k^4\tau} A^{'}_{T{\bf k}}
+ \frac{9}{4k^4\tau^2} A_{T{\bf k}}
-\frac{3}{4k^2} A_{T{\bf k}}
,
\el
\bl\label{Tbeta2bar}
\beta^{(2)}_{T\bf k}
=&
Q_2 \frac{2\sqrt3 \,i}{k^2\tau}e^{-ik\tau/\sqrt3}
-Q_3 \frac{2\sqrt3 \,i}{k^2\tau}e^{ik\tau/\sqrt3}
        \nn\\
        &
-\frac{2}{k}\int^\tau
\l[
Q_2 e^{-ik\tau'/\sqrt3}
+Q_3 e^{ik\tau'/\sqrt3}
\r]
\frac{d\tau'}{\tau'}
\nn\\
&
-\frac{ \ln\tau}{k }\int^\tau \frac{
    3(k^2 \tau'^2+6)}{k^3 \tau'} Z_{T\mathbf k}(\tau')d\tau'
+\int^\tau\frac{3(k^2 \tau'^2+6)\ln\tau'+9}{k^4 \tau'} Z_{T\mathbf k}(\tau')d\tau'
    \nn\\
&
-\frac{2\sqrt3}{k^2\tau} \sin (\frac{k \tau}{\sqrt{3}})
\int^\tau
\bigg(
9\cos (\frac{k \tau'}{\sqrt{3}})
+3\sqrt{3}k \tau' \sin (\frac{k \tau'}{\sqrt{3}})
\bigg)
\frac{  Z_{T\mathbf k}(\tau') }{ k^3 \tau'} \, d\tau'
\nn\\
&
+\frac{2\sqrt3}{k^2\tau} \cos (\frac{k \tau}{\sqrt{3}})
    \int^\tau
    \bigg( 9 \sin (\frac{k \tau'}{\sqrt{3}})
- 3\sqrt{3} k \tau' \cos (\frac{k \tau'}{\sqrt{3}})\bigg)
    \frac{Z_{T\mathbf k}(\tau')}{k^3 \tau'}d\tau'
\nn\\
&
+\int^\tau
\bigg[
\frac{6}{k^3\tau''} \cos (\frac{k \tau''}{\sqrt{3}})
\int^{\tau''}
\bigg(
3\cos (\frac{k \tau'}{\sqrt{3}})
+\sqrt{3}k \tau' \sin (\frac{k \tau'}{\sqrt{3}})
\bigg)
\frac{  Z_{T\mathbf k}(\tau') }{ k \tau'} \, d\tau'
\bigg]d\tau''
\nn\\
&
+\int^\tau
\bigg[
\frac{6}{k^3\tau''} \sin (\frac{k \tau''}{\sqrt{3}})
    \int^{\tau''}
    \bigg(
    3 \sin (\frac{k \tau'}{\sqrt{3}})
- \sqrt{3} k \tau' \cos (\frac{k \tau'}{\sqrt{3}})\bigg)
    \frac{Z_{T\mathbf k}(\tau')}{k \tau'}d\tau'
\bigg]d\tau''
        \nn\\
        &
+\frac{3}{2k^4} E_{T{\bf k}}
- \frac{9}{4k^4\tau} A_{T{\bf k}}(\tau)
-\frac{3}{4k^2}\int^\tau A_{T{\bf k}}(\tau') d\tau'
,
\el
\be\label{Td2bar2}
d^{(2)}_{T\,i}
=
\frac{ \nabla^{-2}q_{4ij}^{,\,j}}{\tau}
 + 2\int^\tau \frac{d\tau'}{\tau^{'2}}
    \int^{\tau'}\tau^{''2}\,\nabla^{-2}V_{T\,ij}^{,j}({\bf x},\tau'')d\tau''
,
\ee
}
where $E_{T}$, $A_{T}$, $Z_{T}$, and $V_{T\,ij}$ are given by
Eqs.(\ref{TES1RD}), (\ref{TAS}), (\ref{TZSall}), and
 (\ref{TSourceCurl1RD}).

Plugging Eqs.(\ref{Tphi2SkSolphy}),
 (\ref{Talpha2bar}), and (\ref{Tbeta2bar})
into (\ref{Tpsi2PChi2}) and (\ref{Tphi2PChi2})
with $a(\tau)\propto\tau$,
one obtains the 2nd-order metric perturbations in the Poisson gauge as
\be\label{Tpsi2Psol}
\psi^{(2)}_{P_{T}}=
\int  \frac{d^3k}{(2\pi)^{3/2}}\Big[
-\alpha^{(2)'}_{T\bf k}
-\frac{1}{\tau}\alpha^{(2)}_{T\bf k}
\Big]e^{i{\bf k\cdot x}}
,
\ee
where
{\allowdisplaybreaks
\bl\label{TalphaTaualpha}
&
-\alpha^{(2)'}_{T\bf k}
-\frac{1}{\tau}\alpha^{(2)}_{T\bf k}
\nn\\
=&
Q_2 (
    -\frac{2\sqrt3 \,i}{k^2\tau^3}
    +\frac{2}{k\tau^2})e^{-ik\tau/\sqrt3}
+Q_3 (
   -\frac{2\sqrt3 \,i}{k^2\tau^3}
    +\frac{2 }{k\tau^2})e^{ik\tau/\sqrt3}
    \nn\\
    &
+\Big(\frac{2\sqrt3}{k^2\tau^3} \sin (\frac{k \tau}{\sqrt{3}})
-\frac{2}{k\tau^2} \cos (\frac{k \tau}{\sqrt{3}})\Big)
\int^\tau
\bigg(
9\cos (\frac{k \tau'}{\sqrt{3}})
+3\sqrt{3}k \tau' \sin (\frac{k \tau'}{\sqrt{3}})
\bigg)
\frac{  Z_{T\mathbf k}(\tau') }{ k^3 \tau'} \, d\tau'
\nn\\
&
+\Big(
-\frac{2\sqrt3}{k^2\tau^3} \cos (\frac{k \tau}{\sqrt{3}})
-\frac{2}{k\tau^2} \sin (\frac{k \tau}{\sqrt{3}})
\Big)
    \int^\tau
    \bigg( 9 \sin (\frac{k \tau'}{\sqrt{3}})
- 3\sqrt{3} k \tau' \cos (\frac{k \tau'}{\sqrt{3}})\bigg)
    \frac{Z_{T\mathbf k}(\tau')}{k^3 \tau'}d\tau'
        \nn\\
        &
+\frac{3}{k^2} Z_{T\mathbf k}
+\frac{9}{k^4\tau}  Z^{\,'}_{T\mathbf k}
-\frac{3}{2k^4\tau} E^{\,'}_{T{\bf k}}
-\frac{3}{2k^4} E^{\,''}_{T{\bf k}}
+ \frac{9}{4k^4\tau} A^{''}_{T{\bf k}}
+ \frac{9}{4k^4\tau^3} A_{T{\bf k}}
\nn\\
&
- \frac{9}{4k^4\tau^2} A^{'}_{T{\bf k}}
+\frac{3}{4k^2\tau} A_{T{\bf k}}
+\frac{3}{4k^2} A^{'}_{T{\bf k}}
,
\el
}
and
\be\label{Tphi2Psol}
\phi^{(2)}_{P_{T}}=
\int  \frac{d^3k}{(2\pi)^{3/2}}\Big[
\phi^{(2)}_{T\bf k}
+\frac{1}{\tau}\alpha^{(2)}_{T\bf k}
-\frac{k^2}{3}\beta^{(2)}_{T\bf k}
\Big]e^{i{\bf k\cdot x}}
,
\ee
where
{\allowdisplaybreaks
\bl
&
\phi^{(2)}_{T\bf k}
+\frac{1}{\tau}\alpha^{(2)}_{T\bf k}
-\frac{k^2}{3}\beta^{(2)}_{T\bf k}
\nn\\
=&
Q_2
\l(-\frac{2\sqrt3 \,i}{k^2\tau^3}
+\frac{2}{k\tau^2}
\r)e^{-ik\tau/\sqrt3}
+Q_3\l(
\frac{2\sqrt3 \,i}{k^2\tau^3}
+\frac{2}{k\tau^2}
\r)e^{ik\tau/\sqrt3}
\nn\\
&
+\bigg(
\frac{2\sqrt3}{k^2\tau^3} \sin (\frac{k \tau}{\sqrt{3}})
-\frac{2}{k\tau^2} \cos (\frac{k \tau}{\sqrt{3}})
\bigg)
\int^\tau
\bigg(
9\cos (\frac{k \tau'}{\sqrt{3}})
+3\sqrt{3}k \tau' \sin (\frac{k \tau'}{\sqrt{3}})
\bigg)
\frac{  Z_{T\mathbf k}(\tau') }{ k^3 \tau'} \, d\tau'
    \nn\\
&
+\bigg(
-\frac{2\sqrt3}{k^2\tau^3} \cos (\frac{k \tau}{\sqrt{3}})
-\frac{ 2}{k\tau^2} \sin (\frac{k \tau}{\sqrt{3}})
\bigg)
    \int^\tau
    \bigg( 9 \sin (\frac{k \tau'}{\sqrt{3}})
- 3\sqrt{3} k \tau' \cos (\frac{k \tau'}{\sqrt{3}})\bigg)
    \frac{Z_{T\mathbf k}(\tau')}{k^3 \tau'}d\tau'
\nn\\
&
-\frac{9}{k^4\tau^2}  Z_{T\mathbf k}
- \frac{9}{4k^4\tau^2} A^{'}_{T{\bf k}}
+ \frac{9}{4k^4\tau^3} A_{T{\bf k}}
+\frac{3}{2k^4\tau} E^{\,'}_{T{\bf k}}
-\frac{1}{2k^2} E_{T{\bf k}}
.
\el
}
Thus, we see  that in the Poisson gauge,
the two 2nd-order scalar modes are not equal.
This can be checked by using  (\ref{Tpsi2PChi2}), (\ref{Tphi2PChi2}),
(\ref{Twi2PncurlSol1}), and (\ref{Tchi2PscalarSol1}),
\[
\phi^{(2)}_{P_{T}}-\psi^{(2)}_{P_{T}}=
\frac{1}{2}\chi^{||(2)\,''}_{T}
+\frac{1}{\tau}\chi^{||(2)\,'}_{T}
+\frac{1}{6}\nabla^2\chi^{||(2)}_{T}
+\phi^{(2)}_{T}
=\frac{3}{2}\nabla^{-2}\nabla^{-2}\bar S_{T\,lm}^{\, ,\,lm},
\]
according to the evolution equation (\ref{TEvo2ndSsChi1RD}).
A similar relation holds for the scalar-tensor and scalar-scalar cases.

The vector mode is given by (\ref{Twi2PChi2})
and (\ref{Talpha2bar})--(\ref{Td2bar2}) as
\bl\label{Twi2Psol}
w^{(2)}_{P_{T}\,i}
=&
\frac{ \nabla^{-2}q_{4ij}^{,\,j}}{\tau^2}
-\frac{2}{\tau^{2}}
    \int^{\tau}\tau^{'2}\,\nabla^{-2}V_{T\,ij}^{,j}({\bf x},\tau')d\tau'
,
\el
which is nonzero
and sourced by
 the 1st-order couplings $ V_{T\,ij}$ in (\ref{TSourceCurl1RD}).

The tensor
$\chi^{\top(2)}_{P_{T}\,ij}$ is given from (\ref{Tchi2PChi2}) as
\be
\chi^{\top(2)}_{P_{T}\,ij}=
\chi^{\top(2)}_{Tij}
,
\ee
which is the same as in the synchronous coordinate.

Next,
we calculate the 2nd-order density in the Poisson coordinate.
We write (\ref{stf2Trans}) in terms of the 2nd-order density contrast
contributed by tensor-tensor couplings as
\be
\delta^{(2)}_{P_{T}}
=
\delta^{(2)}_{T}
+\l[
-2\frac{a''(\tau)}{a'(\tau)}
+4\frac{a'(\tau)}{a(\tau)}
\r]\alpha^{(2)}_{T} .
\ee
For the RD stage,
using the solution
$\delta_{T{\bf k}}^{(2)}$ in (\ref{TdeltaV2soluphy})
and
$\alpha^{(2)}_{T\bf k}$ in (\ref{Talpha2bar}),
the above becomes
\be
\delta^{(2)}_{P_{T}}
=
\int  \frac{d^3k}{(2\pi)^{3/2}}
\Big[
\delta_{T{\bf k}}^{(2)}
+\frac{4}{\tau}\alpha^{(2)}_{T\bf k}
\Big]
e^{i{\bf k\cdot x}} ,
\ee
where
{\allowdisplaybreaks
\bl
&
\delta_{T{\bf k}}^{(2)}
+\frac{4}{\tau}\alpha^{(2)}_{T\bf k}
\nn\\
=&
Q_2
\l(
-\frac{8\sqrt3 \,i}{k^2\tau^3}
+\frac{8}{k\tau^2}
+\frac{8 i}{\sqrt3\,\tau}
-\frac{4 k}{3}
\r)e^{-ik\tau/\sqrt3}
\nn\\
&
+Q_3 \l(
\frac{8\sqrt3 \,i}{k^2\tau^3}
+\frac{8}{k\tau^2}
-\frac{8 i}{\sqrt3\,\tau}
-\frac{4 k}{3}
\r)e^{ik\tau/\sqrt3}
        \nn\\
        &
+\bigg(
\frac{8\sqrt3}{k^2\tau^3} \sin (\frac{k \tau}{\sqrt{3}})
-\frac{8}{k\tau^2} \cos (\frac{k \tau}{\sqrt{3}})
-\frac{8}{\sqrt3\,\tau} \sin (\frac{k \tau}{\sqrt{3}})
\nn\\
&
+\frac{4k}{3}\cos (\frac{k \tau}{\sqrt{3}})
\bigg)
\int^\tau
\bigg(
9\cos (\frac{k \tau'}{\sqrt{3}})
+3\sqrt{3}k \tau' \sin (\frac{k \tau'}{\sqrt{3}})
\bigg)
\frac{  Z_{T\mathbf k}(\tau') }{ k^3 \tau'} \, d\tau'
\nn\\
&
+\bigg(
-\frac{8\sqrt3}{k^2\tau^3} \cos (\frac{k \tau}{\sqrt{3}})
-\frac{ 8}{k\tau^2} \sin (\frac{k \tau}{\sqrt{3}})
+\frac{ 8}{\sqrt{3}\,\tau} \cos (\frac{k \tau}{\sqrt{3}})
\nn\\
&
+\frac{4k}{3} \sin (\frac{k\tau}{\sqrt{3}})
\bigg)
    \int^\tau
    \bigg( 9 \sin (\frac{k \tau'}{\sqrt{3}})
- 3\sqrt{3} k \tau' \cos (\frac{k \tau'}{\sqrt{3}})\bigg)
    \frac{Z_{T\mathbf k}(\tau')}{k^3 \tau'}d\tau'
\nn\\
&
-\frac{36}{k^4\tau^2}  Z_{T\mathbf k}
+\frac{6}{k^4\tau} E^{\,'}_{T{\bf k}}
- \frac{9}{k^4\tau^2} A^{'}_{T{\bf k}}
+ \frac{9}{k^4\tau^3} A_{T{\bf k}}
-\frac{3}{k^2\tau} A_{T{\bf k}} .
\el
}
It is seen that only $\xi^{(2)\mu}$
contributes to $\delta^{(2)}_{P_{T}}$,
yet $\xi^{(1)\mu}$ does not contribute.

The 2nd-order 4-velocity in the Poisson coordinate is derived
from (C11) in Ref.\cite{WangZhang2ndRD2018}
and from (\ref{U0element}), (\ref{Uielement}),
(\ref{stalpha1Po}), and (\ref{stbeta1Po}) as
\be
U^{(2)0}_{P_{T}}
        =
-\frac{1}{\tau}\int  \frac{d^3k}{(2\pi)^{3/2}}\Big[
-\alpha^{(2)'}_{T\bf k}
-\frac{1}{\tau}\alpha^{(2)}_{T\bf k}
\Big]e^{i{\bf k\cdot x}}
\,,
\ee
with $-\alpha^{(2)'}_{T\bf k}
-\frac{1}{\tau}\alpha^{(2)}_{T\bf k}$
given in (\ref{TalphaTaualpha}),
and
\be\label{TvP2i4}
v^{(2)}_{P_{T}\,i}
    =
v^{(2)}_{Ti}
+\beta^{(2)'}_{T,i}
+d^{(2)'}_{Ti}
\,,
\ee
with  $U^{(2)i}_{P_{T}}=a^{-1}v^{(2)i}_{P_{T}}$.
Only $\xi^{(2)\mu}$
contributes to $U^{(2)\mu}_{P_{T}}$.
By writing $v^{(2)}_{Ti}=v^{||(2)}_{T,i}+v^{\perp(2)}_{Ti}$
with $v^{\perp(2),i}_{Ti}=0$,
[$\nabla^{-2}\partial^i$(\ref{TvP2i4})]
gives the noncurl part of $v^{(2)}_{P_{T}\,i}$ as
\bl\label{TvP||2i2}
v^{||(2)}_{P_{T}}
    =&
\int  \frac{d^3k}{(2\pi)^{3/2}}\Big[
v^{||(2)}_{T{\bf k}}
+\beta^{(2)'}_{T{\bf k}}
\Big]e^{i{\bf k\cdot x}}
\,,
\el
where
{\allowdisplaybreaks
\bl
v^{||(2)}_{T{\bf k}}
+\beta^{(2)'}_{T{\bf k}}
=&
Q_2 \l(
    -\frac{2\sqrt3 \,i}{k^2\tau^2}
    +\frac{2}{k\tau}
    +\frac{i}{\sqrt3}
\r)    e^{-ik\tau/\sqrt3}
+Q_3 \l(
     \frac{2\sqrt3 \,i}{k^2\tau^2}
    +\frac{2}{k\tau}
    -\frac{i}{\sqrt3}
    \r)e^{ik\tau/\sqrt3}
\nn\\
&
+\left(
\frac{2\sqrt3}{k^2\tau^2} \sin (\frac{k \tau}{\sqrt{3}})
-\frac{2}{k\tau} \cos (\frac{k \tau}{\sqrt{3}})
-\frac{1}{\sqrt{3}}\sin (\frac{k \tau}{\sqrt{3}})
\right)
\int^\tau
\bigg(
9\cos (\frac{k \tau'}{\sqrt{3}})
\nn\\
&
+3\sqrt{3}k \tau' \sin (\frac{k \tau'}{\sqrt{3}})
\bigg)
\frac{  Z_{T\mathbf k}(\tau') }{ k^3 \tau'} \, d\tau'
\nn\\
&
+\left(
-\frac{2\sqrt3}{k^2\tau^2} \cos (\frac{k \tau}{\sqrt{3}})
-\frac{2}{k\tau} \sin (\frac{k \tau}{\sqrt{3}})
+\frac{1}{\sqrt{3}} \cos (\frac{k\tau}{\sqrt{3}})
\right)
    \int^\tau
    \bigg( 9 \sin (\frac{k \tau'}{\sqrt{3}})
\nn\\
&
- 3\sqrt{3} k \tau' \cos (\frac{k \tau'}{\sqrt{3}})\bigg)
    \frac{Z_{T\mathbf k}(\tau')}{k^3 \tau'}d\tau'
        \nn\\
        &
-\frac{9}{k^4\tau}  Z_{T\mathbf k}
+\frac{3}{2k^4} E^{\,'}_{T{\bf k}}
- \frac{9}{4k^4\tau} A^{'}_{T{\bf k}}
+ \frac{9}{4k^4\tau^2} A_{T{\bf k}}
-\frac{3}{4k^2} A_{T{\bf k}}
,
\el
}
with $E_{T}$,  $A_{T}$, and $Z_{T}$ given by
Eqs.(\ref{TES1RD}), (\ref{TAS}), and (\ref{TZSall}).
The curl part of $v^{(2)}_{P_{T}\,i}$ is given by
[(\ref{TvP2i4})$-\partial_i$(\ref{TvP||2i2})] as follows:
\bl
v^{\perp(2)}_{P_{T}\,i}
    =&
\frac{q_{4ij}^{\,,\,j}({\bf x})}{8}
-\frac{ \nabla^{-2}q_{4ij}^{\,,\,j}}{\tau^2}
- \frac{1}{4}
    \int^{\tau}\tau^{'2}\,V_{T\,ij}^{,j}({\bf x},\tau')d\tau'
\nn\\
&
+ \frac{2}{\tau^{2}}
    \int^{\tau}\tau^{'2}\,\nabla^{-2}V_{T\,ij}^{,j}({\bf x},\tau')d\tau'
+\frac{\tau^2}{4}\l(
M_{T\,i}
-\partial_i\nabla^{-2}M_{T\,k}^{,\,k}
\r)
\,,
\el
where  $(M_{T\,i}-\partial_i\nabla^{-2}M_{T\,k}^{,k})$,
and $V_{T\,ij}$ are given in
(\ref{TMSiCurlRD}) and (\ref{TSourceCurl1RD})
respectively.
The  2nd-order curl part of 3-velocity is nonzero
even if $v^{\perp(1)}_{P\,i}=0$.

The 2nd-order perturbations with scalar-scalar couplings
in Poisson coordinate has not been given in Ref.\cite{WangZhang2ndRD2018}.
Using the solutions  in synchronous coordinate in Ref.\cite{WangZhang2ndRD2018},
by similar gauge transformation procedures to the above
we obtain the results as the following (the computation details are skipped).

With the source terms $E_S$, $F^{||}_S$, $A_S$, $Z_S$, and $V_{S\,ij}$
given in Eqs.(4.2), (4.28), (4.24), (5.14),
and (4.19) of Ref.\cite{WangZhang2ndRD2018},
the 2nd-order metric perturbations in the Poisson coordinates are
\bl\label{sspsi2Psol}
\psi^{(2)}_{P_S}=&
\int  \frac{d^3k}{(2\pi)^{3/2}}\Big[
-\alpha^{(2)'}_{S\bf k}
-\frac{1}{\tau}\alpha^{(2)}_{S\bf k}
\Big]e^{i{\bf k\cdot x}}
+\frac{1}{4}\chi^{||(1)'''}\chi^{||(1)'}
+\frac{5}{4\tau}\chi^{||(1)''}\chi^{||(1)'}
\nn\\
&
+\frac{1}{4\tau^2}\chi^{||(1)'}\chi^{||(1)'}
+\frac{1}{4}\chi^{||(1)''}_{,\,l}\chi^{||(1),\,l}
+\frac{1}{4\tau}\chi^{||(1)'}_{,\,l}\chi^{||(1),\,l}
+\frac{1}{2}\chi^{||(1)''}\chi^{||(1)''}
,
\el
where
{\allowdisplaybreaks
\bl\label{ssalphaTaualpha}
&
-\alpha^{(2)'}_{S\bf k}
-\frac{1}{\tau}\alpha^{(2)}_{S\bf k}
\nn\\
=&
G_2 (
    -\frac{2\sqrt3 \,i}{k^2\tau^3}
    +\frac{2}{k\tau^2})e^{-ik\tau/\sqrt3}
+G_3 (
   -\frac{2\sqrt3 \,i}{k^2\tau^3}
    +\frac{2 }{k\tau^2})e^{ik\tau/\sqrt3}
    \nn\\
    &
+\Big(\frac{2\sqrt3}{k^2\tau^3} \sin (\frac{k \tau}{\sqrt{3}})
-\frac{2}{k\tau^2} \cos (\frac{k \tau}{\sqrt{3}})\Big)
\int^\tau
\bigg(
9\cos (\frac{k \tau'}{\sqrt{3}})
+3\sqrt{3}k \tau' \sin (\frac{k \tau'}{\sqrt{3}})
\bigg)
\frac{  Z_{S\mathbf k}(\tau') }{ k^3 \tau'} \, d\tau'
\nn\\
&
+\Big(
-\frac{2\sqrt3}{k^2\tau^3} \cos (\frac{k \tau}{\sqrt{3}})
-\frac{2}{k\tau^2} \sin (\frac{k \tau}{\sqrt{3}})
\Big)
    \int^\tau
    \bigg( 9 \sin (\frac{k \tau'}{\sqrt{3}})
- 3\sqrt{3} k \tau' \cos (\frac{k \tau'}{\sqrt{3}})\bigg)
    \frac{Z_{S\mathbf k}(\tau')}{k^3 \tau'}d\tau'
        \nn\\
        &
+\frac{3}{k^2} Z_{S\mathbf k}
+\frac{9}{k^4\tau}  Z^{\,'}_{S\mathbf k}
-\frac{3}{2k^4\tau} E^{\,'}_{S\,{\bf k}}
-\frac{3}{2k^4} E^{\,''}_{S\,{\bf k}}
-\frac{27}{4k^4\tau}F^{||'''}_{S\,{\bf k}}
+\frac{135}{4k^4\tau^3}F^{||'}_{S\,{\bf k}}
-\frac{27}{4k^4\tau^2}F^{||''}_{S\,{\bf k}}
        \nn\\
        &
-\frac{54}{k^4\tau^4} F^{||}_{S\,{\bf k}}
-\frac{9}{4k^2\tau}F^{||'}_{S\,{\bf k}}
-\frac{9}{4k^2}F^{||''}_{S\,{\bf k}}
+ \frac{9}{4k^4\tau} A^{''}_{S\,{\bf k}}
+ \frac{9}{4k^4\tau^3} A_{S\,{\bf k}}
- \frac{9}{4k^4\tau^2} A^{'}_{S\,{\bf k}}
\nn\\
&
+\frac{3}{4k^2\tau} A_{S\,{\bf k}}
+\frac{3}{4k^2} A^{'}_{S\,{\bf k}}
-\frac{1}{2\tau}Y^{||\,'}_{S\bf k}
-\frac{1}{2}Y^{||\,''}_{S\bf k}
+\frac{1}{\tau}W^{||}_{S\bf k}
+W^{||\,'}_{S\bf k}
,
\el
}
and
{\allowdisplaybreaks
\bl\label{ssphi2Psol}
\phi^{(2)}_{P_S}=&
\int  \frac{d^3k}{(2\pi)^{3/2}}\Big[
\phi^{(2)}_{S\bf k}
+\frac{1}{\tau}\alpha^{(2)}_{S\bf k}
-\frac{k^2}{3}\beta^{(2)}_{S\bf k}
\Big]e^{i{\bf k\cdot x}}
-\phi^{(1)'}\chi^{||(1)'}
-\frac{2}{\tau}\phi^{(1)}\chi^{||(1)'}
\nn\\
&
-\frac{1}{4\tau^2}\chi^{||(1)'}\chi^{||(1)'}
-\frac{1}{4\tau}\chi^{||(1)''}\chi^{||(1)'}
-\frac{2}{3}\phi^{(1)}\nabla^2\chi^{||(1)}
-\frac{1}{12}\chi^{||(1)'}\nabla^2\chi^{||(1)'}
\nn\\
&
-\frac{1}{12}\chi^{||(1),\,l}\nabla^2\chi^{||(1)}_{,\,l}
-\frac{1}{3\tau}\chi^{||(1)'}\nabla^2\chi^{||(1)}
-\frac{1}{9}\nabla^2\chi^{||(1)}\nabla^2\chi^{||(1)}
-\chi^{||(1),\,l}\phi^{(1)}_{,\,l}
\nn\\
&
-\frac{1}{4\tau}\chi^{||(1),\,l}\chi^{||(1)'}_{,\,l}
+\frac{1}{6}\chi^{||(1)}_{,\,lm}\chi^{||(1),\,lm}
,
\el
}
where
{\allowdisplaybreaks
\bl
&
\phi^{(2)}_{S\bf k}
+\frac{1}{\tau}\alpha^{(2)}_{S\bf k}
-\frac{k^2}{3}\beta^{(2)}_{S\bf k}
\nn\\
=&
G_2
\l(-\frac{2\sqrt3 \,i}{k^2\tau^3}
+\frac{2}{k\tau^2}
\r)e^{-ik\tau/\sqrt3}
+G_3\l(
\frac{2\sqrt3 \,i}{k^2\tau^3}
+\frac{2}{k\tau^2}
\r)e^{ik\tau/\sqrt3}
\nn\\
&
+\bigg(
\frac{2\sqrt3}{k^2\tau^3} \sin (\frac{k \tau}{\sqrt{3}})
-\frac{2}{k\tau^2} \cos (\frac{k \tau}{\sqrt{3}})
\bigg)
\int^\tau
\bigg(
9\cos (\frac{k \tau'}{\sqrt{3}})
+3\sqrt{3}k \tau' \sin (\frac{k \tau'}{\sqrt{3}})
\bigg)
\frac{  Z_{S\mathbf k}(\tau') }{ k^3 \tau'} \, d\tau'
    \nn\\
&
+\bigg(
-\frac{2\sqrt3}{k^2\tau^3} \cos (\frac{k \tau}{\sqrt{3}})
-\frac{ 2}{k\tau^2} \sin (\frac{k \tau}{\sqrt{3}})
\bigg)
    \int^\tau
    \bigg( 9 \sin (\frac{k \tau'}{\sqrt{3}})
- 3\sqrt{3} k \tau' \cos (\frac{k \tau'}{\sqrt{3}})\bigg)
    \frac{Z_{S\mathbf k}(\tau')}{k^3 \tau'}d\tau'
\nn\\
&
-\frac{9}{k^4\tau^2}  Z_{S\mathbf k}
- \frac{9}{4k^4\tau^2} A^{'}_{S\,{\bf k}}
+ \frac{9}{4k^4\tau^3} A_{S\,{\bf k}}
+\frac{3}{2k^4\tau} E^{\,'}_{S\,{\bf k}}
+\frac{27}{4k^4\tau^2}F^{||''}_{S\,{\bf k}}
+\frac{27}{4k^4\tau^3}F^{||'}_{S\,{\bf k}}
        \nn\\
        &
-\frac{27}{k^4\tau^4} F^{||}_{S\,{\bf k}}
-\frac{1}{2k^2} E_{S\,{\bf k}}
-\frac{9}{2k^2\tau^2} F^{||}_{S\,{\bf k}}
-\frac{k^2}{6}Y^{||}_{S\bf k}
+\frac{1}{2\tau}Y^{||\,'}_{S\bf k}
-\frac{1}{\tau}W^{||}_{S\bf k}
,
\el
}
and
\bl\label{sswi2Psol}
w^{(2)}_{P_S\,i}
=&
\frac{ \nabla^{-2}c_{2ij}^{,\,j}}{\tau^2}
-\frac{2}{\tau^{2}}
    \int^{\tau}\tau^{'2}\,\nabla^{-2}V_{S\,ij}^{,j}({\bf x},\tau')d\tau'
-Y^{\perp'}_{S\,i}
+W^{\perp}_{S\,i}
,
\el
and the tensor
$\chi^{\top(2)}_{P\,ij}$ is
\be
\chi^{\top(2)}_{P_S\,ij}=
\chi^{\top(2)}_{Sij}
+Y^\top_{S\,ij}
,
\ee
with $\chi^{\top(2)}_{Sij}$ in (5.1) of Ref.\cite{WangZhang2ndRD2018}
and $W^{||}_S$, $Y^{||}_S$, $W^{\perp}_{S\,i}$, $Y^{\perp}_{S\,i}$,
$Y^{\top}_{S\,ij}$  as
\bl\label{ssW||}
W^{||}_S=&
\nabla^{-2}\partial^iW_i
\nn\\
=&
\nabla^{-2}\Big[
2\phi^{(1),\,l}\chi^{||(1)'}_{,\,l}
+2\phi^{(1)}\nabla^2\chi^{||(1)'}
-\frac{1}{2}\chi^{||(1)'',\,l}\chi^{||(1)'}_{,\,l}
-\frac{1}{2}\chi^{||(1)''}\nabla^2\chi^{||(1)'}
\nn\\
&
-\frac{1}{6}\chi^{||(1)'}_{,\,l}\nabla^2\chi^{||(1),\,l}
+\frac{1}{3}\nabla^2\chi^{||(1)'}\nabla^2\chi^{||(1)}
-\frac{1}{2}\chi^{||(1)'}_{,\,lm}\chi^{||(1),\,lm}
\Big]
.
\el
The curl part of $W_{S\,i}$ is
\bl\label{ssWperp}
W^{\perp}_{S\,i}=&
W_{S\,i}-\partial_i W^{||}_S
\nn\\
=&
2\phi^{(1)}\chi^{||(1)'}_{,\,i}
-\frac{1}{2}\chi^{||(1)''}\chi^{||(1)'}_{,\,i}
+\frac{1}{3}\chi^{||(1)'}_{,\,i}\nabla^2\chi^{||(1)}
-\frac{1}{2}\chi^{||(1)'}_{,\,l}\chi^{||(1),\,l}_{,\,i}
\nn\\
&
+\partial_i\nabla^{-2}\Big[
-2\phi^{(1),\,l}\chi^{||(1)'}_{,\,l}
-2\phi^{(1)}\nabla^2\chi^{||(1)'}
+\frac{1}{2}\chi^{||(1)'',\,l}\chi^{||(1)'}_{,\,l}
+\frac{1}{2}\chi^{||(1)''}\nabla^2\chi^{||(1)'}
\nn\\
&
+\frac{1}{6}\chi^{||(1)'}_{,\,l}\nabla^2\chi^{||(1),\,l}
-\frac{1}{3}\nabla^2\chi^{||(1)'}\nabla^2\chi^{||(1)}
+\frac{1}{2}\chi^{||(1)'}_{,\,lm}\chi^{||(1),\,lm}
\Big]
.
\el
The scalar, vector, and tensor modes of $Y_{S\,ij}$ are
{\allowdisplaybreaks
\bl\label{ssY||}
Y^{||}_S&=
\frac{3}{2}\nabla^{-2}\nabla^{-2}Y_{S\,lm}^{,\,lm}
\nn\\
=&
\frac{1}{8}\chi^{||(1)'}\chi^{||(1)'}
-\frac{1}{8}\chi^{||(1),\,l}\chi^{||(1)}_{,\,l}
\nn\\
&
+\nabla^{-2}\Big[
-2\phi^{(1)}\nabla^2\chi^{||(1)}
+\frac{1}{6}\nabla^2\chi^{||(1)}\nabla^2\chi^{||(1)}
-\frac{5}{8}\chi^{||(1)',\,l}\chi^{||(1)'}_{,\,l}
-\frac{3}{8}\chi^{||(1),\,lm}\chi^{||(1)}_{,\,lm}
\Big]
\nn\\
&
+\nabla^{-2}\nabla^{-2}\Big[
6\phi^{(1),\,lm}\chi^{||(1)}_{,\,lm}
+12\phi^{(1),\,l}\nabla^2\chi^{||(1)}_{,\,l}
+6\phi^{(1)}\nabla^2\nabla^2\chi^{||(1)}
-\frac{3}{4}\chi^{||(1)',\,l}\nabla^2\chi^{||(1)'}_{,\,l}
\nn\\
&
-\frac{3}{4}\chi^{||(1)'}\nabla^2\nabla^2\chi^{||(1)'}
-\frac{1}{4}\chi^{||(1),\,lm}\nabla^2\chi^{||(1)}_{,\,lm}
-\frac{1}{4}\chi^{||(1),\,l}\nabla^2\nabla^2\chi^{||(1)}_{,\,l}
\Big]
,
\el
}
{\allowdisplaybreaks
\bl\label{ssYperp}
Y^{\perp}_{S\,j}&=
\nabla^{-2}Y_{S\,lj}^{,\,l}
-\frac{2}{3}Y^{||}_{S,j}
\nn\\
=&
\nabla^{-2}\Big[
4\phi^{(1),\,l}\chi^{||(1)}_{,\,lj}
+4\phi^{(1)}\nabla^2\chi^{||(1)}_{,j}
-\frac{1}{2}\chi^{||(1)'}\nabla^2\chi^{||(1)'}_{,j}
-\frac{1}{6}\chi^{||(1),\,l}\nabla^2\chi^{||(1)}_{,\,lj}
\Big]
\nn\\
&
+\partial_j\nabla^{-2}\nabla^{-2}\Big[
-4\phi^{(1),\,lm}\chi^{||(1)}_{,\,lm}
-8\phi^{(1),\,l}\nabla^2\chi^{||(1)}_{,\,l}
-4\phi^{(1)}\nabla^2\nabla^2\chi^{||(1)}
+\frac{1}{2}\chi^{||(1)',\,l}\nabla^2\chi^{||(1)'}_{,\,l}
\nn\\
&
+\frac{1}{2}\chi^{||(1)'}\nabla^2\nabla^2\chi^{||(1)'}
+\frac{1}{6}\chi^{||(1),\,lm}\nabla^2\chi^{||(1)}_{,\,lm}
+\frac{1}{6}\chi^{||(1),\,l}\nabla^2\nabla^2\chi^{||(1)}_{,\,l}
\Big]
,
\el
\bl\label{ssYtop1}
Y^{\top}_{S\,ij}&=
Y_{S\,ij}-D_{ij}Y^{||}_S-2Y^{\perp}_{S\,(i,j)}
\nn\\
=&
4\phi^{(1)}\chi^{||(1)}_{,ij}
-2\phi^{(1)}\nabla^2\chi^{||(1)}\delta_{ij}
+\frac{2}{3}\chi^{||(1)}_{,ij}\nabla^2\chi^{||(1)}
-\frac{1}{6}\nabla^2\chi^{||(1)}\nabla^2\chi^{||(1)}\delta_{ij}
\nn\\
&
-\frac{3}{4}\chi^{||(1)'}\chi^{||(1)'}_{,ij}
+\frac{1}{4}\chi^{||(1)'}\nabla^2\chi^{||(1)'}\delta_{ij}
-\frac{1}{4}\chi^{||(1)'}_{,i}\chi^{||(1)'}_{,j}
-\frac{1}{8}\chi^{||(1)',\,l}\chi^{||(1)'}_{,\,l}\delta_{ij}
\nn\\
&
-\frac{1}{4}\chi^{||(1),\,l}\chi^{||(1)}_{,\,lij}
+\frac{1}{12}\chi^{||(1),\,l}\nabla^2\chi^{||(1)}_{,\,l}\delta_{ij}
-\frac{3}{4}\chi^{||(1)}_{,\,il}\chi^{||(1),\,l}_{,j}
+\frac{1}{8}\chi^{||(1),\,lm}\chi^{||(1)}_{,\,lm}\delta_{ij}
\nn\\
&
+\delta_{ij}\nabla^{-2}\Big[
2\phi^{(1),\,lm}\chi^{||(1)}_{,\,lm}
+4\phi^{(1),\,l}\nabla^2\chi^{||(1)}_{,\,l}
+2\phi^{(1)}\nabla^2\nabla^2\chi^{||(1)}
-\frac{1}{4}\chi^{||(1)',\,l}\nabla^2\chi^{||(1)'}_{,\,l}
\nn\\
&
-\frac{1}{4}\chi^{||(1)'}\nabla^2\nabla^2\chi^{||(1)'}
-\frac{1}{12}\chi^{||(1),\,lm}\nabla^2\chi^{||(1)}_{,\,lm}
-\frac{1}{12}\chi^{||(1),\,l}\nabla^2\nabla^2\chi^{||(1)}_{,\,l}
\Big]
\nn\\
&
+\partial_i\nabla^{-2}\Big[
-4\phi^{(1),\,l}\chi^{||(1)}_{,\,lj}
-4\phi^{(1)}\nabla^2\chi^{||(1)}_{,j}
+\frac{1}{2}\chi^{||(1)'}\nabla^2\chi^{||(1)'}_{,j}
+\frac{1}{6}\chi^{||(1),\,l}\nabla^2\chi^{||(1)}_{,\,lj}
\Big]
\nn\\
&
+\partial_j\nabla^{-2}\Big[
-4\phi^{(1),\,l}\chi^{||(1)}_{,\,li}
-4\phi^{(1)}\nabla^2\chi^{||(1)}_{,\,i}
+\frac{1}{2}\chi^{||(1)'}\nabla^2\chi^{||(1)'}_{,i}
+\frac{1}{6}\chi^{||(1),\,l}\nabla^2\chi^{||(1)}_{,\,li}
\Big]
\nn\\
&
+\partial_i\partial_j\nabla^{-2}\Big[
2\phi^{(1)}\nabla^2\chi^{||(1)}
-\frac{1}{6}\nabla^2\chi^{||(1)}\nabla^2\chi^{||(1)}
+\frac{5}{8}\chi^{||(1)',\,l}\chi^{||(1)'}_{,\,l}
+\frac{3}{8}\chi^{||(1),\,lm}\chi^{||(1)}_{,\,lm}
\Big]
\nn\\
&
+\partial_i\partial_j\nabla^{-2}\nabla^{-2}\Big[
2\phi^{(1),\,lm}\chi^{||(1)}_{,\,lm}
+4\phi^{(1),\,l}\nabla^2\chi^{||(1)}_{,\,l}
+2\phi^{(1)}\nabla^2\nabla^2\chi^{||(1)}
-\frac{1}{4}\chi^{||(1)',\,l}\nabla^2\chi^{||(1)'}_{,\,l}
\nn\\
&
-\frac{1}{4}\chi^{||(1)'}\nabla^2\nabla^2\chi^{||(1)'}
-\frac{1}{12}\chi^{||(1),\,lm}\nabla^2\chi^{||(1)}_{,\,lm}
-\frac{1}{12}\chi^{||(1),\,l}\nabla^2\nabla^2\chi^{||(1)}_{,\,l}
\Big]
.
\el
}
Note that Ref.\cite{Lu2008} also
gave  the 2nd-order vector mode of the metric perturbation in Poisson coordinates.
However, by detailed checking of  Eq.(17) of Ref.[52],
we find that in their source  $\sum _{\, lm}$ the asymmetric,  last  term
$$2 \mathcal{H}
\left(\partial_{l} \Phi\right)
\left(\partial_{m} \Phi^{\prime}\right)$$
should be be replaced by the symmetrized term
$$\mathcal{H}\left(\partial_{l} \Phi\right)\left(\partial_{m} \Phi^{\prime}\right)
+ \mathcal{H}\left(\partial_{l} \Phi^{\prime}\right)\left(\partial_{m} \Phi\right)$$
because   $\sum _{\, lm}$ as
the source in the Einstein equation must be symmetric,
and the solution $S ({\bf k},\tau)$ of (20) in Ref.\cite{Lu2008}
should also be revised   accordingly.
We project the inhomogeneous part of $w_{P_Si}$ of  (\ref{sswi2Psol})
into    the following
\bl\label{wPjej2}
 e^jw^{(2)}_{P_S\,j}
=&
 e^j
\bigg\{
-\frac{2}{\tau^{2}}
    \int^{\tau}\tau^{'2}
\nabla^{-2}\bigg[
\frac{\tau'^2}{2}\chi^{||(1)'''}_{,j}\nabla^2\chi^{||(1)'''}
+\tau'\chi^{||(1)''}_{,j}\nabla^2\chi^{||(1)'''}
\nn\\
&
+\tau'\chi^{||(1)'''}_{,j}\nabla^2\chi^{||(1)''}
+\frac{1}{\tau'^2}\chi^{||(1)'}_{,j}\nabla^2\chi^{||(1)'}
+\frac{1}{\tau'}\chi^{||(1)'}_{,j}\nabla^2\chi^{||(1)''}
\nn\\
&
+\frac{1}{\tau'}\chi^{||(1)''}_{,\,j}\nabla^2\chi^{||(1)'}
+3\chi^{||(1)''}_{,j}\nabla^2\chi^{||(1)''}
\bigg]  d \tau'
\bigg\}
.
\el
where  $ e_i$  is  a polarization vector
 orthogonal to the wave vector.
The Fourier  mode
of our result  (\ref{wPjej2})  agrees  with
the  corrected   $  S ({\bf k},\tau)$  in Eq.(20) of Ref.\cite{Lu2008},
where the relation $\Phi=-\frac{1}{2}\chi^{||(1)''}-\frac{1}{2\tau}\chi^{||(1)'}$
is used.

The 2nd-order density contrast is
\bl
\delta^{(2)}_{P_S}
=&
\int  \frac{d^3k}{(2\pi)^{3/2}}
\Big[
\delta_{S\,{\bf k}}^{(2)}
+\frac{4}{\tau}\alpha^{(2)}_{S\bf k}
\Big]
e^{i{\bf k\cdot x}}
+\frac{4}{\tau}\delta^{(1)}\chi^{||(1)'}
-\delta^{(1)'}\chi^{||(1)'}
-\delta^{(1)}_{,\, l}\chi^{||(1),\,l}
\nn\\
&
-\frac{1}{\tau}\chi^{||(1)''}\chi^{||(1)'}
-\frac{1}{\tau}\chi^{||(1)'}_{,\, l}\chi^{||(1),\,l}
+\frac{5}{\tau^2}\chi^{||(1)'}\chi^{||(1)'} .
\el
where
{\allowdisplaybreaks
\bl
&
\delta_{S\,{\bf k}}^{(2)}
+\frac{4}{\tau}\alpha^{(2)}_{S\bf k}
\nn\\
=&
G_2
\l(
-\frac{8\sqrt3 \,i}{k^2\tau^3}
+\frac{8}{k\tau^2}
+\frac{8 i}{\sqrt3\,\tau}
-\frac{4 k}{3}
\r)e^{-ik\tau/\sqrt3}
\nn\\
&
+G_3 \l(
\frac{8\sqrt3 \,i}{k^2\tau^3}
+\frac{8}{k\tau^2}
-\frac{8 i}{\sqrt3\,\tau}
-\frac{4 k}{3}
\r)e^{ik\tau/\sqrt3}
        \nn\\
        &
+\bigg(
\frac{8\sqrt3}{k^2\tau^3} \sin (\frac{k \tau}{\sqrt{3}})
-\frac{8}{k\tau^2} \cos (\frac{k \tau}{\sqrt{3}})
-\frac{8}{\sqrt3\,\tau} \sin (\frac{k \tau}{\sqrt{3}})
\nn\\
&
+\frac{4k}{3}\cos (\frac{k \tau}{\sqrt{3}})
\bigg)
\int^\tau
\bigg(
9\cos (\frac{k \tau'}{\sqrt{3}})
+3\sqrt{3}k \tau' \sin (\frac{k \tau'}{\sqrt{3}})
\bigg)
\frac{  Z_{S\mathbf k}(\tau') }{ k^3 \tau'} \, d\tau'
\nn\\
&
+\bigg(
-\frac{8\sqrt3}{k^2\tau^3} \cos (\frac{k \tau}{\sqrt{3}})
-\frac{ 8}{k\tau^2} \sin (\frac{k \tau}{\sqrt{3}})
+\frac{ 8}{\sqrt{3}\,\tau} \cos (\frac{k \tau}{\sqrt{3}})
\nn\\
&
+\frac{4k}{3} \sin (\frac{k\tau}{\sqrt{3}})
\bigg)
    \int^\tau
    \bigg( 9 \sin (\frac{k \tau'}{\sqrt{3}})
- 3\sqrt{3} k \tau' \cos (\frac{k \tau'}{\sqrt{3}})\bigg)
    \frac{Z_{S\mathbf k}(\tau')}{k^3 \tau'}d\tau'
\nn\\
&
-\frac{36}{k^4\tau^2}  Z_{S\mathbf k}
+\frac{6}{k^4\tau} E^{\,'}_{S\,{\bf k}}
+\frac{27}{k^4\tau^2}F^{||''}_{S\,{\bf k}}
+\frac{27}{k^4\tau^3}F^{||'}_{S\,{\bf k}}
-\frac{108}{k^4\tau^4} F^{||}_{S\,{\bf k}}
+\frac{9}{k^2\tau}F^{||'}_{S\,{\bf k}}
        \nn\\
        &
+3 F^{||}_{S\,{\bf k}}
- \frac{9}{k^4\tau^2} A^{'}_{S\,{\bf k}}
+ \frac{9}{k^4\tau^3} A_{S\,{\bf k}}
-\frac{3}{k^2\tau} A_{S\,{\bf k}}
+\frac{2}{\tau}Y^{||\,'}_{S\bf k}
-\frac{4}{\tau}W^{||}_{S\bf k}.
\el
}
The 0-component of the 4-velocity is
\bl
U^{(2)0}_{P_S}
        =&
-\frac{1}{\tau}\int  \frac{d^3k}{(2\pi)^{3/2}}\Big[
-\alpha^{(2)'}_{S\bf k}
-\frac{1}{\tau}\alpha^{(2)}_{S\bf k}
\Big]e^{i{\bf k\cdot x}}
+\frac{1}{\tau}v^{||(1),\,l}v^{||(1)}_{,\,l}
+\frac{1}{\tau}\chi^{||(1)'}_{,\,l}v^{||(1),\,l}
\nn\\
&
+\frac{1}{2\tau^3}\chi^{||(1)'}\chi^{||(1)'}
+\frac{1}{4\tau^2}\chi^{||(1)''}\chi^{||(1)'}
-\frac{1}{4\tau}\chi^{||(1)'''}\chi^{||(1)'}
+\frac{1}{4\tau}\chi^{||(1)''}\chi^{||(1)''}
\nn\\
&
-\frac{1}{4\tau^2}\chi^{||(1)'}_{,\,l}\chi^{||(1),\,l}
-\frac{1}{4\tau}\chi^{||(1)''}_{,\,l}\chi^{||(1),\,l}
+\frac{1}{4\tau}\chi^{||(1)',\,l}\chi^{||(1)'}_{,\,l}
\,,
\el
where $-\alpha^{(2)'}_{S\bf k}
-\frac{1}{\tau}\alpha^{(2)}_{S\bf k}$
is given in (\ref{ssalphaTaualpha}).
By the definition  $U^{(2)i}_{P_S}=a^{-1}v^{(2)i}_{P_S}$,
one has the $i$-component as
\bl\label{ssvP2i4}
v^{(2)}_{P_S\,i}
    =&
v^{(2)}_{Si}
+\beta^{(2)'}_{S,i}
+d^{(2)'}_{Si}
+\frac{1}{\tau}v^{||(1)}_{,i}\chi^{||(1)'}
-v^{||(1)'}_{,i}\chi^{||(1)'}
-v^{||(1)}_{,\,li}\chi^{||(1),\,l}
\nn\\
&
+v^{||(1),\,l}\chi^{||(1)}_{,\,li}
+\frac{1}{2\tau}\chi^{||(1)'}_{,i}\chi^{||(1)'}
-\frac{1}{4}\chi^{||(1)''}_{,i}\chi^{||(1)'}
+\frac{1}{4}\chi^{||(1)'}_{,i}\chi^{||(1)''}
\nn\\
&
-\frac{1}{4}\chi^{||(1)'}_{,\,li}\chi^{||(1),\,l}
+\frac{1}{4}\chi^{||(1)}_{,\,li}\chi^{||(1)',\,l}
\,.
\el
By writing $v^{(2)}_{Si}=v^{||(2)}_{S,i}+v^{\perp(2)}_{Si}$
with $v^{\perp(2),i}_{Si}=0$,
[$\nabla^{-2}\partial^i$(\ref{ssvP2i4})]
gives the noncurl part of $v^{(2)}_{P_S\,i}$ as
\bl\label{ssvP||2i2}
v^{||(2)}_{P_S}
    =&
\int  \frac{d^3k}{(2\pi)^{3/2}}\Big[
v^{||(2)}_{S{\bf k}}
+\beta^{(2)'}_{S{\bf k}}
\Big]e^{i{\bf k\cdot x}}
-v^{||(1)}_{,\,l}\chi^{||(1),\,l}
+\frac{1}{4\tau}\chi^{||(1)'}\chi^{||(1)'}
-\frac{1}{4}\chi^{||(1)''}\chi^{||(1)'}
\nn\\
&
-\frac{1}{4}\chi^{||(1)'}_{,\,l}\chi^{||(1),\,l}
+\nabla^{-2}\Big[
\frac{1}{\tau}v^{||(1)}_{,\,l}\chi^{||(1)',\,l}
+\frac{1}{\tau}\chi^{||(1)'}\nabla^2v^{||(1)}
-\chi^{||(1)'}\nabla^2v^{||(1)'}
\nn\\
&
-v^{||(1)'}_{,\,l}\chi^{||(1)',\,l}
+2v^{||(1),\,l}\nabla^2\chi^{||(1)}_{,\,l}
+2v^{||(1),\,lm}\chi^{||(1)}_{,\,lm}
+\frac{1}{2}\chi^{||(1)'}_{,\,l}\chi^{||(1)'',\,l}
\nn\\
&
+\frac{1}{2}\chi^{||(1)''}\nabla^2\chi^{||(1)'}
+\frac{1}{2}\chi^{||(1)',\,l}\nabla^2\chi^{||(1)}_{,\,l}
+\frac{1}{2}\chi^{||(1)}_{,\,lm}\chi^{||(1)',\,lm}
\Big]
\,,
\el
where
{\allowdisplaybreaks
\bl
v^{||(2)}_{S{\bf k}}
+\beta^{(2)'}_{S{\bf k}}
=&
G_2 \l(
    -\frac{2\sqrt3 \,i}{k^2\tau^2}
    +\frac{2}{k\tau}
    +\frac{i}{\sqrt3}
\r)    e^{-ik\tau/\sqrt3}
+G_3 \l(
     \frac{2\sqrt3 \,i}{k^2\tau^2}
    +\frac{2}{k\tau}
    -\frac{i}{\sqrt3}
    \r)e^{ik\tau/\sqrt3}
\nn\\
&
+\left(
\frac{2\sqrt3}{k^2\tau^2} \sin (\frac{k \tau}{\sqrt{3}})
-\frac{2}{k\tau} \cos (\frac{k \tau}{\sqrt{3}})
-\frac{1}{\sqrt{3}}\sin (\frac{k \tau}{\sqrt{3}})
\right)
\int^\tau
\bigg(
9\cos (\frac{k \tau'}{\sqrt{3}})
\nn\\
&
+3\sqrt{3}k \tau' \sin (\frac{k \tau'}{\sqrt{3}})
\bigg)
\frac{  Z_{S\mathbf k}(\tau') }{ k^3 \tau'} \, d\tau'
\nn\\
&
+\left(
-\frac{2\sqrt3}{k^2\tau^2} \cos (\frac{k \tau}{\sqrt{3}})
-\frac{2}{k\tau} \sin (\frac{k \tau}{\sqrt{3}})
+\frac{1}{\sqrt{3}} \cos (\frac{k\tau}{\sqrt{3}})
\right)
    \int^\tau
    \bigg( 9 \sin (\frac{k \tau'}{\sqrt{3}})
\nn\\
&
- 3\sqrt{3} k \tau' \cos (\frac{k \tau'}{\sqrt{3}})\bigg)
    \frac{Z_{S\mathbf k}(\tau')}{k^3 \tau'}d\tau'
        \nn\\
        &
-\frac{9}{k^4\tau}  Z_{S\mathbf k}
+\frac{3}{2k^4} E^{\,'}_{S\,{\bf k}}
+\frac{27}{4k^4\tau}F^{||''}_{S\,{\bf k}}
+\frac{27}{4k^4\tau^2}F^{||'}_{S\,{\bf k}}
-\frac{27}{k^4\tau^3} F^{||}_{S\,{\bf k}}
        \nn\\
        &
+\frac{9}{4k^2}F^{||'}_{S\,{\bf k}}
- \frac{9}{4k^4\tau} A^{'}_{S\,{\bf k}}
+ \frac{9}{4k^4\tau^2} A_{S\,{\bf k}}
-\frac{3}{4k^2} A_{S\,{\bf k}}
+\frac{1}{2}Y^{||'}_{S{\bf k}}
.
\el
}

The curl part of $v^{(2)}_{P_S\,i}$ is given by
[(\ref{ssvP2i4})$-\partial_i$(\ref{ssvP||2i2})] as
{\allowdisplaybreaks
\bl
v^{\perp(2)}_{P_S\,i}
    =&
\frac{c_{2ij}^{\,,j}({\bf x})}{8}
-\frac{ \nabla^{-2}c_{2ij}^{\,,\,j}}{\tau^2}
- \frac{1}{4}
    \int^{\tau}\tau^{'2}\,V_{S\,ij}^{,j}({\bf x},\tau')d\tau'
\nn\\
&
+ \frac{2}{\tau^{2}}
    \int^{\tau}\tau^{'2}\,\nabla^{-2}V_{S\,ij}^{,j}({\bf x},\tau')d\tau'
+\frac{\tau^2}{4}\l(
M_{S\,i}
-\partial_i\nabla^{-2}M_{S\,k}^{,\,k}
\r)
+Y_{S\,i}^{\perp'}
\nn\\
&
+\frac{1}{\tau}v^{||(1)}_{,i}\chi^{||(1)'}
-v^{||(1)'}_{,i}\chi^{||(1)'}
+2v^{||(1),\,l}\chi^{||(1)}_{,\,li}
+\frac{1}{2}\chi^{||(1)'}_{,i}\chi^{||(1)''}
+\frac{1}{2}\chi^{||(1)}_{,\,li}\chi^{||(1)',\,l}
\nn\\
&
+\partial_i\nabla^{-2}\Big[
-\frac{1}{\tau}v^{||(1)}_{,\,l}\chi^{||(1)',\,l}
-\frac{1}{\tau}\chi^{||(1)'}\nabla^2v^{||(1)}
+\chi^{||(1)'}\nabla^2v^{||(1)'}
+v^{||(1)'}_{,\,l}\chi^{||(1)',\,l}
\nn\\
&
-2v^{||(1),\,l}\nabla^2\chi^{||(1)}_{,\,l}
-2v^{||(1),\,lm}\chi^{||(1)}_{,\,lm}
-\frac{1}{2}\chi^{||(1)'}_{,\,l}\chi^{||(1)'',\,l}
-\frac{1}{2}\chi^{||(1)''}\nabla^2\chi^{||(1)'}
\nn\\
&
-\frac{1}{2}\chi^{||(1)',\,l}\nabla^2\chi^{||(1)}_{,\,l}
-\frac{1}{2}\chi^{||(1)}_{,\,lm}\chi^{||(1)',\,lm}
\Big]
\,.
\el
where  $(M_{S\,i}-\partial_i\nabla^{-2}M_{S\,k}^{,k})$ is given in
(4.8) in Ref.\cite{WangZhang2ndRD2018}.
The curl part of the 3-velocity is nonzero in the 2nd order,
even if $v^{\perp(1)}_{P_S\,i}=0$.
}

\section{Conclusion}

As part of a series study \cite{WangZhang2ndRD2018}
of the 2nd-order cosmological perturbations
in the RD stage in synchronous coordinates,
this paper presents the 2nd-order formal solutions
for the scalar-tensor and tensor-tensor couplings
in the integral form.
The cosmic matter content is represented by a relativistic fluid.
The 1st-order vector perturbations and
the 1st-order transverse velocity are assumed to be zero.

From the 2nd-order solutions  we find that in the RD stage
the scalar modes, density contrast and longitudinal velocity
all propagate as a wave at the sound speed $\frac{1}{\sqrt{3}}$,
while the tensor modes are waves at the speed of light.
These  wave features are similar to the 1st-order solutions in the RD stage.
This is in contrast to the MD stage,
in which only tensor modes are waves at the speed of light,
and other types of perturbations evolve as a power law of the comoving time $\tau$
  \cite{WangZhang2017,ZhangQinWang2017}.

A general 2nd-order gauge transformation,
from synchronous to synchronous coordinates,
is implemented by  1st-order and  2nd-order transformation vector fields.
When the gauge-invariant 1st-order solutions are used in the couplings,
only the 2nd-order  transformation vector is effective at
carrying out the 2nd-order transformations.
With this,
we obtain the gauge-invariant 2nd-order formal solutions
in the integral form within a chosen synchronous coordinate.
In addition, we also perform the 2nd-order gauge transformations
of the solutions from  synchronous to Poisson coordinates,
and obtain the 2nd-order formal solutions in Poisson coordinates.
In the Poisson coordinates,
for the 2nd-order vector mode with scalar-scalar couplings,
we find that the last term in Eq.(17) of Ref.\cite{Lu2008}
should be symmetrized.
After correcting  this mistake in Eq.(17) of Ref.[52],
the revised    solution Eq.(20) in Ref.[52] will agree with our Eq.(6.100).

From these 2nd-order solutions,
one will be able to investigate the properties of nonlinearity evolution
not only within one type of metric perturbation,
such as the transfer of perturbation power among various $k$-modes,
the influence by the growing and decaying modes, etc,
but also the transfer of perturbation power between different types of
perturbations, such as those between scalar and tensor modes, etc.
After accumulations during cosmic expansion,
these nonlinear effects might possibly
lead to changes of the  tensor/scalar ratio of metric perturbations
observed in CMB anisotropies
from that of the  primordial metric perturbations generated during inflation.
A detailed investigation of these issues would require the initial conditions
pertinent to the cosmic evolution
and the proper adjoining of perturbations
from the  precedent expansion stages
such as inflation and  a possible period of reheating.

To use these solutions,
one  needs to do three types of $\int d\tau$ integrals
and one type of $\int d^3\bf k$ integral
that involve
the functions $Z_{s(t)\mathbf k} $, $A_{s(t){\bf k}} $,
 $Z_{T{\bf k}}$, $ A_{T{\bf k}}$, etc,
which can be done numerically.
To apply these in the cosmological study,
one needs to specify the initial conditions
which are presented by  $D_2(\mathbf k)$,  $D_3(\mathbf k)$,
$b_1(\mathbf k)$,  $b_2(\mathbf k)$,
$P_2(\mathbf k)$,  $P_3(\mathbf k)$,
 $Q_2(\mathbf k)$,  $Q_3(\mathbf k)$,
 etc.

The results of scalar-tensor and tensor-tensor in this paper,
together with the results of scalar-scalar couplings in Ref.\cite{WangZhang2ndRD2018},
constitute
the full solution of the 2nd-order cosmological perturbations
in the integral form in the RD stage in synchronous coordinates.
These  2nd-order   results of the RD stage,
in conjunction with the 2nd-order results
of the MD stage \cite{WangZhang2017,ZhangQinWang2017},
can be used to study
the nonlinear effects of cosmological perturbations.

\textbf{Acknowledgements}

Y. Z. is supported by
NSFC Grants No. 11421303, No. 11675165, and No. 11633001.

\newpage

\appendix

\numberwithin{equation}{section}

\section{2nd-order perturbed Einstein equation and conservation equations}
\label{Ap:PertEinEq}

In this appendix,
we shall list the 2nd-order perturbed Einstein equation
 with the scalar-tensor and tensor-tensor couplings,
 as well as the 2nd-order covariant conservation equations of the stress tensor
in a general RW spacetime.

First, we present the 2nd-order perturbed equations with scalar-tensor couplings.
The  $(00)$ component of 2nd-order perturbed Einstein equation,
i.e.,
the 2nd-order energy constraint, is
\be  \label{Ein2th003}
-\frac{6a'}{a} \phi^{(2)'}_{s(t)}
+2\nabla^2\phi^{(2)}_{s(t)}
+\frac{1}{3}\nabla^2\nabla^2\chi^{||(2)}_{s(t)}
=
3\l(\frac{a'}{a}\r)^2\delta^{(2)}_{s(t)}
+E_{s(t)}
,
\ee
where
\bl\label{ES1}
E_{s(t)}
\equiv
&
2\phi^{(1),\,lm}\chi^{\top(1)}_{lm}
+\frac{1}{2}\chi^{\top(1)'}_{lm}\chi^{||(1)',\,lm}
+\frac{2a'}a \chi^{\top(1)}_{lm}\chi^{||(1)',\,lm}
+\frac{2a'}a \chi^{\top(1)'}_{lm}\chi^{||(1),\,lm}
    \nn\\
    &
+\frac{1}{3}\chi^{\top(1)}_{lm}\nabla^2\chi^{||(1),\,lm}
-\chi^{||(1),\,lm}\nabla^2\chi^{\top(1)}_{lm}
-\frac{1}{2}\chi^{\top(1),\,l}_{mn}\chi^{||(1),\,mn}_{,\,l}
.
\el
The $(0i)$ component of the 2nd-order perturbed Einstein equation,
i.e.,
the 2nd-order momentum constraint, is
\be\label{MoConstr2ndv3}
2 \phi^{(2)'} _{s(t),\,i}
+\frac{1}{2}D_{ij}\chi^{||(2)',\,j}_{s(t)}
+\frac{1}{2}\chi^{\perp(2)',\,j}_{s(t)ij}
=
-3(1+c_s^2)\l(\frac{a'}{a}\r)^2v^{(2)}_{s(t)i}
+M_{s(t)i} \, ,
\ee
where
\bl\label{MSi1}
M_{s(t)i}\equiv&
-6\l(\frac{a'}{a}\r)^2(1+c_s^2)\chi^{\top(1)}_{il}v^{||(1),\,l}
+\phi^{(1) ,\,l }\chi^{\top(1)' }_{il}
-2\phi^{(1)' ,\,l} \chi^{\top(1) }_{il}
\nn\\
&
-\chi^{\top(1)' }_{lm,  \, i}\chi^{||(1),\,lm}
- \frac{1}{2}\chi^{\top(1)'}_{lm }\chi^{||(1),\, lm}_{,i}
-\frac{1}{2} \chi^{\top(1)}_{lm,\,i}\chi^{||(1)',\,lm}
    \nn\\
&
+ \chi^{\top(1)'}_{il,\,m}\chi^{||(1),\,lm}
-\frac{1}{3}\chi^{\top(1)}_{li}\nabla^2\chi^{||(1)',\,l}
+\frac{2}{3}\chi^{\top (1)' }_{il}\nabla^2\chi^{||(1),\,l} \, .
\el
The longitudinal part of
the momentum constraint (\ref{MoConstr2ndv3}) is
\be \label{MoCons2ndNoCurl2}
2\phi^{(2)'}_{s(t)}
+\frac{1}{3}\nabla^2\chi^{||(2)'}_{s(t)}
=
-3(1+c_s^2)\l(\frac{a'}{a}\r)^2v^{||(2)}_{s(t)}
+\nabla^{-2}M_{s(t)l}^{\,,\,l}
\ ,
\ee
and the transverse part is
\be \label{MoCons2ndCurl1}
\frac{1}{2}\chi^{\perp(2)',\,j}_{s(t)ij}
=
-3(1+c_s^2)\l(\frac{a'}{a}\r)^2v^{\perp(2)}_{s(t)i}
+ \l( M_{s(t)i} -\partial_i\nabla^{-2}M_{s(t)l}^{\,,\,l} \r)
\ ,
\ee
where
{\allowdisplaybreaks
\bl\label{MSkk}
M_{s(t)l}^{\,,\,l}
=&
-6\l(\frac{a'}{a}\r)^2(1+c_s^2)\chi^{\top(1)}_{lm}v^{||(1),\,lm}
+\phi^{(1) ,\,lm}\chi^{\top(1)' }_{lm}
-2\phi^{(1)' ,\,lm} \chi^{\top(1) }_{lm}
\nn\\
&
- \frac{1}{2}\chi^{\top(1)'}_{lm ,n}\chi^{||(1),\, lmn}
+\frac{1}{6}\chi^{\top(1)'}_{lm}\nabla^2\chi^{||(1),\,lm}
-\chi^{||(1),\,lm}\nabla^2\chi^{\top(1)' }_{lm}
    \nn\\
&
-\frac{1}{2} \chi^{\top(1)}_{lm,\,n}\chi^{||(1)',\,lmn}
-\frac{1}{3}\chi^{\top(1)}_{lm}\nabla^2\chi^{||(1)',\,lm}
-\frac{1}{2} \chi^{||(1)',\,lm}\nabla^2\chi^{\top(1)}_{lm}
,
\el
}
and
\bl\label{MSiCurl}
&
\l( M_{s(t)i}-\partial_i\nabla^{-2}M_{s(t)l}^{\,,\,l} \r)
\nn\\
=&
-6\l(\frac{a'}{a}\r)^2(1+c_s^2)\chi^{\top(1)}_{il}v^{||(1),\,l}
+\phi^{(1) ,\,l }\chi^{\top(1)' }_{il}
-2\phi^{(1)' ,\,l} \chi^{\top(1) }_{il}
-\frac{1}{2}\chi^{\top(1)' }_{lm,  \, i}\chi^{||(1),\,lm}
    \nn\\
&
-\frac{1}{2} \chi^{\top(1)}_{lm,\,i}\chi^{||(1)',\,lm}
+ \chi^{\top(1)'}_{il,\,m}\chi^{||(1),\,lm}
-\frac{1}{3}\chi^{\top(1)}_{li}\nabla^2\chi^{||(1)',\,l}
+\frac{2}{3}\chi^{\top (1)' }_{il}\nabla^2\chi^{||(1),\,l}
\nn\\
&
+\partial_i\nabla^{-2}\Big[
6\l(\frac{a'}{a}\r)^2(1+c_s^2)\chi^{\top(1)}_{lm}v^{||(1),\,lm}
-\phi^{(1) ,\,lm}\chi^{\top(1)' }_{lm}
+2\phi^{(1)' ,\,lm} \chi^{\top(1) }_{lm}
\nn\\
&
- \frac{1}{2}\chi^{\top(1)'}_{lm ,n}\chi^{||(1),\, lmn}
-\frac{2}{3}\chi^{\top(1)'}_{lm}\nabla^2\chi^{||(1),\,lm}
+\frac{1}{2}\chi^{||(1),\,lm}\nabla^2\chi^{\top(1)' }_{lm}
    \nn\\
&
+\frac{1}{2} \chi^{\top(1)}_{lm,\,n}\chi^{||(1)',\,lmn}
+\frac{1}{3}\chi^{\top(1)}_{lm}\nabla^2\chi^{||(1)',\,lm}
+\frac{1}{2} \chi^{||(1)',\,lm}\nabla^2\chi^{\top(1)}_{lm}
\Big]  .
\el
The $(ij)$ component of 2nd-order perturbed Einstein equation,
i.e.,
the 2nd-order evolution equation, is
{\allowdisplaybreaks
\bl\label{Evo2ndSs1}
&
2\phi^{(2)''}_{s(t)} \delta_{ij}
+4\frac{{a'}}{a}\phi^{(2)'}_{s(t)} \delta_{ij}
+\phi^{(2)}_{s(t),\,ij}
-\nabla^2\phi^{(2)}_{s(t)} \delta_{ij}
+\l[4\frac{a''}{a}
    +6(c_s^2-\frac{1}{3})(\frac{a'}{a})^2\r]\phi^{(2)}_{s(t)} \delta_{ij}
\nn\\
&
+\frac{1}{2}D_{ij}\chi^{||(2)''}_{s(t)}
+\frac{a'}{a} D_{ij}\chi^{||(2)'}_{s(t)}
 +\l[(1-3c_s^2)(\frac{a'}{a})^2
    -2\frac{a''}a \r]D_{ij}\chi^{||(2)}_{s(t)}
\nn\\
&
+\frac{1}{6}\nabla^2D_{ij}\chi^{||(2)}_{s(t)}
-\frac{1}{9}\delta_{ij}\nabla^2\nabla^2\chi^{||(2) }_{s(t)}
\nn\\
&
+\frac{1}{2}\chi^{\perp(2)''}_{s(t)ij}
+\frac{a'}{a} \chi^{\perp(2)'}_{s(t)ij}
+\l[(1-3c_s^2)(\frac{a'}a )^2
    -2\frac{a''}a \r]\chi^{\perp(2)}_{s(t)ij}
\nn\\
&
+\frac{1}{2}\chi^{\top(2)''}_{s(t)ij}
+\frac{a'}{a} \chi^{\top(2)'}_{s(t)ij}
+\l[(1-3c_s^2)(\frac{a'}a )^2
    -2\frac{a''}a \r]\chi^{\top(2)}_{s(t)ij}
-\frac{1}{2}\nabla^2\chi^{\top(2)}_{s(t)ij}
\nn\\
=&
3c_N^2\l(\frac{a'}{a}\r)^2\delta^{(2)}_{s(t)}\delta_{ij}
+S_{s(t)ij}
,
\el
}
where
{\allowdisplaybreaks
\bl\label{Ss2ndij2}
S_{s(t)ij}\equiv&
6c_L^2\l(\frac{a'}{a}\r)^2\chi^{\top(1)}_{ij}\delta^{(1)}
-6\phi^{(1)''}\chi^{\top(1)}_{ij}
-12\frac{a'}a \phi^{(1)'}\chi^{\top(1)}_{ij}
+4\chi^{\top(1)}_{ij}\nabla^2\phi^{(1)}
-\phi^{(1)'}\chi^{\top(1)'}_{ij}
\nn\\
&
-\phi^{(1),\,l}\chi^{\top(1)}_{l j,\,i}
-\phi^{(1),\,l}\chi^{\top(1)}_{l i,\,j}
+3\phi^{(1),\,l}\chi^{\top(1)}_{ij,\,l}
+2\phi^{(1)}\nabla^2\chi^{\top(1)}_{ij}
-2\phi^{(1),\,l}_{,\,i}\chi^{\top(1)}_{l j}
\nn\\
&
-2\phi^{(1),\,l}_{,\,j}\chi^{\top(1)}_{l i}
-\chi^{\top(1)''}_{lm}\chi^{||(1),\,lm}\delta_{ij}
-\chi^{\top(1)}_{lm}\chi^{||(1)'',\,lm}\delta_{ij}
-\frac{2a'}{a}\chi^{\top(1)'}_{lm}\chi^{||(1),\,lm}\delta_{ij}
\nn\\
&
-\frac{2a'}{a}\chi^{\top(1)}_{lm}\chi^{||(1)',\,lm}\delta_{ij}
+\chi^{\top(1)'}_{l i}\chi^{||(1)',\,l}_{,\,j}
+\chi^{\top(1)'}_{lj}\chi^{||(1)',\,l}_{,\,i}
-\frac{3}{2}\chi^{\top(1)'}_{lm}\chi^{||(1)',\,lm}\delta_{ij}
\nn\\
&
-\frac{1}{3}\chi^{\top(1)}_{li}\nabla^2\chi^{||(1),\,l}_{,\,j}
-\frac{1}{3}\chi^{\top(1)}_{lj}\nabla^2\chi^{||(1),\,l}_{,\,i}
-\chi^{\top(1)}_{lm,\,ij}\chi^{||(1),\,lm}
+\chi^{||(1),\,lm}\nabla^2\chi^{\top(1)}_{lm}\delta_{ij}
\nn\\
&
-\frac{1}{2}\chi^{\top(1)}_{lm,\,j}\chi^{||(1),\,lm}_{,\,i}
-\frac{1}{2}\chi^{\top(1)}_{lm,\,i}\chi^{||(1),\,lm}_{,\,j}
+\frac{1}{2}\chi^{||(1),\,nm l}\chi^{\top(1)}_{nm,\,l}\delta_{ij}
-\frac{2}{3}\chi^{\top(1)'}_{ij}\nabla^2\chi^{||(1)'}
\nn\\
&
+\frac{2}{3}\chi^{\top(1)}_{ij}\nabla^2\nabla^2\chi^{||(1)}
+\frac{1}{3}\nabla^2\chi^{\top(1)}_{ij}\nabla^2\chi^{||(1)}
+\frac{1}{3}\chi^{\top(1)}_{lj,\,i}\nabla^2\chi^{||(1),\,l}
+\frac{1}{3}\chi^{\top(1)}_{li,\,j}\nabla^2\chi^{||(1),\,l}
\nn\\
&
+\chi^{\top(1)}_{lj,\,im}\chi^{||(1),\,lm}
+\chi^{\top(1)}_{li,\,jm}\chi^{||(1),\,lm}
-\chi^{\top(1)}_{ij,\,lm}\chi^{||(1),\,lm}
\ .
\el
}
The trace part of the 2nd-order evolution equation (\ref{Evo2ndSs1}) is
\bl\label{Evo2ndSsTr2}
&
2\phi^{(2)''}_{s(t)}
+4\frac{{a'}}{a}\phi^{(2)'}_{s(t)}
-\frac{2}{3}\nabla^2\phi^{(2)}_{s(t)}
+\l[4\frac{a''}{a}
    +6(c_s^2-\frac{1}{3})(\frac{a'}{a})^2\r]\phi^{(2)}_{s(t)}
-\frac{1}{9}\nabla^2\nabla^2\chi^{||(2) }_{s(t)}
\nn\\
=&
3c_N^2\l(\frac{a'}{a}\r)^2\delta^{(2)}_{s(t)}
+\frac{1}{3}S_{s(t)l}^{\,l}
\,,
\el
where
\bl\label{SijTrace}
S_{s(t)l}^{\,l}
=&
-4\phi^{(1),\,lm}\chi^{\top(1)}_{lm}
-3\chi^{\top(1)''}_{lm}\chi^{||(1),\,lm}
-3\chi^{\top(1)}_{lm}\chi^{||(1)'',\,lm}
-\frac{6a'}{a}\chi^{\top(1)'}_{lm}\chi^{||(1),\,lm}
\nn\\
&
-\frac{6a'}{a}\chi^{\top(1)}_{lm}\chi^{||(1)',\,lm}
-\frac{5}{2}\chi^{\top(1)'}_{lm}\chi^{||(1)',\,lm}
-\frac{2}{3}\chi^{\top(1)}_{lm}\nabla^2\chi^{||(1),\,lm}
+2\chi^{||(1),\,lm}\nabla^2\chi^{\top(1)}_{lm}
\nn\\
&
+\frac{1}{2}\chi^{\top(1)}_{lm,\,n}\chi^{||(1),\,lmn}
\ .
\el
The traceless part of the 2nd-order evolution equation (\ref{Evo2ndSs1}) is
\bl\label{Evo2ndSsNoTr1}
&
D_{ij}\phi^{(2)}_{s(t)}
+\frac{1}{2}D_{ij}\chi^{||(2)''}_{s(t)}
+\frac{a'}{a} D_{ij}\chi^{||(2)'}_{s(t)}
 +\l[(1-3c_s^2)(\frac{a'}{a})^2
    -2\frac{a''}a \r]D_{ij}\chi^{||(2)}_{s(t)}
+\frac{1}{6}\nabla^2D_{ij}\chi^{||(2)}_{s(t)}
\nn\\
&
+\frac{1}{2}\chi^{\perp(2)''}_{s(t)ij}
+\frac{a'}{a} \chi^{\perp(2)'}_{s(t)ij}
+\l[(1-3c_s^2)(\frac{a'}a )^2
    -2\frac{a''}a \r]\chi^{\perp(2)}_{s(t)ij}
\nn\\
&
+\frac{1}{2}\chi^{\top(2)''}_{s(t)ij}
+\frac{a'}{a} \chi^{\top(2)'}_{s(t)ij}
+\l[(1-3c_s^2)(\frac{a'}a )^2
    -2\frac{a''}a \r]\chi^{\top(2)}_{s(t)ij}
-\frac{1}{2}\nabla^2\chi^{\top(2)}_{s(t)ij}
=
\bar S_{s(t)ij}
,
\el
where
{\allowdisplaybreaks
\bl\label{SijTraceless}
\bar S_{s(t)ij}
\equiv&
S_{s(t)ij}-\frac{1}{3}S_{s(t)k}^k\delta_{ij}
\nn\\
=&
6c_L^2\l(\frac{a'}{a}\r)^2\chi^{\top(1)}_{ij}\delta^{(1)}
-6\phi^{(1)''}\chi^{\top(1)}_{ij}
-12\frac{a'}a \phi^{(1)'}\chi^{\top(1)}_{ij}
+4\chi^{\top(1)}_{ij}\nabla^2\phi^{(1)}
-\phi^{(1)'}\chi^{\top(1)'}_{ij}
\nn\\
&
-\phi^{(1),\,l}\chi^{\top(1)}_{l j,\,i}
-\phi^{(1),\,l}\chi^{\top(1)}_{l i,\,j}
+3\phi^{(1),\,l}\chi^{\top(1)}_{ij,\,l}
+2\phi^{(1)}\nabla^2\chi^{\top(1)}_{ij}
-2\phi^{(1),\,l}_{,\,i}\chi^{\top(1)}_{l j}
\nn\\
&
-2\phi^{(1),\,l}_{,\,j}\chi^{\top(1)}_{l i}
+\frac{4}{3}\phi^{(1),\,lm}\chi^{\top(1)}_{lm}\delta_{ij}
+\chi^{\top(1)'}_{l i}\chi^{||(1)',\,l}_{,\,j}
+\chi^{\top(1)'}_{lj}\chi^{||(1)',\,l}_{,\,i}
\nn\\
&
-\frac{2}{3}\chi^{\top(1)'}_{lm}\chi^{||(1)',\,lm}\delta_{ij}
-\frac{1}{3}\chi^{\top(1)}_{li}\nabla^2\chi^{||(1),\,l}_{,\,j}
-\frac{1}{3}\chi^{\top(1)}_{lj}\nabla^2\chi^{||(1),\,l}_{,\,i}
+\frac{2}{9}\chi^{\top(1)}_{lm}\nabla^2\chi^{||(1),\,lm}\delta_{ij}
\nn\\
&
-\chi^{\top(1)}_{lm,\,ij}\chi^{||(1),\,lm}
+\frac{1}{3}\chi^{||(1),\,lm}\nabla^2\chi^{\top(1)}_{lm}\delta_{ij}
-\frac{1}{2}\chi^{\top(1)}_{lm,\,j}\chi^{||(1),\,lm}_{,\,i}
-\frac{1}{2}\chi^{\top(1)}_{lm,\,i}\chi^{||(1),\,lm}_{,\,j}
\nn\\
&
+\frac{1}{3}\chi^{\top(1)}_{lm,\,n}\chi^{||(1),\,lmn}\delta_{ij}
-\frac{2}{3}\chi^{\top(1)'}_{ij}\nabla^2\chi^{||(1)'}
+\frac{2}{3}\chi^{\top(1)}_{ij}\nabla^2\nabla^2\chi^{||(1)}
+\frac{1}{3}\nabla^2\chi^{\top(1)}_{ij}\nabla^2\chi^{||(1)}
\nn\\
&
+\frac{1}{3}\chi^{\top(1)}_{lj,\,i}\nabla^2\chi^{||(1),\,l}
+\frac{1}{3}\chi^{\top(1)}_{li,\,j}\nabla^2\chi^{||(1),\,l}
+\chi^{\top(1)}_{lj,\,im}\chi^{||(1),\,lm}
+\chi^{\top(1)}_{li,\,jm}\chi^{||(1),\,lm}
\nn\\
&
-\chi^{\top(1)}_{ij,\,lm}\chi^{||(1),\,lm}
\ .
\el
}
The scalar part of
the traceless part of the 2nd-order evolution equation (\ref{Evo2ndSsNoTr1}) is
\bl\label{Evo2ndSsChi1}
&
\phi^{(2)}_{s(t)}
+\frac{1}{2}\chi^{||(2)''}_{s(t)}
+\frac{a'}{a}\chi^{||(2)'}_{s(t)}
\nn\\
&
 +\l[(1-3c_s^2)(\frac{a'}{a})^2
    -2\frac{a''}a \r]\chi^{||(2)}_{s(t)}
+\frac{1}{6}\nabla^2\chi^{||(2)}_{s(t)}
=
\frac{3}{2}\nabla^{-2}\nabla^{-2}\bar S_{s(t)kl}^{\, ,\,kl}
,
\el
where
{\allowdisplaybreaks
\bl\label{SijPijbar}
\bar S_{s(t)lm}^{\, ,\,lm}
=&
-\frac{2}{3}\nabla^2\nabla^2\Big[\chi^{||(1),\,lm}\chi^{\top(1)}_{lm}\Big]
+\nabla^2\Big[
\frac{4}{3}\phi^{(1),\,lm}\chi^{\top(1)}_{lm}
+\frac{1}{3}\chi^{\top(1)'}_{lm}\chi^{||(1)',\,lm}
\nn\\
&
+\frac{8}{9} \chi^{\top(1)}_{lm}\nabla^2\chi^{||(1),\,lm}
+\frac{7}{6}\chi^{\top(1)}_{lm,\,n}\chi^{||(1),\,lmn}
\Big]
\nn\\
&
+6c_L^2\l(\frac{a'}{a}\r)^2\chi^{\top(1)}_{lm}\delta^{(1),\,lm}
-6\phi^{(1)'',\,lm}\chi^{\top(1)}_{lm}
-12\frac{a'}a \phi^{(1)',\,lm}\chi^{\top(1)}_{lm}
\nn\\
&
-\phi^{(1)',\,lm}\chi^{\top(1)'}_{lm}
-3\phi^{(1),\,lmn}\chi^{\top(1)}_{lm,\,n}
-\chi^{||(1)',\,lm}\nabla^2\chi^{\top(1)'}_{lm}
+\frac{1}{3}\chi^{\top(1)'}_{lm}\nabla^2\chi^{||(1)',\,lm}
\nn\\
&
-\frac{1}{2}\chi^{\top(1)}_{lm,\,n}\nabla^2\chi^{||(1),\,lmn}
+\nabla^2\chi^{\top(1)}_{lm}\nabla^2\chi^{||(1),\,lm}
+\frac{3}{2}\chi^{||(1),\,lmn}\nabla^2\chi^{\top(1)}_{lm,n}
\ .
\el
}
The vector part of
the traceless part of the 2nd-order evolution equation (\ref{Evo2ndSsNoTr1}) is
\bl\label{Evo2ndSsVec2}
&
\frac{1}{2}\chi^{\perp(2)''}_{s(t)ij}
+\frac{a'}{a} \chi^{\perp(2)'}_{s(t)ij}
+\l[(1-3c_s^2)(\frac{a'}a )^2
    -2\frac{a''}a \r]\chi^{\perp(2)}_{s(t)ij}
\nn\\
& =
\nabla^{-2}\bar S_{s(t)kj,\,i}^{,\,k}
+\nabla^{-2}\bar S_{s(t)ki,j}^{,\,k}
-2\nabla^{-2}\nabla^{-2}\bar S_{s(t)kl,\,ij}^{\, ,\,kl}
,
\el
where
the rhs of the above is
{\allowdisplaybreaks
\bl\label{SourceCurl1}
&
\nabla^{-2}\bar S_{s(t)lj,\,i}^{,\,l}
+\nabla^{-2}\bar S_{s(t)li,j}^{,\,l}
-2\nabla^{-2}\nabla^{-2}\bar S_{s(t)lm,\,ij}^{\, ,\,lm}
\nn\\
=&
\frac{2}{3}\partial_i\partial_j\Big[\chi^{||(1),\,lm}\chi^{\top(1)}_{lm}\Big]
+\partial_i\partial_j\nabla^{-2}\Big[
-\phi^{(1),\,lm}\chi^{\top(1)}_{lm}
-\frac{5}{6} \chi^{\top(1)}_{lm}\nabla^2\chi^{||(1),\,lm}
\nn\\
&
-\frac{1}{6}\chi^{||(1),\,lm}\nabla^2\chi^{\top(1)}_{lm}
-\frac{4}{3}\chi^{\top(1)}_{lm,\,n}\chi^{||(1),\,lmn}
\Big]
+\partial_i\nabla^{-2}\Big[
6c_L^2\l(\frac{a'}{a}\r)^2\chi^{\top(1)}_{lj}\delta^{(1),\,l}
-6\phi^{(1)'',\,l}\chi^{\top(1)}_{lj}
\nn\\
&
-12\frac{a'}a \phi^{(1)',\,l}\chi^{\top(1)}_{lj}
-\phi^{(1)',\,l}\chi^{\top(1)'}_{lj}
+2\chi^{\top(1)}_{lj}\nabla^2\phi^{(1),\,l}
+\phi^{(1),\,l}\nabla^2\chi^{\top(1)}_{lj}
-\phi^{(1),\,lm}_{,j}\chi^{\top(1)}_{lm}
\nn\\
&
+\chi^{\top(1)'}_{lj,\,m}\chi^{||(1)',\,lm}
-\chi^{\top(1)'}_{lm,j}\chi^{||(1)',\,lm}
+\frac{1}{3}\chi^{\top(1)'}_{lj}\nabla^2\chi^{||(1)',\,l}
-\frac{1}{6}\chi^{\top(1)}_{lm}\nabla^2\chi^{||(1),\,lm}_{,j}
\nn\\
&
+\frac{1}{3}\chi^{\top(1)}_{lj}\nabla^2\nabla^2\chi^{||(1),\,l}
+\frac{2}{3}\nabla^2\chi^{\top(1)}_{lj}\nabla^2\chi^{||(1),\,l}
+\chi^{||(1),\,lm}\nabla^2\chi^{\top(1)}_{lj,\,m}
-\frac{1}{2}\chi^{||(1),\,lm}\nabla^2\chi^{\top(1)}_{lm,j}
\Big]
\nn\\
&
+\partial_i\partial_j\nabla^{-2}\nabla^{-2}\Big[
-6c_L^2\l(\frac{a'}{a}\r)^2\chi^{\top(1)}_{lm}\delta^{(1),\,lm}
+6\phi^{(1)'',\,lm}\chi^{\top(1)}_{lm}
+12\frac{a'}a \phi^{(1)',\,lm}\chi^{\top(1)}_{lm}
\nn\\
&
+\phi^{(1)',\,lm}\chi^{\top(1)'}_{lm}
+3\phi^{(1),\,lmn}\chi^{\top(1)}_{lm,\,n}
+\chi^{||(1)',\,lm}\nabla^2\chi^{\top(1)'}_{lm}
-\frac{1}{3}\chi^{\top(1)'}_{lm}\nabla^2\chi^{||(1)',\,lm}
\nn\\
&
+\frac{1}{2}\chi^{\top(1)}_{lm,\,n}\nabla^2\chi^{||(1),\,lmn}
-\nabla^2\chi^{\top(1)}_{lm}\nabla^2\chi^{||(1),\,lm}
-\frac{3}{2}\chi^{||(1),\,lmn}\nabla^2\chi^{\top(1)}_{lm,n}
\Big]
\nn\\
&
+(i\leftrightarrow j)
\ .
\el
}
The tensor part (GW) of the 2nd-order evolution equation (\ref{Evo2ndSsNoTr1}) is
\bl \label{Evo2ndSsTen1}
&
\frac{1}{2}\chi^{\top(2)''}_{s(t)ij}
+\frac{a'}{a} \chi^{\top(2)'}_{s(t)ij}
+\l[(1-3c_s^2)(\frac{a'}a )^2
    -2\frac{a''}a \r]\chi^{\top(2)}_{s(t)ij}
-\frac{1}{2}\nabla^2\chi^{\top(2)}_{s(t)ij}
\nn\\
=&
\bar S_{s(t)ij}
-\frac{3}{2}D_{ij}\nabla^{-2}\nabla^{-2}\bar S_{s(t)kl}^{\, ,\,kl}
-\nabla^{-2}\bar S_{s(t)kj,\,i}^{,\,k}
-\nabla^{-2}\bar S_{s(t)ki,j}^{,\,k}
+2\nabla^{-2}\nabla^{-2}\bar S_{s(t)kl,\,ij}^{\, ,\,kl}
\ ,
\el
where the rhs of the above is
{
\allowdisplaybreaks
\bl\label{2ndTensorSource}
&
\bar S_{s(t)ij}
-\frac{3}{2}D_{ij}\nabla^{-2}\nabla^{-2}\bar S_{s(t)lm}^{\, ,\,lm}
-\nabla^{-2}\bar S_{s(t)lj,\,i}^{,\,l}
-\nabla^{-2}\bar S_{s(t)li,j}^{,\,l}
+2\nabla^{-2}\nabla^{-2}\bar S_{s(t)lm,\,ij}^{\, ,\,lm}
\nn\\
=&
\Big[
6c_L^2\l(\frac{a'}{a}\r)^2\chi^{\top(1)}_{ij}\delta^{(1)}
-6\phi^{(1)''}\chi^{\top(1)}_{ij}
-12\frac{a'}a \phi^{(1)'}\chi^{\top(1)}_{ij}
+4\chi^{\top(1)}_{ij}\nabla^2\phi^{(1)}
-\phi^{(1)'}\chi^{\top(1)'}_{ij}
\nn\\
&
-\phi^{(1),\,l}\chi^{\top(1)}_{l j,\,i}
-\phi^{(1),\,l}\chi^{\top(1)}_{l i,\,j}
+3\phi^{(1),\,l}\chi^{\top(1)}_{ij,\,l}
+2\phi^{(1)}\nabla^2\chi^{\top(1)}_{ij}
-2\phi^{(1),\,l}_{,\,i}\chi^{\top(1)}_{l j}
-2\phi^{(1),\,l}_{,\,j}\chi^{\top(1)}_{l i}
\nn\\
&
+2\phi^{(1),\,lm}\chi^{\top(1)}_{lm}\delta_{ij}
+\chi^{\top(1)'}_{l i}\chi^{||(1)',\,l}_{,\,j}
+\chi^{\top(1)'}_{lj}\chi^{||(1)',\,l}_{,\,i}
-\frac{1}{2}\chi^{\top(1)'}_{lm}\chi^{||(1)',\,lm}\delta_{ij}
\nn\\
&
-\frac{1}{3}\chi^{\top(1)}_{li}\nabla^2\chi^{||(1),\,l}_{,\,j}
-\frac{1}{3}\chi^{\top(1)}_{lj}\nabla^2\chi^{||(1),\,l}_{,\,i}
+\frac{1}{3}\chi^{\top(1)}_{lm}\nabla^2\chi^{||(1),\,lm}\delta_{ij}
-\chi^{\top(1)}_{lm,\,ij}\chi^{||(1),\,lm}
\nn\\
&
-\frac{1}{2}\chi^{\top(1)}_{lm,\,j}\chi^{||(1),\,lm}_{,\,i}
-\frac{1}{2}\chi^{\top(1)}_{lm,\,i}\chi^{||(1),\,lm}_{,\,j}
+\frac{1}{4}\chi^{\top(1)}_{lm,\,n}\chi^{||(1),\,lmn}\delta_{ij}
-\frac{2}{3}\chi^{\top(1)'}_{ij}\nabla^2\chi^{||(1)'}
\nn\\
&
+\frac{2}{3}\chi^{\top(1)}_{ij}\nabla^2\nabla^2\chi^{||(1)}
+\frac{1}{3}\nabla^2\chi^{\top(1)}_{ij}\nabla^2\chi^{||(1)}
+\frac{1}{3}\chi^{\top(1)}_{lj,\,i}\nabla^2\chi^{||(1),\,l}
+\frac{1}{3}\chi^{\top(1)}_{li,\,j}\nabla^2\chi^{||(1),\,l}
\nn\\
&
+\chi^{\top(1)}_{lj,\,im}\chi^{||(1),\,lm}
+\chi^{\top(1)}_{li,\,jm}\chi^{||(1),\,lm}
-\chi^{\top(1)}_{ij,\,lm}\chi^{||(1),\,lm}
\Big]
+\delta_{ij}\nabla^{-2}\Big[
3c_L^2\l(\frac{a'}{a}\r)^2\chi^{\top(1)}_{lm}\delta^{(1),\,lm}
\nn\\
&
-3\phi^{(1)'',\,lm}\chi^{\top(1)}_{lm}
-6\frac{a'}a \phi^{(1)',\,lm}\chi^{\top(1)}_{lm}
-\frac{1}{2}\phi^{(1)',\,lm}\chi^{\top(1)'}_{lm}
-\frac{3}{2}\phi^{(1),\,lmn}\chi^{\top(1)}_{lm,\,n}
\nn\\
&
-\frac{1}{2}\chi^{||(1)',\,lm}\nabla^2\chi^{\top(1)'}_{lm}
+\frac{1}{6}\chi^{\top(1)'}_{lm}\nabla^2\chi^{||(1)',\,lm}
-\frac{1}{4}\chi^{\top(1)}_{lm,\,n}\nabla^2\chi^{||(1),\,lmn}
+\frac{1}{2}\nabla^2\chi^{\top(1)}_{lm}\nabla^2\chi^{||(1),\,lm}
\nn\\
&
+\frac{3}{4}\chi^{||(1),\,lmn}\nabla^2\chi^{\top(1)}_{lm,n}
\Big]
+\partial_i\partial_j\nabla^{-2}\Big[
-\frac{1}{2}\chi^{\top(1)'}_{lm}\chi^{||(1)',\,lm}
+\frac{1}{4}\chi^{\top(1)}_{lm,n}\chi^{||(1),\,lmn}
\Big]
\nn\\
&
+\partial_i\nabla^{-2}\Big[
-6c_L^2\l(\frac{a'}{a}\r)^2\chi^{\top(1)}_{lj}\delta^{(1),\,l}
+6\phi^{(1)'',\,l}\chi^{\top(1)}_{lj}
+12\frac{a'}a \phi^{(1)',\,l}\chi^{\top(1)}_{lj}
+\phi^{(1)',\,l}\chi^{\top(1)'}_{lj}
\nn\\
&
-2\chi^{\top(1)}_{lj}\nabla^2\phi^{(1),\,l}
-\phi^{(1),\,l}\nabla^2\chi^{\top(1)}_{lj}
+\phi^{(1),\,lm}_{,j}\chi^{\top(1)}_{lm}
-\chi^{\top(1)'}_{lj,\,m}\chi^{||(1)',\,lm}
+\chi^{\top(1)'}_{lm,j}\chi^{||(1)',\,lm}
\nn\\
&
-\frac{1}{3}\chi^{\top(1)'}_{lj}\nabla^2\chi^{||(1)',\,l}
+\frac{1}{6}\chi^{\top(1)}_{lm}\nabla^2\chi^{||(1),\,lm}_{,j}
-\frac{1}{3}\chi^{\top(1)}_{lj}\nabla^2\nabla^2\chi^{||(1),\,l}
-\frac{2}{3}\nabla^2\chi^{\top(1)}_{lj}\nabla^2\chi^{||(1),\,l}
\nn\\
&
-\chi^{||(1),\,lm}\nabla^2\chi^{\top(1)}_{lj,\,m}
+\frac{1}{2}\chi^{||(1),\,lm}\nabla^2\chi^{\top(1)}_{lm,j}
\Big]
+\partial_j\nabla^{-2}\Big[
-6c_L^2\l(\frac{a'}{a}\r)^2\chi^{\top(1)}_{li}\delta^{(1),\,l}
\nn\\
&
+6\phi^{(1)'',\,l}\chi^{\top(1)}_{li}
+12\frac{a'}a \phi^{(1)',\,l}\chi^{\top(1)}_{li}
+\phi^{(1)',\,l}\chi^{\top(1)'}_{li}
-2\chi^{\top(1)}_{li}\nabla^2\phi^{(1),\,l}
-\phi^{(1),\,l}\nabla^2\chi^{\top(1)}_{li}
\nn\\
&
+\phi^{(1),\,lm}_{,i}\chi^{\top(1)}_{lm}
-\chi^{\top(1)'}_{li,\,m}\chi^{||(1)',\,lm}
+\chi^{\top(1)'}_{lm,i}\chi^{||(1)',\,lm}
-\frac{1}{3}\chi^{\top(1)'}_{li}\nabla^2\chi^{||(1)',\,l}
+\frac{1}{6}\chi^{\top(1)}_{lm}\nabla^2\chi^{||(1),\,lm}_{,i}
\nn\\
&
-\frac{1}{3}\chi^{\top(1)}_{li}\nabla^2\nabla^2\chi^{||(1),\,l}
-\frac{2}{3}\nabla^2\chi^{\top(1)}_{li}\nabla^2\chi^{||(1),\,l}
-\chi^{||(1),\,lm}\nabla^2\chi^{\top(1)}_{li,\,m}
+\frac{1}{2}\chi^{||(1),\,lm}\nabla^2\chi^{\top(1)}_{lm,i}
\Big]
\nn\\
&
+\partial_i\partial_j\nabla^{-2}\nabla^{-2}\Big[
3c_L^2\l(\frac{a'}{a}\r)^2\chi^{\top(1)}_{lm}\delta^{(1),\,lm}
-3\phi^{(1)'',\,lm}\chi^{\top(1)}_{lm}
-6\frac{a'}a \phi^{(1)',\,lm}\chi^{\top(1)}_{lm}
\nn\\
&
-\frac{1}{2}\phi^{(1)',\,lm}\chi^{\top(1)'}_{lm}
-\frac{3}{2}\phi^{(1),\,lmn}\chi^{\top(1)}_{lm,\,n}
-\frac{1}{2}\chi^{||(1)',\,lm}\nabla^2\chi^{\top(1)'}_{lm}
+\frac{1}{6}\chi^{\top(1)'}_{lm}\nabla^2\chi^{||(1)',\,lm}
\nn\\
&
-\frac{1}{4}\chi^{\top(1)}_{lm,\,n}\nabla^2\chi^{||(1),\,lmn}
+\frac{1}{2}\nabla^2\chi^{\top(1)}_{lm}\nabla^2\chi^{||(1),\,lm}
+\frac{3}{4}\chi^{||(1),\,lmn}\nabla^2\chi^{\top(1)}_{lm,n}
\Big] \, .
\el
}

Next, we present the 2nd-order perturbed Einstein equation with tensor-tensor couplings.
The  $(00)$ component of the 2nd-order perturbed Einstein equation,
i.e.,
the 2nd-order energy constraint, is
\be  \label{TEin2th003}
-\frac{6a'}{a} \phi^{(2)'}_{T}
+2\nabla^2\phi^{(2)}_{T}
+\frac{1}{3}\nabla^2\nabla^2\chi^{||(2)}_{T}
=
3\l(\frac{a'}{a}\r)^2\delta^{(2)}_{T}
+E_{T}
.
\ee
where
\bl\label{TES1}
E_{T}
\equiv
&
\frac{1}{4}\chi^{\top(1)'lm}\chi^{\top(1)'}_{lm}
+\frac{2a'}a \chi^{\top(1)lm}\chi^{\top(1)'}_{lm}
-\chi^{\top(1)lm}\nabla^2\chi^{\top(1)}_{lm}
    \nn\\
    &
-\frac{3}{4}\chi^{\top(1)lm,\,n}\chi^{\top(1)}_{lm,\,n}
+\frac{1}{2}\chi^{\top(1)lm,\,n}\chi^{\top(1)}_{ln,\,m}
.
\el
The $(0i)$ component of 2nd-order perturbed Einstein equation,
i.e.,
the 2nd-order momentum constraint is
\be\label{TMoConstr2ndv3}
2 \phi^{(2)'} _{T,\,i}
+\frac{1}{2}D_{ij}\chi^{||(2)',\,j}_{T}
+\frac{1}{2}\chi^{\perp(2)',\,j}_{T\,ij}
=
-3(1+c_s^2)\l(\frac{a'}{a}\r)^2v^{(2)}_{T\,i}
+M_{T\,i}
\ee
where  $M_{T\,i}$ is the same as (\ref{TMSi1RD}).
The longitudinal part of the momentum constraint (\ref{TMoConstr2ndv3}) is
\be \label{TMoCons2ndNoCurl2}
2\phi^{(2)'}_{T}
+\frac{1}{3}\nabla^2\chi^{||(2)'}_{T}
=
-3(1+c_s^2)\l(\frac{a'}{a}\r)^2v^{||(2)}_{T}
+\nabla^{-2}M_{T\,l}^{\,,\,l}
\ ,
\ee
and the transverse part:
\be \label{TMoCons2ndCurl1}
\frac{1}{2}\chi^{\perp(2)',\,j}_{T\,ij}
=
-3(1+c_s^2)\l(\frac{a'}{a}\r)^2v^{\perp(2)}_{T\,i}
+ \l( M_{T\,i} -\partial_i\nabla^{-2}M_{T\,l}^{\,,\,l} \r)
\ ,
\ee
where
$\nabla^{-2}M_{T\,l}^{\,,\,l}$
and
$\l( M_{T\,i}-\partial_i\nabla^{-2}M_{T\,l}^{,\,l} \r)$
are the same as (\ref{TMSkkScalarRD}) and (\ref{TMSiCurlRD}).

The $(ij)$ component of 2nd-order perturbed Einstein equation,
i.e.,
the 2nd-order evolution equation, is
\bl\label{TEvo2ndSs1}
&
2\phi^{(2)''}_{T} \delta_{ij}
+4\frac{{a'}}{a}\phi^{(2)'}_{T} \delta_{ij}
+\phi^{(2)}_{T,\,ij}
-\nabla^2\phi^{(2)}_{T} \delta_{ij}
+\l[4\frac{a''}{a}
    +6(c_s^2-\frac{1}{3})(\frac{a'}{a})^2\r]\phi^{(2)}_{T} \delta_{ij}
\nn\\
&
+\frac{1}{2}D_{ij}\chi^{||(2)''}_{T}
+\frac{a'}{a} D_{ij}\chi^{||(2)'}_{T}
 +\l[(1-3c_s^2)(\frac{a'}{a})^2
    -2\frac{a''}a \r]D_{ij}\chi^{||(2)}_{T}
\nn\\
&
+\frac{1}{6}\nabla^2D_{ij}\chi^{||(2)}_{T}
-\frac{1}{9}\delta_{ij}\nabla^2\nabla^2\chi^{||(2) }_{T}
\nn\\
&
+\frac{1}{2}\chi^{\perp(2)''}_{T\,ij}
+\frac{a'}{a} \chi^{\perp(2)'}_{T\,ij}
+\l[(1-3c_s^2)(\frac{a'}a )^2
    -2\frac{a''}a \r]\chi^{\perp(2)}_{T\,ij}
\nn\\
&
+\frac{1}{2}\chi^{\top(2)''}_{T\,ij}
+\frac{a'}{a} \chi^{\top(2)'}_{T\,ij}
+\l[(1-3c_s^2)(\frac{a'}a )^2
    -2\frac{a''}a \r]\chi^{\top(2)}_{T\,ij}
-\frac{1}{2}\nabla^2\chi^{\top(2)}_{T\,ij}
\nn\\
=&
3c_N^2\l(\frac{a'}{a}\r)^2\delta^{(2)}_{T}\delta_{ij}
+S_{T\,ij}
,
\el
where
\bl\label{TSs2ndij2}
S_{Tij}\equiv&
-2\frac{a'}{a}\chi^{\top(1)lm}\chi^{\top(1)'}_{lm}\delta_{ij}
-\chi^{\top(1)lm}\chi^{\top(1)}_{lm,\,ij}
+\chi^{\top(1)lm}\nabla^2\chi^{\top(1)}_{lm}\delta_{ij}
\nn\\
&
-\frac{1}{2}\chi^{\top(1)lm}_{,\,i}\chi^{\top(1)}_{lm,\,j}
-\chi^{\top(1)}_{li,\,m}\chi^{\top(1)l,\,m}_{j}
+\frac34\chi^{\top(1)lm,\,n}\chi^{\top(1)}_{lm,\,n}\delta_{ij}
\nn\\
&
+\chi^{\top(1)}_{l i,\,m}\chi^{\top(1)m,\,l}_{j}
-\frac{1}{2}\chi^{\top(1)lm,\,n}\chi^{\top(1)}_{ln,\,m}\delta_{ij}
+\chi^{\top(1)'l}_{i}\chi^{\top(1)'}_{lj}
-\frac34\chi^{\top(1)'\,lm}\chi^{\top(1)'}_{lm}\delta_{ij}
\nn\\
&
-\chi^{\top(1)lm}\chi^{\top(1)''}_{lm}\delta_{ij}
+\chi^{\top(1)lm}\chi^{\top(1)}_{lj,\,im}
+\chi^{\top(1)lm}\chi^{\top(1)}_{li,\,jm}
-\chi^{\top(1)lm}\chi^{\top(1)}_{ij,\,lm}
\ .
\el
The trace part of the 2nd-order evolution equation (\ref{TEvo2ndSs1}) is
\bl\label{TEvo2ndSsTr2}
&
2\phi^{(2)''}_{T}
+4\frac{{a'}}{a}\phi^{(2)'}_{T}
-\frac{2}{3}\nabla^2\phi^{(2)}_{T}
+\l[4\frac{a''}{a}
    +6(c_s^2-\frac{1}{3})(\frac{a'}{a})^2\r]\phi^{(2)}_{T}
-\frac{1}{9}\nabla^2\nabla^2\chi^{||(2) }_{T}
\nn\\
=&
3c_N^2\l(\frac{a'}{a}\r)^2\delta^{(2)}_{T}
+\frac{1}{3}S_{Tl}^{\,l}
\,,
\el
where
$S_{Tl}^{\,l}$ is the same as (\ref{TSijTraceRD}).
The traceless part of the 2nd-order evolution equation (\ref{TEvo2ndSs1}) is
{\allowdisplaybreaks
\bl\label{TEvo2ndSsNoTr1}
&
D_{ij}\phi^{(2)}_{T}
+\frac{1}{2}D_{ij}\chi^{||(2)''}_{T}
+\frac{a'}{a} D_{ij}\chi^{||(2)'}_{T}
 +\l[(1-3c_s^2)(\frac{a'}{a})^2
    -2\frac{a''}a \r]D_{ij}\chi^{||(2)}_{T}
+\frac{1}{6}\nabla^2D_{ij}\chi^{||(2)}_{T}
\nn\\
&
+\frac{1}{2}\chi^{\perp(2)''}_{T\,ij}
+\frac{a'}{a} \chi^{\perp(2)'}_{T\,ij}
+\l[(1-3c_s^2)(\frac{a'}a )^2
    -2\frac{a''}a \r]\chi^{\perp(2)}_{T\,ij}
\nn\\
&
+\frac{1}{2}\chi^{\top(2)''}_{T\,ij}
+\frac{a'}{a} \chi^{\top(2)'}_{T\,ij}
+\l[(1-3c_s^2)(\frac{a'}a )^2
    -2\frac{a''}a \r]\chi^{\top(2)}_{T\,ij}
-\frac{1}{2}\nabla^2\chi^{\top(2)}_{T\,ij}
=
\bar S_{T\,ij}
,
\el
}
where
$\bar S_{T\,ij}$ is the same as (\ref{TSijTracelessRD}).
The scalar part of
the traceless part of 2nd-order evolution equation (\ref{TEvo2ndSsNoTr1}) is
\bl\label{TEvo2ndSsChi1}
&
\chi^{||(2)''}_{T}
+\frac{2a'}{a}\chi^{||(2)'}_{T}
\nn\\
&
+2\l[(1-3c_s^2)(\frac{a'}{a})^2
    -2\frac{a''}a \r]\chi^{||(2)}_{T}
+\frac{1}{3}\nabla^2\chi^{||(2)}_{T}
+2\phi^{(2)}_{T}
=
3\nabla^{-2}\nabla^{-2}\bar S_{T\,lm}^{\, ,\,lm}
,
\el
where
$\bar S_{T\,lm}^{\, ,\,lm}$ is the same as (\ref{TSijPijbarRD}).
The vector part of
the traceless part   (\ref{TEvo2ndSsNoTr1}) is
\bl\label{TEvo2ndSsVec2}
&
\frac{1}{2}\chi^{\perp(2)''}_{T\,ij}
+\frac{a'}{a} \chi^{\perp(2)'}_{T\,ij}
+\l[(1-3c_s^2)(\frac{a'}a )^2
    -2\frac{a''}a \r]\chi^{\perp(2)}_{T\,ij}
\nn\\
& =
\nabla^{-2}\bar S_{T\,kj,\,i}^{,\,k}
+\nabla^{-2}\bar S_{T\,ki,j}^{,\,k}
-2\nabla^{-2}\nabla^{-2}\bar S_{T\,kl,\,ij}^{\, ,\,kl}
,
\el
where
$\nabla^{-2}\bar S_{T\,lj,\,i}^{,\,l}
+\nabla^{-2}\bar S_{T\,li,j}^{,\,l}
-2\nabla^{-2}\nabla^{-2}\bar S_{T\,lm,\,ij}^{\, ,\,lm}$
is the same as (\ref{TSourceCurl1RD}).
The tensor part   of   (\ref{TEvo2ndSsNoTr1}) is
\bl \label{TEvo2ndSsTen1}
&
\frac{1}{2}\chi^{\top(2)''}_{T\,ij}
+\frac{a'}{a} \chi^{\top(2)'}_{T\,ij}
+\l[(1-3c_s^2)(\frac{a'}a )^2
    -2\frac{a''}a \r]\chi^{\top(2)}_{T\,ij}
-\frac{1}{2}\nabla^2\chi^{\top(2)}_{T\,ij}
\nn\\
=&
\bar S_{T\,ij}
-\frac{3}{2}D_{ij}\nabla^{-2}\nabla^{-2}\bar S_{T\,kl}^{\, ,\,kl}
-\nabla^{-2}\bar S_{T\,kj,\,i}^{,\,k}
-\nabla^{-2}\bar S_{T\,ki,j}^{,\,k}
+2\nabla^{-2}\nabla^{-2}\bar S_{T\,kl,\,ij}^{\, ,\,kl}
\ ,
\el
where the rhs of the above is
 the same as (\ref{T2ndTensorSourceRD}).

Finally,
we present the 2nd-order equations of
the covariant conservation
of the energy-momentum tensor,
which are also needed for solving the 2nd-order perturbations.
The  conservation equations with all the couplings  for a general RW spacetime
are given in (\ref{EnConsv2}) and (\ref{MoConsv2}).
Keeping only the scalar-tensor terms,
we have
\bl\label{enCons2ndstRD}
&
\delta^{(2)'}_{s(t)}
+2a''(a')^{-1}\delta^{(2)}_{s(t)}
+(-1+3c_N^2)a'a^{-1}\delta^{(2)}_{s(t)}
+(1+c_s^2)\nabla^2v^{||(2)}_{s(t)}
-3(1+c_s^2)\phi^{(2)'}_{s(t)}
\nn\\
&
-(1+c_s^2)\chi^{\top(1)'}_{lm}\chi^{||(1),\,lm}
-(1+c_s^2)\chi^{\top(1)}_{lm}\chi^{||(1)',\,lm}
=0
\ ,
\el
\bl\label{MoCons2ndstRD}
&
c_N^2 \delta^{(2)}_{s(t),\,i}
+2(1+c_s^2)a''(a')^{-1} v^{||(2)}_{s(t),i}
+(1+c_s^2) v^{||(2)'}_{s(t),i}
+2(1+c_s^2)a''(a')^{-1} v^{\perp(2)}_{s(t)i}
\nn\\
&
+(1+c_s^2) v^{\perp(2)'}_{s(t)i}
-2c_L^2\delta^{(1),\,l}\chi^{\top(1)}_{li}
+2(1+c_s^2)v^{||(1),\,l}\chi^{\top(1)'}_{li}=0
\ .
\el
Keeping only  the tensor-tensor coupling terms,
we have
\bl\label{TenCons2ndstRD}
&
\delta^{(2)'}_{T}
+2a''(a')^{-1}\delta^{(2)}_{T}
+(-1+3c_N^2)a'a^{-1}\delta^{(2)}_{T}
+(1+c_s^2)\nabla^2v^{||(2)}_{T}
-3(1+c_s^2)\phi^{(2)'}_{T}
\nn\\
&
-(1+c_s^2)\chi^{\top(1)'}_{lm}\chi^{\top(1)lm}
=0
\ ,
\el
\bl\label{TMoCons2ndstRD}
&
c_N^2 \delta^{(2)}_{T,\,i}
+2(1+c_s^2)a''(a')^{-1} v^{||(2)}_{T,i}
+(1+c_s^2) v^{||(2)'}_{T,i}
+2(1+c_s^2)a''(a')^{-1} v^{\perp(2)}_{Ti}
\nn\\
&
+(1+c_s^2) v^{\perp(2)'}_{Ti}=0
\ .
\el

\end{document}